\documentclass[final,times,5p,sort&compress]{elsarticle}

\usepackage{placeins}
\usepackage[english]{babel}
\usepackage[utf8]{inputenc}
\usepackage{amsmath}
\usepackage{amssymb}
\usepackage{amsthm}
\usepackage{eurosym}
\usepackage{color}
\usepackage[colorlinks]{hyperref}  %
\usepackage{cleveref}
\usepackage{caption}  %
\usepackage{longtable}
\usepackage{booktabs}
\usepackage{multicol}  %
\usepackage{multirow}  %
\usepackage{nomencl}
\usepackage{lipsum}
\usepackage{subfig}
\usepackage[nohyperlinks, nolist]{acronym}  %
\usepackage[export]{adjustbox} %

\makenomenclature
\newcommand{\mycircle}[1]{\raisebox{.5pt}{\normalsize \textcircled{\raisebox{-.9pt} {\small #1}}}}

\usepackage{etoolbox}
\renewcommand\nomgroup[1]{
	\item[\bfseries \itshape
	\ifstrequal{#1}{A}{Acronyms / abbreviations}{
		\ifstrequal{#1}{Y}{Symbols}{
			\ifstrequal{#1}{I}{Indices / sets}{}}}
		]}

\usepackage{framed}  %

\usepackage[final]{pdfpages}  %

\usepackage{microtype}  %

\allowdisplaybreaks

\begin{document}

\captionsetup[figure]{labelfont={bf},labelformat={default},labelsep=period,name={Fig.}}
\captionsetup[table]{labelfont={bf},labelsep=newline}

\hypersetup{
	pdftitle={The effect of price-based demand response on carbon emissions in European electricity markets},
	pdfauthor={Markus Fleschutz, Markus Bohlayer, Marco Braun, Gregor Henze, Michael D. Murphy}
}

\begin{frontmatter}
\title{The effect of price-based demand response on carbon emissions in European electricity markets: The importance of adequate carbon prices}

\author[HSKA,CIT]{Markus Fleschutz}
\author[HSKA,FAU]{Markus Bohlayer}
\author[HSKA]{Marco Braun}
\author[UCB,NREL,RASEI]{Gregor Henze}
\author[CIT]{Michael D. Murphy\corref{cor1}}
\cortext[cor1]{Corresponding author: MichaelD.Murphy@cit.ie}

\address[HSKA]{Institute of Refrigeration, Air-Conditioning, and Environmental Engineering, Karlsruhe University of Applied Sciences, Moltkestraße 30, 76133 Karlsruhe, Germany}
\address[CIT]{Department of Process, Energy and Transport Engineering, Munster Technological University, Bishopstown, Cork, Ireland}
\address[FAU]{School of Business and Economics, Friedrich-Alexander-Universität Erlangen-Nürnberg, Lange Gasse 20, 90403 Nürnberg, Germany}
\address[UCB]{Department of Civil, Environmental and Architectural Engineering, University of Colorado, Boulder, Colorado 80309, USA}
\address[NREL]{National Renewable Energy Laboratory, Golden, Colorado 80401, USA}
\address[RASEI]{Renewable and Sustainable Energy Institute, Boulder, Colorado 80309, USA}
	
\date{\today}

\begin{abstract}
\noindent Price-based demand response (PBDR) has recently been attributed great economic but also environmental potential.
However, the determination of its short-term effects on carbon emissions requires the knowledge of marginal emission factors (MEFs), which compared to grid mix emission factors (XEFs), are cumbersome to calculate due to the complex characteristics of national electricity markets.
This study, therefore, proposes two merit order-based methods to approximate hourly MEFs and applies it to readily available datasets from 20 European countries for the years 2017--2019.
Based on the resulting electricity prices, MEFs, and XEFs, standardized daily load shifts were simulated to quantify their effects on marginal costs and carbon emissions.
Finally, by repeating the load shift simulations for different carbon price levels, the impact of the carbon price on the resulting carbon emissions was analyzed.
Interestingly, the simulated price-based load shifts led to increases in operational carbon emissions for 8 of the 20 countries and to an average increase of 2.1\% across all 20 countries.
Switching from price-based to MEF-based load shifts reduced the corresponding carbon emissions to a decrease of 35\%, albeit with 56\% lower monetary cost savings compared to the price-based load shifts.
Under specific circumstances, PBDR leads to an increase in carbon emissions, mainly due to the economic advantage fuel sources such as lignite and coal have in the merit order.
However, as the price of carbon is increased, the correlation between the carbon intensity and the marginal cost of the fuels substantially increases.
The Spearman correlation coefficient between carbon intensity and marginal cost along the German merit order is -0.13 for the current carbon price of 24.9\,\euro/t, 0 for 42.6\,\euro/t, and 0.4 for 100.0\,\euro/t.
Therefore, with adequate carbon prices, PBDR can be an effective tool for both economical and environmental improvement.
\end{abstract}

	
\end{frontmatter}

\nomenclature[A]{PBDR}{Price-Based Demand Response}
\nomenclature[A]{DR}{Demand Response}
\nomenclature[A]{GHG}{Greenhouse Gas}
\nomenclature[A]{ETS}{Emissions Trading System}
\nomenclature[A]{CEF}{Carbon Emission Factor}
\nomenclature[A]{XE}{Grid Mix Emission}
\nomenclature[A]{ME}{Marginal Emission}
\nomenclature[A]{MEF}{Marginal (power plant) Emission Factor}
\nomenclature[A]{XEF}{Grid Mix Emission Factor}
\nomenclature[A]{CC}{Combined-Cycle}
\nomenclature[A]{ETP}{ENTSO-E Transparency Platform}
\nomenclature[A]{PP}{\underline{P}ower \underline{P}lant method to calculate carbon emission factors}
\nomenclature[A]{PWL}{\underline{P}iece\underline{w}ise-\underline{L}inear approximation method to calculate carbon emission factors}
\nomenclature[A]{PWLv}{Validation variant of PWL method}
\nomenclature[A]{CONV}{Conventional energy sources}
\nomenclature[A]{vRES}{variable Renewable Energy Sources}
\nomenclature[A]{RES}{Renewable Energy Sources}
\nomenclature[A]{EDRM}{Empirical Data \& Relationship Models}
\nomenclature[A]{PSOM}{Power System Optimization Models}

\nomenclature[Y]{$\varepsilon$}{Carbon emissions}
\nomenclature[Y]{$\eta_p^\mathrm{el}$}{Generation efficiency per power plant~$p$}
\nomenclature[Y]{$P_p^\mathrm{inst}$}{Installed power plant capacity per power plant~$p$ in MW}
\nomenclature[Y]{$E_{t,f}^\mathrm{gen}$}{Generated energy per fuel type~$f$ and time step~$t$ in MWh}
\nomenclature[Y]{$\gamma_{t,p}^\mathrm{m}$}{Binary indicator of marginal unit}
\nomenclature[Y]{$\gamma_{t,p}^\mathrm{x}$}{Capacity utilization rate per time step~$t$ and power plant~$p$}
\nomenclature[Y]{$\Delta t$}{Time-step-width in hours}
\nomenclature[Y]{$\eta^\mathrm{T}$}{Constant transmission efficiency considering all transmission and distribution losses}

\nomenclature[I]{$f \in F$}{Generation fuel types}
\nomenclature[I]{$t \in T$}{Timesteps}
\nomenclature[I]{$p \in P$}{Power plants}

\begin{acronym}
	\acro{CC}{combined-cycle}
	\acro{CEF}{carbon emission factor}
	\acro{DR}{demand response}
	\acro{ETP}{ENTSO-E Transparency Platform}
	\acro{ETS}{Emissions Trading System}
	\acro{EU}{European Union}
	\acro{GHG}{greenhouse gas}
	\acro{PBDR}{price-based demand response}
	\acro{RES}{renewable energy sources}
	\acro{vRES}{variable renewable energy sources}
	\acro{ME}{marginal emission}
	\acro{MEF}{marginal (power plant) emission factor}
	\acro{XE}{grid mix emission}
	\acro{XEF}{average electricity mix emission factor}
\end{acronym}

\section{Introduction}

\subsection{Motivation}
High penetrations of \ac{vRES} in national electricity grids are essential to achieve the global aims of the Paris agreement.
Price-based demand response \acused{PBDR} (\ac{PBDR}) is a promising approach, especially in smart grids, to provide the operational flexibility, that is needed for the integration of \ac{vRES}.
Due to the characteristics of the electricity market, available approaches used for carbon emissions accounting lead to misleading results when applied in the context of \ac{DR}.
Specifically, they overestimate the environmental potential of \ac{PBDR} by ignoring the nature of the electricity markets and their special phenomenon called the merit order dilemma of emissions.
This phenomenon refers to the fact that, for electricity generation, certain emission-intensive fuel types are, due to their low marginal costs, preferred to relatively lower emission-intensive technologies.
At the time of writing, e.g., highly efficient combined-cycle gas power plants, are, according to the merit-order dispatch principle, only considered after more carbon-intensive lignite-fired power plants.
Adequate approaches based on the idea of \acp{ME}, however, require detailed data, which are usually not available.
As a consequence, as long as the merit order of an energy market does not correlate with the carbon emission factors, \ac{PBDR} cannot exploit the full carbon reduction potential of load shifting.
Increasing the carbon price to an appropriate level is crucial to exploit this potential by internalizing the external cost of climate change and creating financial incentives for sustainable development~\cite{TheWorldBank.2020}.

\subsection{Background}

\begin{figure*}[!ht]
	\centering
	\includegraphics[width=1.0\linewidth]{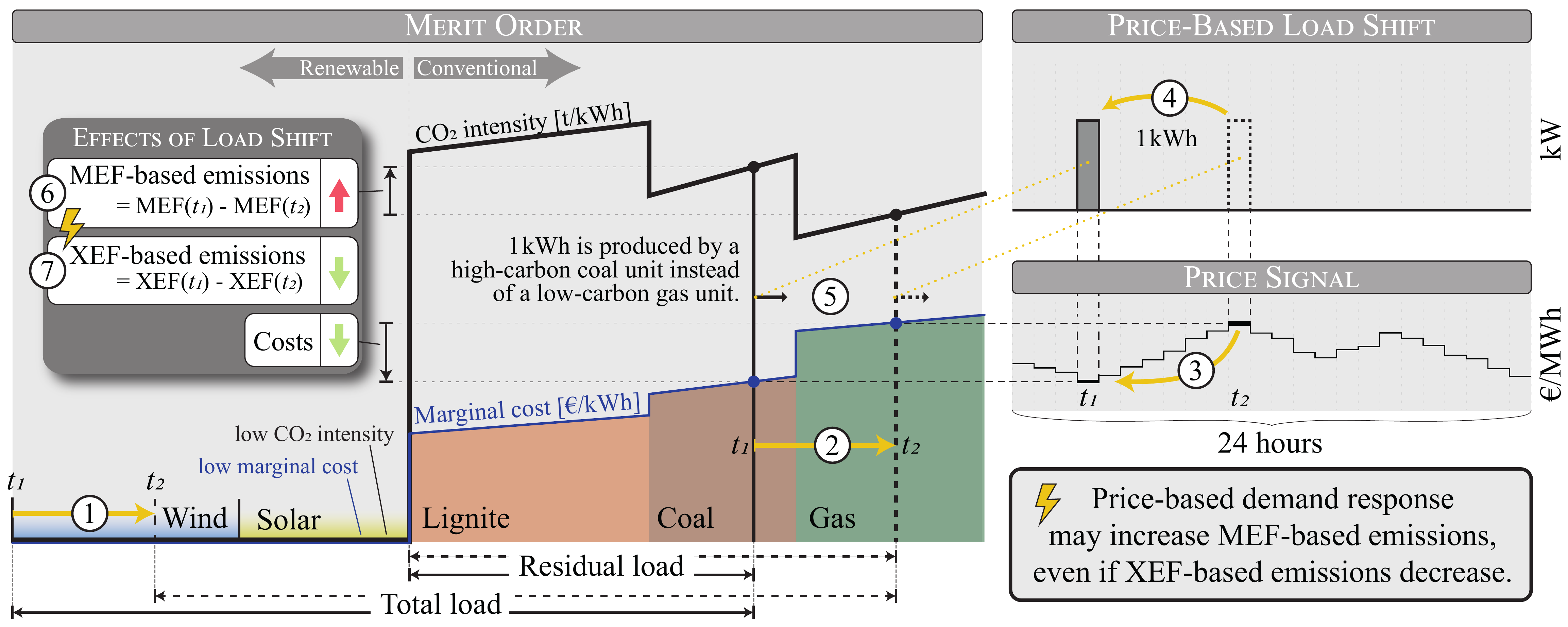}
	\vspace{-0.8cm}
	\caption{Exemplary case where price-based or XEF-based load shifting leads to increased total emissions:
		We assume two time points $t_1$ and $t_2$. $t_2$ differs from $t_1$ only by a decreased wind supply (\mycircle{1}).
		Since the total load is identical in both time points, the wind decrease leads to a higher residual load in $t_2$ and therefore to a shift of the operating point in the merit order curve from the area of coal-fired power plants to the area of gas-fired power plants (\mycircle{2}).
		The different marginal prices of the two marginal power plant types lead to a spot market price spread between $t_1$ and $t_2$ (\mycircle{3}).
		This price spread incentivizes a load shift of 1\,kWh from $t_2$ to $t_1$ (\mycircle{4}).
		However, the shifted load causes an additional production of 1\,kWh at a carbon-intensive coal-fired power plant in $t_1$, whereas in the case without load shift, the load causes an additional production of 1\,kWh at a less carbon-intensive gas-fired power plant (\mycircle{5}).
		Since the marginal emissions of the coal power plant (MEF($t_1$)) are higher than the marginal emissions of the gas power plant (MEF($t_2$)) the load shift causes increased MEF-based emissions (\mycircle{6}).
		In contrast, XEF($t_1$) is lower than XEF($t_2$) since in $t_2$, the reduced wind power is compensated by coal and gas.
		if effects on emissions are evaluated using \acp{XEF}, a load shift from $t_2$ to $t_1$ suggests a decrease in emissions, since the \ac{XEF} at $t_1$ is lower than at $t_2$ (\mycircle{7}).}
	\label{fig:scheme_merit_order_dilemma}
\end{figure*}

The integration of additional \ac{vRES} requires an increase in the operational flexibility of the electricity grids to compensate for the increasing fluctuation of the residual load, which defines the total load minus the production from \ac{vRES}.
Operational flexibility in this context means that it is not based on structural changes in the electricity system such as power station commissioning / decommissioning or fuel price changes~\cite{Hawkes.2014}.
This flexibility can be provided by flexible generation, interconnection, energy storage, and demand-side resources~\cite{R.A.Verzijlbergh.2017}.
\ac{PBDR} is seen as an essential and promising approach for unlocking the flexibility potential of demand-side resources in a cost-efficient way~\cite{DNVGL.2014}.
Thus, \ac{PBDR} has been the subject of many recent studies across the residential, commercial, and industrial sectors, as shown in the review articles~\cite{Jordehi.2019,Good.2017,Paterakis.2017,Siano.2014}.
Also, \ac{PBDR} can be combined with incentive-based DR as demonstrated in~\cite{Bohlayer.2020}.

For the quantification of electricity-related carbon emissions, the temporal granularity of \acp{CEF} is important.
Annual average \acp{CEF}, which are still commonly used, lead to inaccurate results because of the high variance of emission factors of different fuel types.
This holds all the more with the evaluation of emissions due to load changes such as with \ac{PBDR}.
Several studies, therefore, suggest the usage of time-varying \acp{CEF} instead of yearly average \acp{CEF} for short and long term decision making~\cite{Khan.2018b,Baumgartner.2019,KopsakangasSavolainen.2017}.

Dynamic \acp{XEF} are useful for calculating carbon emission balances of energy consumers.
Since usually it is not possible to trace electricity from a specific producer to consumer, the average carbon intensity of the entire generation system is attributed to each customer~\cite{Jiusto.2006b}.
However, if \acp{XEF} are used to assess or reduce the real effect of \ac{DR} on carbon emission, the results may be misleading~\cite{Hawkes.2010,SilerEvans.2012}.
The reason for this is that not all power plants react proportionally to a change in demand~\cite{Hawkes.2010}.
In theory, the electricity requested will come from the power plant with the lowest marginal cost and spare capacity, the so-called marginal power plant~\cite{Corradi.2019}.
Dynamic \acp{MEF} estimate the carbon intensity of demand changes as the carbon intensity of the marginal power plant for each time step~\cite{Hawkes.2010}.
This is why, if the real impact of \ac{DR} on operational carbon emissions is to be determined or even minimized, \acp{MEF} should be used where possible.
However, the calculation of hourly \acp{MEF} requires a very detailed database, which is why hourly \acp{MEF} are not available for most areas~\cite{SilerEvans.2012}.
As a consequence and despite a growing body of literature that recognizes the necessity of \acp{MEF} for assessing the environmental effects of \ac{PBDR}~\cite{Hawkes.2010, SilerEvans.2012, Regett.2018, Baumgartner.2019, Tranberg.2019, Rinne.2013, Pareschi.2017, Pean.2018, Khan.2019}, \acp{XEF} are still used in this context.

Through \ac{PBDR}, price incentives and potential savings that arise in an energy spot market by varying supply and demand of electricity are passed on to the energy consumer.
However, energy spot markets lead, under perfect competitions, to a cost-minimizing dispatch, which, depending on the correlation between prices and emissions along the merit order, may lead to a suboptimal dispatch in an environmental sense.
This phenomenon is known as the merit order emission dilemma as illustrated in \cite{Regett.2018}.

The fluctuating feed-in of \ac{vRES} significantly affects the carbon emissions of the electricity supplied to consumers.
In times of high \ac{vRES} shares, the carbon emissions per produced unit of electricity are usually lower than in times of lower \ac{vRES} shares since less fossil-based power plants per produced unit of electricity are in operation.
This phenomenon leads to the hypothesis that a shift of energy consumption from an hour with a low \ac{vRES} share to an hour with a high \ac{vRES} share leads to a decrease in carbon emissions.
In fact, a calculation according to \acp{XEF}, which describes the current generation mix of the electricity system supports this hypothesis in many cases, e.g.,~\cite{Summerbell.2017, Stoll.2014, KopsakangasSavolainen.2017}.
However, price-based or \ac{XEF}-based load shifting might lead to increased emissions as illustrated in \Cref{fig:scheme_merit_order_dilemma}.

One option to minimize carbon emissions through load shifting is the usage of \acp{MEF} instead of prices as a primary DR incentive signal, as Leerbeck et al.~\cite{Leerbeck.2020} suggest.
Real-time \acp{MEF} with resolutions between 5 and 15 minutes are already commercially available via application programming interfaces from companies such as WattTime~\cite{WattTime.2020} or Tomorrow~\cite{ElecMapApi.2020}.
However, since prices and emissions along the merit order are not fully correlated, cost minimization and emission minimization are conflicting.
As a second option, this cost-emission conflict can be resolved by internalizing the external costs of climate change by placing an adequate price on carbon emissions.
Both options will be analyzed in this study.

In summary, fuel type specific, temporally resolved national generation data (e.g., from the \ac{ETP}~\cite{ENTSOE.2020}) allows for the calculation of temporally resolved \acp{XEF} which are necessary for carbon emissions accounting of electric consumers.
However, for the environmental evaluation of \ac{PBDR} activities like load shifting, dynamic \acp{MEF} are needed, since they reflect the effects of system changes, which \acp{XEF} do not.
Due to the lack of power plant specific electricity generation data, the purely empirical identification of the marginal power plant is not straightforward~\cite{Doucette.2011}.
As a consequence, dynamic historic \acp{MEF} are not available free of charge for most European countries.

\begin{table*}[htb]
	\let\tsc\textsuperscript
	\let\mc\multicolumn
	\centering
	\caption{Classification of studies that propose methods to calculate electricity \acp{CEF} according to type, methodology, scope, and temporal resolution. The studies are clustered in three groups and within these sorted by years. The two characteristics of methodology refer to Empirical Data \& Relationship Models (EDRMs) and Power System Optimization Models (PSOMs) proposed by Ryan et. al~\cite{Ryan.2016}.}
	\label{tab:lit_review}
	\resizebox{1.0\linewidth}{!}{\begin{tabular}{lllrcccllll}
			\toprule
			& \mc3c{Study}&\mc2c{CEF Type}&\mc2c{Methodology}&\mc2c{Scope}&Temporal\\
			\cmidrule(r){2-4} \cmidrule(r){5-6} \cmidrule(lr){7-8} \cmidrule(lr){9-10}
			&Year&Name&Ref.&XEF&MEF&EDRM&PSOM&Temporal&Geographic&resolution of CEFs\\
			\midrule
			\multirow{10}{*}{\rotatebox{90}{Group A}}&2014&Stoll et al.&\cite{Stoll.2014}&$\bullet$&&$\bullet$\tsc{a}&&historic: 2011, 2012&UK, SE, US&hourly\\
			&2014&Messagie et al.&\cite{Messagie.2014}&$\bullet$&&$\bullet$\tsc{a}&&historic: 2011&BE&hourly\\
			&2016&Roux et al.&\cite{Roux.2016}&$\bullet$&&$\bullet$\tsc{a}&&historic: 2013&FR&hourly\\
			&2017&Kono et al.&\cite{Kono.2017}&$\bullet$&&$\bullet$\tsc{a}&&historic: 2011--2015&DE&hourly\\
			&2017&Kopsakangas-S. et al.&\cite{KopsakangasSavolainen.2017}&$\bullet$&&$\bullet$\tsc{a}&&historic: 2011&FI&hourly\\
			&2017&Summerbell et al.&\cite{Summerbell.2017}&$\bullet$&&$\bullet$\tsc{a}&& historic: 2015&UK&half-hourly\\ 
			&2018&Khan et al.&\cite{Khan.2018b}&$\bullet$&&$\bullet$\tsc{a}&&historic: 2015&NZ&half-hourly\\
			&2019&Clauß et al.&\cite{Clau.2019b}&$\bullet$&&$\bullet$\tsc{a}&&historic: 2015&NO, SE, DK, FI&hourly\\
			&2019&Munné-Collado et al.&\cite{MunneCollado.2019}&$\bullet$&&$\bullet$\tsc{a}&&historic: 2018&5 European countries &hourly\\
			&2019&Tranberg et al.&\cite{Tranberg.2019}&$\bullet$&&$\bullet$\tsc{b}&&historic: 2017&27 European countries&hourly\\
			\midrule
			\multirow{7}{*}{\rotatebox{90}{Group B}}&2010&Hawkes&\cite{Hawkes.2010}&$\bullet$&$\bullet$&$\bullet$\tsc{c}&&historic: 2002--2009&UK&time-of-day/year\\ 
			&2010&Greensfelder et al.&\cite{Greensfelder.2010}&$\bullet$&$\bullet$&$\bullet$\tsc{c}&&historic: 2008&US (4 regions)&hourly\\
			&2012&Siler-Evans et al.&\cite{SilerEvans.2012}&$\bullet$&$\bullet$&$\bullet$\tsc{c}&&historic: 2006--2011&US (8 regions)&time-of-day/year\\
			&2014&Thomson&\cite{Thomson.2014}&$\bullet$&$\bullet$&$\bullet$\tsc{c}&&historic: 2008--2013&UK&time-of-day/year\\
			&2017&Pareschi et al.&\cite{Pareschi.2017}&$\bullet$&$\bullet$&$\bullet$\tsc{c}&&historic: 2015&AT, CH, DE, FR&annual\\
			&2017&Thomson et al.&\cite{Thomson.2017}&$\bullet$&$\bullet$&$\bullet$\tsc{c}&&historic: 2009--2014&UK&time-of-day/year\\
			&2018&Péan et al.&\cite{Pean.2018}&$\bullet$&$\bullet$&$\bullet$\tsc{c}&&historic: 2016&ES&hourly\\
			\midrule
			\multirow{5}{*}{\rotatebox{90}{Group C}}&2006&Bettle et al.&\cite{Bettle.2006}&$\bullet$&$\bullet$&&$\bullet$\tsc{d}&historic: 2000&UK&half-hourly\\
			&2018&Regett et al.&\cite{Regett.2018}&$\bullet$&$\bullet$&&$\bullet$\tsc{d}&future: 2030&DE&hourly\\
			&2019&Baumgärtner et al.&\cite{Baumgartner.2019}&$\bullet$&$\bullet$&&$\bullet$\tsc{d}&historic: 2016&DE&hourly\\
			&2019&Böing \& Regett&\cite{Boing.2019}&$\bullet$&$\bullet$&&$\bullet$\tsc{d}&future: 2020--2050&DE&hourly\\
			&\mc3l{~------~~~Our study~~~------~~}&$\bullet$&$\bullet$&&$\bullet$\tsc{d}&historic: 2017--2019&20 European countries&hourly\\
			\bottomrule
			\mc{11}{l}{\tsc{a} Simple Emission Factors, \tsc{b} Flow Tracing, \tsc{c} Statistical Relationship Model, \tsc{d} Economic Dispatch Model}
	\end{tabular}}
\end{table*}

\subsection{Literature review}

In the literature, essentially there are two dimensions to classify electricity grid \acp{CEF}.
The methodology dimension classifies the methods behind the \acp{CEF} into Empirical Data \& Relationship Models (EDRM) such as regression analyses and Power System Optimization Models (PSOM) such as economic dispatch models~\cite{Ryan.2016}.
The other dimension divides the \acp{CEF} according to their \ac{CEF}-type into \acp{XEF} and \acp{MEF}.
\acp{XEF} describe the current generation mix of the electricity system and account for \ac{RES} shares while \acp{MEF} quantify marginal system effects~\cite{Regett.2018}.

The methodology behind \acp{XEF} is an attributional approach where all emissions of the electricity grid in a particular time frame are shared across all electricity consumers in proportion to their demand~\cite{Baumgartner.2019}.
This prevents double counting of emissions and is therefore often used in the context of carbon accounting~\cite{Roux.2016}.
The \acp{XEF} are determined by weighting the power plant type specific carbon emissions with its share of total electricity consumption during each hour.
Therefore, it reflects the current state of the electricity mix, e.g., the share of renewable and conventional electricity production~\cite{Regett.2018}.

The marginal power plant method to calculate \acp{MEF} focuses on the element of the power plant mix that will actually be affected by changes on the demand side and reflect the consequences on carbon emissions of the electricity system~\cite{Harmsen.2013}.
In uniform-pricing markets, power plant owners are incentivized to bid at their individual marginal cost.
This merit order sorts all power plants according to their marginal costs.
A load reduction or increase within a specific hour is now compensated by an increased power output of the marginal power plant.
The \ac{MEF} is therefore equal to the specific emission factor of the marginal power plant.

Ryan et al.~\cite{Ryan.2016} provides a useful guideline for selecting the most appropriate method.
For a review of carbon emission accounting approaches in electricity generation systems, the reader is referred to the work of Khan~\cite{Khan.2019}.

Within the literature, different approaches for the determination of \acp{CEF} can be found.
\Cref{tab:lit_review} shows an overview of studies clustering them by type, methodology, temporal and geographic scope, and temporal resolution into three groups.

Studies in Group A calculated historic, hourly or half-hourly \acp{XEF} with EDRMs for different countries.
Most of these studies \cite{Stoll.2014, Messagie.2014, Roux.2016, Kono.2017, KopsakangasSavolainen.2017, Summerbell.2017, Khan.2018b, Clau.2019b, MunneCollado.2019} focus on calculating simple emission factors for either individual or a few countries by weighting the historic generation data with the according carbon emission factors per fuel type. %

Tranberg et al.~\cite{Tranberg.2019} proposed a more sophisticated approach.
Using flow-tracing which allows the tracking of power flows on the transmission network from the region of generation to the region of consumption they calculated consumption-based \acp{XEF} for 27 European countries.
The results show significant deviations between the consumption-based and the generation-based \acp{XEF} and suggest to include cross-border flows for carbon emission accounting of electricity.
This method is demonstrated in the electricityMap~\cite{ElecMap.2020}, which is a real-time visualization of the carbon emission footprint of electricity consumption.
However, they do not focus on \acp{MEF} as we do.

For Group B, Hawkes~\cite{Hawkes.2010} set the foundation in 2010.
Using half-hourly data from the UK for 2002--2009, he calculated the first difference of system carbon emissions and system load, respectively, and determined the average \ac{MEF} by the slope of the regression line of the two difference vectors.
Compared to purely merit-order based methods, this has the advantage that it implicitly takes into account the trading decisions of the players in the market, the logistical constraints of power plant operation, and transmission and distribution restrictions.
However, the nature of statistical relationship models restricts these methods in the temporal resolution of the resulting \acp{MEF} which is significant for assessing \ac{DR} measures.
The regression can be performed repeatedly on subsets of the data to obtain, e.g., time-of-day or time-of-year \acp{MEF} as also shown by subsequent studies following Hawkes' approach by applying it to different geographic regions~\cite{SilerEvans.2012, Thomson.2014, Pareschi.2017, Thomson.2017} but this still neglects important information such as the fluctuating \ac{RES} share.
Through the comparison of \acp{XEF} and \acp{MEF}, SilerEvans et al.~\cite{SilerEvans.2012} found that \acp{XEF} may misestimate the emissions that can be avoided from an demand-side intervention.

Pean et al.~\cite{Pean.2018} extended Hawkes' approach by clustering the data of Spain in the year 2016 according to system load and \ac{RES} share and then realizing a linear regression on every cluster.
By fitting a quadratic function of \ac{RES} share and system load to the results of the regression they were able to compute hourly \acp{MEF}.
Although this approach seems to be promising for our aim of comparing the environmental effect of \ac{DR} for different countries, it involves constant emission factors per fuel type that disregards the efficiency differences of power plants within a fuel type which is of high significance for national energy systems that base on only one or two fuel types as, e.g., in Lithuania or Serbia.

Greensfelder et al.~\cite{Greensfelder.2010} studied the relationship between load and emissions for the four US regions IL, NY, TX, and CA.
The availability of power plant-specific emission and load time series allowed for the utilization of the flexibility weighted hourly average emissions rate method, which determines the power plants that are most likely to be operating at the margin.
However, due to the lack of comparable data, this method is not applicable to Europe. 

Finally, studies in Group C computed historic or future \acp{MEF} in hourly or higher resolution using PSOMs, more precisely Economic Dispatch Models.
Bettle et al.~\cite{Bettle.2006} used historic generation data per power plant of the UK for the year 2000 to calculate half-hourly \acp{MEF} that indicated up to 50\% higher carbon emission savings than the \acp{XEF}.
However, this method is not applicable to most of the European countries where power plant specific generation data are not readily available.
Furthermore, they used fix power plant efficiencies per fuel type.
Regett et al.~\cite{Regett.2018} computed future, hourly \acp{XEF} and \acp{MEF} for Germany and found that they are negatively correlated with each other meaning that they can lead to opposing results, which highlights the importance of the choice of \ac{CEF} type.
Based on \cite{Regett.2018}, Böing \& Regett~\cite{Boing.2019} proposed an emission accounting method that determined dynamic \acp{XEF} and \acp{MEF} for different energy carriers in multi-energy systems.
However, their focus laid on the future energy system for an individual country and not on the comparison of multiple countries based on historic data.
Baumgärtner et al.~\cite{Baumgartner.2019} ran an economic dispatch model on German data of the year 2016 that resulted in hourly \acp{XEF} and \acp{MEF} which fed into a subsequent multi-objective synthesis problem of a low-carbon utility system.
They used power plant specific efficiencies that were approximated by a logarithmic size-dependent regression of real power plants.
However, the power plant sizes are not available for all countries and efficiencies also depend on the year of construction.

In summary, the literature contains multiple studies that calculate \acp{CEF} for different countries and years with most studies focusing on \acp{XEF} (see Group A of \Cref{tab:lit_review}).
Only a few studies calculate \acp{MEF} with hourly resolution and none of them compares hourly \acp{MEF} between different countries.

\begin{figure*}[ht]
	\centering
	\subfloat[][\textbf{Net electricity generation ($t, y, c, f$)}. Here: Calculated shares of annual generation in \%.]{
		\includegraphics[width=1.0\linewidth]{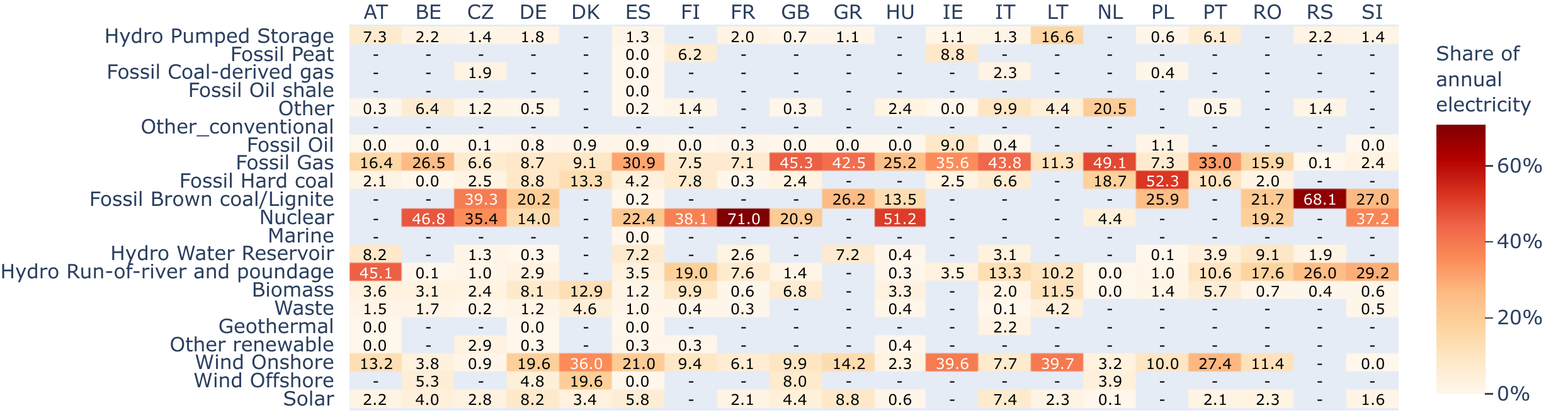}
		\label{fig:gen}}\\
	\subfloat[][\textbf{Installed generation capacity ($y, c, f$)}. Here: Share of total capacity in \%.]{
		\includegraphics[width=1.0\linewidth]{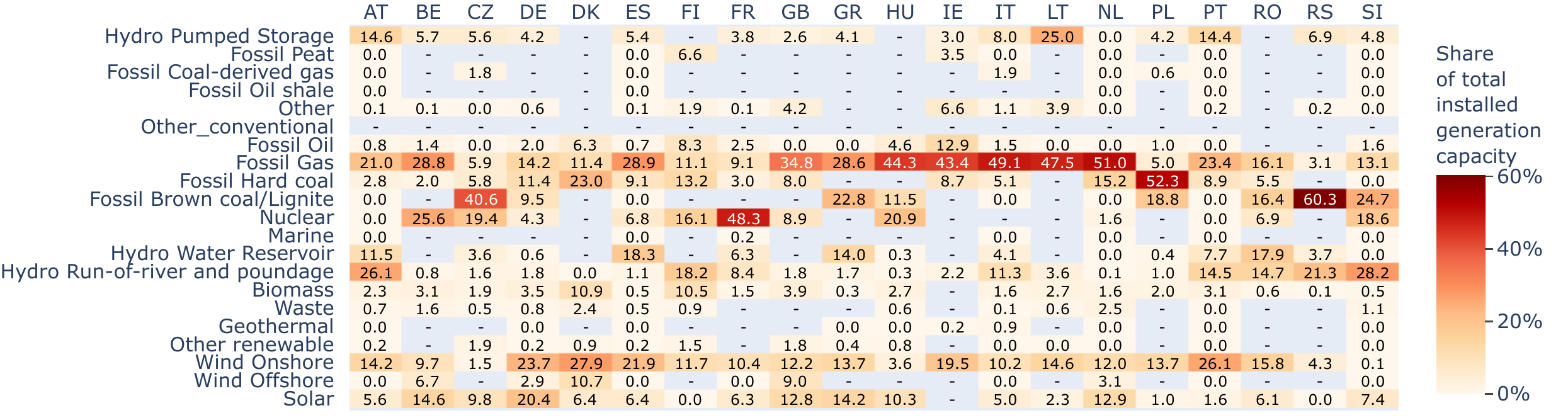}
		\label{fig:capa}}
	\caption{Overview of used data from \ac{ETP}~\cite{ENTSOE.2020} per fuel type and country after preprocessing for the year 2019: \protect\subref{fig:gen} Net electricity generation and \protect\subref{fig:capa} Installed generation capacity.
		Blue fields with "-" indicate missing data.}
	\label{fig:generation_rel_heatmap_annotated}
\end{figure*}

\subsection{Contribution}
Within this work, we quantitatively compare the environmental effects of PBDR between 20 European countries.
Since MEFs are currently not readily available and XEFs are not appropriate for measuring the effects of PBDR activities, we propose two approximating methods (Power plant (PP) and PieceWise Linear (PWL)) to calculate time-dependent CEFs with readily available data for European countries.

The aims of this paper are:
\begin{itemize} %
	\item Propose and validate a method that approximates \acp{MEF} and \acp{XEF} from readily available datasets.
	\item Apply the method to 20 European countries for the years 2017--2019 to compare resulting \acp{CEF}.
	\item Simulate load shifting based on prices, XEFs, and MEFs to assess its impact on carbon emissions.
	\item Evaluate the impact of carbon prices on the marginal cost--emission correlation, the merit order, and on the effects of load shifting.
\end{itemize}

The resulting hourly \acp{CEF}, marginal costs, and marginal fuel types for 2017--2019 can be downloaded as CSV files under a GitHub repository\footnote{\url{https://github.com/mfleschutz/marginal-emission-factors}}.

\subsection{Paper organization}
The remainder of this paper is structured as followed.
First, all data sets used for the calculation of the CEFs are described in \Cref{sec:data}.
Then, the methods for the CEFs calculation, PP and PWL, are detailed in \Cref{sec:methods}.
In \Cref{sec:validation}, the PP method is applied to the power plant resolved dataset for Germany in 2019 to evaluate the approximation error of the PWL method.
In \Cref{sec:results}, the PWL method is applied to the data of 20 European countries.
First, the resulting merit order and \acp{CEF} are described, \Cref{sec:merit-orders-and-cefs-for-european-countries}.
In \Cref{sec:correlation-analysis-of-emissions-and-prices}, cost--emissions correlations are analyzed.
In \Cref{sec:load-shift-analysis} and \Cref{sec:cef-based-load-shift-analysis}, load shifts based on prices, MEFs, and XEFs are simulated and the results discussed.
In \Cref{sec:six-example-countries}, the phenomenon of increasing carbon emission as effect of \ac{PBDR} is explained using the results of six exemplary countries.
In \Cref{sec:impact-of-carbon-price}, a sensitivity analysis of carbon emissions on the merit order is conducted that demonstrates the impact of the carbon price on the merit order and the mitigation of the merit order dilemma of emissions.
\Cref{sec:discussion-summary} summarizes the results and discussions and in \Cref{sec:conclusions}, we conclude the findings of this paper.

\section{Data sources for this study}\label{sec:data}
This section describes the datasets that are used in this study grouped by data source.

\subsection{ENTSO-E Transparency Platform (ETP)}

\subsubsection{Net generation}\label{sec:data-generation}
For the calculation of the temporally resolved residual and total load of a national electricity system, historic electricity generation time series data (“Aggregated Generation per Type”) for each fuel type~$f$ and country~$c$ from the \ac{ETP}~\cite{ENTSOE.2020} were used and accessed via the ENTSOE-py client~\cite{ENTSOEPY.2020}.
For the years 2017--2019 and the 20 countries used in this paper, 10.2 Million data points were processed.
Across all European countries, 20 different fuel types are used for electricity generation.
The temporal resolution of the raw data varies depending on the country (15, 30, and 60 minute intervals).
A comprehensive review of the data composition of the \ac{ETP} can be found in \cite{Hirth.2018}.
Within the data preprocessing for the simulation, all time series were downsampled to hourly resolution.
In the case of missing values and outliers, the last available data points were considered instead.
The percentage of missing values that were filled were 1.9\% for 2017, 1.3\% for 2018, and 0.4\% for 2019.
However, 67\% of the missing values appear in only two fuel types ('Fossil Hard coal' and 'Other') for the Netherlands for the years 2017 and 2018.
Nine outliers were detected by a combination of Z-score analysis and treated as missing values, see Supplementary Material D.
To save space, the following renaming was conducted: Biomass $\to$ biomass, Fossil Brown coal/Lignite $\to$ lignite, Fossil Oil $\to$ oil, Fossil Gas $\to$ gas, Fossil Hard coal $\to$ coal, Hydro Run-of-river and poundage $\to$ hydro, Nuclear $\to$ nuclear.
\Cref{fig:gen} shows the share of annual generation per fuel type and country in relation to the total generation for 2019 after preprocessing.
Supplementary Material A and B contain basic analyses of available generation data from the \ac{ETP}~\cite{ENTSOE.2020} for Europe and Germany, respectively.

\subsubsection{Installed generation capacity}\label{sec:data-installed-generation-capacity}
Installed electricity generation capacities per fuel type~$f$ were also obtained from the \ac{ETP}~\cite{ENTSOE.2020}.
The same renaming of fuel types as in \Cref{sec:data-generation} was conducted.
\Cref{fig:capa} shows all preprocessed data for 2019 in a concise way.
Supplementary Material A contains an overview of available installed generation capacity data on the \ac{ETP}~\cite{ENTSOE.2020}.

\subsection{Open Power System Data (OPSD)}

\subsubsection{OPSD Power plant list}\label{sec:data-OPSD-PPlist}
For the PP method and to verify the PWL method with a higher granularity of generation capacity, the power plant list from the Open Power System Data~\cite{OpenPowerSystemData.2018} was used.
This data consists of 893 power plants for the electricity market region Germany, Austria and Luxembourg with 38 features each.
The most relevant features were country, capacity, date for commissioning and shutdown, type, as well as efficiency estimate.
In this paper, for the sake of simplicity, individually controllable power plant units are referred to as power plants.
In \Cref{fig:regression}, e.g., the nine light brown horizontally distributed dots near 20~GW represent the nine generation units of the Jänschwalde power plant which have similar characteristics.
For the construction of the merit order, only active power plants were considered by constraining the year of simulation after date of commissioning and before shutdown.
The distribution of efficiency and capacity can be seen in Supplementary Material C.

\subsection{Global Energy Observatory (GEO)}

\subsubsection{GEO Power plant list}\label{sec:data-GEO-PPlist}
Due to the particularly strong influence of power plant efficiency, power plants are further distinguished according to their technology type in open cycle gas turbines (indicated with gas) and combined-cycle gas turbines (indicated with gas\_cc).
The division into gas and gas\_cc was based on $k^\mathrm{cc}$ which is the proportion of combined-cycle gas turbines in all gas power plants for each analyzed country.
These shares were calculated from the GEO power plant list~\cite{GEO.2020} with two exceptions:
For Germany, the share was calculated from the OPSD list in favour of more detailed data.
For Serbia, no gas power plants were listed on~\cite{GEO.2020} at the time of data retrieval.
Independent investigation identified its share to be $k^\mathrm{cc}=0\%$.
All used values for $k^\mathrm{cc}$ are listed in \Cref{tab:share_cc}.

\begin{table}[htb]
	\centering
	\caption{Assumption of $k^\mathrm{cc}$ (share of combined-cycle gas turbines across all gas fuelled power plants  in \%).}
	\label{tab:share_cc}
	\resizebox{1.0\linewidth}{!}{\begin{tabular}{lr|lr|lr|lr}
			\toprule
			AT &   99.57 \textsuperscript{b} &  ES &   94.73 \textsuperscript{b} &  HU &  32.80 \textsuperscript{b} &  PL &  100.00 \textsuperscript{b} \\
			BE &   80.92 \textsuperscript{b} &  FI &  100.00 \textsuperscript{b} &  IE &  71.76 \textsuperscript{b} &  PT &   93.26 \textsuperscript{b} \\
			CZ &  100.00 \textsuperscript{b} &  FR &   83.50 \textsuperscript{b} &  IT &  99.78 \textsuperscript{b} &  RO &   46.24 \textsuperscript{b} \\
			DE &   53.52 \textsuperscript{a} &  GB &   42.31 \textsuperscript{b} &  LT &   0.00 \textsuperscript{b} &  RS &    0.00 \textsuperscript{c} \\
			DK &   28.12 \textsuperscript{b} &  GR &   79.90 \textsuperscript{b} &  NL &  85.96 \textsuperscript{b} &  SI &    0.00 \textsuperscript{b} \\
			\bottomrule
			\multicolumn{8}{l}{\textsuperscript{a} from the OPSD power plant list of the PP method \cite{OpenPowerSystemData.2018}} \\
			\multicolumn{8}{l}{\textsuperscript{b} GEO list \cite{GEO.2020}} \\
			\multicolumn{8}{l}{\textsuperscript{c} \cite{WIKIPEDIA.2020}} \\
	\end{tabular}}
\end{table}

\subsection{Other data sources}

\subsubsection{Carbon emission certificate prices}\label{sec:data-carbon-emission-certificate-prices}
\begin{figure}[htb]
	\centering
	\includegraphics[width=1.0\linewidth]{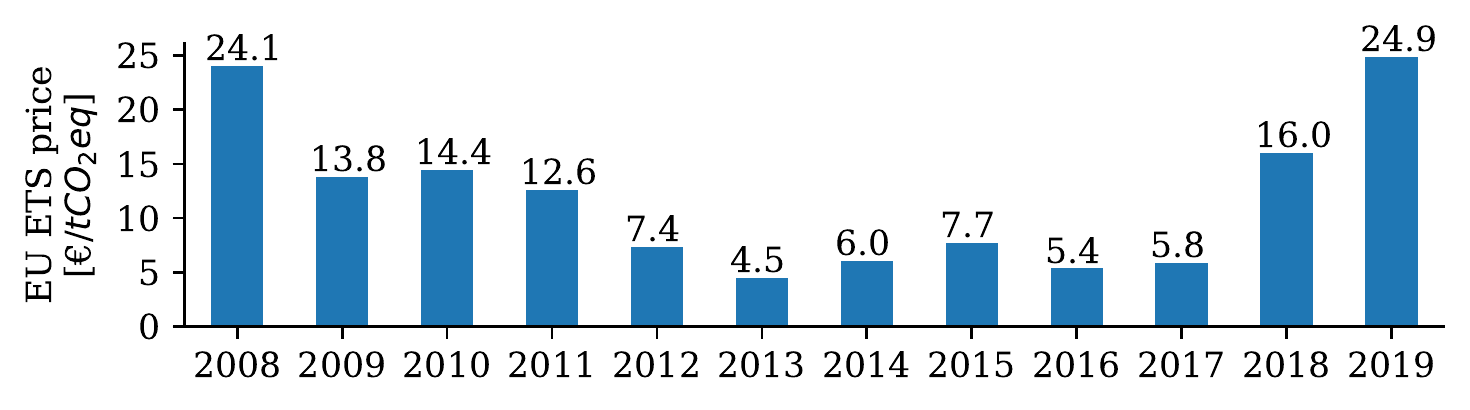}
	\caption{Historic annual average $\text{CO}_2$-emission prices of the EU ETS for the years 2008--2019. Data source: \cite{EEX.2020}.}
	\label{fig:ETS_price}
\end{figure}

In order to calculate the carbon-related marginal costs for the merit order, European Emission Allowances (EUA) prices of the European Union \ac{ETS} are used from the European Energy Exchange (EEX)~\cite{EEX.2020} (downloaded via~\cite{Sandbag.2019}).
The data was downsampled from weekly to annual resolution with average method and used in mean annual form as $c^\mathrm{GHG}$, see \Cref{fig:ETS_price}.

\subsubsection{Fuel type specific emissions and costs}\label{sec:data-fuel-type-specific-emissions-and-costs}
\Cref{tab:input_params} shows an overview of the fuel type~$f$ specific input parameters used in this study.
The fuel type specific cost were used depending on the year and in the case of natural gas, also depending on the country.
As fuel type specific $\text{CO}_2$-intensity $\varepsilon_f$, operational net emission factors from \cite{Quaschning.2019} were used which, in contrast to life cycle assessment approaches, do not include emissions embodied in infrastructure.
This is consistent with the short-term nature of \ac{PBDR}.

\begin{table}[htb]
	\centering
	\caption{Overview of the fuel type~$f$ specific input parameters for the merit order and its data sources.}
	\label{tab:input_params}
	\begin{tabular}{lll}
		\toprule
		Fuel type & $ \text{CO}_2 $-intensity & Fuel price\\ 
		$ f $  & $\varepsilon_f$ & $c_f$\\
		& t\textsubscript{CO\textsubscript{2}eq}/MWh & \euro/MWh \\ 
		\midrule
		oil & 0.28 \cite{Quaschning.2019}  & 54.31\textsuperscript{y} \cite{Konstantin.2017}, \cite{StatistischeBundesamt.2020}, \cite{MineralFranken.2003} \\
		gas & 0.25 \cite{Quaschning.2019}  & 26.10\textsuperscript{y,c} \cite{StatistischeBundesamt.2020} \\
		coal & 0.34 \cite{Quaschning.2019}  & 14.58\textsuperscript{y} \cite{Konstantin.2017}, \cite{StatistischeBundesamt.2020} \\
		lignite & 0.36 \cite{Quaschning.2019}  & 6.18\textsuperscript{y} \cite{Konstantin.2017}, \cite{StatistischeBundesamt.2020} \\
		nuclear & 0.0 \cite{Quaschning.2019}  & 4.18\textsuperscript{y} \cite{Konstantin.2017} \\
		\bottomrule
		\multicolumn{3}{l}{\textsuperscript{y} Year-specific values were used. Here: 2019.} \\
		\multicolumn{3}{l}{\textsuperscript{c} Country-specific values were used. Here: DE.} \\
	\end{tabular} 
\end{table}

\subsubsection{Transmission efficiency}\label{sec:data-transmission-efficiency}
Based on the methods employed in \cite{Kono.2017} and \cite{Baumgartner.2019} we employed a constant transmission efficiency $ \eta^\mathrm{T} $ to consider all losses for transmission and distribution.
\Cref{tab:transmissions_efficiency} shows the used average value over the last four published years (2010--2014) for each country~\cite{WorldBank.2020}.

\begin{table}[htb]
	\centering
	\caption{Transmission efficiencies $\eta^\mathrm{T}$ used in simulations. Data source:~\cite{WorldBank.2020}.}
	\label{tab:transmissions_efficiency}
	\begin{tabular}{lrlrlrlr}
		\toprule
		AT & 94.9\% &  ES & 90.8\% &  HU & 88.8\% &  PL & 93.3\% \\
		BE & 95.1\% &  FI & 96.2\% &  IE & 92.3\% &  PT & 90.6\% \\
		CZ & 95.1\% &  FR & 93.6\% &  IT & 93.0\% &  RO & 88.4\% \\
		DE & 96.1\% &  GB & 92.3\% &  LT & 79.5\% &  RS & 84.7\% \\
		DK & 93.7\% &  GR & 94.2\% &  NL & 95.2\% &  SI & 94.6\% \\
		\bottomrule
	\end{tabular}
\end{table}

\begin{figure*}[htb]
	\centering
	\includegraphics[width=\linewidth]{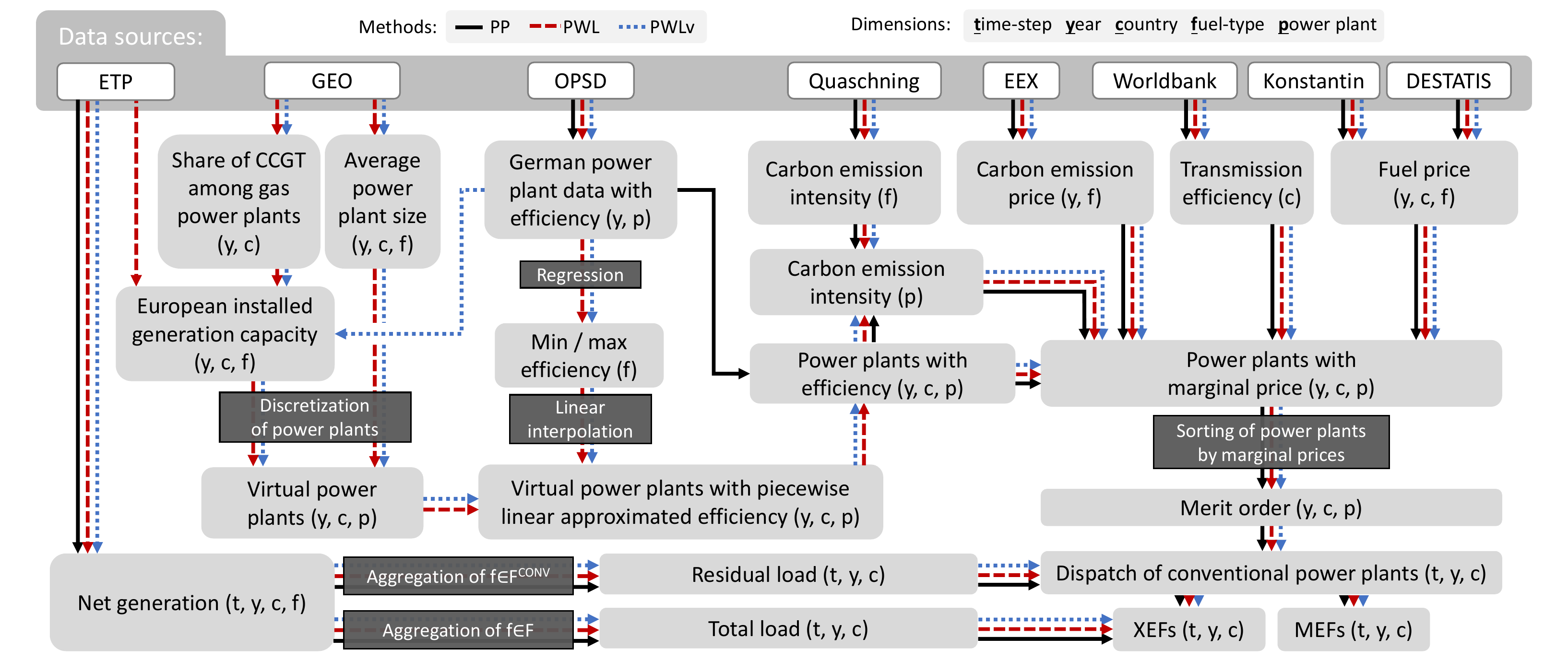}
	\caption{Schematic depiction of data sources, data subsets and special calculations of the PP, PWL, and PWLv method to calculate \ac{CEF}. White boxes indicate the following data sets: ETP\,\cite{ENTSOE.2020}, GEO\,\cite{GEO.2020}, OPSD\,\cite{OpenPowerSystemData.2018}, Quaschning\,\cite{Quaschning.2019}, EEX\,\cite{EEX.2020}, Konstantin\,\cite{Konstantin.2017}, DESTATIS\,\cite{DESTATIS.2020}. Light grey boxes indicate subsets or modified data. The dimension of each dataset is denoted by the characters t, y, f, and p in brackets. Dark grey boxes indicate special mathematical operations.}
	\label{fig:scheme_CEF_calculation}
\end{figure*}

\section{Methods}\label{sec:methods}
In this study, we propose two approximating methods -- the Power Plant (PP) method and the PieceWise-Linear (PWL) method -- to calculate dynamic \acp{MEF} and \acp{XEF} from the available data.
Both methods (PP and PWL) include the application of a country and year specific merit order on historic residual load data to simulate the time-dependent dispatch of conventional power plants and to identify the marginal power plant per time step~$t$.
In other words, the prices, XEFs, and MEFs from the simulation depend on the available conventional power plants, the time-dependent \ac{RES} shares, and the total load, which was proxied by the national total generation.

The PP method uses a power plant list with given efficiencies.
Since power plant specific data is not readily available for most European countries, we also propose the PWL method, which does not rely on given power plant specific efficiency and capacity.
Instead, national installed generation capacities per fuel type were used to discretize the fuel type specific generation capacity into virtual power plants.
\Cref{fig:scheme_CEF_calculation} shows a schematic depiction of the data sources and calculation steps.
Besides the two main methods (PP and PWL), it also shows the PWL validation mode (PWLv), which is detailed in \Cref{sec:validation}.

For the sake of simplicity, from here onwards, the term fuel types also include gas\_cc which technically is a special combination of a fuel type and a technology.
The remainder of this section describes the calculation of \acp{XEF} and \acp{MEF} using the PP and the PWL method.

\subsection{Calculation of MEFs}

Regardless of the merit order calculation method, the power plant specific emissions $\varepsilon_p$ are given by

\begin{align}
& \varepsilon_p = \frac{\varepsilon_f}{\eta_p^\mathrm{el}}  \label{eq:marg_emissions}
\end{align}
where $\varepsilon_f$ is the carbon emission intensity per fuel type~$f$ and 
$\eta_p^\mathrm{el}$ is the generation efficiency per power plant~$p$.

The time-dependent dispatch is then given by applying the merit order on the residual load.
The time-dependent marginal emission factor $\mathrm{MEF}_t$ is given by the power plant specific emission intensity $\varepsilon_p$ of the marginal power plant in given time step~$t$ divided by the transmission efficiency $\eta^\mathrm{T}$:

\begin{equation}
\begin{aligned}
\mathrm{MEF}_t = \frac{\sum\limits_{p \in P}\varepsilon_p \gamma_{t,p}^\mathrm{m}}{\eta^\mathrm{T}}
\label{eq:CALC_mefs}
\end{aligned}
\end{equation}
where $ \gamma_{t,p}^\mathrm{m} $ is a binary variable:

\begin{equation}
\gamma_{t,p}^\mathrm{m} =
\begin{cases}
1, & \text{if}~ \sum_{i=1}^{p-1}{P_{i}^\mathrm{inst}} < P_t^\mathrm{resi} \leq \sum_{i=1}^{p}{P_{i}^\mathrm{inst}},\\
0, & \text{else.}
\end{cases}
\label{eq:CALC_mefs2}
\end{equation}

\subsection{Calculation of XEFs}
In all methods (PP, PWL, PWLv), the grid mix emission factor $\mathrm{XEF}_t$ is calculated with the following equation:

\begin{equation}
\begin{aligned}
\mathrm{XEF}_t = \frac{\text{electricity supply specific emissions (t)}}{\text{electricity consumption (t)}}. \\
\label{eq:CALC_xefs1}
\end{aligned}
\end{equation}

We simplify \Cref{eq:CALC_xefs1} to:

\begin{equation}
\begin{aligned}
\mathrm{XEF}_t = \frac{\sum\limits_{p \in P}{\varepsilon_p \gamma_{t,p}^\mathrm{x} P_p^\mathrm{inst} \Delta t}}{\eta^\mathrm{T} \sum\limits_{f \in F}{E_{t,f}^\mathrm{gen}}}
\label{eq:CALC_xefs2}
\end{aligned}
\end{equation}
where $\varepsilon_p$ is the carbon emission intensity per power plant~$p$,
$P_p^\mathrm{inst}$ is the installed power plant capacity,
$\gamma_{t,p}^\mathrm{x}$ is the capacity utilization rate defined in \Cref{eq:CALC_xefs4},
$\Delta t$ is the time-step-width,
$\eta^{T}$ is the constant transmission efficiency considering all transmission and distribution losses,
$E_{t,f}^\mathrm{gen}$ is the generated energy per fuel type~$f$ and time step~$t$,
and $F$ is the set of all generation fuel types.

\begin{equation}
\gamma_{t,p}^\mathrm{x} =
\begin{cases}
1, & \text{if}~ \sum_{i=1}^{p}{P_{i}^\mathrm{inst}} < P_t^\mathrm{resi} , \\
0, & \text{if}~ \sum_{i=1}^{p-1}{P_{i}^\mathrm{inst}} \geq P_t^\mathrm{resi}, \\
\frac{P_t^\mathrm{resi} - \sum_{i=1}^{p-1} P_{i}^\mathrm{inst}}{P_p^\mathrm{inst}}, & \text{else.}\\
\end{cases}
\label{eq:CALC_xefs4}
\end{equation}

\begin{figure}[tb]
	\centering
	\includegraphics[width=1.0\linewidth,trim=0 0.7cm 0 0,clip]{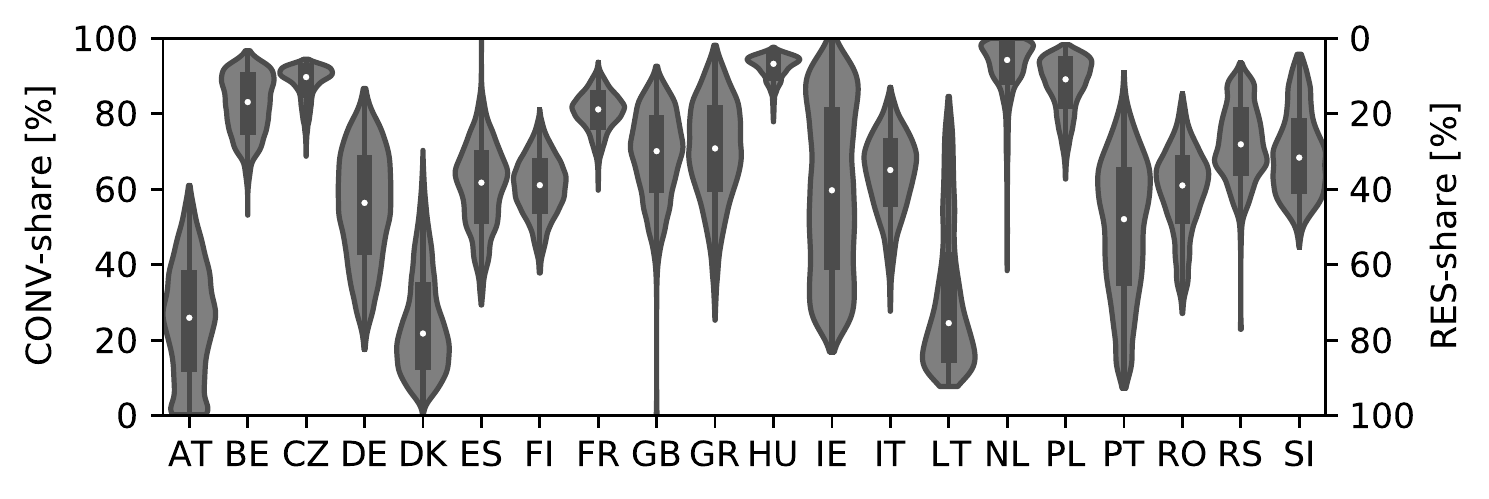}
	\includegraphics[width=0.93\linewidth,trim=0 0 -1.1cm 0.2,clip]{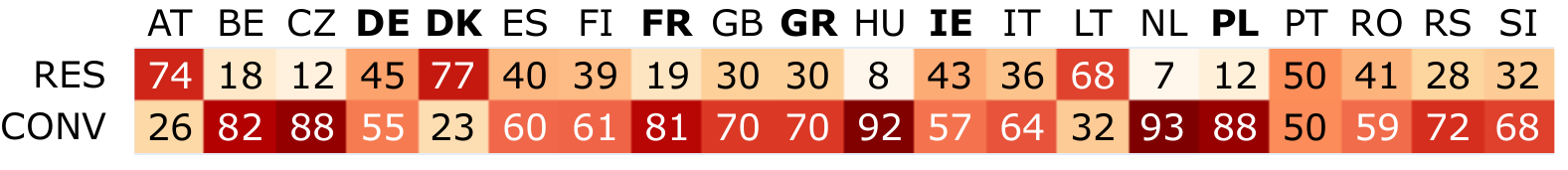}
	\caption{Top: Distribution of the CONV share (left) and \ac{RES} share (right) of the total national generation. White dots represent medians. Bottom: Corresponding average values in \%.}
	\label{fig:violin_share_conv_2019}
\end{figure}

\subsection{Residual load}\label{sec:residual-load}
Since fuel type specific generation data $P_{t,f}^\mathrm{gen}$ is available in most European countries, the residual load $P_t^\mathrm{resi}$ is approximated by the sum of energy generated by the conventional power plants for all methods (PP, PWL, PWLv) and both \ac{CEF} types (\ac{XEF} and \ac{MEF}):

\begin{equation}
\begin{aligned}
P_t^\mathrm{resi} = \sum_{f \in F^\mathrm{CONV}}{P_{t,f}^\mathrm{gen}}
\label{eq:residual_load}
\end{aligned}
\end{equation}
where $F^\mathrm{CONV}$ is the set of all available conventional fuel types (first 11 fuel types in \Cref{fig:gen}).

\Cref{fig:violin_share_conv_2019} shows the distributions and average values of the time-dependent CONV share and \ac{RES} share of the total national generation for the year 2019 for different countries.
The average \ac{RES} share varies from 7\% in the Netherlands (NL) to 77\% in Denmark (DK).
Given \Cref{eq:residual_load} and the limitation to national electricity supply, in this analysis, the CONV share in \Cref{fig:violin_share_conv_2019} is also the residual load share of the total national load.

\subsection{Merit order calculation with the PP method}\label{sec:pp-method}
The PP method can be used to calculate the merit order if power plant specific efficiencies are available.
The merit order results from sorting all active power plants $p$ according to their ascending marginal costs~$c_p^\mathrm{m}$, which are given by:

\begin{align}
& c_p^\mathrm{m} = \underbrace{\frac{x_f}{\eta_p^\mathrm{el}}}_\text{fuel costs} + \underbrace{\frac{\varepsilon_f}{\eta_p^\mathrm{el}}  c^\mathrm{GHG}}_\text{carbon-related costs} \label{eq:marg_costs}
\end{align}
where $\eta_p^\mathrm{el}$ is the efficiency per power plant~$p$ (data from \Cref{sec:data-OPSD-PPlist}),
$c^\mathrm{GHG}$ is the carbon emission price (data from \Cref{sec:data-carbon-emission-certificate-prices}),
and $\varepsilon_f$ and $x_f$ are the fuel type specific emission intensities and prices, respectively (data from \Cref{sec:data-fuel-type-specific-emissions-and-costs}).

\Cref{fig:merit_orders}a) exemplarily shows the resulting merit order with power plant specific marginal costs $c_p^\mathrm{m}$ and emissions $\varepsilon_p^\mathrm{m}$ for Germany, 2019.

\begin{figure}[tb]
	\centering
	\includegraphics[width=1.0\linewidth]{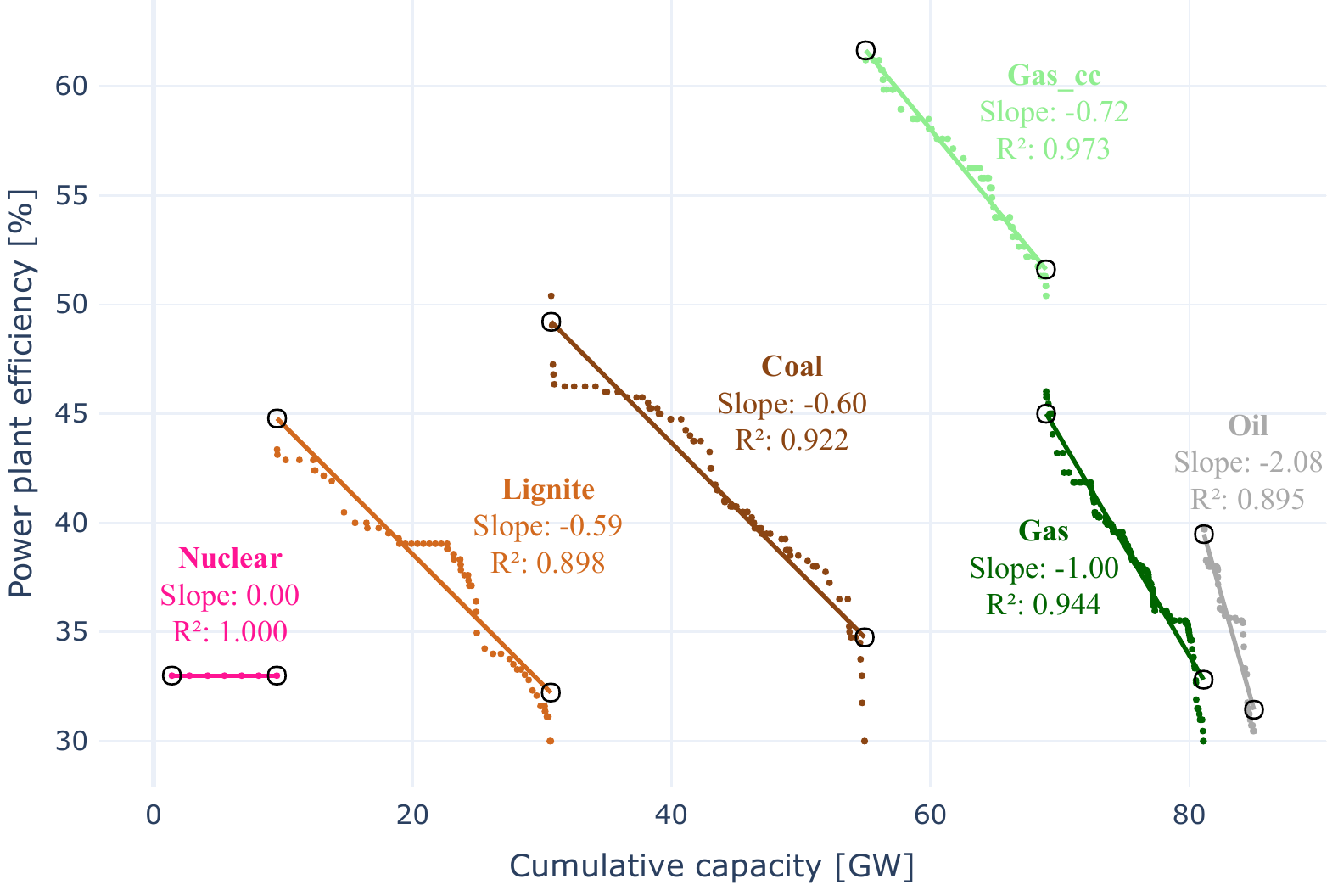}
	\caption{Ordinary least squares regressions on efficiencies of the power plant list to identify fuel type specific minimal and maximal values (indicated with black open circles). Linear interpolation between these minimal and maximal values result in a piecewise linear function. Source: Calculations based on \cite{OpenPowerSystemData.2018}.}
	\label{fig:regression}
\end{figure}

\begin{figure}[tb]
	\centering
	\includegraphics[width=1.0\linewidth]{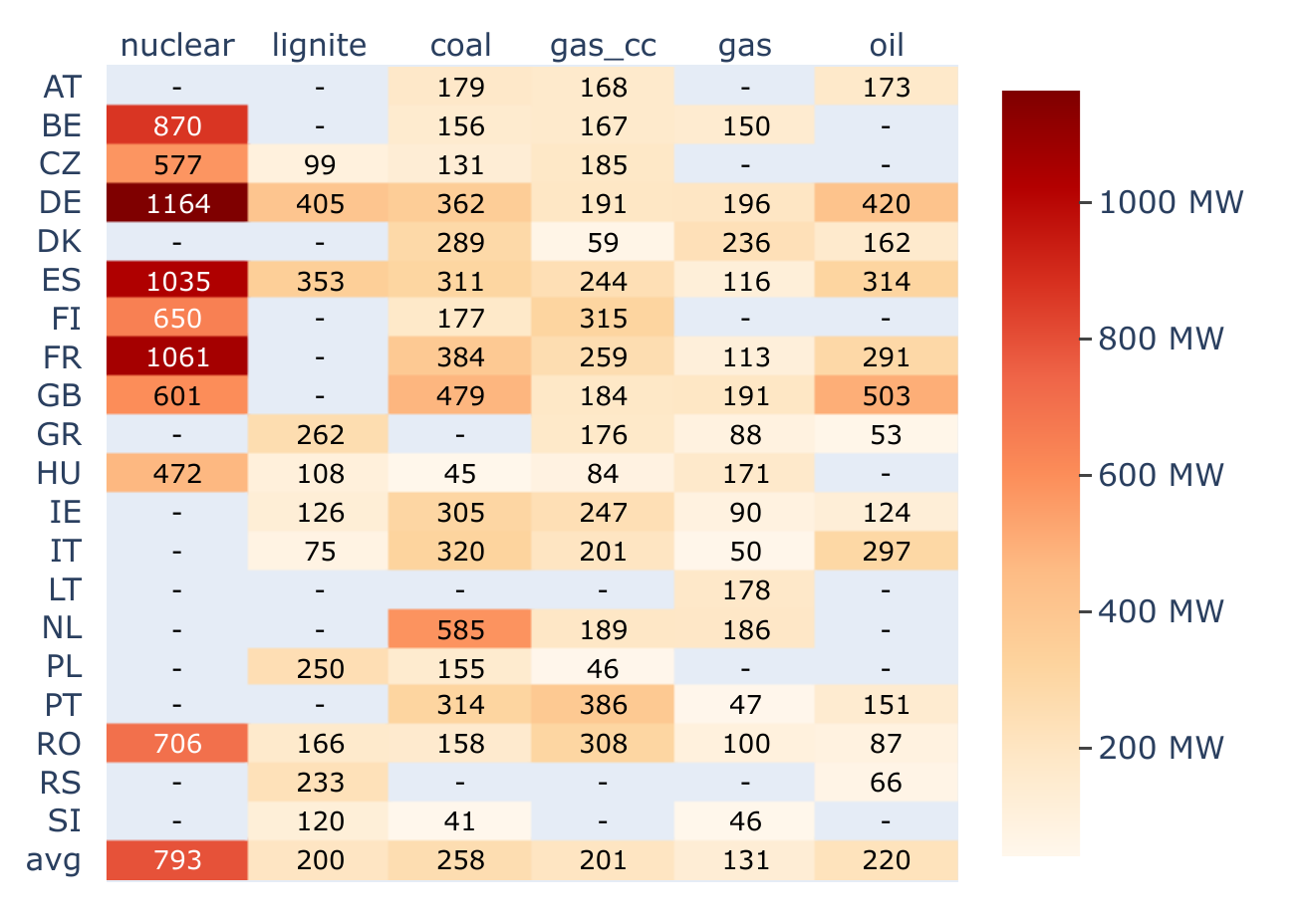}
	\caption{Average power plant sizes used in PWL method. The average values at the bottom were used where no data were available (indicated with "-"). Data source:~\cite{GEO.2020}.}
	\label{fig:pp_sizes_heatmap_annotated}
\end{figure}

\subsection{Merit order calculation with the PWL method}\label{sec:pwl-method}
The PWL method is a piecewise-linear approximation approach that can be used to approximate the country and year specific merit order if power plant specific capacity and efficiency data are unavailable.
For most European countries only fuel type specific data are provided, hence the PP method which is based on the power plant list including power plant efficiencies is not applicable in these countries.
A very simple approach to approximate \acp{MEF} is the usage of fuel type specific efficiencies.
In this PWL method, instead, we assume that all countries have the same maximum and minimum efficiency per fuel type and that within each fuel type the capacity over the range of occurring efficiencies is uniformly distributed.
These assumptions allow us to approximate the efficiencies of the power plants by a piecewise-linear function.
For each fuel type, the minimum and maximum efficiencies were determined by the minimum and maximum values of an ordinary least squares regression of the power plant efficiencies of the German power plant list, see \Cref{fig:regression}.
To enable the non-disjunct structure of the merit order, which can be observed in reality, the total generation capacities per fuel type were discretized into discrete equally sized virtual power plants.
In our case, country and fuel type specific values shown in \Cref{fig:pp_sizes_heatmap_annotated} were used.
They were computed from the GEO power plant list~\cite{GEO.2020}, which contains capacity data but no efficiency data.

\subsection{Limitations}\label{sec:limitations}
The resulting \acp{MEF} depend on the determination of marginal power plants, which is subject to some uncertainty since the merit order is calculated based on an approximation of marginal costs and installed generation capacity.
Therefore, the error generated through the improper calculation of the merit order has a substantially stronger effect on \acp{MEF} than on \acp{XEF}.

As did all previous studies presented in \Cref{tab:lit_review}, we do not consider transnational power flows in the calculation of \acp{MEF} since there is no existing method to do so.
Commercial entities like Tomorrow~\cite{ElecMapApi.2020,Corradi.2018} provide estimated \acp{MEF} considering cross border flows, however their method is not published.

Due to these two limitations, the \acp{MEF} calculated in this study are subject to an unquantified level of uncertainty.
However, the authors deem this to be reasonably low due to the high number of empirically derived parameters factored into this studies.

\section{Validation of the PWL method}\label{sec:validation}
\begin{figure}[htb]
	\centering
	\includegraphics[width=1.0\linewidth]{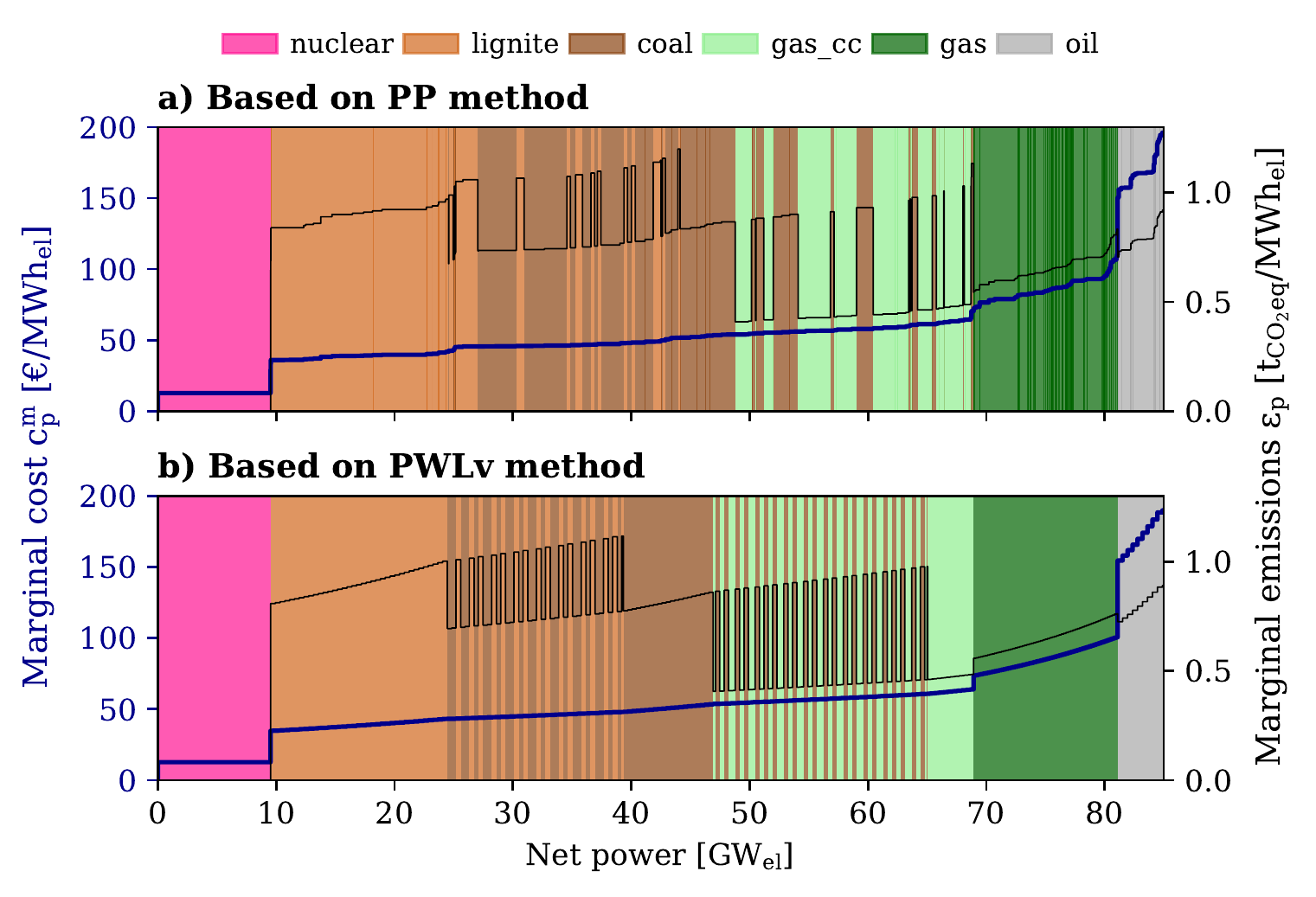}
	\vspace{-0.8cm}
	\caption{Merit orders for Germany, 2019 based on a) PP method and b) PWLv method, respectively.
		The background colors correspond to the fuel types. The x-axis is shared.}
	\label{fig:merit_orders}
	\vspace{0.3cm}
	\includegraphics[width=1.0\linewidth]{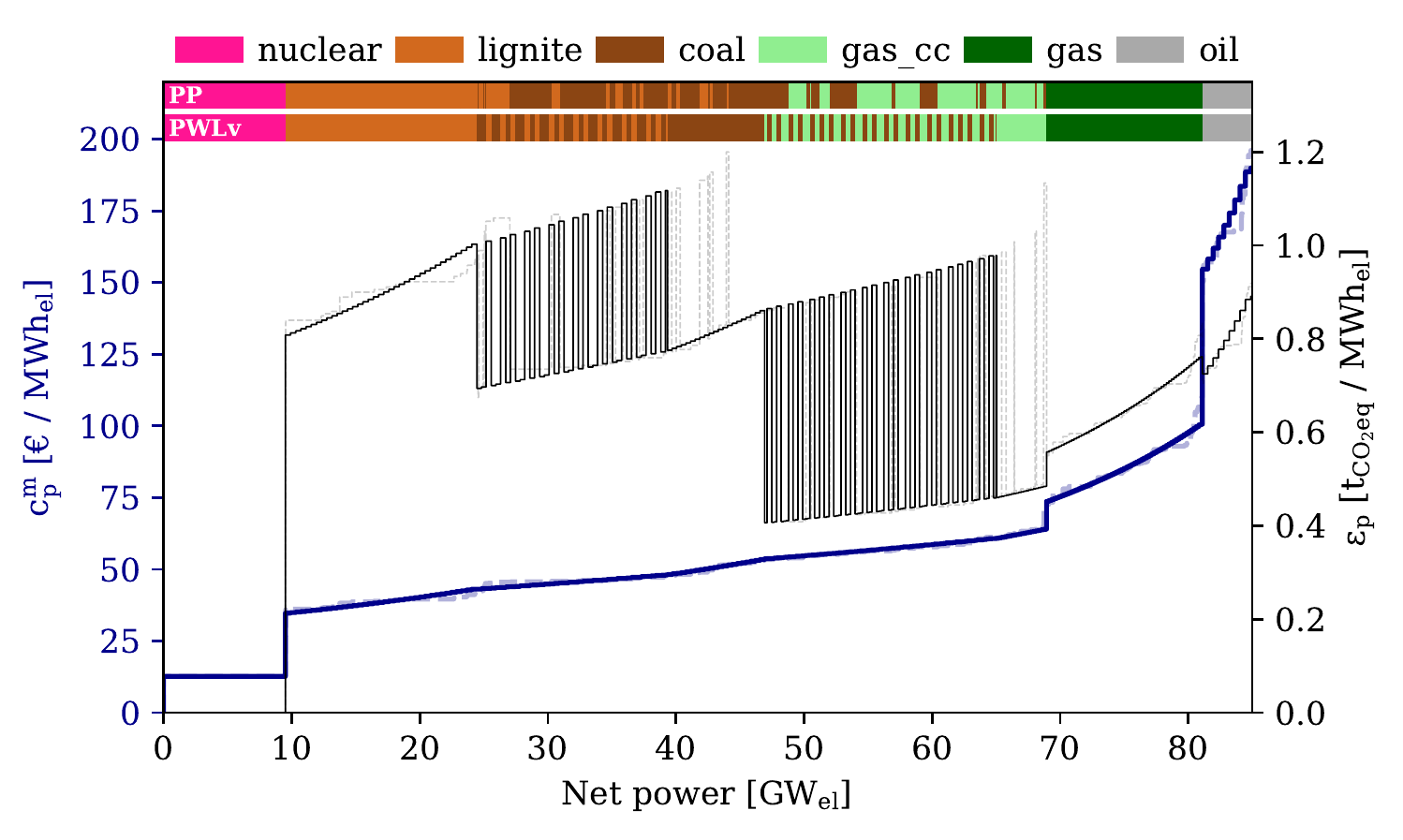}
	\vspace{-0.8cm}
	\caption{Direct comparison of PP vs. PWLv method based merit orders.
		The colors of the additional horizontal bars at the top correspond to the fuel types.}
	\label{fig:pp_and_pwl_merit_order_in_one}
\end{figure}

\begin{figure}[htb]
	\centering
	\includegraphics[width=1.0\linewidth]{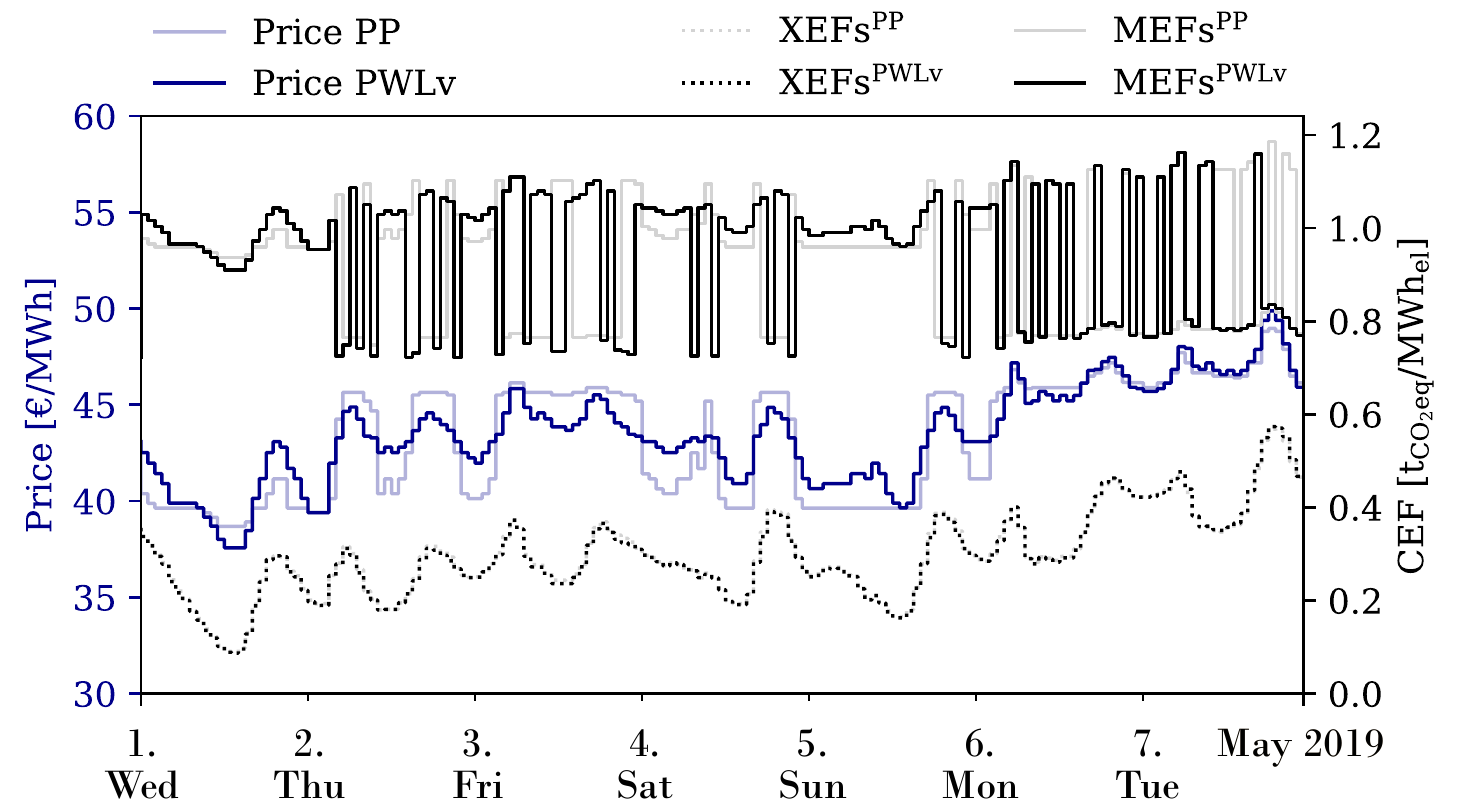}
	\vspace{-0.5cm}
	\caption{Comparison between XEFs, MEFs, and marginal prices for the methods PP and PWLv in hourly resolution for the first week in May 2019 (Wednesday--Tuesday) in Germany.}
	\label{fig:comparison_XEF_MEF}
	\vspace{0.3cm}
	\includegraphics[width=1.0\linewidth]{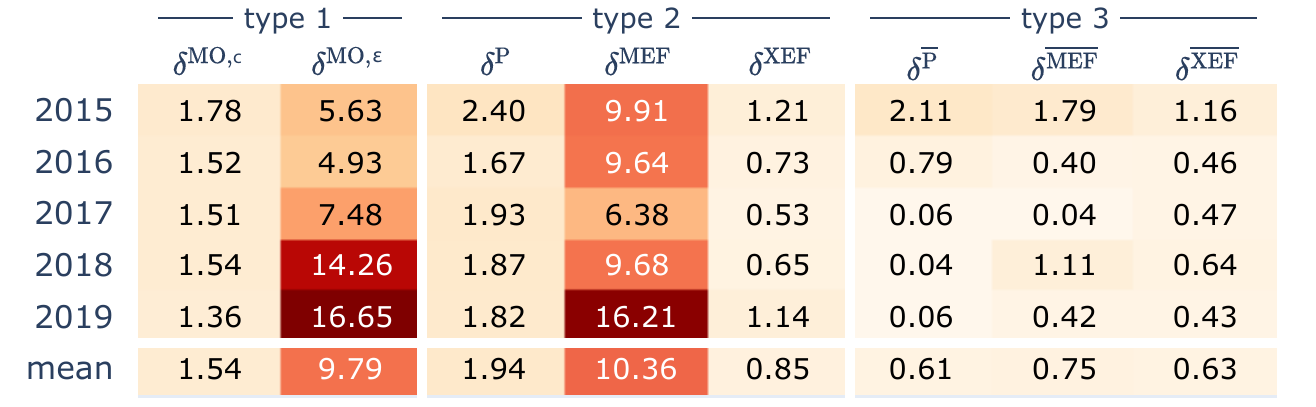}
	\vspace{-0.5cm}
	\caption{Relative approximation errors of the PWL method vs. PP method for Germany and the years 2015--2019 in \% grouped in three error types.}
	\label{fig:errors_heatmap_annotated}
	\vspace{0.3cm}
	\includegraphics[width=1.0\linewidth]{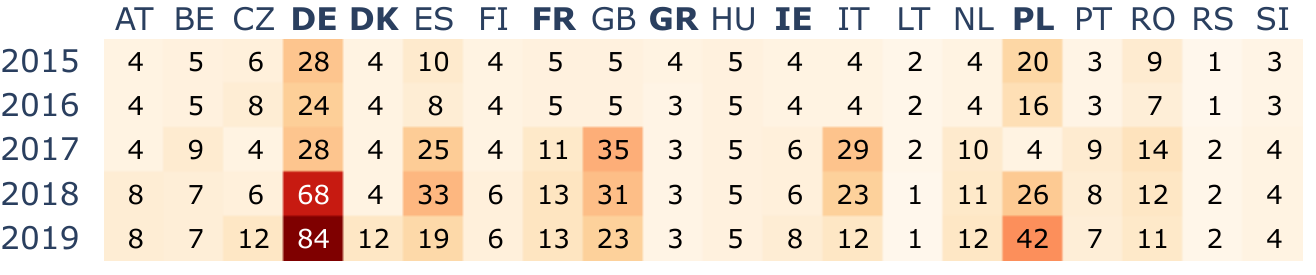}
	\vspace{-0.5cm}
	\caption{Number of fuel type changes along PWL-based merit orders for the years 2015--2019.}
	\label{fig:heatmap_no_of_fuel_type_changes_of_merit_order}
\end{figure}

To evaluate the approximation error of the PWL method, the PP method was applied to the detailed dataset for Germany described in \Cref{sec:data-OPSD-PPlist} for the years 2015--2019.
For this, we introduce the PWLv method as a validation variant of the PWL method which ensures consistency of data sources between the two main methods (PP and PWL).
In \Cref{fig:scheme_CEF_calculation}, it can be seen that in contrast to the PWL method, the PWLv method used installed generation capacity data from the German OPSD power plant list to ensure the same data basis with the PP method.
The PWLv-based merit order for Germany, 2019 is shown in \Cref{fig:merit_orders}b); and for better comparison, \Cref{fig:pp_and_pwl_merit_order_in_one} shows the same merit order together with the PP-based merit order.
\Cref{fig:comparison_XEF_MEF} shows a comparison of the two CEF-types and the two calculation methods of Germany for the first week in May 2019.
It can be seen that the PWL method succeeds to represent the \acp{MEF} well.
A whole-year comparison between the resulting $\mathrm{MEF}_t^\mathrm{PP}$ and $\mathrm{MEF}_t^\mathrm{PWLv}$ can be found in Supplementary Material F.

\Cref{fig:errors_heatmap_annotated} shows the values of the three different relative error types that were calculated to evaluate the approximation quality of the PWL method compared to the PP method:
\begin{enumerate}
	\item The error of power plant specific marginal costs~$c_p^\mathrm{m}$ and emission intensities~$\varepsilon_p$ along the merit order ($\delta^\mathrm{MO,c}$, $\delta^\mathrm{MO,\varepsilon}$).
	$\delta^\mathrm{MO,c}$ and $\delta^\mathrm{MO,\varepsilon}$ were calculated as averaged relative errors of marginal costs and emission intensities, respectively, between the PWLv method and the PP method along the merit order.
	To be able to deal with the different cumulative power values in the merit orders, the merit order was discretized by  10\,MW-elements which are more than ten times smaller than the average power plant.
	\item The error of time-dependent prices, MEFs, and XEFs ($\delta^\mathrm{P}$, $\delta^\mathrm{MEF}$, $\delta^\mathrm{XEF}$).
	\item The error of yearly aggregated prices, MEFs, and XEFs ($\delta^\mathrm{\overline{P}}$, $\delta^\mathrm{\overline{MEF}}$, $\delta^\mathrm{\overline{XEF}}$).
\end{enumerate}

From \Cref{fig:errors_heatmap_annotated}, one can see that while all other error types are below 2.5\%, $\delta^\mathrm{MO,\varepsilon}$ and  $\delta^\mathrm{MEF}$ are above.
However, due to the high number of lignite and coal power plants with similar marginal costs, Germany has by far the most fuel type changes along the merit order, see \Cref{fig:heatmap_no_of_fuel_type_changes_of_merit_order}.
And since $\delta^\mathrm{MO,\varepsilon}$ and $\delta^\mathrm{MEF}$ correlate with the number of fuel type changes, they are expected to be smaller for all other countries.\footnote{The Pearson correlation coefficients for $\delta^\mathrm{MO,\varepsilon}$ and $\delta^\mathrm{MEF}$ are 0.99 and 0.77, respectively.}

\section{Analyses and discussions for European countries}\label{sec:results}
In the following analysis, we apply the previously developed methods (PP and PWL) presented in \Cref{sec:methods} to data described in \Cref{sec:data}, see also \Cref{fig:scheme_CEF_calculation}.
More specifically, we apply the PP method to the German power plant list described in \Cref{sec:data-OPSD-PPlist} and the PWL method to data of 20 European countries described in \Cref{sec:data-installed-generation-capacity}.
The remaining European countries were removed from the analysis due to poor data quality or insufficient electricity demand (small countries).
The rejection criteria are detailed in Supplementary Material F.

\begin{figure*}[htb]
	\centering
	\includegraphics[width=1.0\linewidth]{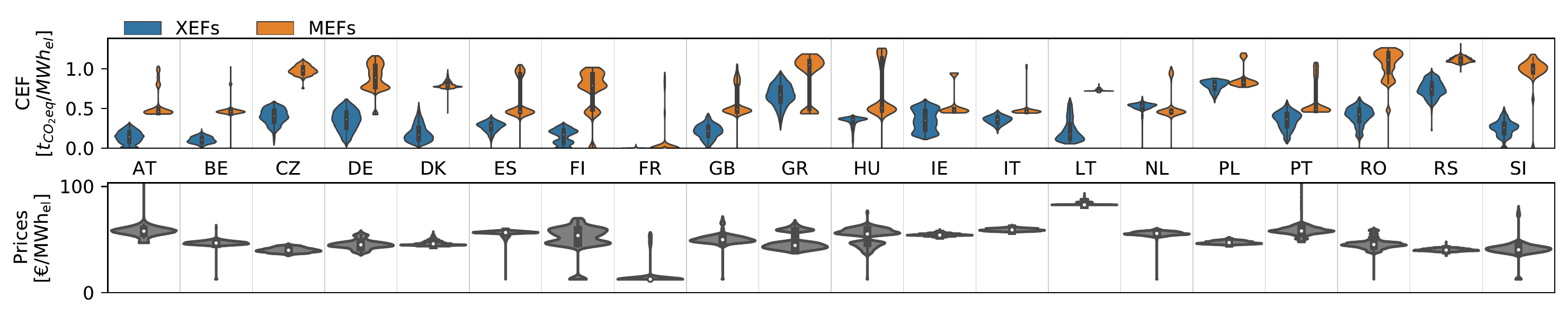}
	\vspace{-0.8cm}
	\caption{Distribution of resulting CEFs and marginal prices for 2019. Note that 1.2\% of AT's values (104--190 \euro/MWh\textsubscript{el}) were trimmed.}
	\label{fig:cef_and_prices_violins_2019}
\end{figure*}

\subsection{Merit orders and CEFs for European countries}\label{sec:merit-orders-and-cefs-for-european-countries}
\begin{figure}[htb]
	\centering
	\includegraphics[width=1.0\linewidth]{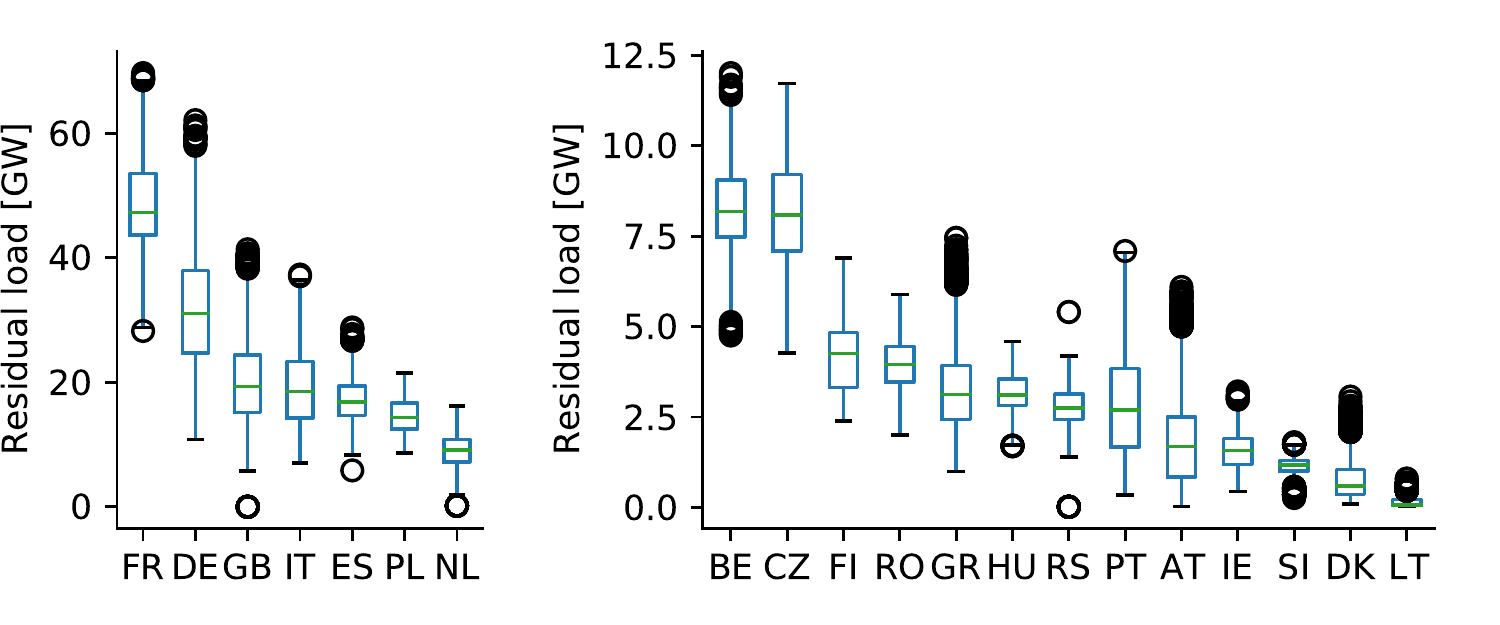}
	\vspace{-0.8cm}
	\caption{Distribution of residual loads for countries under study in descending order of median. Plots were splitted to enhance readability.}
	\label{fig:all_resi_box}
\end{figure}

Of the analyzed countries, \Cref{fig:all_resi_box} shows the distribution of the residual load (the load that has to be provided by conventional power plants) described in \Cref{sec:residual-load}.
The median value for France (FR), 47.31\,GW, is 739 times greater than that of Lithuania (LT), 0.06\,GW.
The respective maximum values are 69.67\,GW and 0.80\,GW.
Detailed depictions of the resulting merit orders for the years 2017--2019 for the 20 analyzed countries are contained in Supplementary Material G.

In \Cref{fig:cef_and_prices_violins_2019}, the distributions of XEFs, MEFs, and marginal prices resulting from the simulations are depicted.
One can see that the MEFs tend to be higher than the XEFs.
Only in the Netherlands (NL) is the median of XEFs higher than the median of MEFs.
The reasons are a \ac{RES} share of only 7\%, a coal baseload mostly, and efficient gas\_cc marginal power plants.
However, the average values of MEFs are higher than those of XEFs for all countries.
This means that the marginal power plant, on average, emits more emissions per unit of electricity than the national generation mix.
From \Cref{fig:cef_and_prices_violins_2019}, we can divide the 20 countries into three groups.
The first group contains Poland (PL) and France (FR), where fluctuations of the CEFs are low compared to the other countries and the medians are almost identical.
The reason for this could be that the main marginal power plant's fuel type is also the main energy source, e.g., coal for PL, nuclear for FR.
The second group includes Austria (AT) and Czech Republic (CZ), where the value sets of XEFs and MEFs are disjointed due to dominant low carbon energy sources, such as hydro in Austria with 26\% of generation, or nuclear and solar in Czechia (CZ) with 10.4\% and 9.8\% generation, respectively.
All other countries are in the third group, where the value sets of XEFs and MEFs intersect, but the medians differ substantially.
An explanation could be that the marginal power plant's fuel type is only one of many energy sources, including \ac{RES}.

\begin{figure}[htb]
	\centering
	\includegraphics[width=1.0\linewidth]{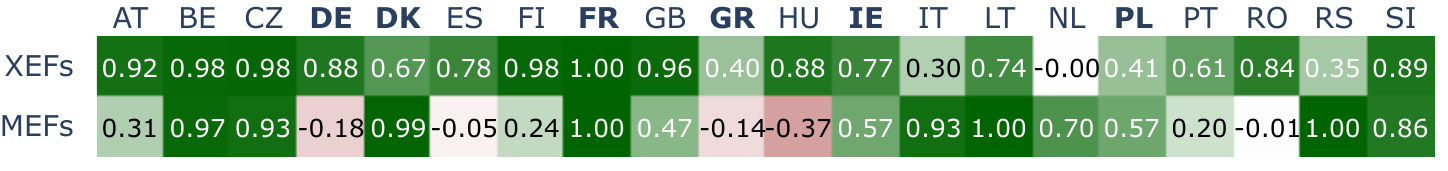}
	\vspace{-0.5cm}
	\caption{Spearman correlation coefficients $r$ between \acp{CEF} (MEFs, XEFs) and marginal prices with the PWL method for the year 2019.}
	\label{fig:bas_corr_c_vs_CEF_2019}
\end{figure}

\subsection{Correlation analysis of emissions and prices}\label{sec:correlation-analysis-of-emissions-and-prices}
For the correlation analysis of the \acp{CEF}, we follow \cite{Stoll.2014} in using the Spearman rank correlation coefficient $r$ as the relationship between \acp{CEF} and marginal costs is not expected to be linear~\cite{Stoll.2014}.
This coefficient $r$ quantifies how well the relationship between two variables can be described using a monotonic function.

In the Supplementary Data H, the correlation between electricity prices (simulated and historic) and the calculated \acp{CEF} (XEFs\textsuperscript{PP}, XEFs\textsuperscript{PWLv}, MEFs\textsuperscript{PP}, MEFs\textsuperscript{PWLv}) for Germany for the years 2017--2019 are presented in scatter plots.
It can be seen that the \acp{XEF} correlate positively with the simulated and the historic prices for all three years 2017--2019 and for the methods PP and PWLv ($r$-values range between 0.62 and 0.88).
In contrast, the \acp{MEF} have negative $r$-values (between -0.16 and -0.42) for all mentioned combinations, which is a first indicator of the phenomenon where \ac{PBDR} leads to a carbon emission increase.

\Cref{fig:bas_corr_c_vs_CEF_2019} shows $r$-values for all analyzed countries using the PWL method.
The correlation values are positive for 16 and negative for 4 of the 20 countries.

\subsection{Price-based load shift analysis}\label{sec:load-shift-analysis}
For environmental evaluation of \ac{PBDR}, only the source and sink hours are relevant, i.e., the hours where energy is shifted from and to.
The analysis of the emissions without the consideration of the electricity price, which is the driving signal for \ac{PBDR}, might be misleading.
Thus, we quantify and compare the incentives and effects of \ac{PBDR} for the years 2017--2019 and the 20 analyzed countries with a simulated load shift: Every day, an hourly load of 1\,kWh is shifted from the most expensive to the cheapest hour of that day.
Annual simulations were conducted for the years 2017--2019 for the 20 analyzed countries.
The electricity prices relate to the marginal costs of the previous simulation.

Since, \acp{MEF} quantify the marginal system effects, the \ac{ME} changes are the total emission changes, i.e., if \acp{ME} increase by 1\,t, total emissions increase by 1\,t, too.
In the remainder of the paper, we refer to the total emission changes as \ac{ME} changes only to distinguish them from \ac{XE} changes, which are the change in emissions calculated from the \ac{XEF}.

\begin{figure}[htb]
	\centering
	\includegraphics[width=1.0\linewidth]{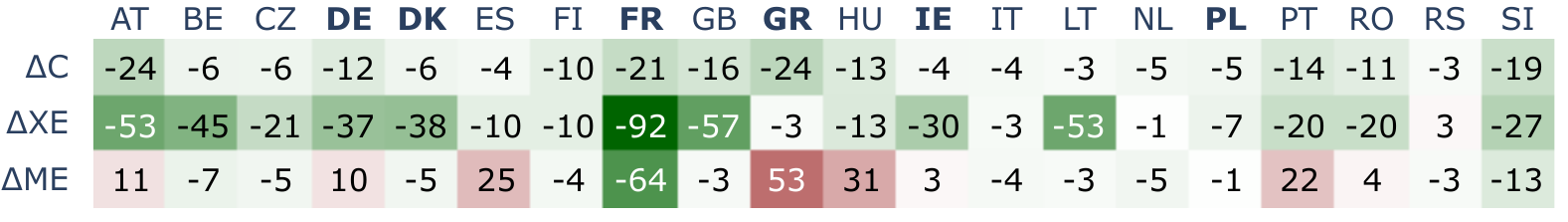}
	\vspace{-0.5cm}
	\caption{Relative annual changes of cost and carbon emissions resulting from load shift simulation for 2019 in \%. The grid mix emissions are based on $\mathrm{XEF}_t^\mathrm{PWL}$, the marginal carbon emissions are based on $\mathrm{MEF}_t^\mathrm{PWL}$.}
	\label{fig:load_shift_effects_heatmap}
\end{figure}

In \Cref{fig:load_shift_effects_heatmap}, the relative changes of cost and carbon emissions of the shifted energy due to load shifts are shown.
Since we consider \ac{PBDR}, the cost reductions are the incentives and the changes in carbon emissions are the effects.
It can be seen that, in contrast to the costs, which decreased for all countries as expected, the carbon emissions increased for some countries.
\acp{ME} increased for the eight countries Austria (AT), Germany (DE), Spain (ES), Greece (GR), Hungary (HU), Ireland (IE), Portugal (PT), and Romania (RO).

\begin{figure}[htb]
	\centering
	\includegraphics[width=0.49\linewidth]{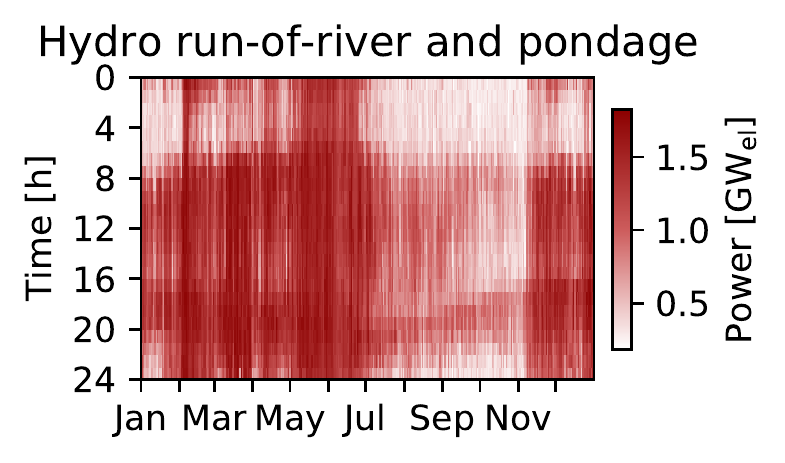}
	\includegraphics[width=0.46\linewidth]{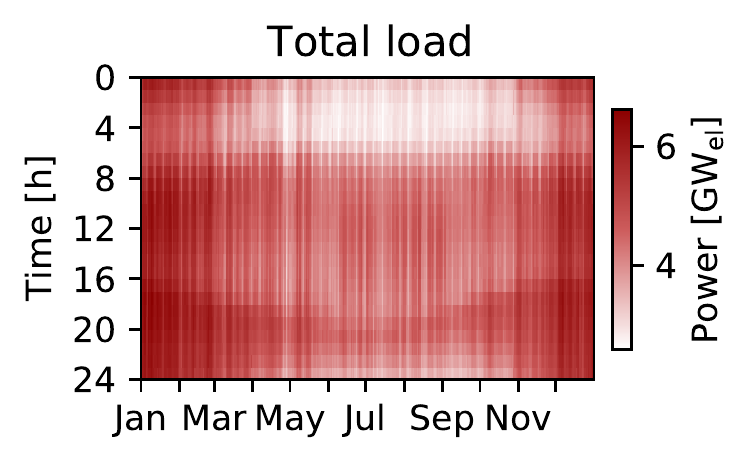}
	\vspace{-0.3cm}
	\caption{Historic data for the year 2019 for Serbia: Hydro run-of-river and pondage (left) and total load (right). Data source:\cite{ENTSOE.2020}.}
	\label{fig:RS_hydro}
\end{figure}

\acp{XE} are usually reduced since \ac{PBDR} leads to load shifts from high-XEF-hours to low-XEF-hours.
The only exception of the 20 countries is Serbia (RS), where the load shifts increased the total \acp{XE} by 3\%.
The reason might stem from the fact that Serbia's power supply system is dominated by hydro (run-of-river and pondage) and lignite.
In the simulation, the marginal power plant for Serbia is always a lignite power plant (see also \Cref{fig:load_shift_fuel_types} and the merit order in Supplementary Material G).
With the aid of the pondage, hydro power plants have storage capacities enabling them to shift electricity generation from low-load-hours (0:00--6:00) to peak load hours (8:00, 19:00), see \Cref{fig:RS_hydro}.
While this also happens in other countries, in Serbia the pondage effect has more weight as there are no significant shares of \ac{vRES} such as wind or solar.
A similar case, where \ac{PBDR} led to an increase in \acp{XE} is reported by \cite{Clau.2019b} for Norway in the year 2015.

\Cref{fig:spreads_time_violin} shows the source and sink time of the load shifts.
These are the times with the highest and lowest prices of the day.
For most countries, the distribution of source time has two humps: One for 5:00--10:00 and another for 16:00--20:00.
This result was expected by the authors since it reflects the double-hump pattern of historic price curves.
The load shift sink is mainly at night (21:00--07:00).
Only in Germany and Italy does the sink occasionally occur around midday (11:00--15:00).
In France, source and sink hours are both just after midnight.
This is because France's sole conventional energy source is usually nuclear which is evaluated with a constant efficiency by the data source~\cite{OpenPowerSystemData.2018}.
Therefore, the daily price spreads were zero on most days, and load shifts were only carried out in cold winter months where coal and gas\_cc power plants additionally stepped in to cover the increased load.

\Cref{fig:load_shift_fuel_types} gives insights on the marginal fuel type combination of the conducted load shifts.
For the small country of Lithuania (LT), it can be seen that the national energy supply with gas as the only energy source and a low ratio between load and average power plant size provides only little incentive for load shifting.

\begin{figure}[htb]
	\centering
	\includegraphics[width=1.0\linewidth]{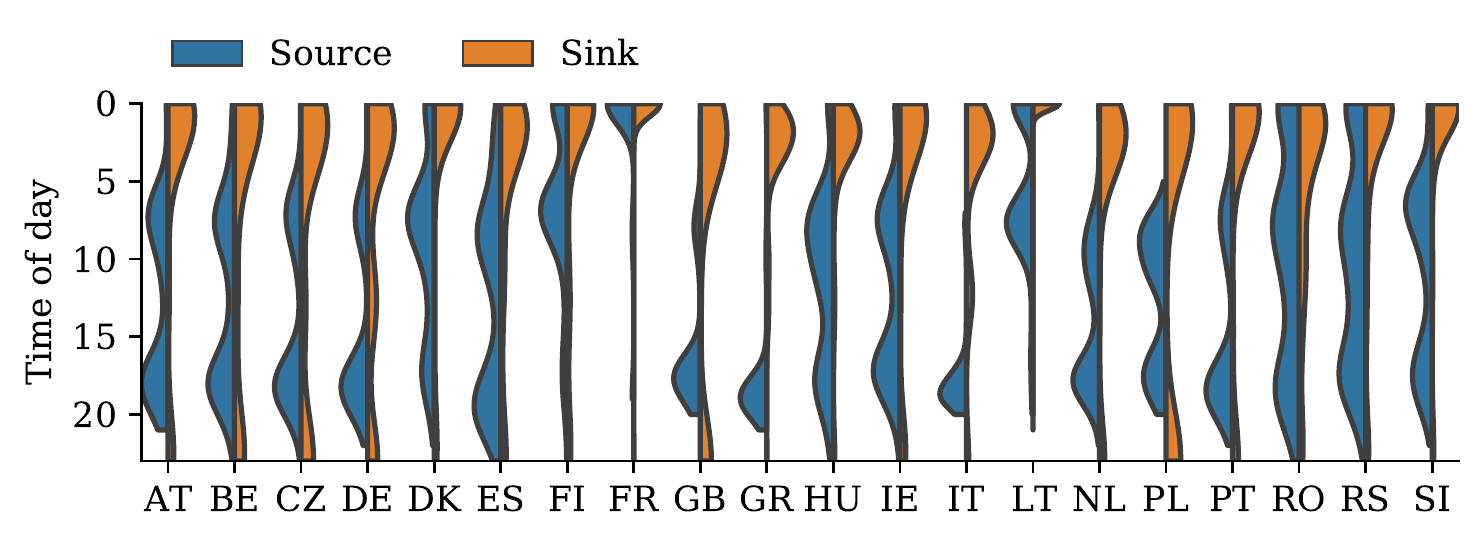}
	\vspace{-0.8cm}
	\caption{Distribution of source time and sink time of simulated load shifts for 2019.}
	\label{fig:spreads_time_violin}
	\vspace{0.3cm}
	\includegraphics[width=1.\linewidth]{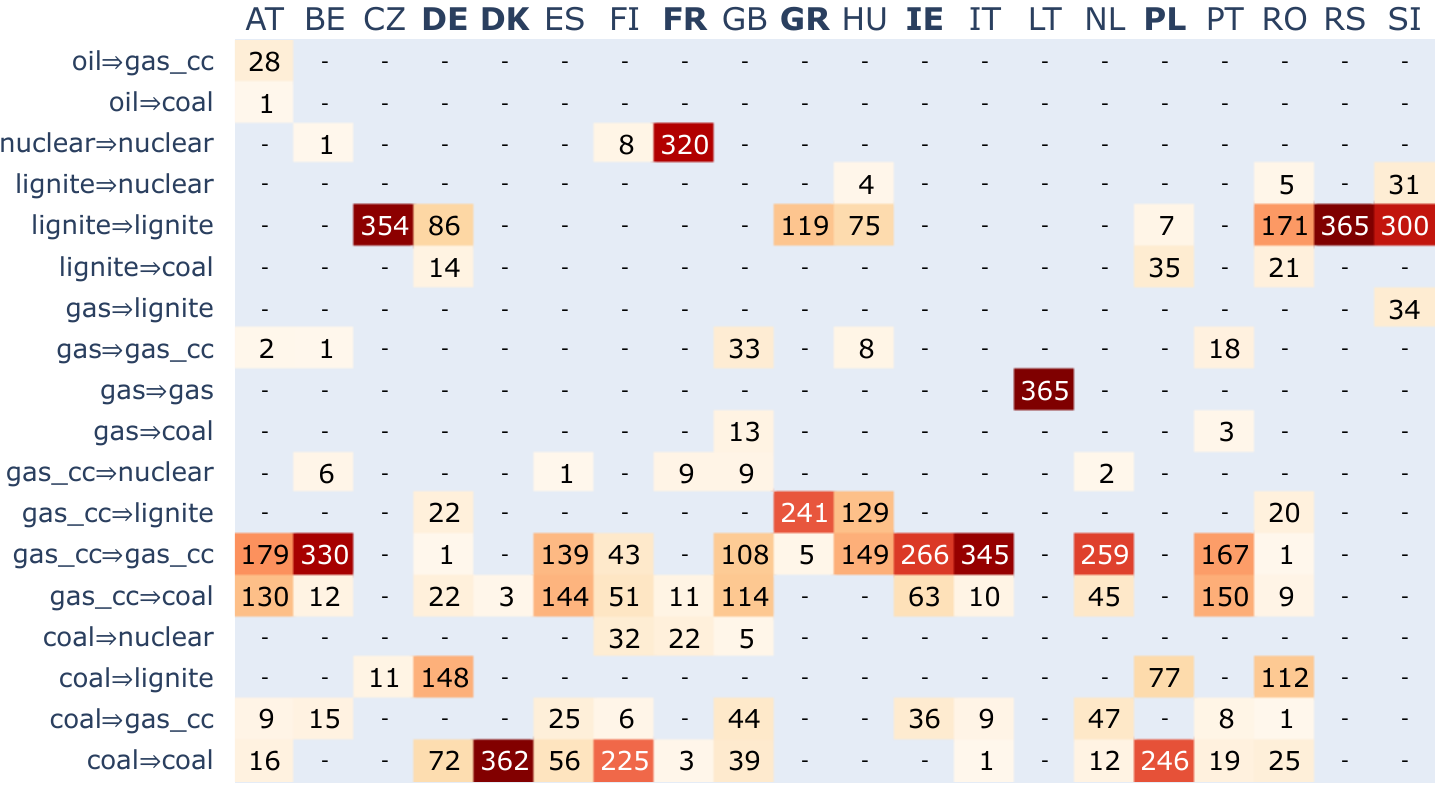}
	\vspace{-0.5cm}
	\caption{Frequency of load shift events per country and marginal fuel type combination (denoted in the form $f^\mathrm{source} \Rightarrow f^\mathrm{sink}$) for the year 2019.}
	\label{fig:load_shift_fuel_types}
\end{figure}

\subsection{CEF-based load shift analysis}\label{sec:cef-based-load-shift-analysis}
\begin{figure}[htb]
	\centering
	\includegraphics[width=1.0\linewidth,right]{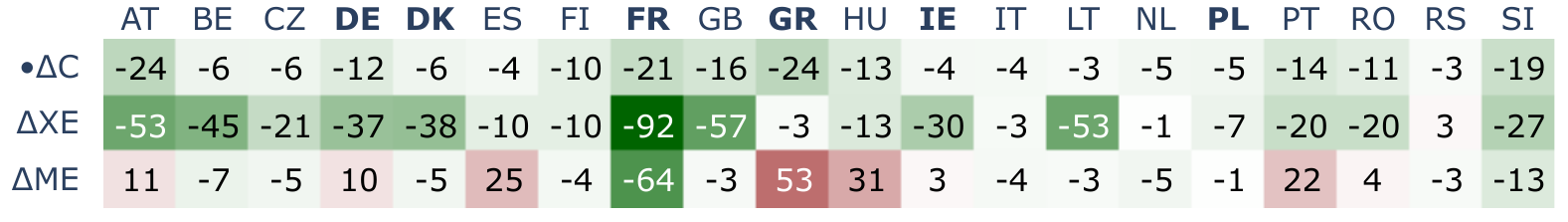}
	\includegraphics[width=1.0\linewidth,right,trim=0 0 0 0.4cm,clip]{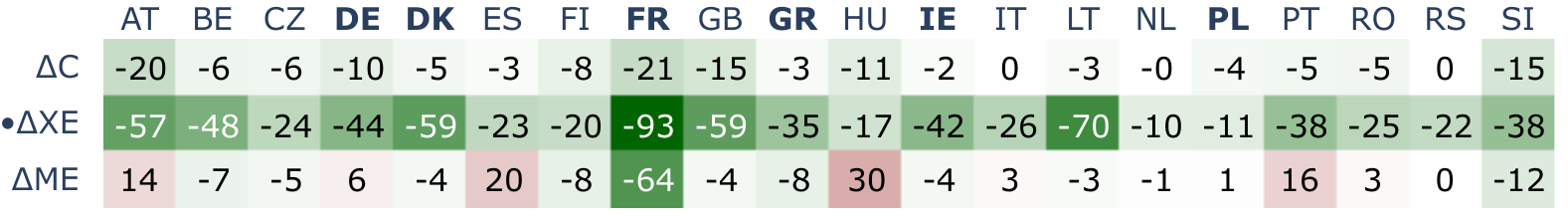}
	\includegraphics[width=1.0\linewidth,right,trim=0 0 0 0.4cm,clip]{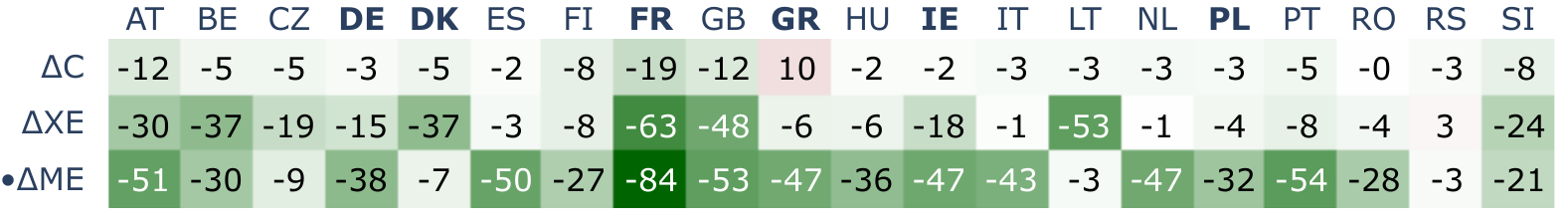}
	\vspace{-0.5cm}
	\caption{Relative annual changes of Cost (C), Grid Mix Emissions (XE), and Marginal Emissions (ME) resulting from load shift simulations for 2019 in \% with ``$\bullet$'' indicating the optimized criteria. The \acp{XE} are based on $\mathrm{XEF}_t^\mathrm{PWL}$, the \acp{ME} are based on $\mathrm{MEF}_t^\mathrm{PWL}$.}
	\label{fig:load_shift_effects_heatmap_XEFs_MEFs}
\end{figure}

For \ac{PBDR}, the time-dependent electricity price is the incentive signal that determines the source and sink hours.
However, conducting \ac{DR} based on \acp{XEF} or \acp{MEF} could be an option for consumers who want to minimize their carbon emissions.
In this analysis, we rerun the previous load shift simulation with \acp{XEF} or \acp{MEF} as incentive signals, so 1 kWh is shifted every day from the hour with the highest \ac{XEF} or \ac{MEF} of that day to the lowest, respectively.
In \Cref{fig:load_shift_effects_heatmap_XEFs_MEFs}, the relative changes of cost and carbon emissions of the shifted energy due to load shifts are shown; note that the results from \Cref{sec:load-shift-analysis} are repeated for better comparison.
The resulting effects of the XEF-based load shifts are similar to that of the cost-based load shifts: Both lead to increased \acp{ME} in 8 countries.
In contrast, the \ac{MEF}-based load shifts lead to \ac{ME} savings between 3 and 84\% with an average of 35\%, albeit reducing the average cost-saving potential by 56\% compared to cost-based load shift.

\subsection{Detailed discussion of six sample countries}\label{sec:six-example-countries}
In the following, the reasons and conditions under which \ac{PBDR} lead to increased carbon emissions will be discussed using detailed results of the load shift analyses of six exemplary countries for the year 2019, shown in \Cref{fig:mols2019}.
For corresponding analyses of the remaining 14 countries, see Supplementary Material J.
The six countries were selected because first, they are representative in terms of country size and the range of relative annual changes from \Cref{fig:load_shift_effects_heatmap}, including its maximum (GR) and minimum (FR) of \ac{ME} changes, and second, they have different interesting attributes, e.g., the dominance of nuclear (FR), high wind share (DK), the dominance of lignite/coal (PL), or being a small country very reliant on wind and gas\_cc (IE).
For each country, \Cref{fig:mols2019} shows 
a) the merit order, 
b) the histogram of $P_t^\mathrm{resi}$ as the range of possible load shifts, 
c) the load shifts with its sources and sinks, and 
d) their inverted load duration curves.
For according plots for all 20 countries for the years 2017--2019, see Supplementary Material I.
In \Cref{fig:mols2019}, the three countries in the first row (DE, GR, IE) have increased \acp{ME} in \Cref{fig:load_shift_effects_heatmap}, the countries in the second row (DK, FR, PL) do not.
(To guide the reader's eye, the six sample countries have been marked with bold letters for several previous diagrams.)

\begin{figure*}[!h]
	\centering
	\includegraphics[width=0.33\linewidth,trim=0 0.7cm 0 0,clip]{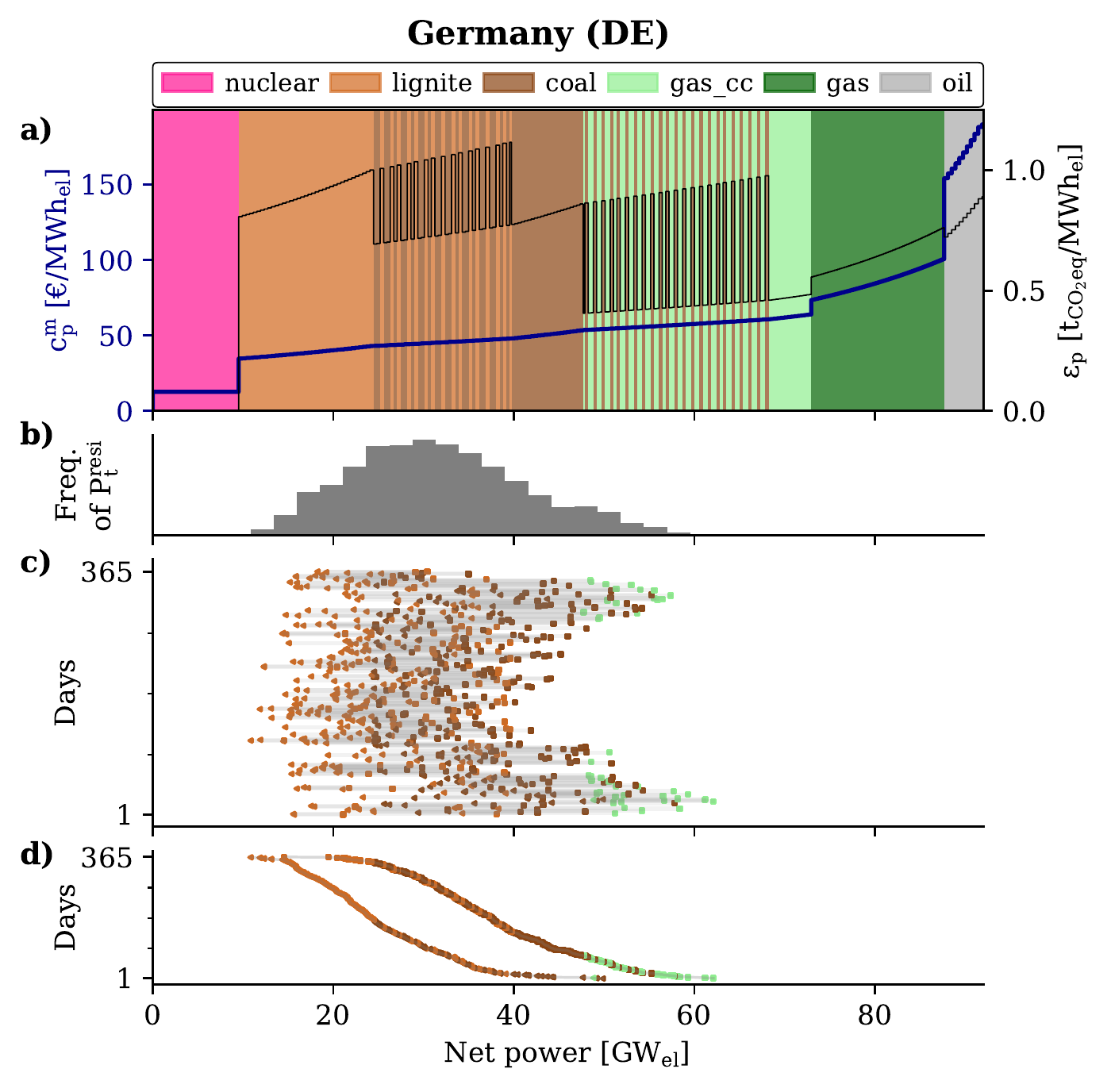}
	\includegraphics[width=0.33\linewidth,trim=0 0.7cm 0 0,clip]{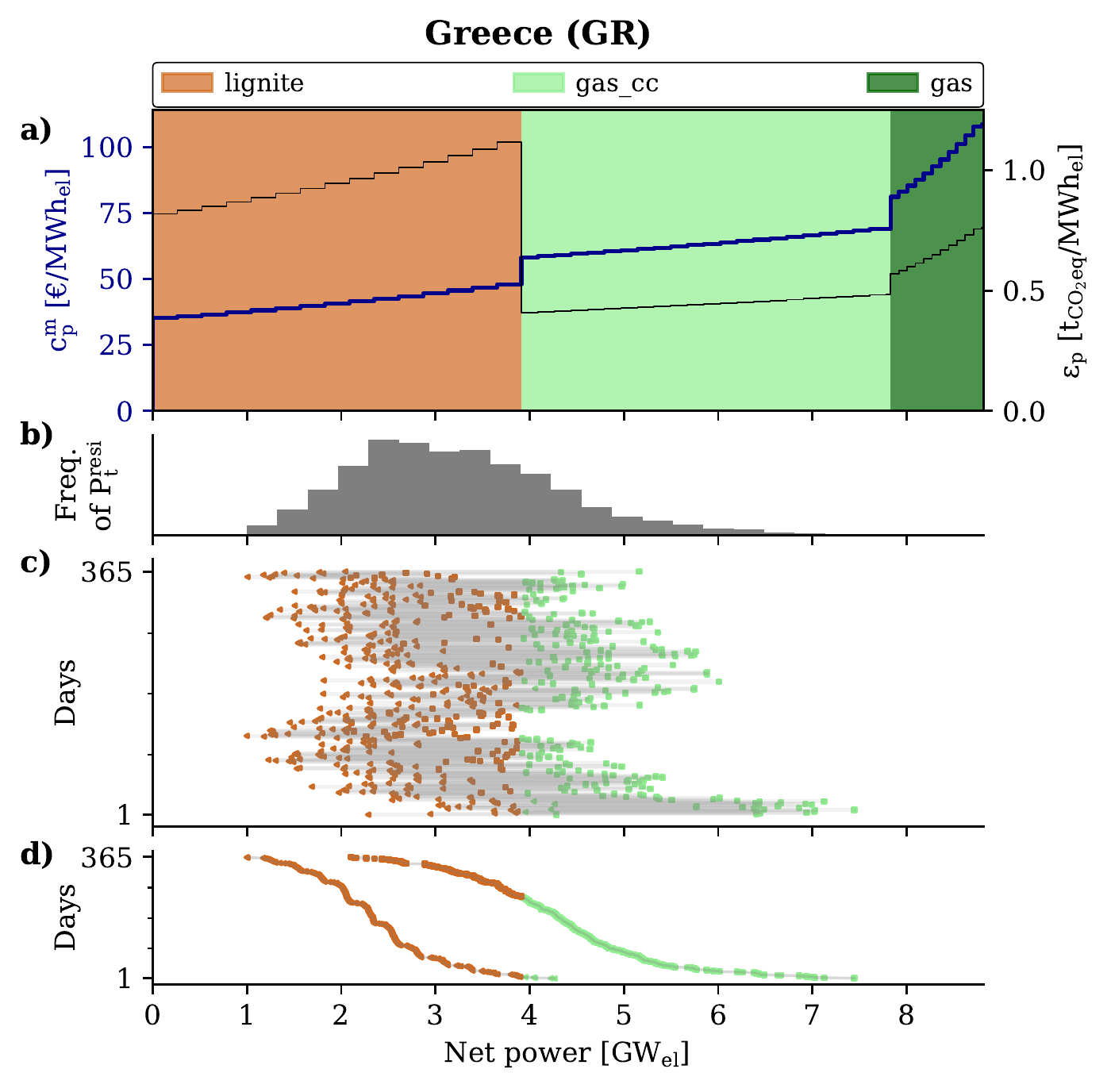}
	\includegraphics[width=0.33\linewidth,trim=0 0.7cm 0 0,clip]{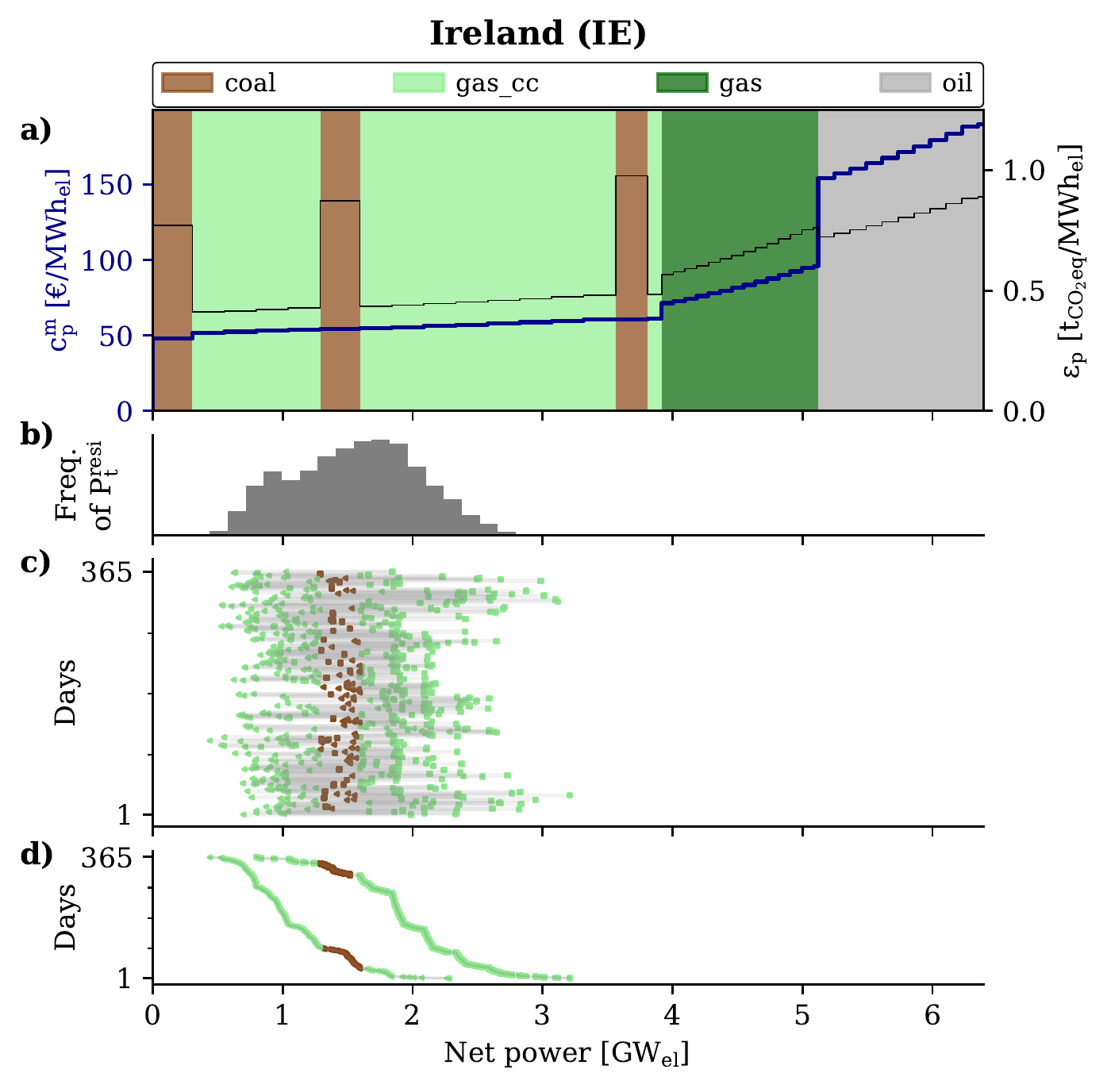}
	\includegraphics[width=0.33\linewidth]{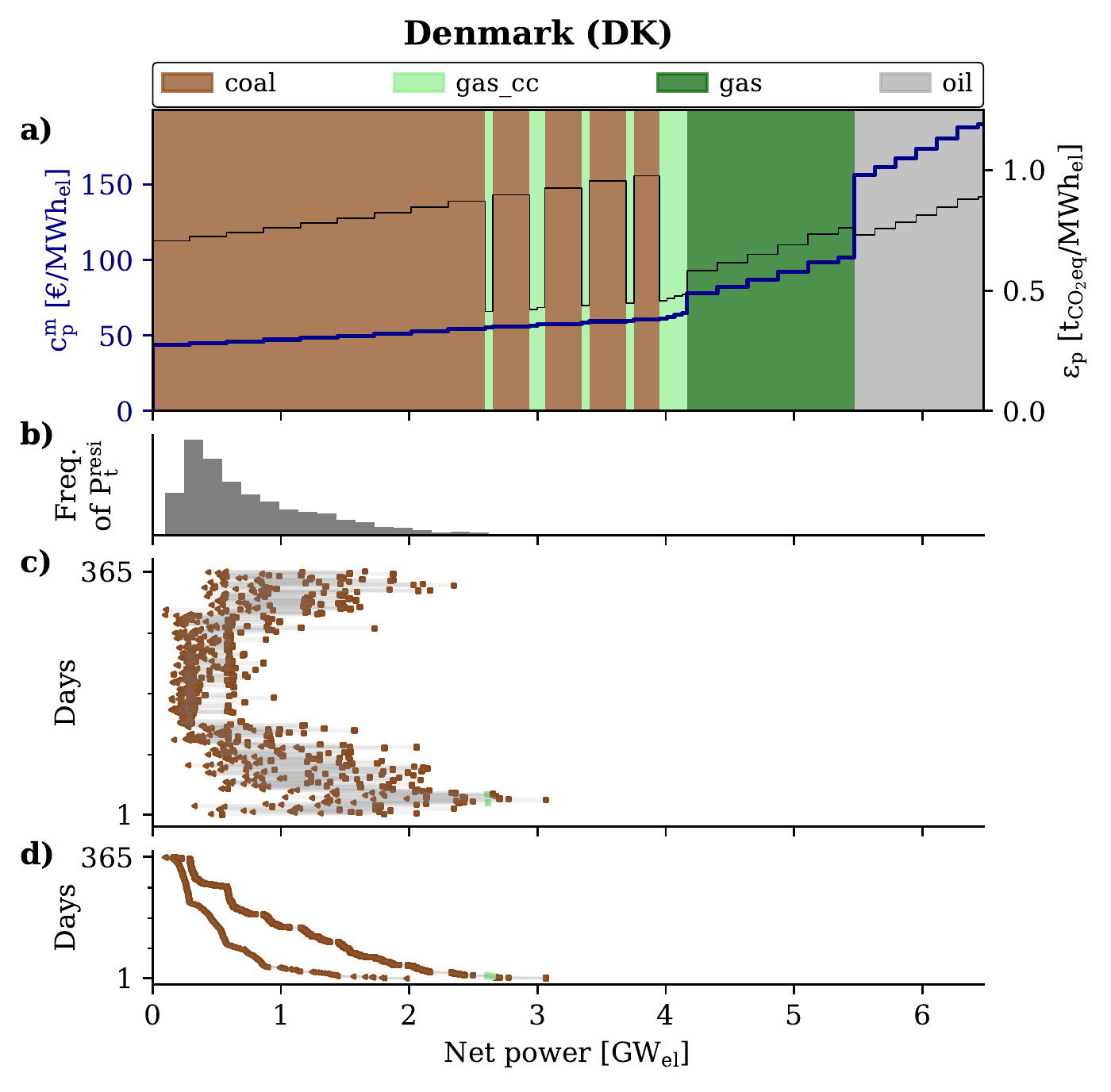}
	\includegraphics[width=0.33\linewidth]{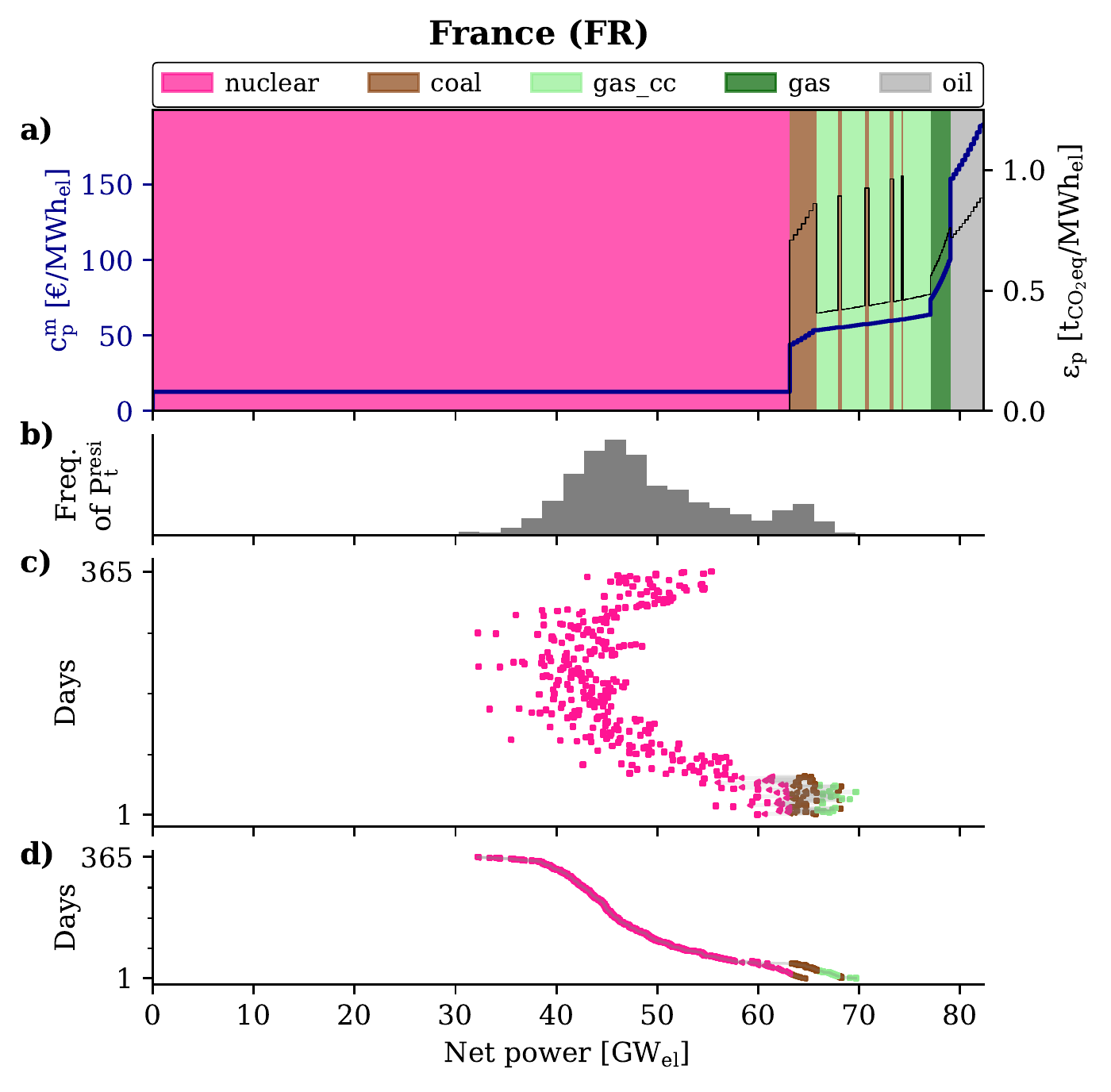}
	\includegraphics[width=0.33\linewidth]{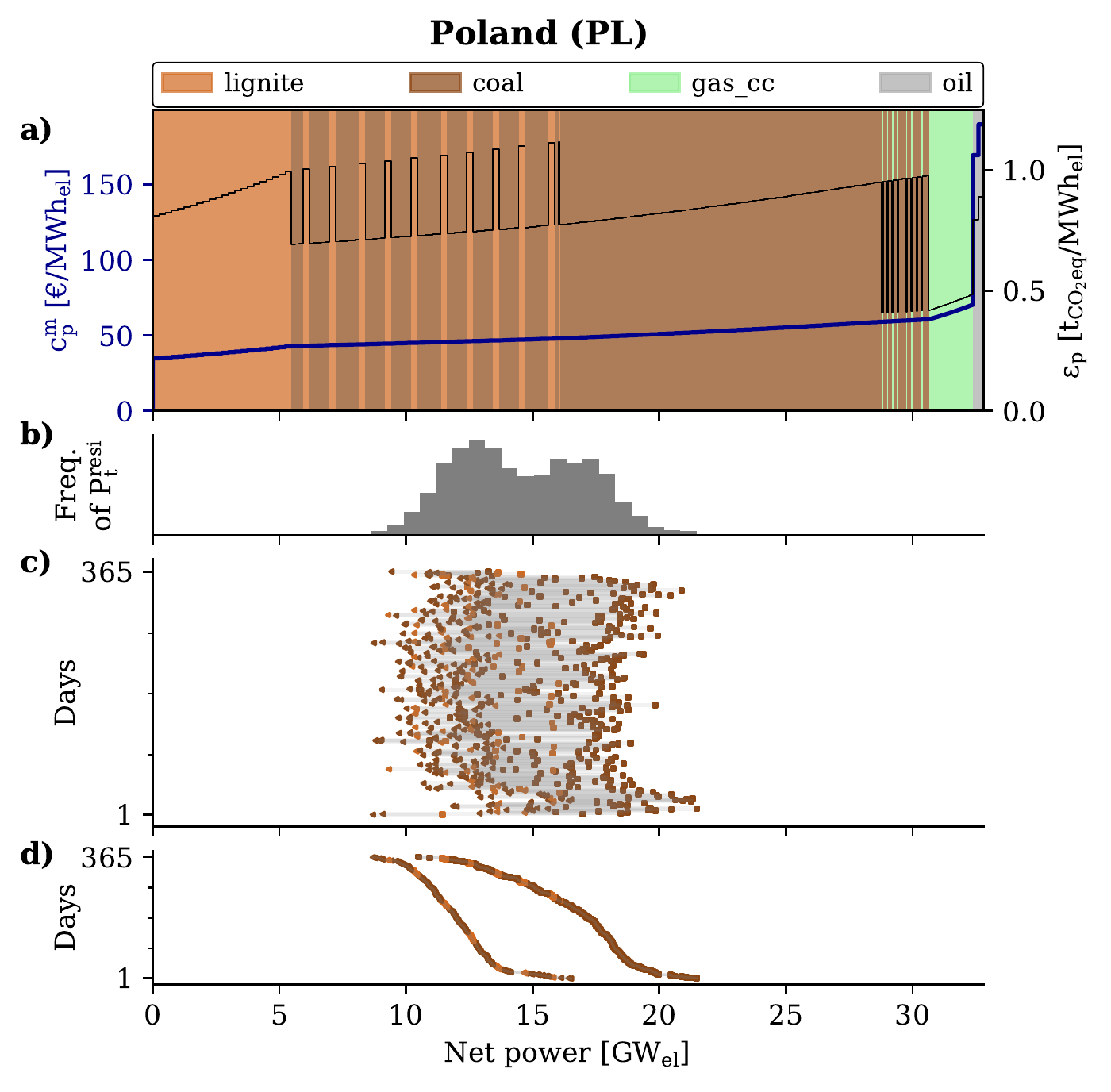}
	\vspace{-0.3cm}
	\caption{Load shift analyses of 6 European countries for the year 2019 based on the PWL method: \textbf{a)} Merit order with marginal costs $c_p^m$ (left) and emissions intensities $\varepsilon_p$ (right), \textbf{b)} histogram of residual load $P_t^\mathrm{resi}$, \textbf{c)} load shifts of all days of the year (squares indicating sources and triangles indicating sinks), and \textbf{d)} inverted load duration curves of sources (right) and sinks (left)). The x-axis (net generation power in GW\textsubscript{el}) is shared across the four subplots.}
	\label{fig:mols2019}
\end{figure*}

\begin{itemize}
\item{\textbf{Germany (DE):}}
Even though Germany has a comparatively diversified energy supply, the main marginal fuel types are lignite, coal, and gas\_cc, which are also the main causers of the merit order dilemma of emissions.
The results show 159 load shifts in which the marginal fuel type does not change.
In these cases, the load is generated by a generation unit with the same fuel type but with a higher efficiency, which leads to reductions in cost and \acp{ME}.
With 192 cases, more than half of the German load shifts display a reduction in \acp{ME} (148 coal-to-lignite, 22 gas\_cc-to-coal, and 22 gas\_cc-to-lignite). 
Only 14 energy units were shifted towards a situation with a greener marginal fuel type (lignite-to-coal).

\item{\textbf{Greece (GR):}}
In Greece, the increase of \acp{ME} due to load shifting is particularly strong with 53\%.
The reason is that the residual load oscillates most days around the capacity limit between lignite and gas\_cc at around 4~GW leading to 241 gas\_cc-to-lignite load shifts, which are the most disadvantageous occurring load shifts regarding the unwanted effect of increasing \acp{ME}.
Due to Greece's still moderately developed expansion of renewable energies with 30\% \ac{RES} share, the potential for reducing \acp{XE} is only 3\%.

\item{\textbf{Ireland (IE):}}
With a wind share of 40\%, Irelands RES share of 43\% is similar to that of Germany even with hardly any photovoltaic power.
This drives the \ac{XE} saving to up to 30\%.
However, \ac{ME} changes are slightly positive with 3\% due to 63 gas\_cc-to-coal load shifts.
Only the 36 coal-to-gas\_cc-load shifts imply changing to a greener marginal fuel type while the vast majority of load shifts (266) stay within the dominant fuel type gas\_cc.

\item{\textbf{Denmark (DK):}}
In Denmark, the load shifts lead to 40\% \ac{XE} reductions.
This is mainly caused by the high RES share of 76\% of which wind\_onshore contributes almost half.
The national residual load $P_t^\mathrm{resi}$ is subject to strong seasonal fluctuations.
Causing factors could be winterly heating power demand and reduced imports of German excess solar power.
The \acp{ME} are moderately reduced by 6\%.
More than 99\% of the load shifts are within the fuel type of coal, thus, yield only emission reductions through higher power plant efficiencies.

\item{\textbf{France (FR):}}
In France, 71\% of electricity is produced by nuclear power plants making the country's national power supply system heavily conventional-based, yet low in carbon emissions.
The national power supply gives only little economic incentive for load shifting since the residual load is predominantly in the range of marginal nuclear power.
Only in the cold winter months, when the residual load exceeds the nuclear power capacity limits, incentives for intraday load shifting are created.
There are 31 load shifts into hours with nuclear as marginal fuel type: 9 shifted from gas\_cc and 22 from coal.
These few cases do not lead to particularly high CO\textsubscript{2} savings in absolute terms.
In relative terms, however, due to the low emission baseline level of France's power system, they contribute to the highest savings of all analyzed countries: 92\% \ac{XE} and 64\% \ac{ME} savings.

\item{\textbf{Poland (PL):}}
Poland has a low RES share.
Predominant fuel types are coal (52\%) and lignite (26\%).
In the merit order, they are intertwined due to similar marginal cost levels -- in contrast to Greece where the fuel types lignite and gas\_cc form continuous blocks in the merit order.
This leads to all combinations of load shifts between the two fuel types: 35 lignite-to-coal-shifts, 77 coal-to-lignite-shifts, 7 lignite-to-lignite-shifts, and 246 coal-to-coal-shifts.
Together these result in 7\% \ac{ME} savings mainly stemming from power plant efficiency gains of the coal-to-coal load shifts.

\end{itemize}

\begin{figure*}[!h]
	\centering
	\includegraphics[width=1.0\linewidth]{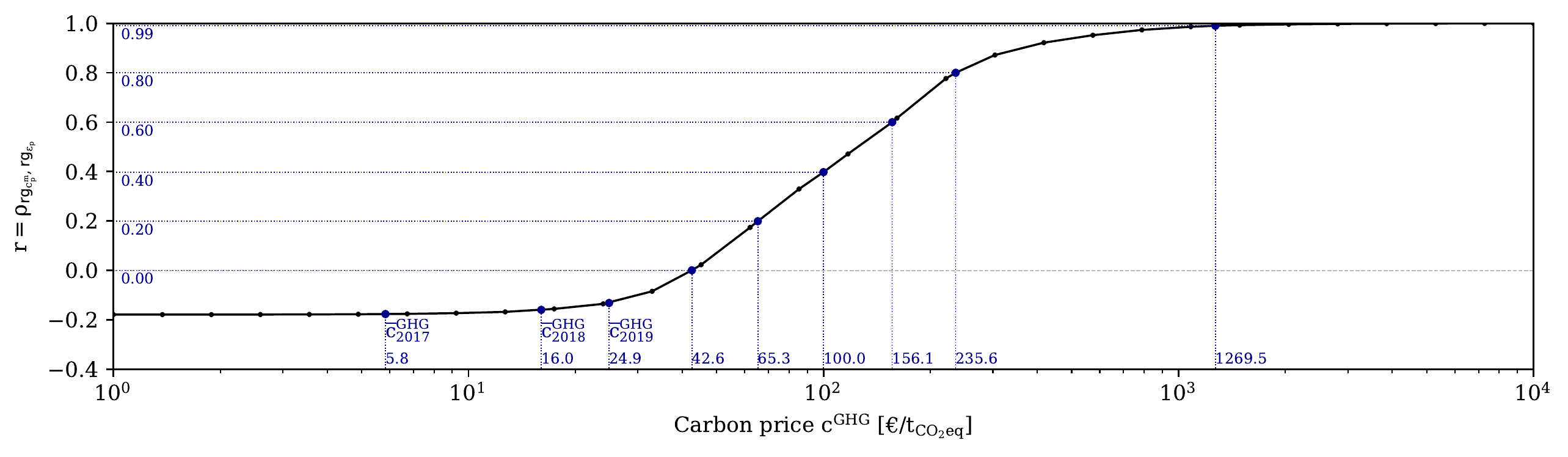}
	\vspace{-0.8cm}
	\caption{Sensitivity analysis of the carbon price $c^\mathrm{GHG}$ and the Spearman correlation coefficient $r$ between marginal costs $c_p^\mathrm{m}$ and marginal emissions $\varepsilon_p^\mathrm{m}$ with German power plant data in 2019.
		Indicated in blue color, the average carbon prices $\overline{c}^\mathrm{GHG}_*$ for the years 2017--2019 and additional interesting carbon prices and r-values are added.}
	\label{fig:spearman_merit_order_cGHG_sa}
\end{figure*}

\begin{figure*}[htb]
	\centering 
	\includegraphics[width=0.9\linewidth]{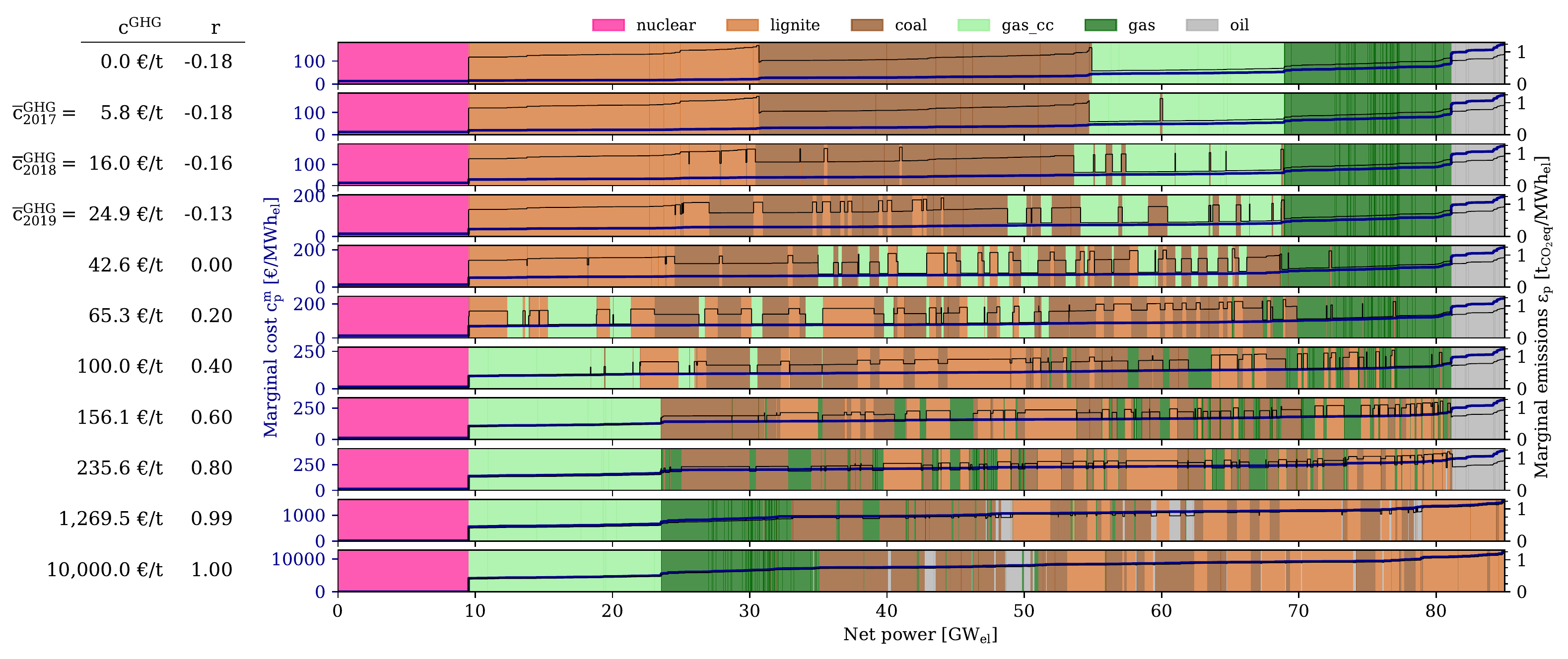} %
	\vspace{-0.4cm}
	\caption{How the German merit order in 2019 and its Spearman correlation coefficient $r$ between marginal costs $c_p^\mathrm{m}$ and marginal emissions $\varepsilon_p^\mathrm{m}$ would change with carbon price $c^\mathrm{GHG}$.}
	\label{fig:merit_order_cGHG_sa}
\end{figure*}

\subsection{Impact of carbon price}\label{sec:impact-of-carbon-price}
A promising measure to solve the merit order emission dilemma is to set a price for carbon emissions in the form of a carbon price or carbon tax to an appropriate level.
This increases the correlation between marginal costs and carbon intensities in the merit order.

\subsubsection{Impact on marginal costs--emissions correlation }
To demonstrate this, a sensitivity analysis was carried out in which the merit order of the German power plants for the year 2019 was determined for different hypothetical carbon prices.
To quantify the effect on the merit order, the Spearman correlation coefficient $r$ between the marginal prices $c_p^m$ and the power plant-specific carbon intensities $\varepsilon_p$ along the merit order was calculated for each scenario.
The quantitative results can be seen in \Cref{fig:spearman_merit_order_cGHG_sa}.
It shows negative $r$ values for carbon prices below 42.58\,\euro/t.
For comparison, the average carbon prices for the years 2017--2019, $\overline{c}^\mathrm{GHG}_{2017}$, $\overline{c}^\mathrm{GHG}_{2018}$, and $\overline{c}^\mathrm{GHG}_{2019}$, were 5.8\,\euro/t, 16.0\,\euro/t and 24.9\,\euro/t, respectively.
$r=0$ means that \acp{CEF} and marginal costs are decoupled.
$r=0.2$ was calculated for $c^\mathrm{GHG}=65.3$\,\euro/t, $r=0.4$ for $c^\mathrm{GHG}=100.0$\,\euro/t, $r=0.6$ for $c^\mathrm{GHG}=156.1$\,\euro/t, and  $r=0.8$ for $c^\mathrm{GHG}=235.6$\,\euro/t.
In comparison, the German Federal Environment Agency suggests the usage of climate damage costs of 180\,\euro$_{2016}$/t for the year 2016, 205\,\euro$_{2016}$/t for 2030, and 240\,\euro$_{2016}$/t for 2050~\cite{GermanEnvironmentAgency.2019}.
For $r \gtrapprox 0.8$ or $c^\mathrm{GHG} \gtrapprox 235.6$\,\euro/t, the marginal gains of $r$ decrease.
In order to reach $r=0.99$, a carbon price of 1269.5\,\euro/t is needed.

\subsubsection{Impact on the merit order}
\Cref{fig:merit_order_cGHG_sa} shows the according effects on the merit order.
One can see that with increasing carbon prices, the low-emission gas\_cc power plants gain a comparative economic advantage over lignite and coal and shift to the left side of the merit order.
With $c^\mathrm{GHG}=100$\,\euro/t, almost all gas\_cc power plants are directly behind nuclear.
The same happens to the gas power plants, coal power plants, and for $c^\mathrm{GHG}$ somewhere above 236\,\euro/t even to oil power plants.
These values mostly align with the simulation results in~\cite{Boing.2019b}.
For $c^\mathrm{GHG}=10\,000$\,\euro/t, where $r$ reaches the value 1.00, the carbon emission intensity $\varepsilon_p$ (black line) is, except for a few small power plants, monotonically increasing and fully correlated with the marginal price $c_\mathrm{p}^{m}$ (blue line).

\begin{figure}[htb]
	\centering
	\includegraphics[width=1.0\linewidth]{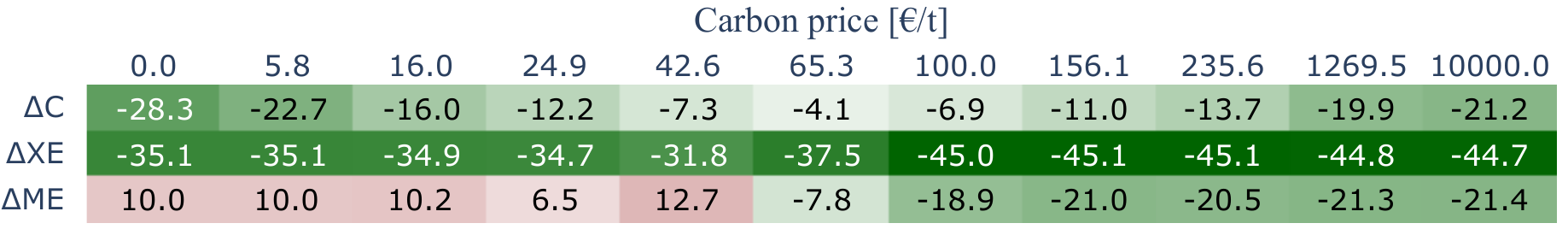}
	\includegraphics[width=1.0\linewidth]{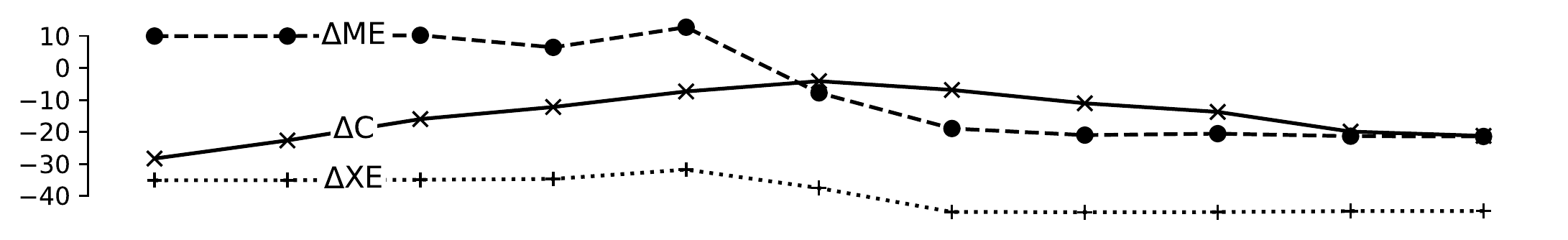}
	\vspace{-0.5cm}
	\caption{Relative annual changes of Cost (C), Grid Mix Emissions (XE), and Marginal Emissions (ME) resulting from load shift simulation for 2019 in \% for different carbon prices. The grid mix emissions are based on $\mathrm{XEF}_t^\mathrm{PP}$, the marginal carbon emissions are based on $\mathrm{MEF}_t^\mathrm{PP}$.}
	\label{fig:load_shift_effects_heatmap_tax} 
\end{figure}

\subsubsection{Impact on load shifting}
In a final analysis, we calculated the load shift effects on \acp{ME}, \acp{XE}, and costs for the different carbon prices for the case of Germany in 2019.
The results in \Cref{fig:load_shift_effects_heatmap_tax} are counterintuitive.
E.g., while \acp{ME} decrease to 6.5\% with a carbon price raise to 24.9\,\euro/t, they increase again by 6.2 percentage points to 12.7\% when the carbon price is further increased to 42.6\,\euro/t.
This is because with $c^\mathrm{GHG}=42.6$\,\euro/t, gas\_cc power plants move into the 35\,GW area, around where the residual load fluctuates (see \Cref{fig:mols2019}b for Germany) increasing the number of gas\_cc-to-lignite and gas\_cc-to-coal load shifts, see Figure S20 in Supplementary Material I.
At $c^\mathrm{GHG}=65.3$\,\euro/t, gas\_cc power plants already arrived on the lower half of the residual load range and therefore act more frequently as load shift sink.
$c^\mathrm{GHG}=65.3$\,\euro/t is already sufficient to achieve 58\% of the possible achievable \ac{ME} savings, and $c^\mathrm{GHG}=100.0$\,\euro/t yields 93\% of the possible achievable \ac{ME} reduction of 21.4\%.
For carbon prices around 235.6\,\euro/t, the aforementioned counterintuitive increase occurs again, however, with lower impact and through increased gas-to-coal load shifts.

The effects on \acp{XE} are similar to the effects on \acp{ME}.
Counterintuitive increases occur for the same reasons but with lower amplitude, due to the damping influence of averaging.

The effect curve on costs in \Cref{fig:load_shift_effects_heatmap_tax} is concave.
For $c^\mathrm{GHG}\leq65.3$\,\euro/t, the cost savings decrease with increasing carbon prices.
The reducing daily price spreads are caused by the convergence of marginal costs of power plants in the relevant residual load range (around 35\,GW) as the high carbon intensities of formerly cheap coal and lignite power plants are effectively penalized by the increasing carbon price.
For $c^\mathrm{GHG}>65.3$\,\euro/t, the carbon-related marginal costs (cf. \Cref{eq:marg_costs}) increasingly outweighs the fuel costs and ultimately makes the fuel costs insignificant.
Through the formation of coherent technology blocks -- now in the new ascending order of carbon intensity -- steps in the marginal costs curve are shaped (e.g., at 23\,GW) which in turn lead to higher price spreads.

\subsection{Discussion summary}\label{sec:discussion-summary}
While \acp{MEF} are essential for quantifying the carbon differential of load shifts, \acp{XEF} are more suitable for calculating the carbon emissions of a static electricity load profile.
Also, \acp{XEF} can be determined with less uncertainty and in a straightforward approach which is reflected in their high availability.

European national electricity supply systems differ widely in both size (0.13--49\,GW residual load) and composition (7--77\% RES share), which is reflected in the varying prices and CEFs that resulted from the simulation, \Cref{fig:cef_and_prices_violins_2019}.
The differences between the European countries became even clearer when running yearly simulations of daily load shifts on the basis of the calculated prices and CEFs.
The electricity cost-saving potentials ranged between 3\% for Lithuania (LT) and Serbia (RS) and 24\% for Austria (AT) and Greece (GR).
The resulting changes in \acp{ME} varied between 64\% decrease for France (FR) and 53\% increase for Greece (GR), with increases in eight countries (AT, DE, ES, GR, HU, IE, PT, RO).
The changes of \acp{XE} varied between -92\% for France (FR) and +3\% for Serbia (RS), with Serbia being the only country where \acp{XE} increase.
Averaged over all countries, the costs decreased by 10.4\%, the \acp{XE} decreased by 26.9\%, but the \acp{ME} increased by 2.1\%.
While XEF-based load shifts, like the price-based load shifts, led to \ac{ME} increases in eight countries, MEF-based load shifts resulted in average emission savings of 35\%, albeit with 56\% lower cost savings.
A final sensitivity analysis regarding the carbon price brought the following salient findings:
(1) For Germany, a carbon price of 42.6\,\euro/t was necessary to decouple emissions from prices, i.e., where $r$, the Spearman correlation coefficient of emissions and prices along the merit order, is zero.
(2) A carbon price increase from 0 to 156.1\,\euro/t led to a switch between gas\_cc and lignite/coal in the German merit order and flipped effects of the according price-based load shifts on carbon emissions from a 10\% increase to a 21\% decrease.
(3) Increases of the carbon price beyond 156.1\,\euro/t led to insignificant changes.

\section{Conclusions}\label{sec:conclusions}
The aim of this paper was the quantification and discussion of the effects of \ac{PBDR} on operational carbon emissions for European countries.
Straightforward approaches based on the calculation of \acp{XEF} are not suitable for this purpose due to the characteristics of electricity markets.
More adequate methods based on the knowledge of marginal power plants require detailed data, thus \ac{MEF} values are not readily available for European countries.
In this paper, we therefore proposed a method (PWL) to approximate \acp{MEF} with readily available datasets and validated it with another method (PP) using power plant-specific efficiency data from Germany.
We then applied the PWL method to 20 European countries for the years 2017--2019 to calculate prices, \acp{MEF}, and, for comparison purposes, \acp{XEF}.
The resulting prices and \acp{CEF} served as basis for subsequently conducted load shift simulations, to evaluate its effects on carbon emissions.
Starting from the so-called merit order dilemma of emissions, the results were discussed for six representative countries.
In a final analysis, the impact of carbon pricing was analyzed by calculating the Spearman correlation coefficient between prices and emissions along the merit order for different carbon prices.

The key findings of the paper are:
\begin{enumerate}
	\item The great diversity of European countries in terms of the composition and the size of their national electricity supply systems is reflected in the \acp{XEF} and \acp{MEF}.
	\item Price-based and XEF-based load shifts led to increases in operational carbon emissions in 8 of 20 European countries.
	\item MEF-based load shifts led to average carbon emission savings of 35\%, however decreasing the cost-saving potential by 56\% compared to price-based load shifts.
	\item Emissions and prices along the German merit order for the year 2019 decoupled with a carbon price of 42.6\,\euro/t.
	\item An carbon price increase from 0 to 156.1\,\euro/t led to a switch between gas\_cc and lignite/coal in the German merit order of 2019 and improved the carbon emission effect of the according price-based load shifts for European countries from a 10\% increase to a 21\% decrease.
\end{enumerate}

Despite the limitations outlined in \Cref{sec:limitations}, the following main conclusions can be drawn:
\begin{enumerate}
\item Currently, DR could increase operational carbon emissions if spot-market prices are used as an incentive signal (under the current carbon price).
\item This phenomenon does not occur
	\begin{enumerate}
	\item if dynamic \acp{MEF} are used as DR incentive signal or
	\item if an adequate carbon price is set.
	\end{enumerate}
\end{enumerate}

While \ac{PBDR} leads to negative environmental effects under specific circumstances, it is a very promising method of reducing operating cost and carbon emissions when adequate carbon prices are implemented.
To exploit the full positive environmental potential of PBDR, high correlations between carbon intensity and marginal cost in the merit order need to be ensured either by adequate carbon prices or other market interventions.

In future research, the interconnectivity of individual countries to form a large interconnected network should be investigated, as this is an increasingly important aspect, which was however outside the scope of this paper.
Furthermore, dynamic \acp{MEF} may be used for the assessment of the environmental potential of \ac{PBDR} in real case studies considering technical, organizational, and process-related constraints in a realistic way.

\printnomenclature

\section*{Acknowledgments}
This research was performed as part of the MeSSO Research Group at the Cork Institute of Technology (CIT) and in relation to the project WIN4Climate as part of the National Climate Initiative financed by the Federal Ministry of the Environment, Nature Conservation and Nuclear Safety (BMU) according to a decision of the German Federal Parliament (No. 03KF0094A).
It was additionally funded by the CIT Risam scholarship scheme.

\section{Conflicts of interest}
The authors declare no conflict of interest.

\section{CRediT authorship contribution statement}
\textit{\textbf{Markus Fleschutz:}} Conceptualization, Methodology, Software, Validation, Data Curation, Writing - Original Draft, Writing - Review \& Editing, Visualization.
\textit{\textbf{Markus Bohlayer:}} Conceptualization, Methodology, Writing - Review \& Editing.
\textit{\textbf{Marco Braun:}} Supervision, Project administration, Funding acquisition.
\textit{\textbf{Gregor Henze:}} Writing - Review \& Editing.
\textbf{Michael D. Murphy:} Supervision, Conceptualization, Writing - Review \& Editing, Funding acquisition.

\section*{References}
\bibliographystyle{elsarticle-num}
\bibliography{references,ref2}

\end{document}


\savegeometry{mydefaultgeometry}
\newgeometry{a4paper,lmargin=3cm,rmargin=3cm,tmargin=5cm,bmargin=2cm}
\maketitle
\tableofcontents
\loadgeometry{mydefaultgeometry}

\newpage

\section{ENTSOE data}
\begin{figure}[!h]
	\centering
	\includegraphics[width=1.0\linewidth]{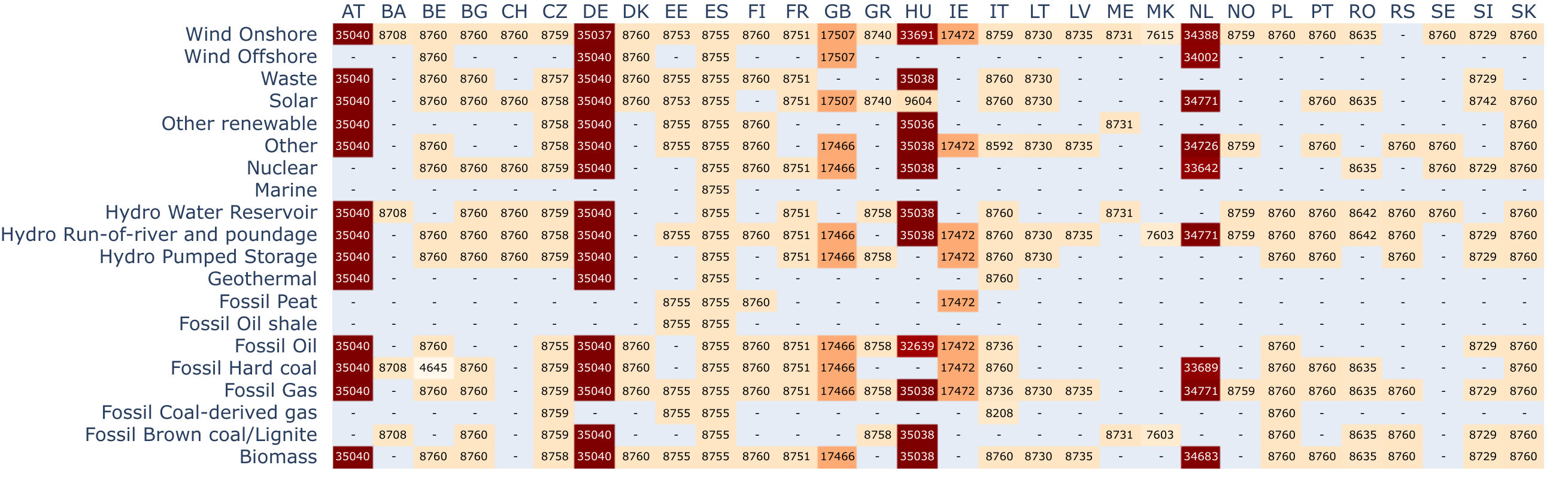}
	\caption{Number of available national generation data points available on the ENTSOE Transparency Platform~\cite{ENTSOE.2020} for European countries in 2019.}
	\label{fig:gen_avail_heatmap_ploty_raw}
\end{figure}

\begin{figure}[!h]
	\centering
	\includegraphics[width=1.0\linewidth]{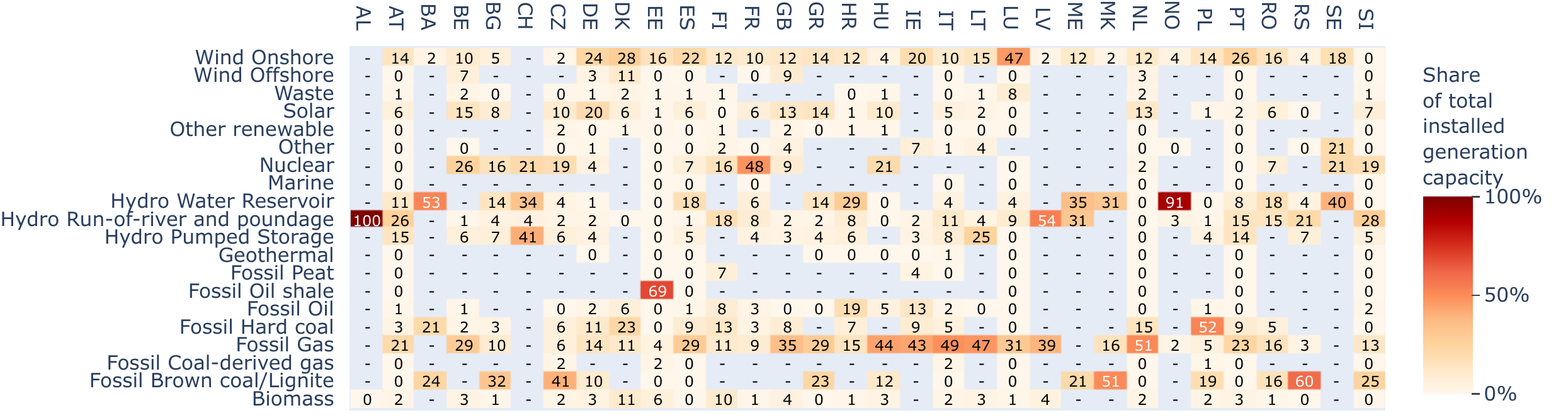}
	\caption{Available national installed generation capacity data available on the ENTSOE Transparency Platform~\cite{ENTSOE.2020} for European countries in 2019.}
	\label{fig:heatmap_installed_capa}
\end{figure}

To save space for the following diagrams, the following renaming and aggregation was conducted: biomass=Biomass, lignite=Fossil Brown coal/Lignite, oil=Fossil Oil, gas=Fossil Gas, coal=Fossil Hard coal, pumped\_hydro=Hydro Pumped Storage, hydro=Hydro Run-of-river and poundage, nuclear=Nuclear, solar=Solar, wind\_offshore=Wind Offshore, wind\_onshore=Wind Onshore, other\_CONV = Other + other\_conventional + Fossil Coal-derived gas + Fossil Oil shale + Fossil Peat, other\_RES = Waste + Geothermal + Hydro Water Reservoir + Marine + Other renewable + other renewables.

\begin{figure}[htb]
	\centering
	\includegraphics[width=0.7\linewidth]{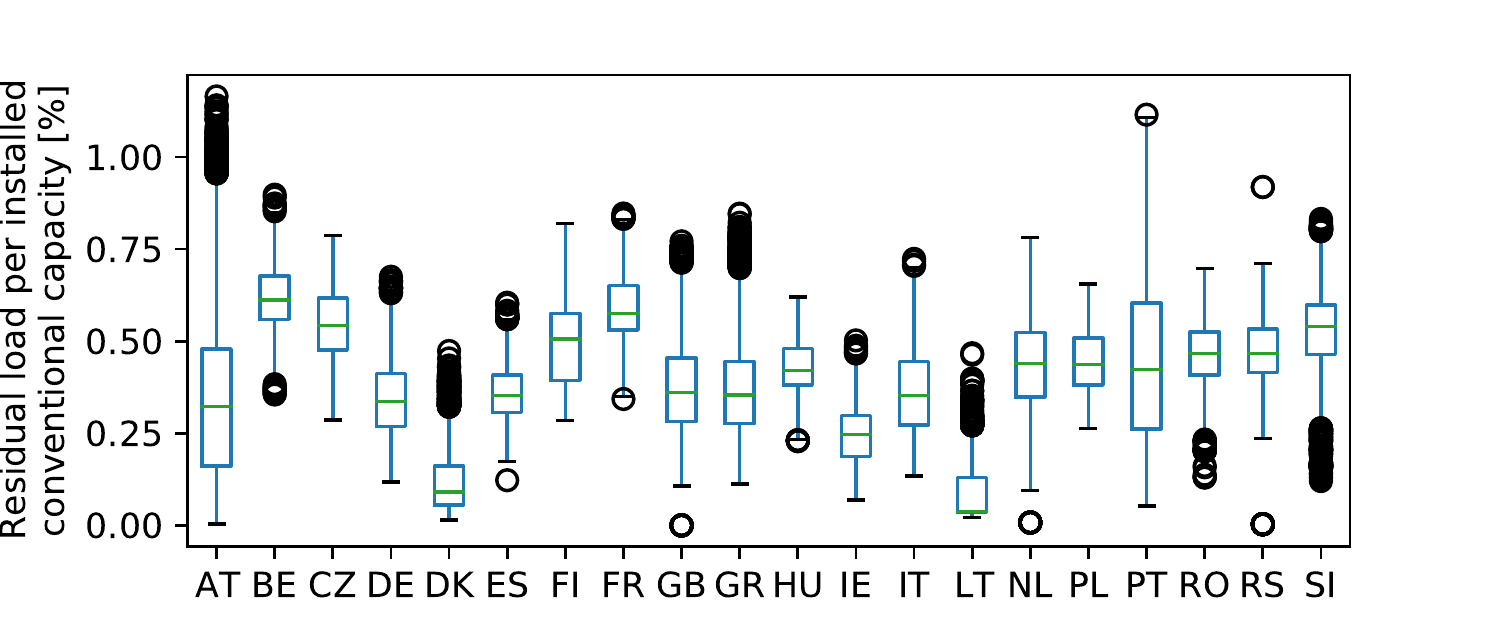}
	\caption{Residual load per installed conventional capacity in 2019 for analyzed countries.}
	\label{fig:resi_relative_to_installed_capa_box_}
\end{figure}

\begin{figure}[htb]
	\centering
	\includegraphics[width=0.7\linewidth]{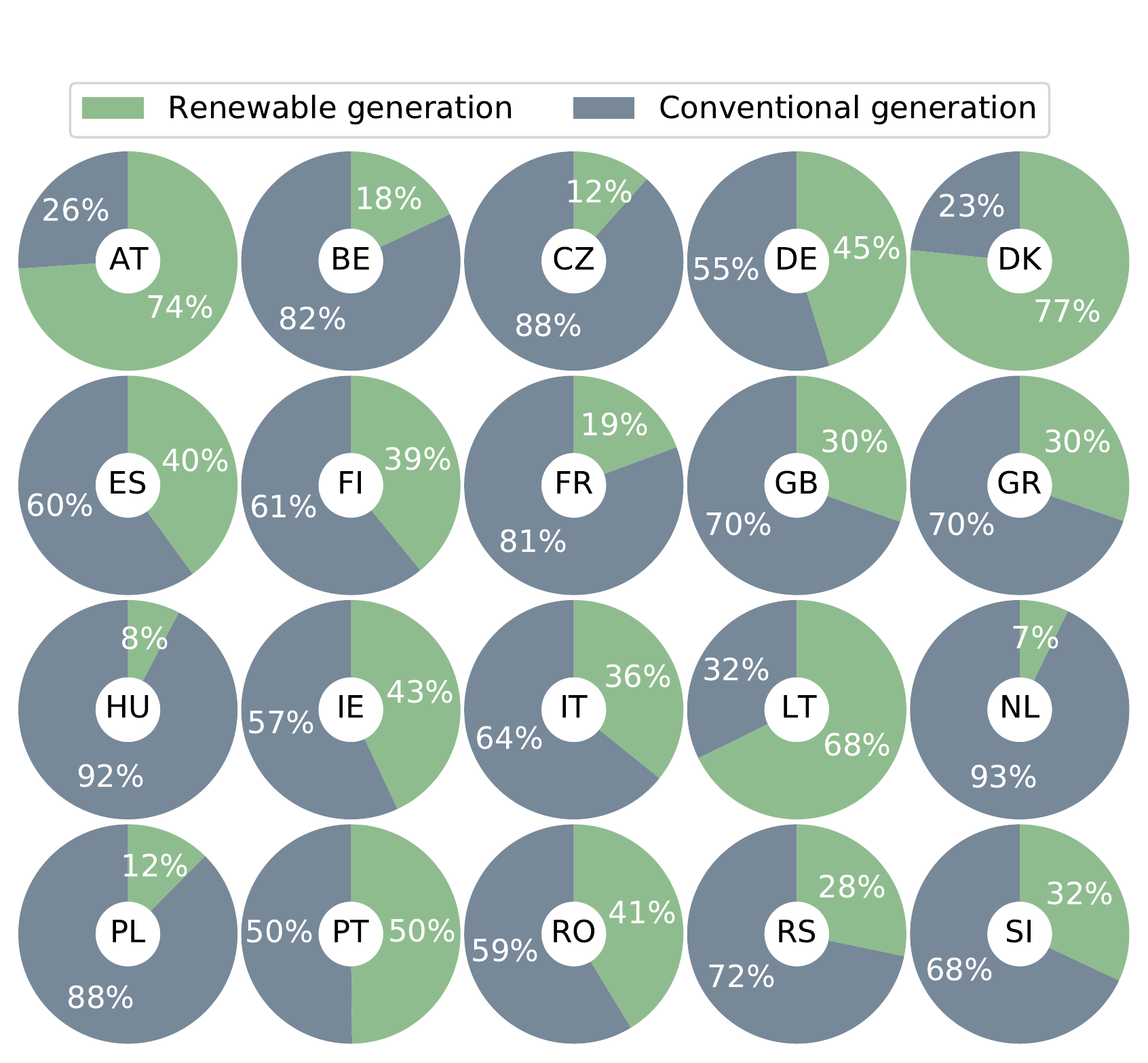}
	\caption{Shares of renewable vs. conventional generation in 2019 for analyzed countries.}
	\label{fig:pie_share_res_conv}
\end{figure}

\clearpage
\section{Analysis of German generation data}

\begin{figure}[!h]
	\centering
	\includegraphics[scale=0.70]{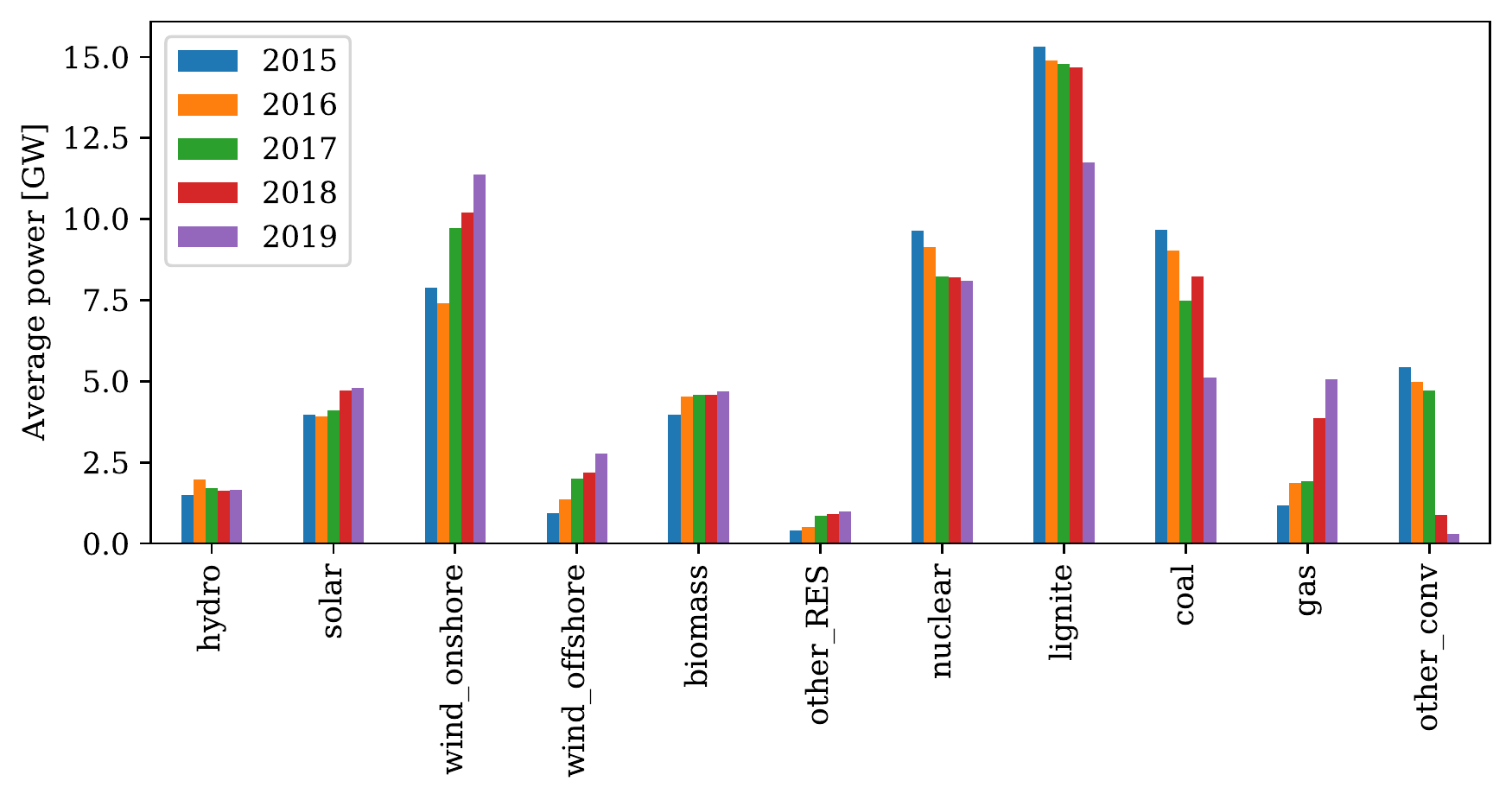}
	\caption{Average power generation per fuel type and year in Germany. Data: ENTSOE Transparency Platform~\cite{ENTSOE.2020}.}
	\label{fig:national_generation_last_years}
\end{figure}

\begin{figure}[!h]
	\centering
	\includegraphics[width=1.0\linewidth]{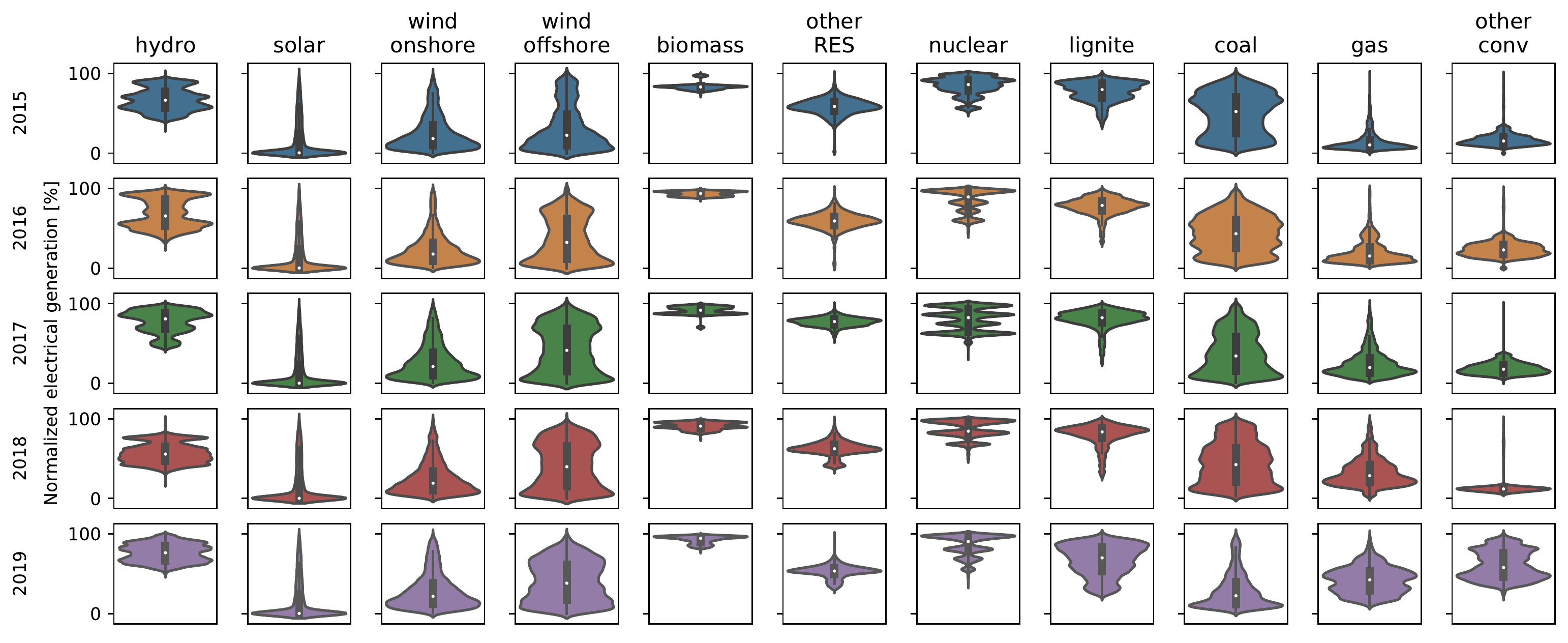}
	\caption{Normalized violin plots of German electricity generation technologies for the years 2015--2019. Data: ENTSOE Transparency Platform~\cite{ENTSOE.2020}.}
	\label{fig:gen_violins}
\end{figure}

\begin{figure}[!h]
	\centering
	\includegraphics[width=0.8\linewidth]{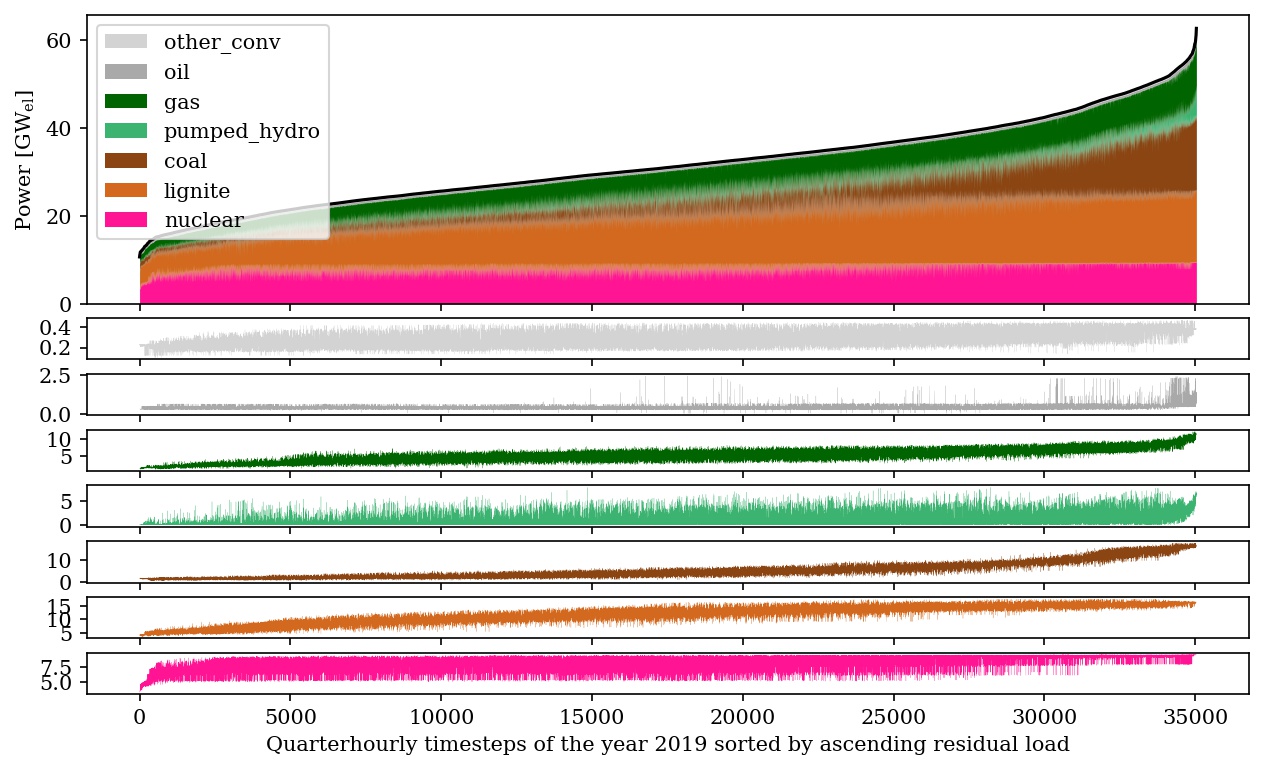}
	\caption{Top: Stacked-area plot of German conventional electricity generation for the year 2019 sorted by ascending residual load.
		Bottom: Individual line plots for each conventional technology.}
	\label{fig:sorted_gen}
\end{figure}

\begin{figure}[!h]
	\centering
	\includegraphics[width=0.8\linewidth]{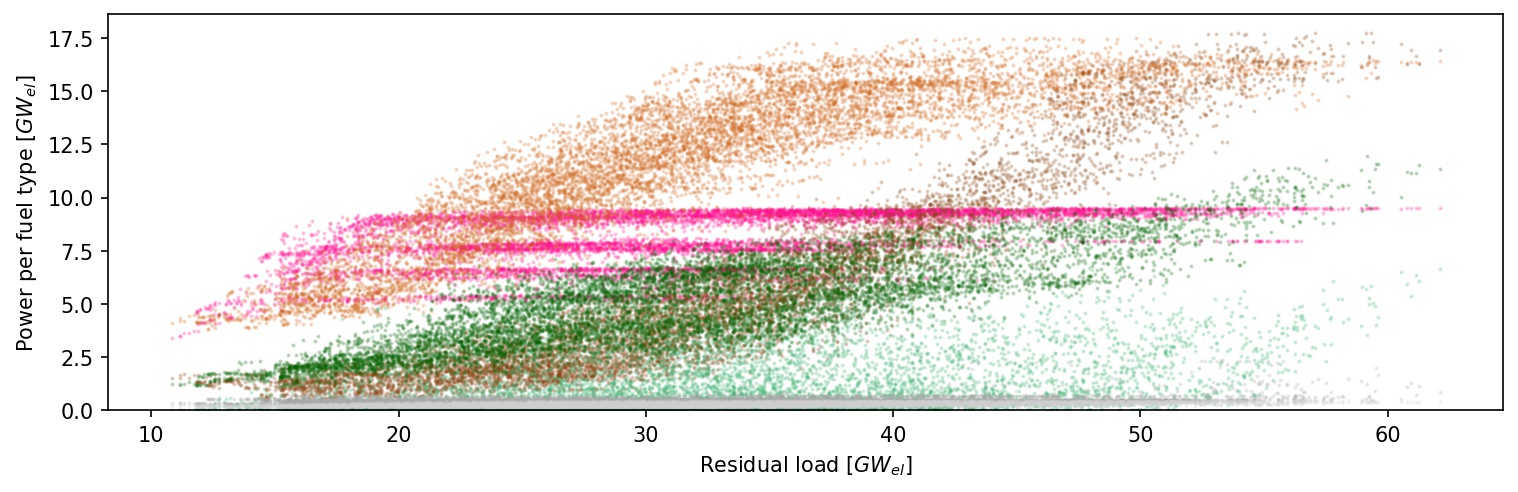}
	\includegraphics[width=0.8\linewidth]{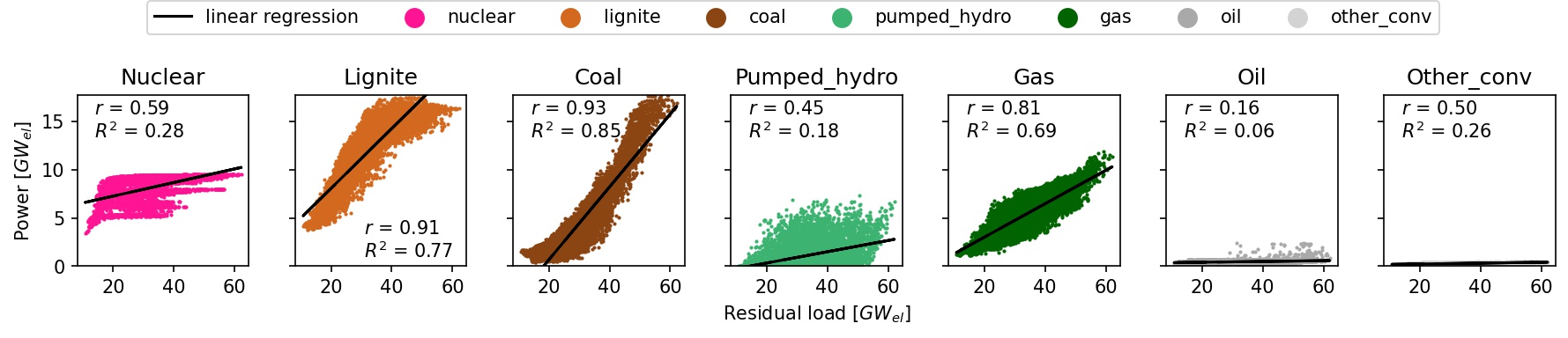}
	\caption{Top: German conventional electricity generation for the year 2019. Bottom: Linear regression analysis for each generation technology where $r$ is the spearman correlation coefficient and $R^2$ is the coefficient of determination.}
	\label{fig:scatter_res_vs_conv}
\end{figure}

\clearpage
\section{OPSD power plant list}

\begin{figure}[!h]
	\centering
	\includegraphics[width=0.7\linewidth]{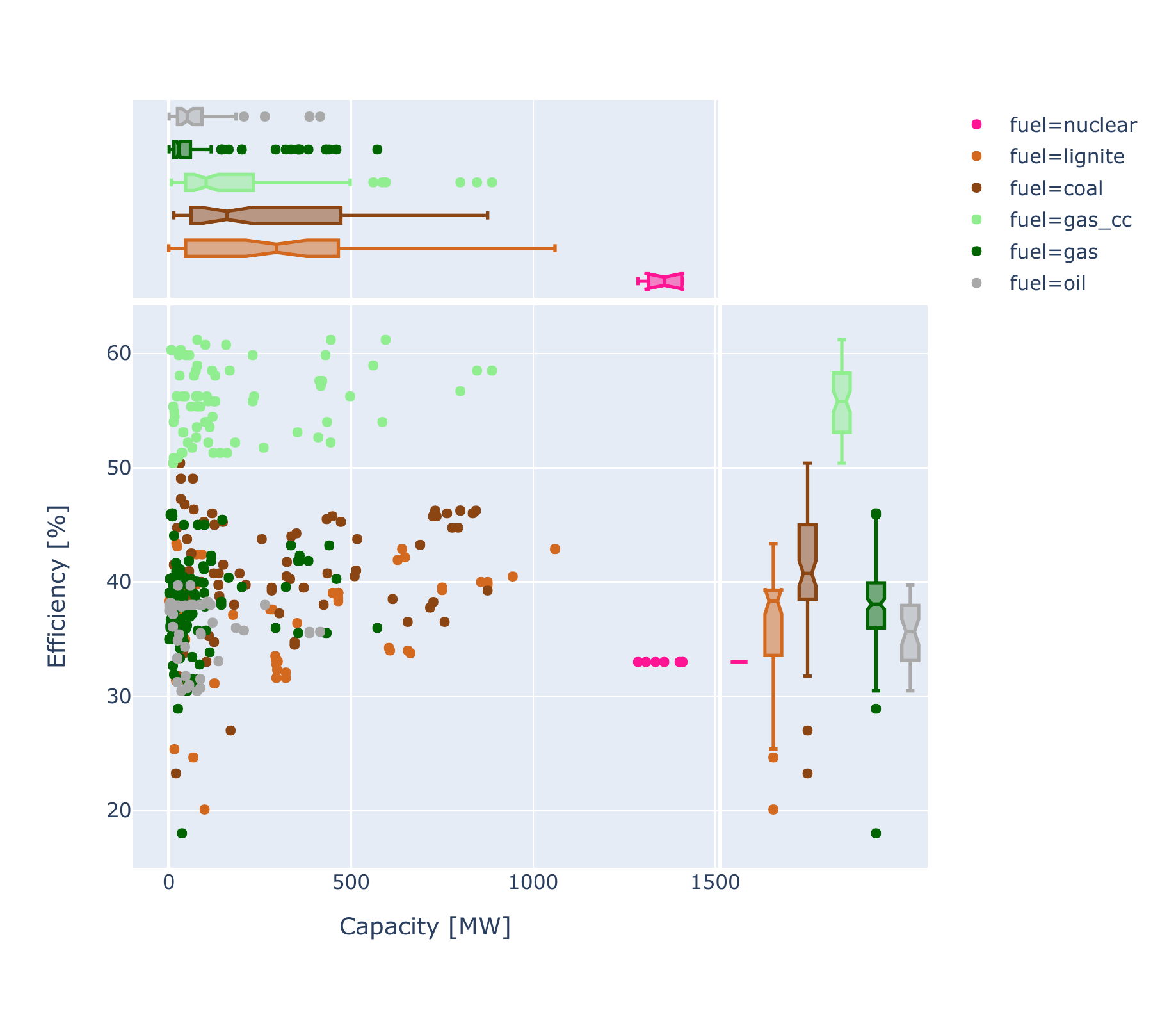}
	\caption{Net generation capacity and power plant efficiency distribution of all in 2019 available conventional power plants from OPSD~\cite{OpenPowerSystemData.2018}.}
	\label{fig:px_scatter_capa_vs_eff}
\end{figure}

\newpage
\section{Outliers and special missing data}

\begin{table}[htb]
	\centering
	\caption{Handling of outliers and special missing data in national electricity generation data.}
	\label{tab:missing_data}
	\begin{tabular}{llp{5cm}rll}
		\toprule
		Country&Fuel type&Timestamp(s)&z-Score&Type&Treatment\\
		\midrule
		CZ&Hydro Run-of-river and poundage&2018-05-24 12:00&$>$40&outlier&as missing value\\
		DK&Waste&2019-11-04 08:00&$>$90&outlier&as missing value\\
		HU&Other renewable&2019-12-04 13:00, 2019-12-04 13:15&$>$18&outlier&as missing value\\
		LT&Other&2018-01-07 08:00&$>$90&outlier&as missing value\\
		LT&Solar&2018-10-24 15:00, 2018-10-26 11:00, 2018-10-26 12:00&$>$12&outlier&as missing value\\
		SI&Waste&2019-11-04 08:00&$>$90&outlier&as missing value\\
		\midrule
		NL&All&2018-12-31 23:00&&no data&data of previous hour is used.\\
		NL&All&2019-01-01 (whole day)&&no data&data of next sunday is used.\\
		\bottomrule
	\end{tabular}
\end{table}

\section{Reasons for exclusion of countries from analyses}

\begin{table*}[!h]
	\centering
	\caption{Exclusion reasons for European countries that were excluded from the analyses.}
	\label{tab:exclusion_reasons}
	\resizebox{1.0\linewidth}{!}{\begin{tabular}{p{5cm}|p{12cm}}
			Reason & Countries \\
			\hline
			No generation data available & AD, AL, AM, AZ, BY, CY, FO, GE, GI, HR, IS, KZ, LI, LU, MC, MD, MT, RU, SM, TR, UA, VA, XK \\
			Only one fossil fuel-type available & coal: BA; nuclear: CH, SE; gas: NO, LV;\newline lignite: MK, ME; other\_conv: EE \\
			No installed capacity for 2019 available & BG, SK \\
			\hline
	\end{tabular}}
\end{table*}

\newpage
\section{Validation}

\begin{figure*}[!h]
	\centering 
	\includegraphics[width=0.8\linewidth]{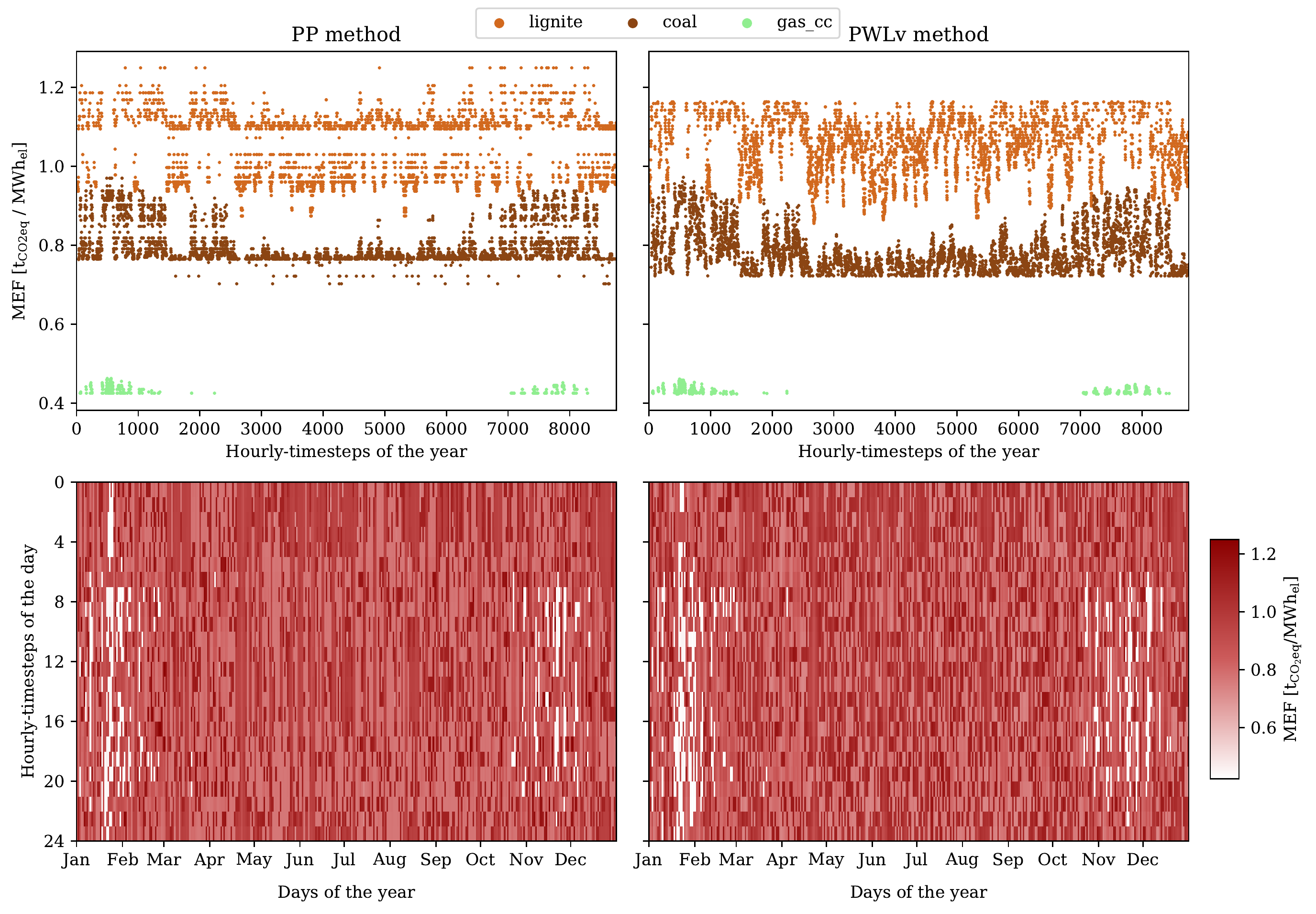}
	\caption{Comparison of methods to calculate MEFs. For each method (PP method left and PWLv method right) the hourly MEFs for Germany for the year 2019 are shown as scatter plot (top) and as heatmap (bottom). While in the scatter plot the range of values of the two resulting MEFs can be compared well, the heatmaps clearly show the similar daily and seasonal patterns, especially for times in winter where gas\_cc power plants kick in due to high residual load.}
	\label{fig:scatter_and_heatmap}
\end{figure*}

\clearpage
\section{Merit orders for the years 2017--2019}

\begin{figure}[!h]
	\centering
	\textbf{2017}\par\medskip
	\includegraphics[width=0.85\linewidth]{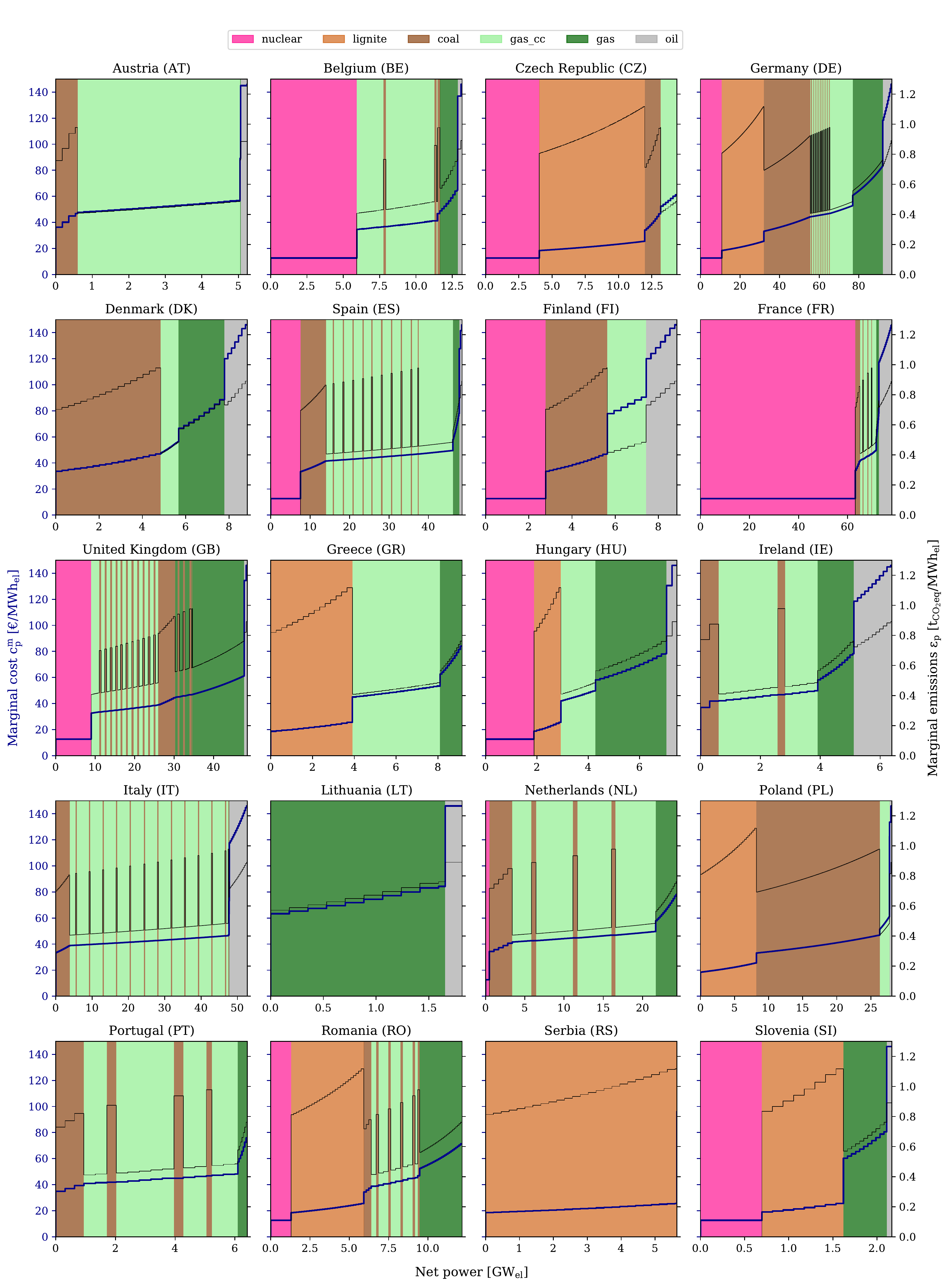}
	\caption{Merit orders of for analyzed European countries resulting from PWL-method for the year 2017.}
	\label{fig:all_merit_orders_2017}
\end{figure}

\begin{figure}[t]
	\centering
	\vspace{0.8cm}
	\textbf{2018}\par\medskip
	\includegraphics[width=0.85\linewidth]{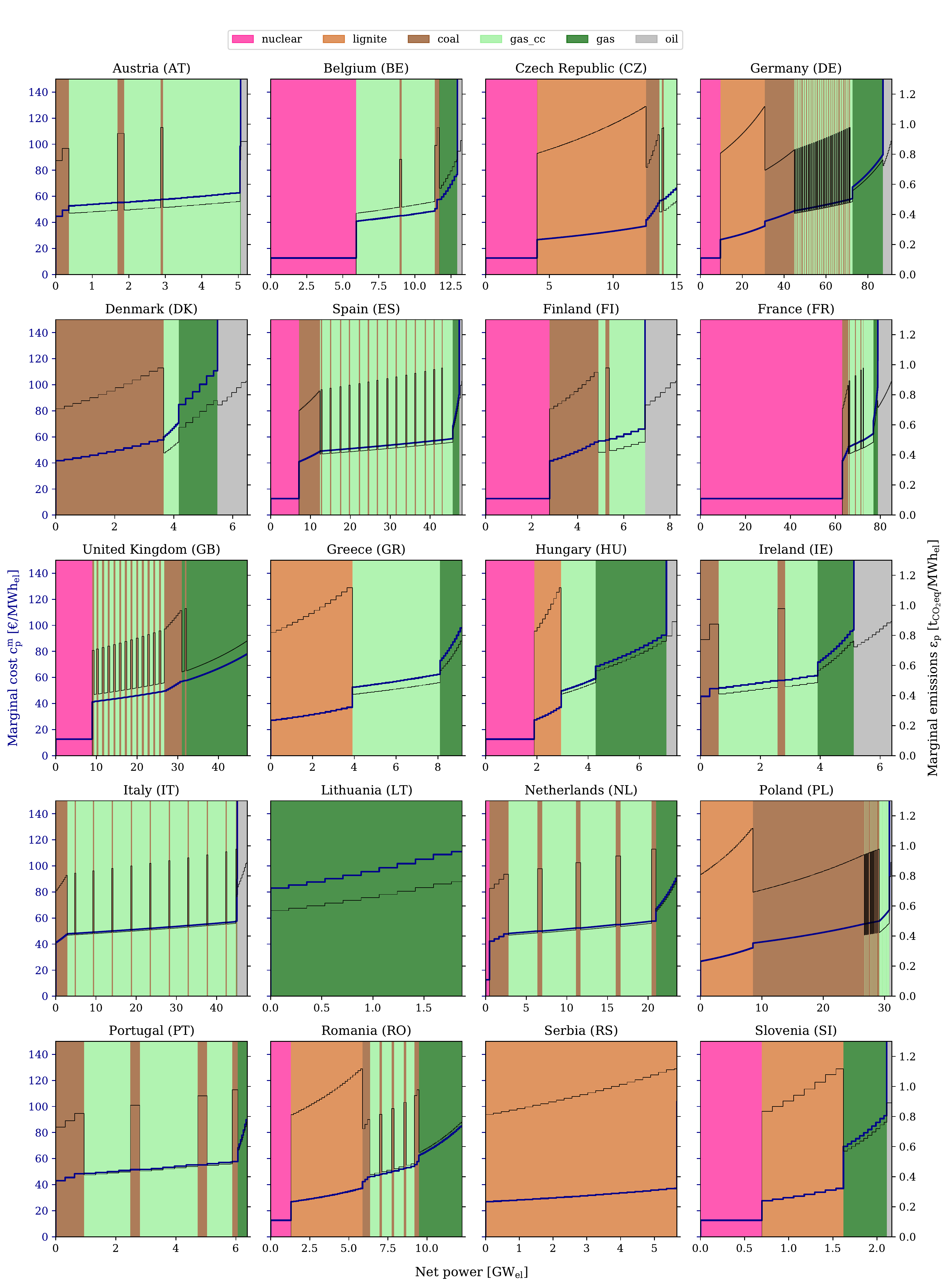}
	\caption{Merit orders of for analyzed European countries resulting from PWL-method for the year 2018.}
	\label{fig:all_merit_orders_2018}
\end{figure}

\begin{figure}[t]
	\centering
	\vspace{0.8cm}
	\textbf{2019}\par\medskip
	\includegraphics[width=0.85\linewidth]{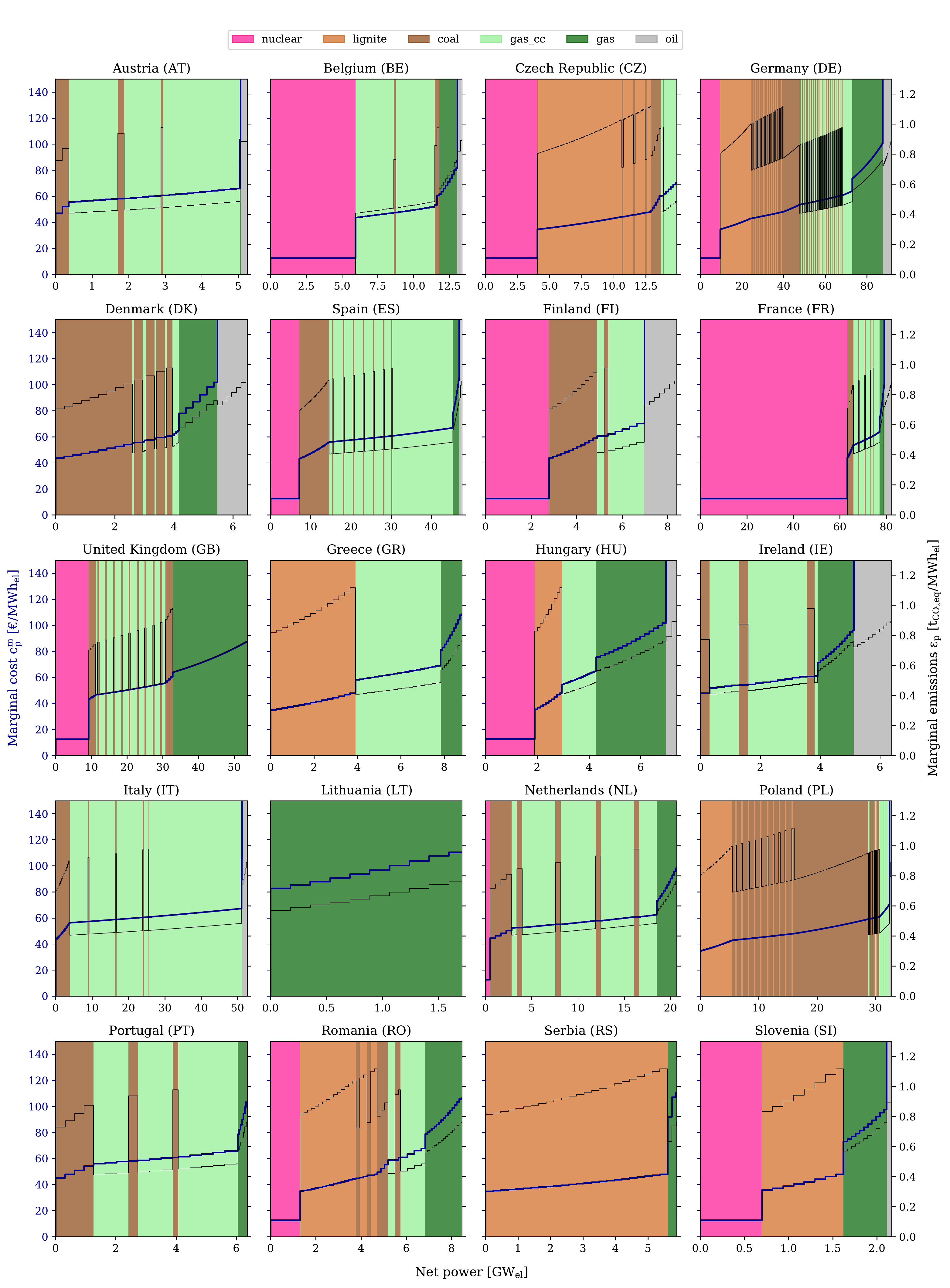}
	\caption{Merit orders of for analyzed European countries resulting from PWL-method for the year 2019.}
	\label{fig:all_merit_orders_2019}
\end{figure}

\section{Correlation of prices and CEFs for the years 2017--2019}

\begin{figure}[!h]
	\centering
	\textbf{2017}\par\medskip
	\includegraphics[width=1.0\linewidth]{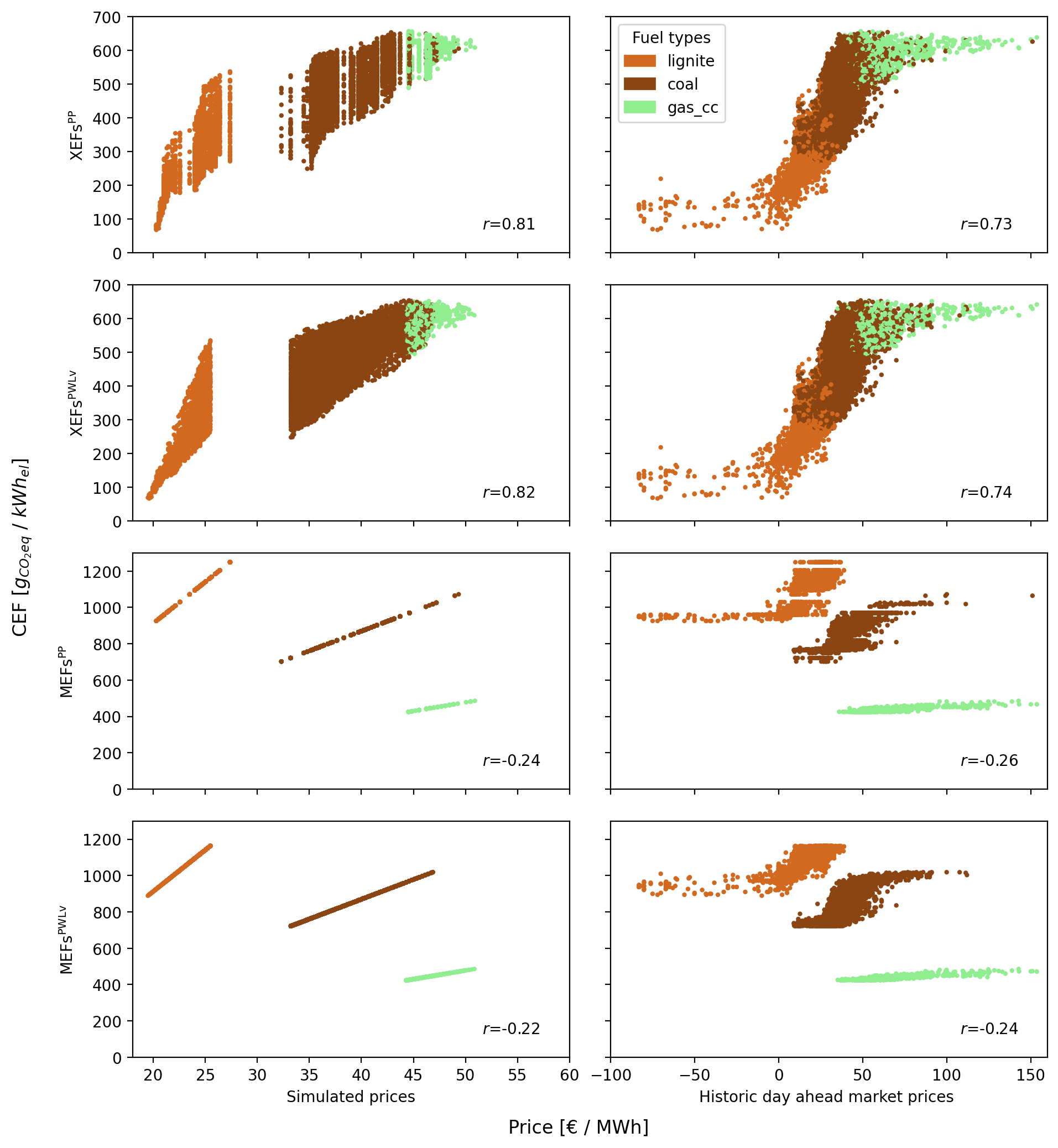}
	\caption{Scatter plots of German CEFs (from top to bottom: XEFs\textsuperscript{PP}, XEFs\textsuperscript{PWLv}, MEFs\textsuperscript{PP}, MEFs\textsuperscript{PWLv}) and prices (left: marginal cost from simulation, right: historic day ahead market prices from the ENTSOE Transparency Platform~\cite{ENTSOE.2020}) for the year 2017. The Spearman correlation coefficient $r$ is given in the lower right corner.}
	\label{fig:2017_c_ce_correl}
\end{figure}

\begin{figure}[!t]
	\centering
	\vspace{1.09cm}
	\textbf{2018}\par\medskip
	\includegraphics[width=1.0\linewidth]{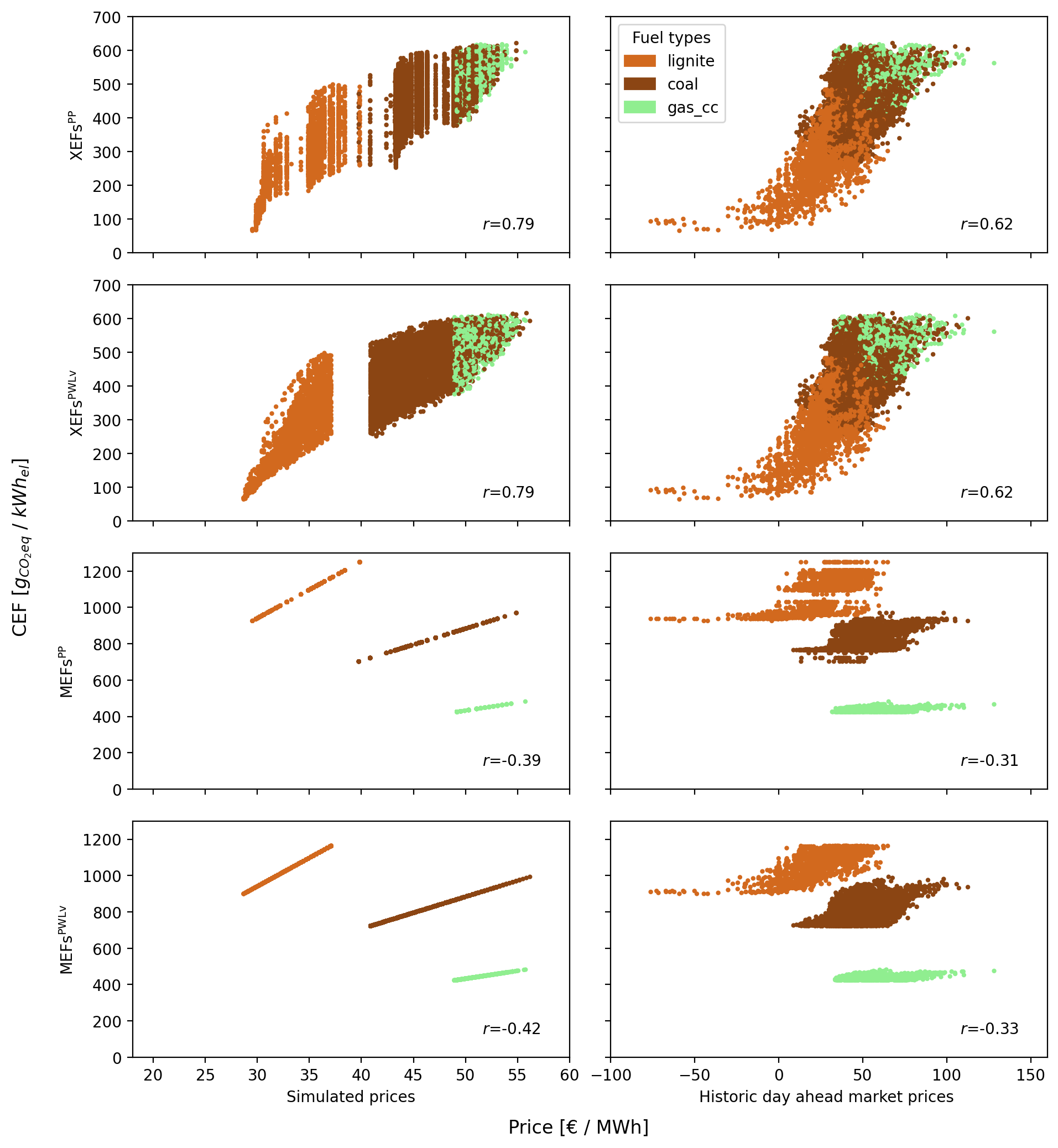}
	\caption{Scatter plots of German CEFs (from top to bottom: XEFs\textsuperscript{PP}, XEFs\textsuperscript{PWLv}, MEFs\textsuperscript{PP}, MEFs\textsuperscript{PWLv}) and prices (left: marginal cost from simulation, right: historic day ahead market prices from the ENTSOE Transparency Platform~\cite{ENTSOE.2020}) for the year 2018. The Spearman correlation coefficient $r$ is given in the lower right corner.}
	\label{fig:2018_c_ce_correl}
\end{figure}

\begin{figure}[!t]
	\centering
	\vspace{1.09cm}
	\textbf{2019}\par\medskip
	\includegraphics[width=1.0\linewidth]{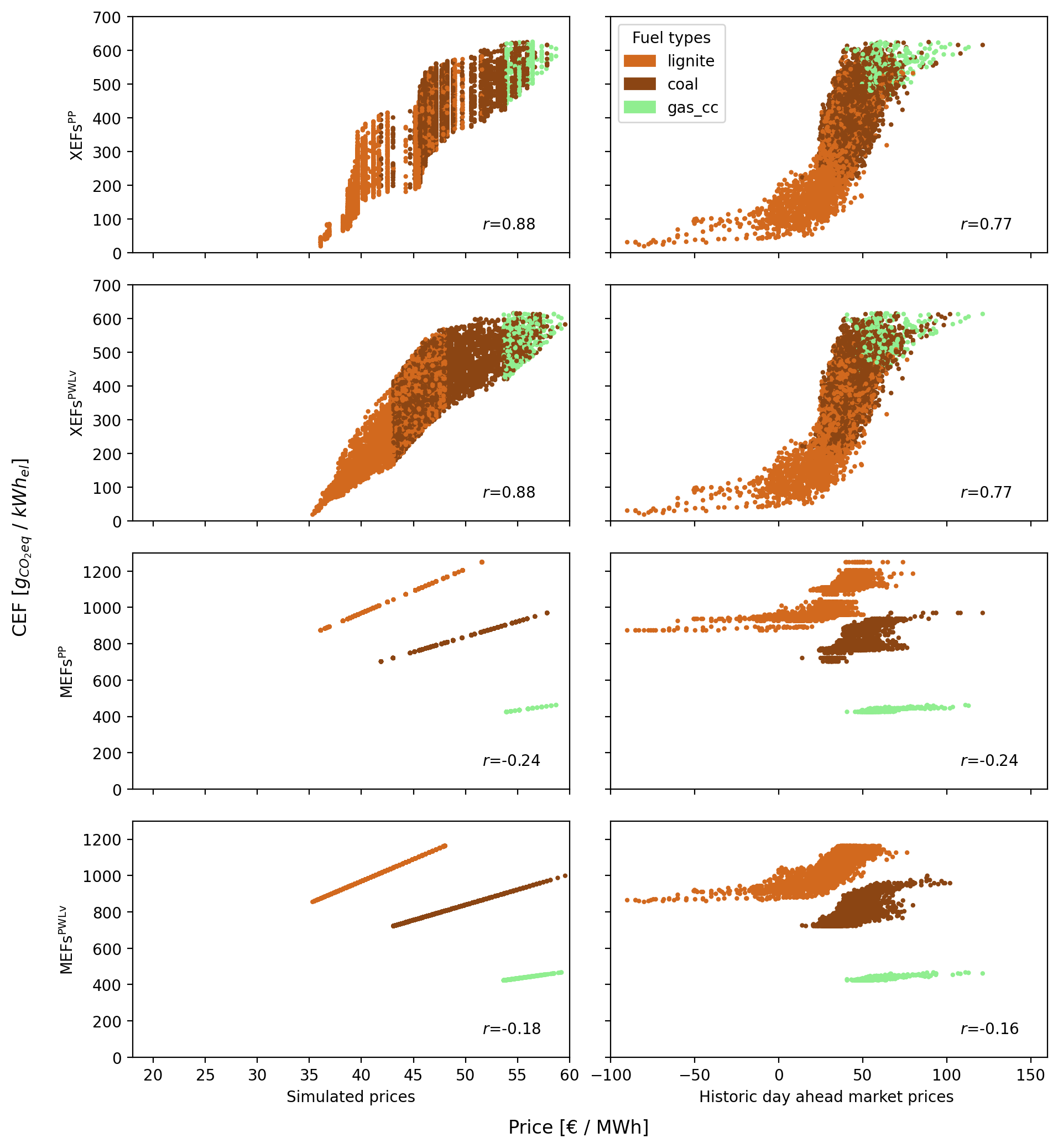}
	\caption{Scatter plots of German CEFs (from top to bottom: XEFs\textsuperscript{PP}, XEFs\textsuperscript{PWLv}, MEFs\textsuperscript{PP}, MEFs\textsuperscript{PWLv}) and prices (left: marginal cost from simulation, right: historic day ahead market prices from the ENTSOE Transparency Platform~\cite{ENTSOE.2020}) for the year 2019. The Spearman correlation coefficient $r$ is given in the lower right corner.}
	\label{fig:2019_c_ce_correl}
\end{figure}

\section{Load shift analysis}

\begin{figure}[!h]
	\centering
	\includegraphics[scale=0.60]{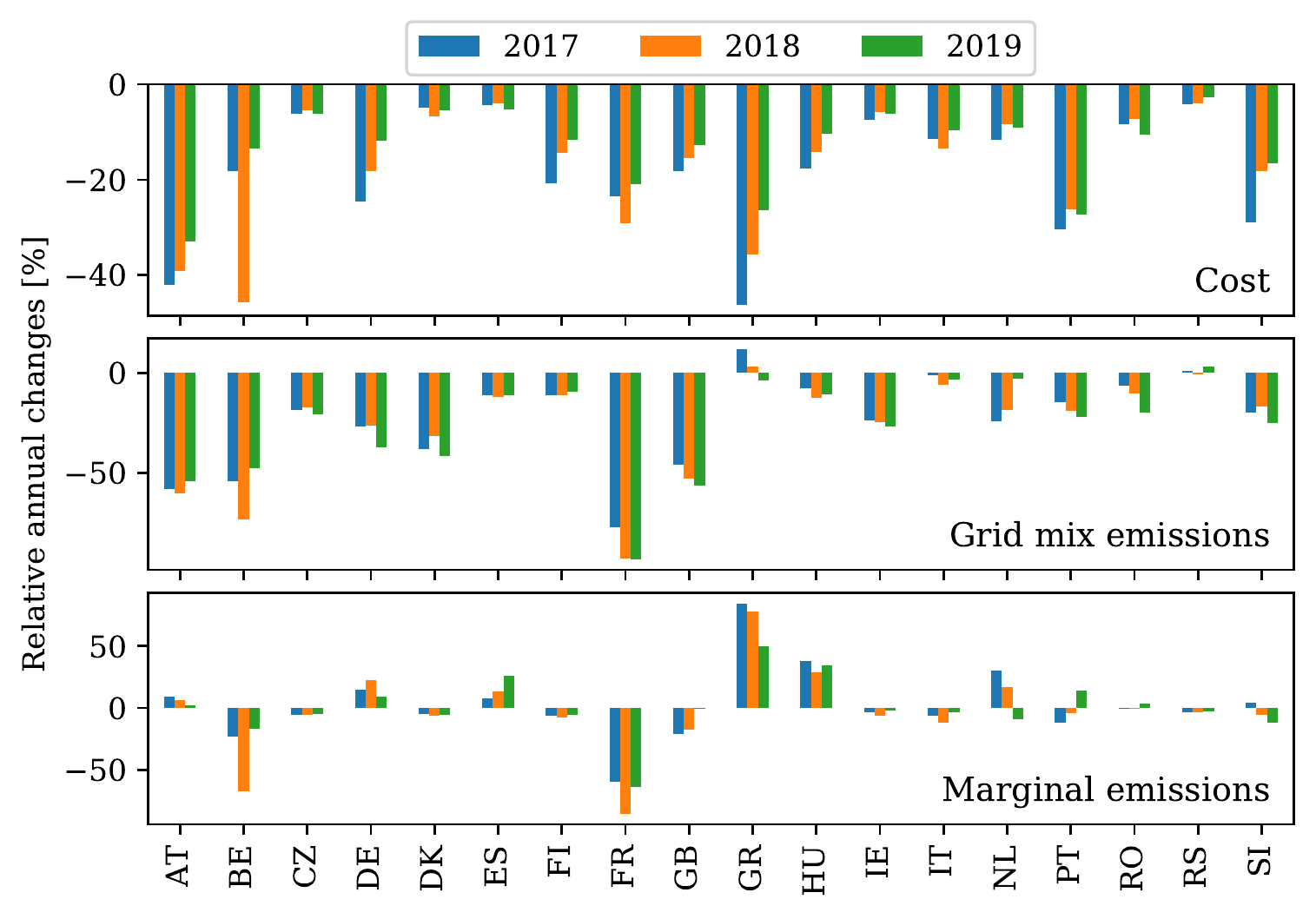}
	\caption{Relative changes in cost and carbon emissions of the shifted energy for 2017--2019 and different countries. The grid mix and marginal carbon emissions are based on XEF\textsuperscript{PWL} and MEF\textsuperscript{PWL} respectively.}
	\label{fig:spreads_effects_bar_allYears}
\end{figure}

\begin{figure}[htb]
	\centering
	\includegraphics[width=0.6\linewidth]{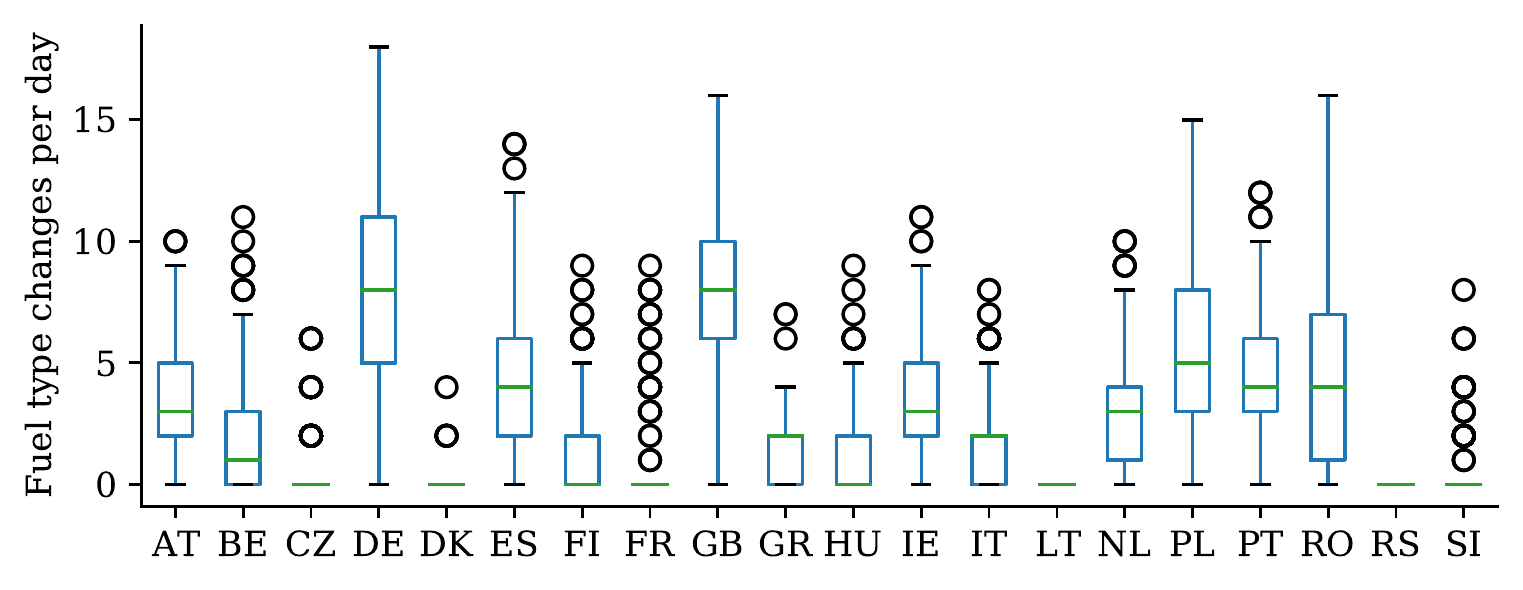}
	\caption{Distribution of the number of daily marginal fuel type changes for 2019.}
	\label{fig:fueltypeChanges_box}
\end{figure}

\begin{figure}[htb]
	\centering
	\includegraphics[width=0.6\linewidth]{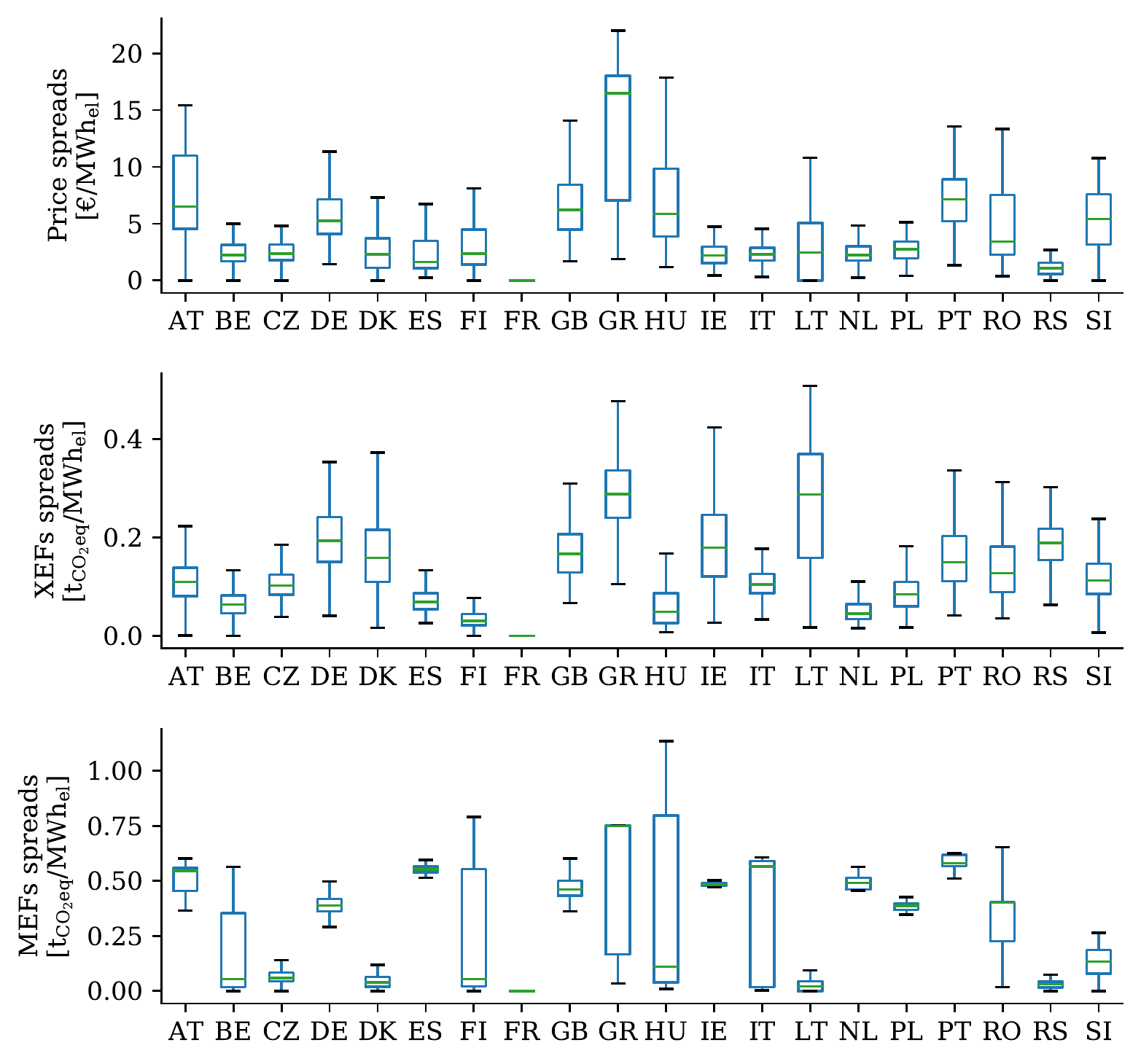}
	\caption{Boxplots of the daily prices and CEFs for 2019. Outliers were removed for this plot for better readability.}
	\label{fig:box_c_ce_daySpreads}
\end{figure}

\begin{figure}[htb]
	\centering
	\includegraphics[width=.7\linewidth]{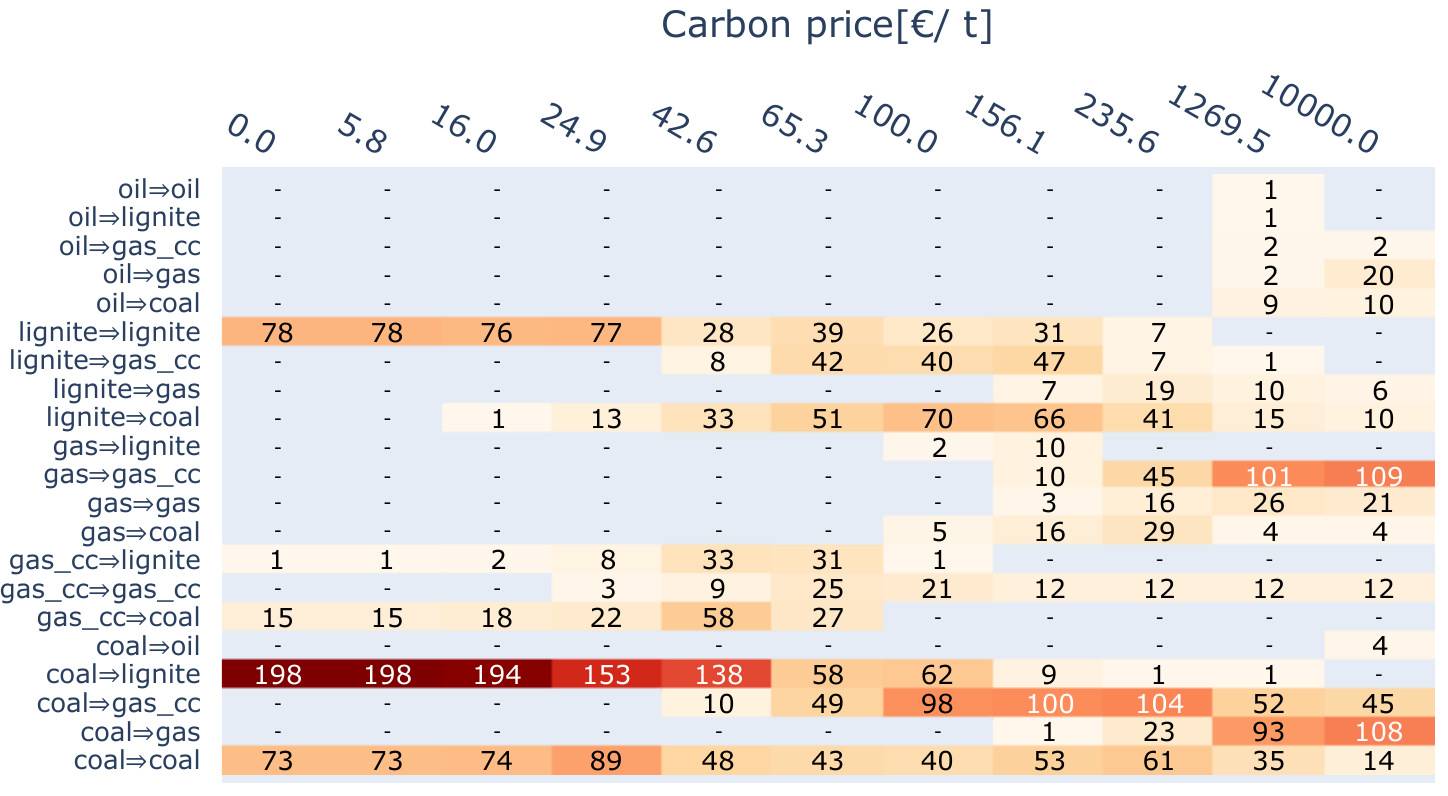}
	\caption{Frequency of load shift events per carbon price and marginal fuel type combination (denoted in the form $f^\mathrm{source} \Rightarrow f^\mathrm{sink}$) for Germany in 2019.}
	\label{fig:load_shift_fuel_types}
\end{figure}

\begin{figure}[!h]
	\centering
	\textbf{2017}\par\medskip
	\includegraphics[width=0.245\linewidth]{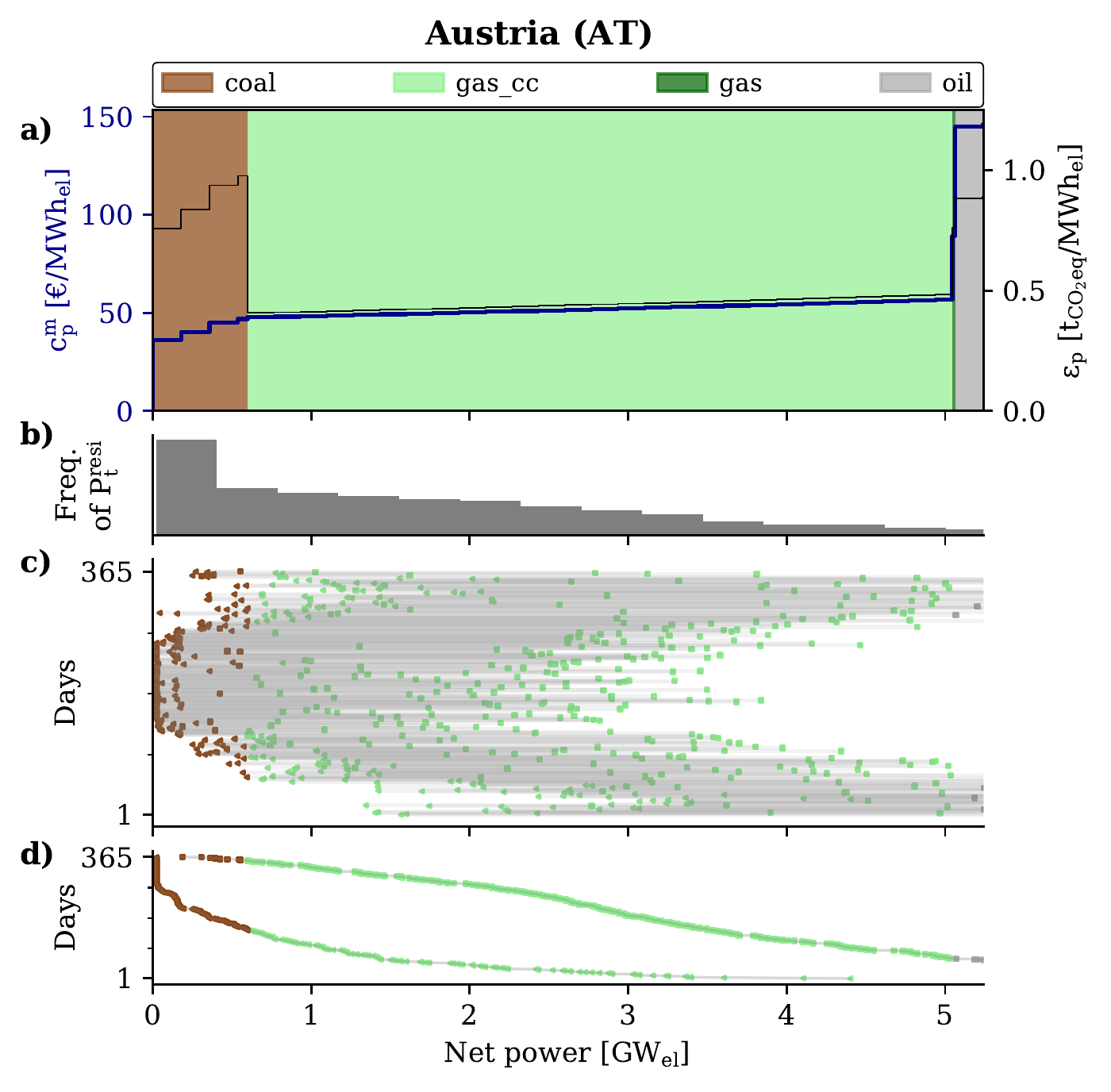}
	\includegraphics[width=0.245\linewidth]{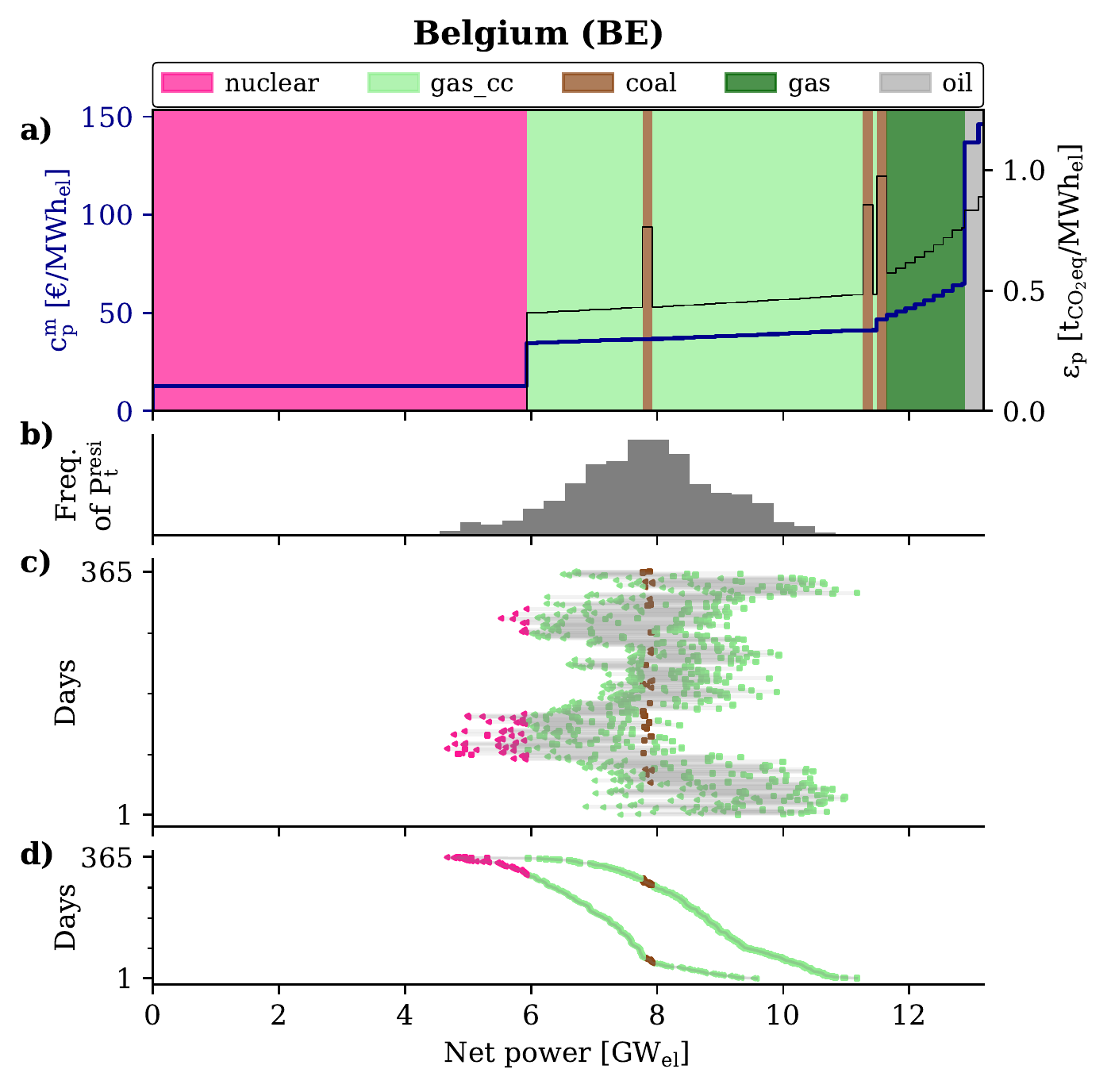}
	\includegraphics[width=0.245\linewidth]{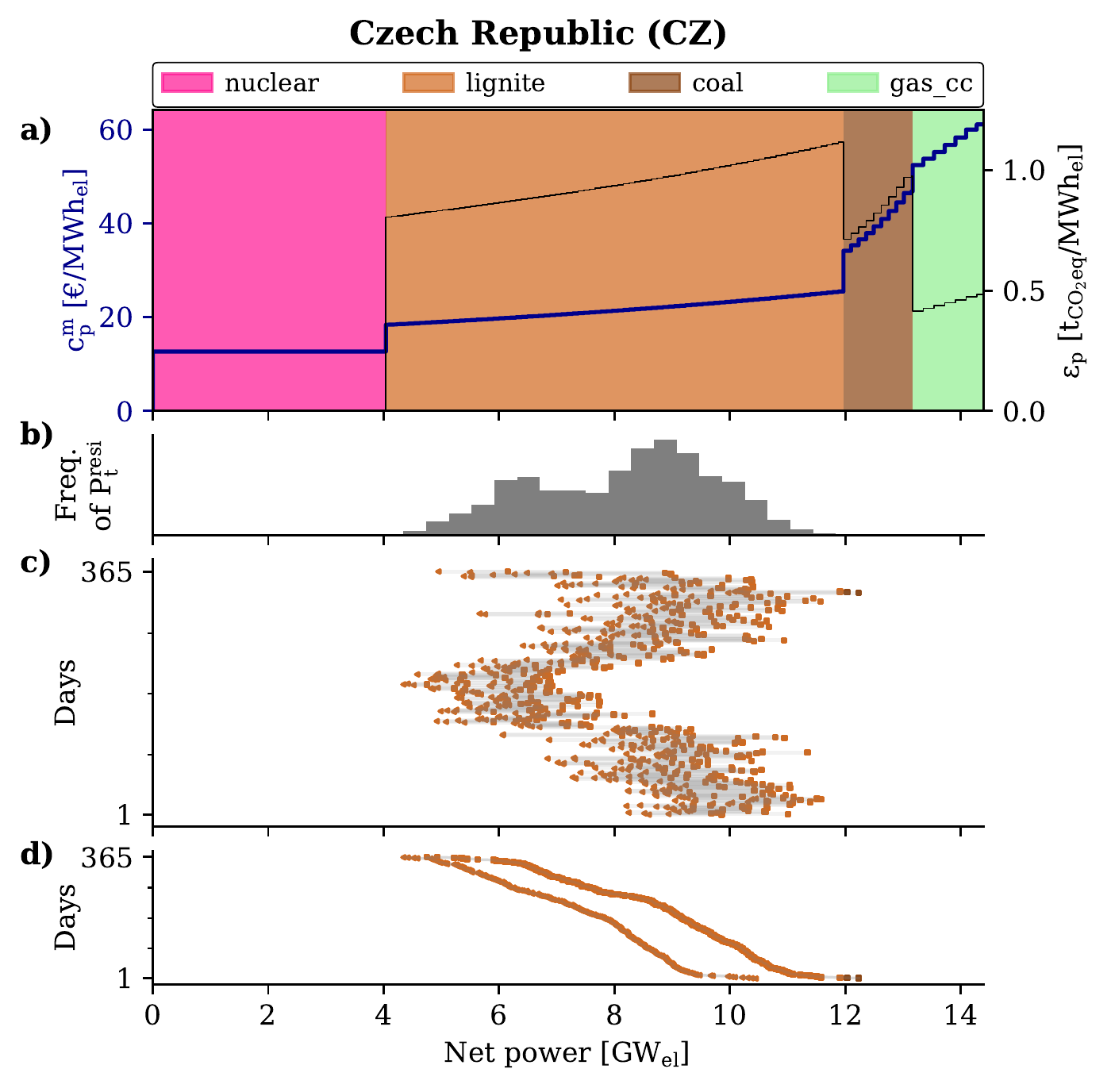}
	\includegraphics[width=0.245\linewidth]{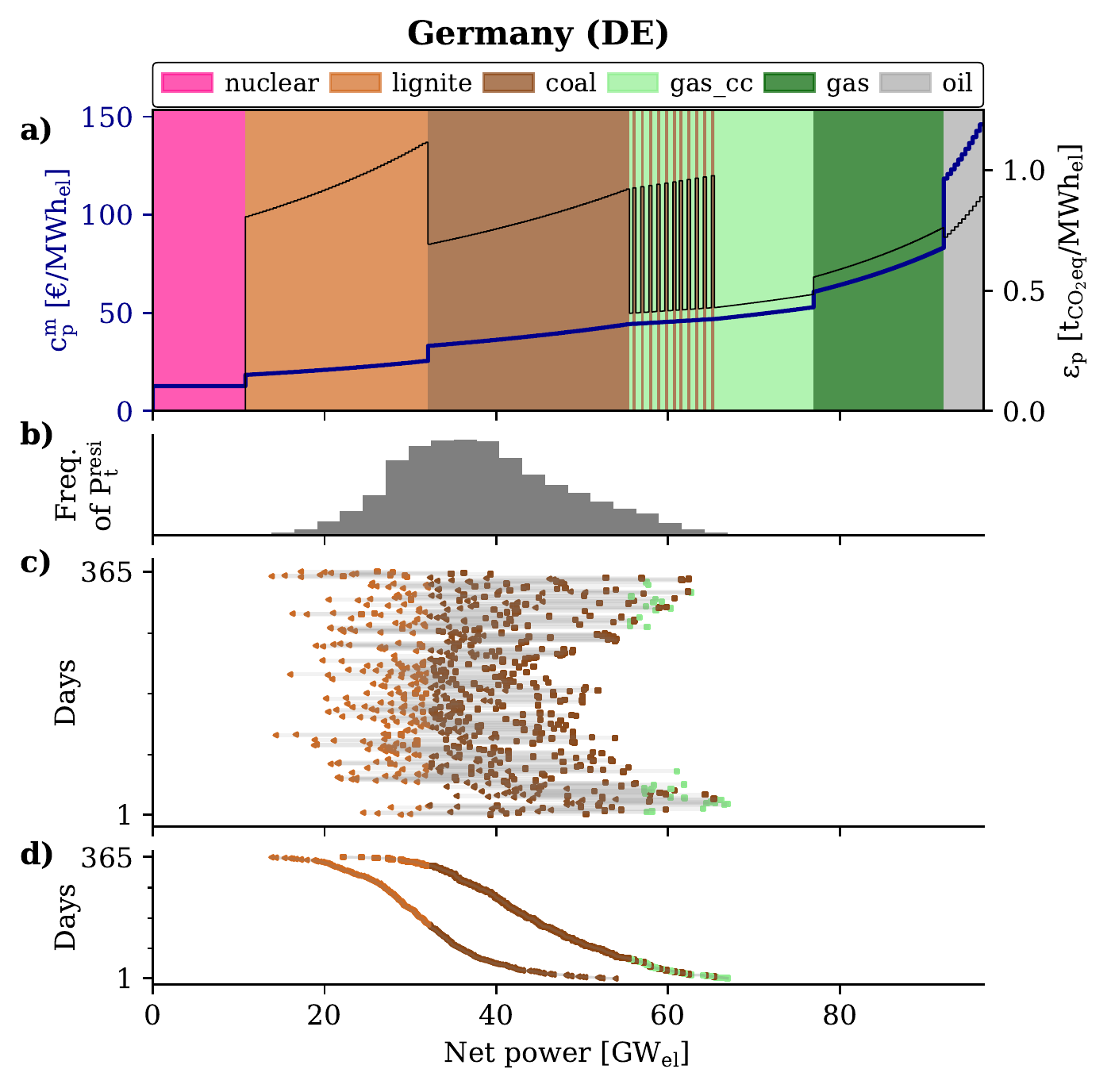}
	\includegraphics[width=0.245\linewidth]{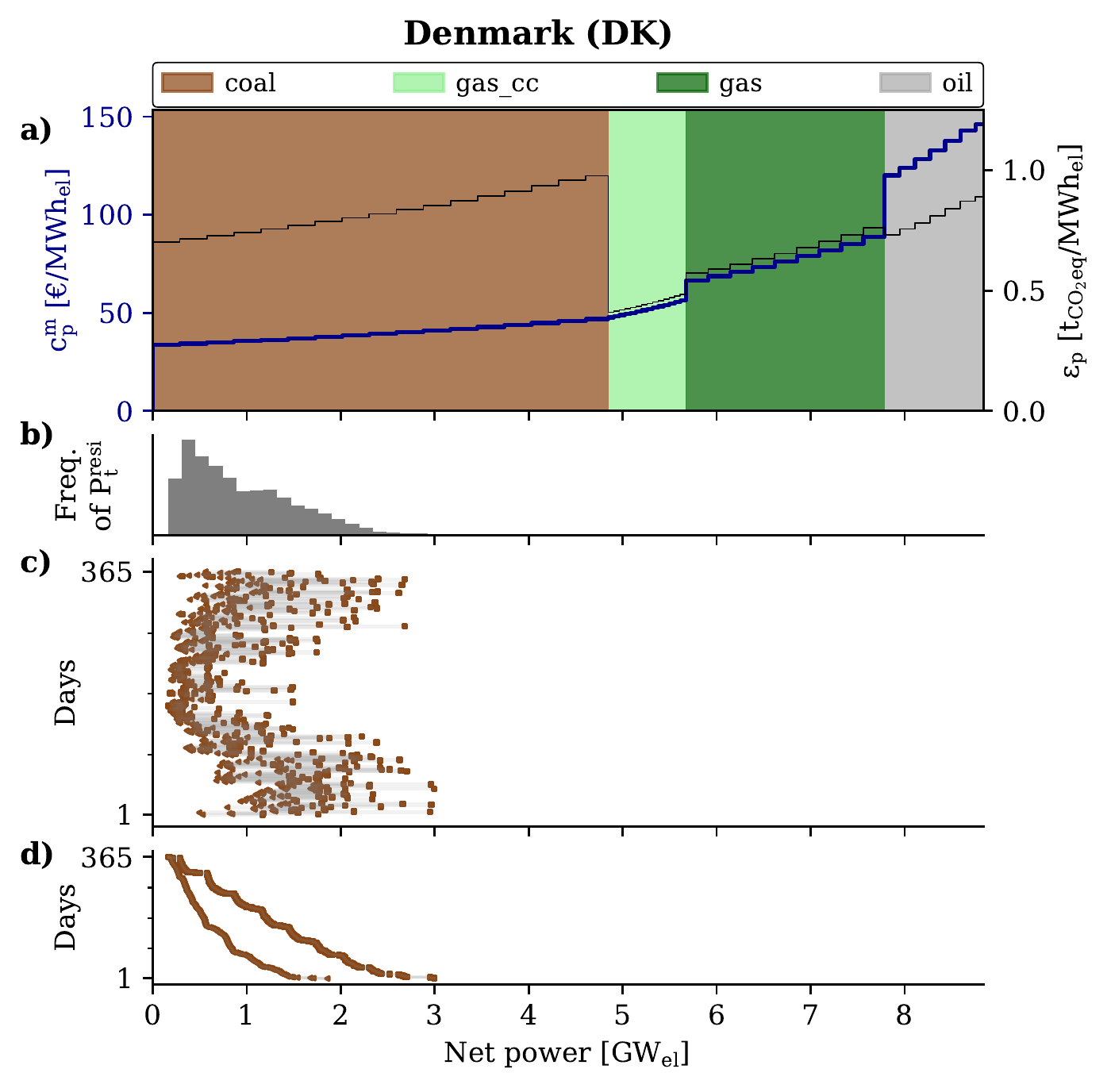}
	\includegraphics[width=0.245\linewidth]{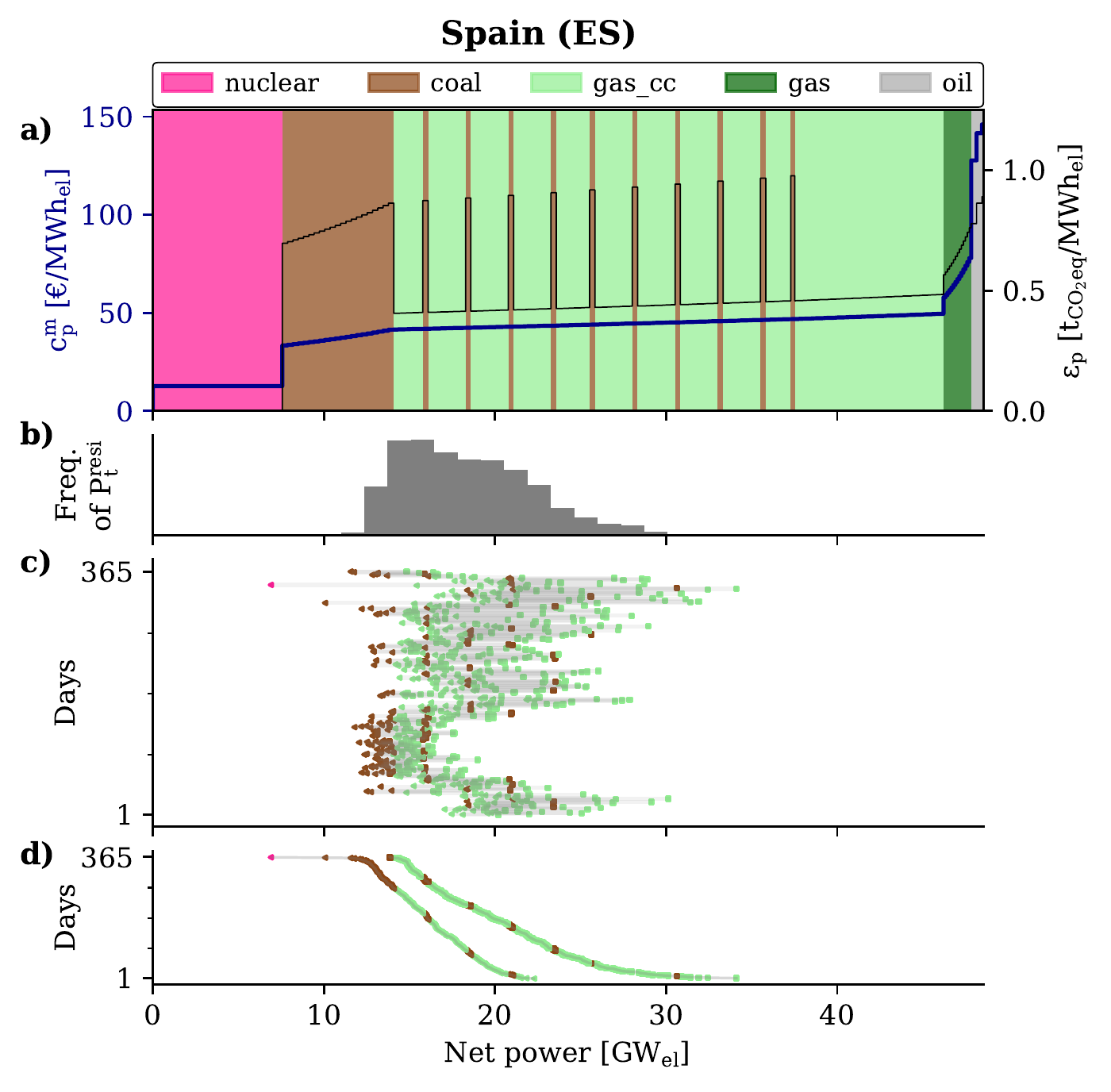}
	\includegraphics[width=0.245\linewidth]{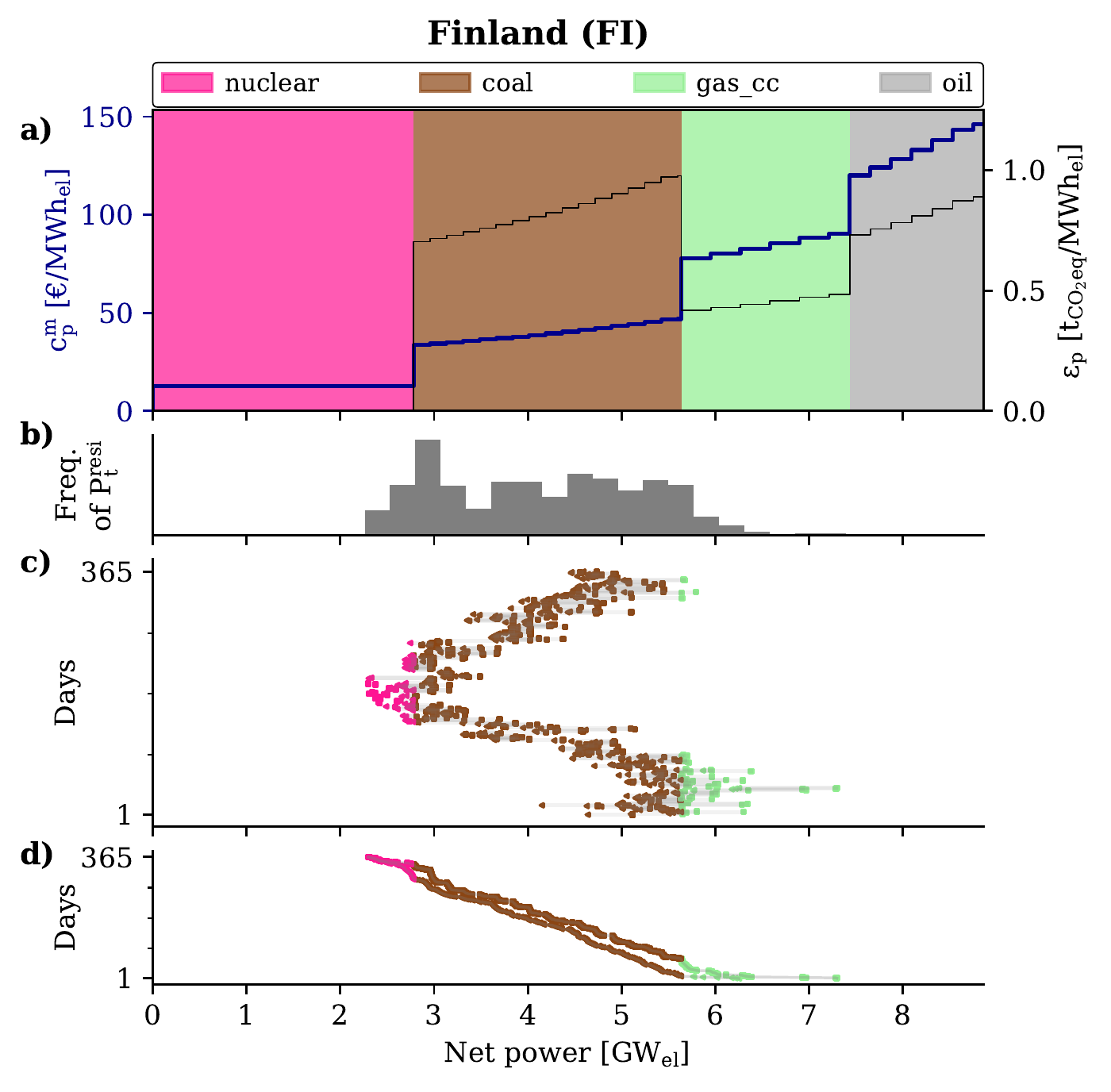}
	\includegraphics[width=0.245\linewidth]{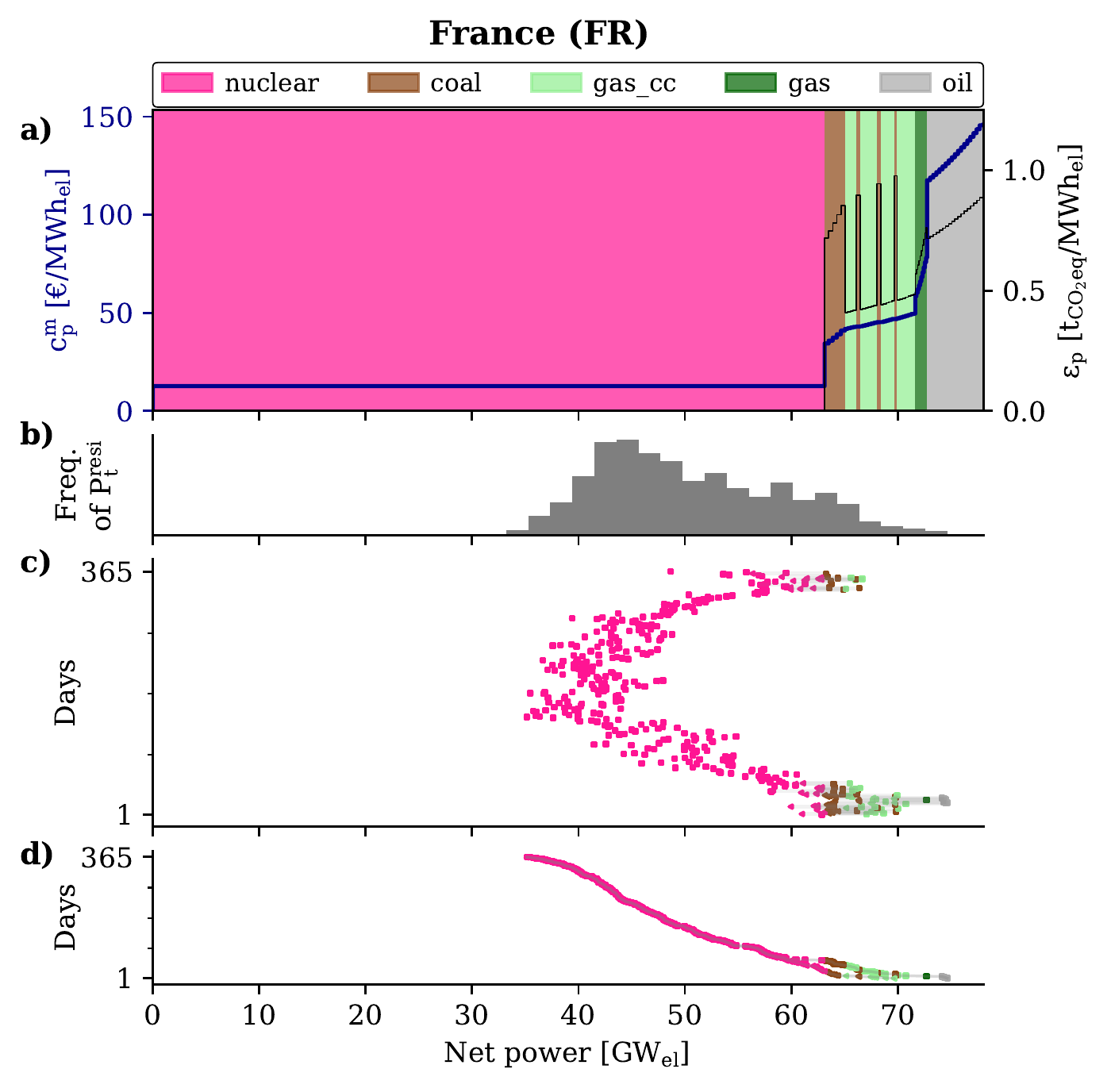}
	\includegraphics[width=0.245\linewidth]{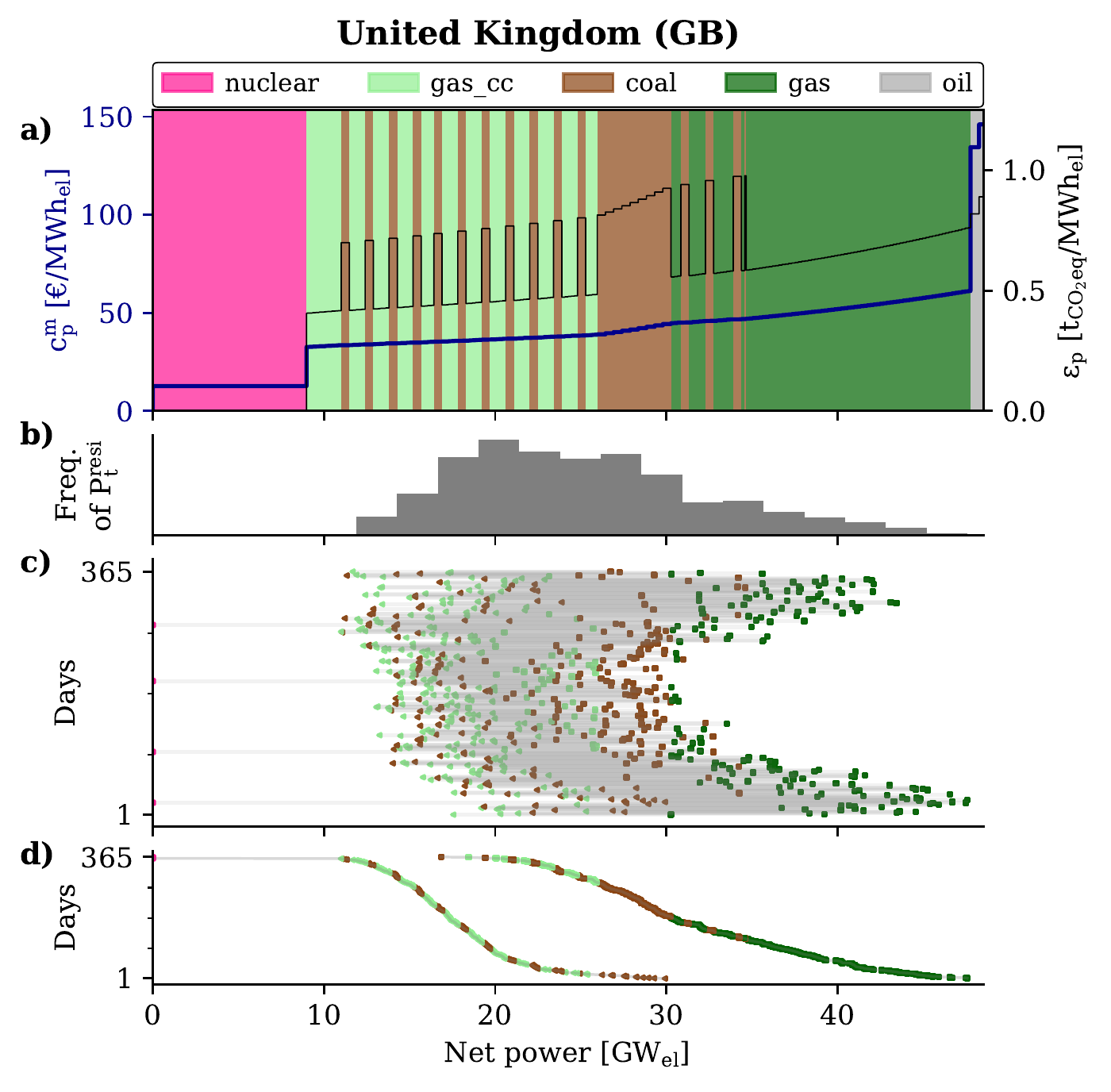}
	\includegraphics[width=0.245\linewidth]{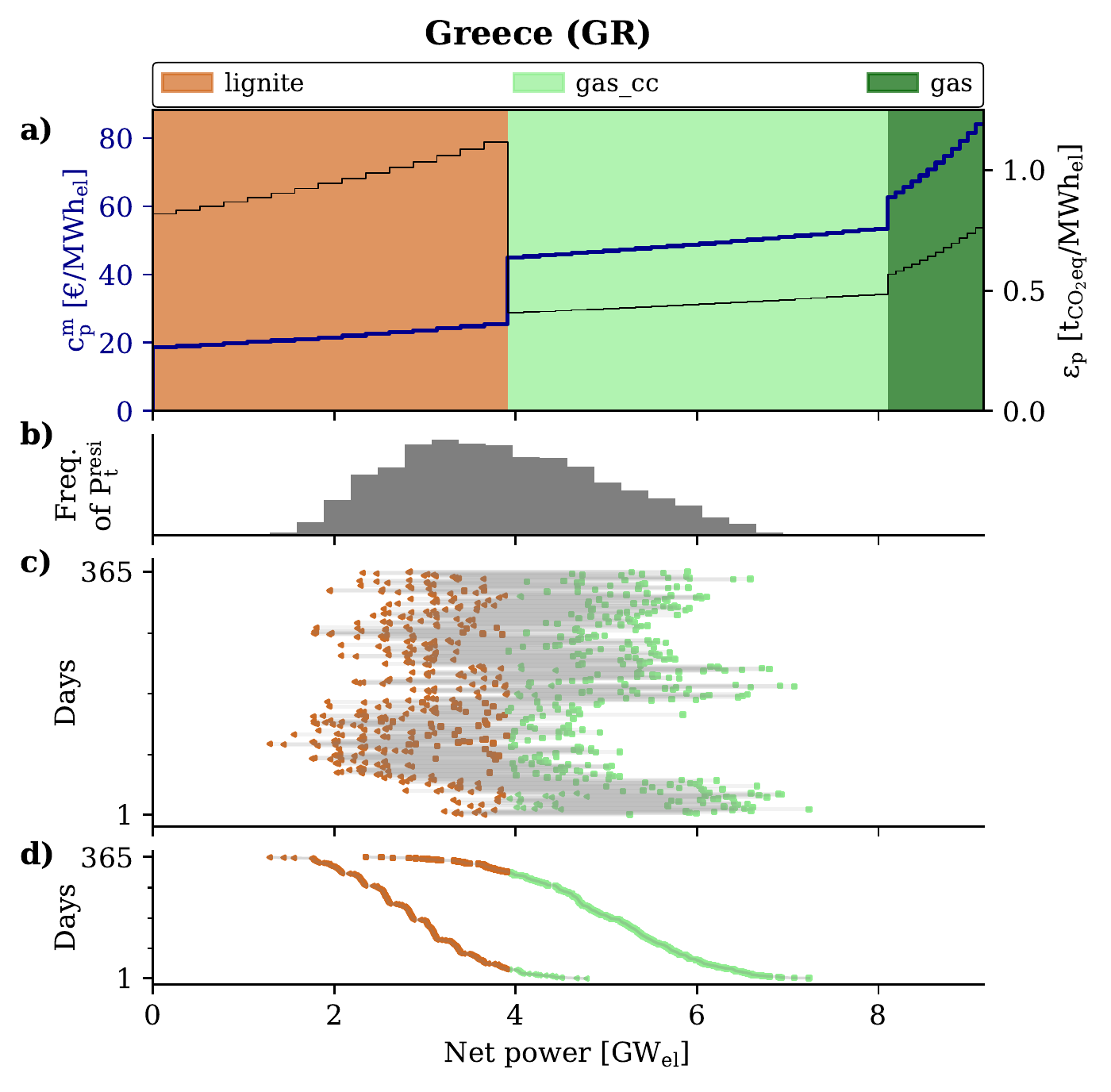}
	\includegraphics[width=0.245\linewidth]{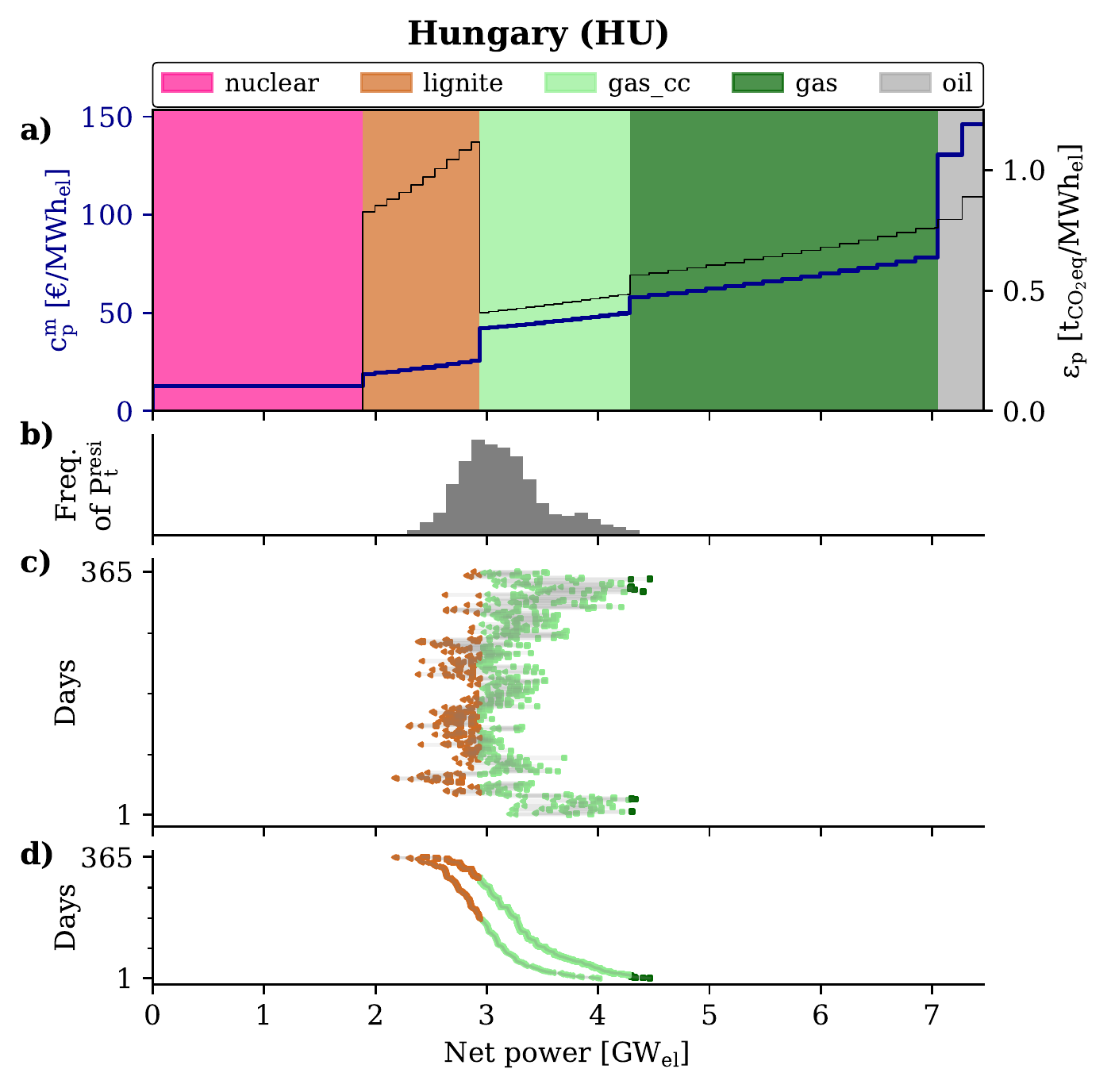}
	\includegraphics[width=0.245\linewidth]{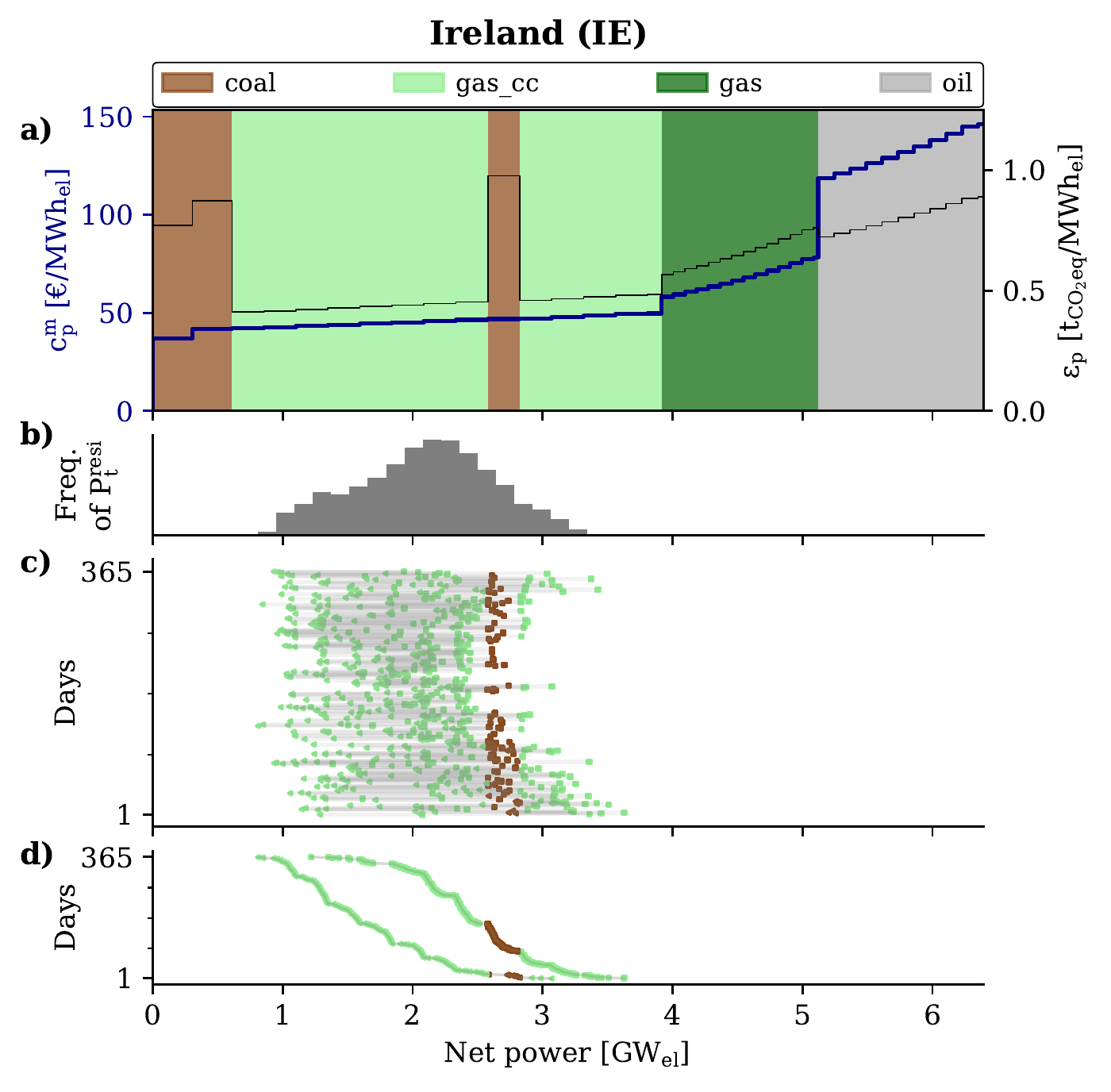}
	\includegraphics[width=0.245\linewidth]{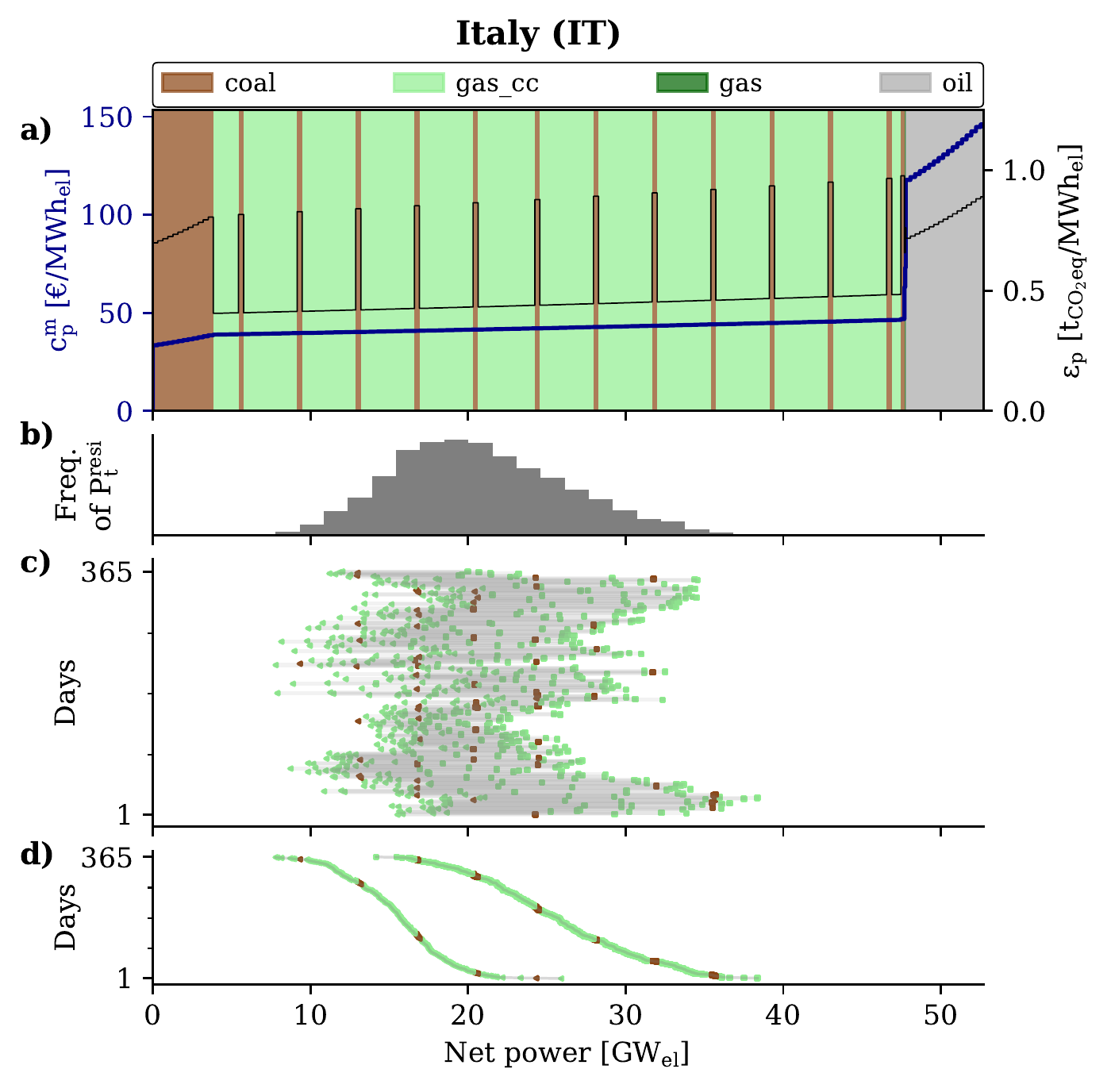}
	\includegraphics[width=0.245\linewidth]{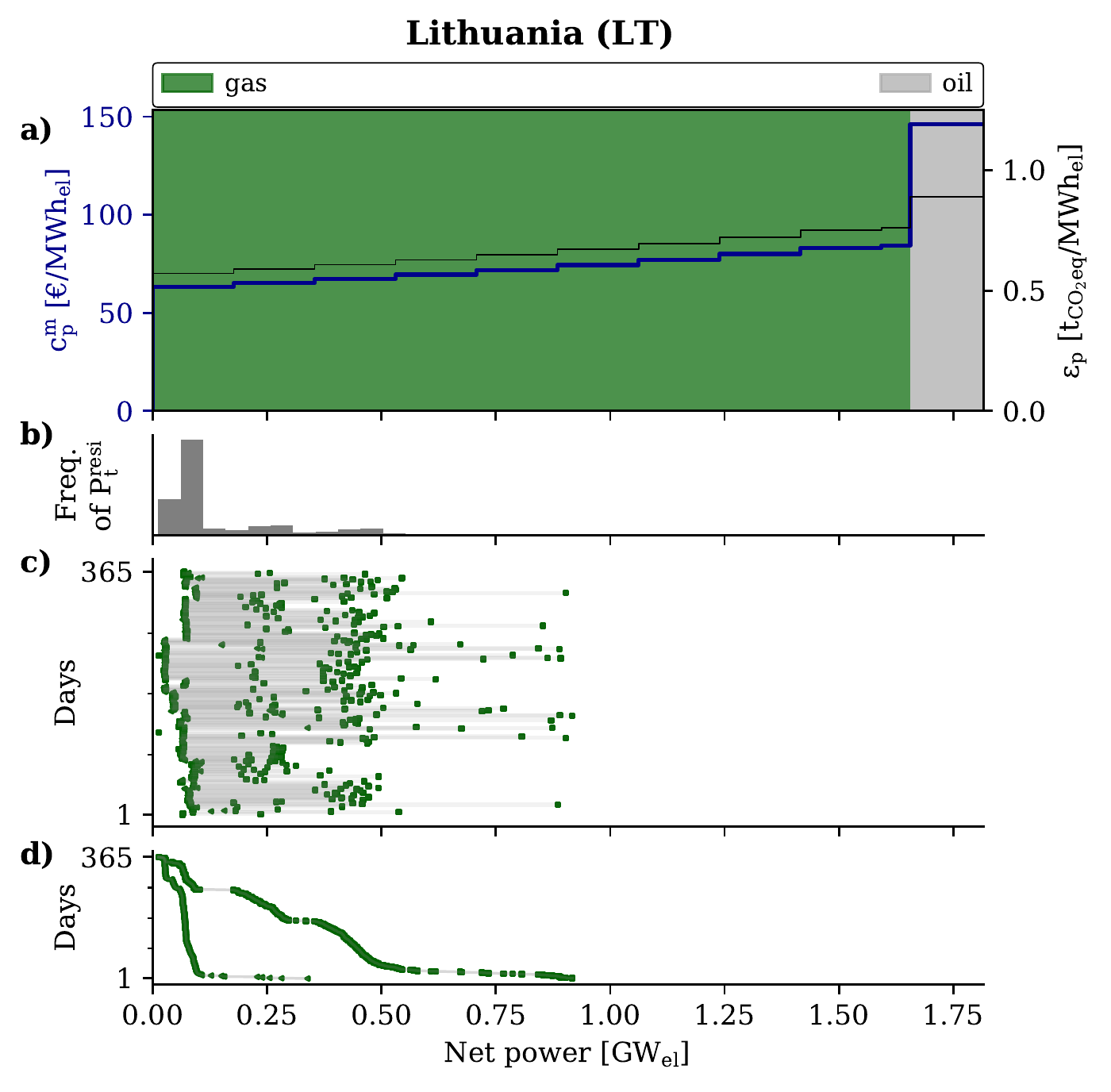}
	\includegraphics[width=0.245\linewidth]{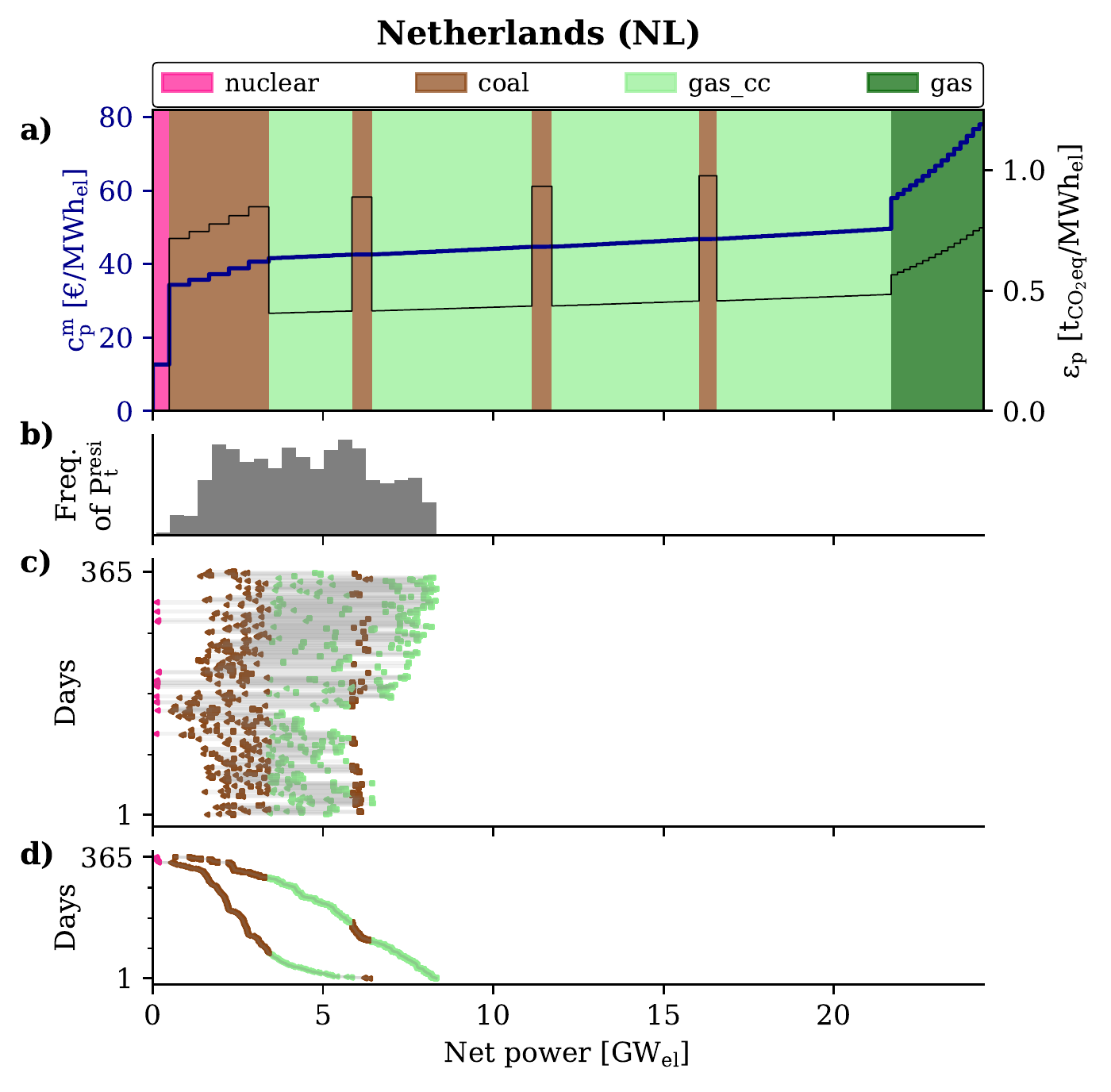}
	\includegraphics[width=0.245\linewidth]{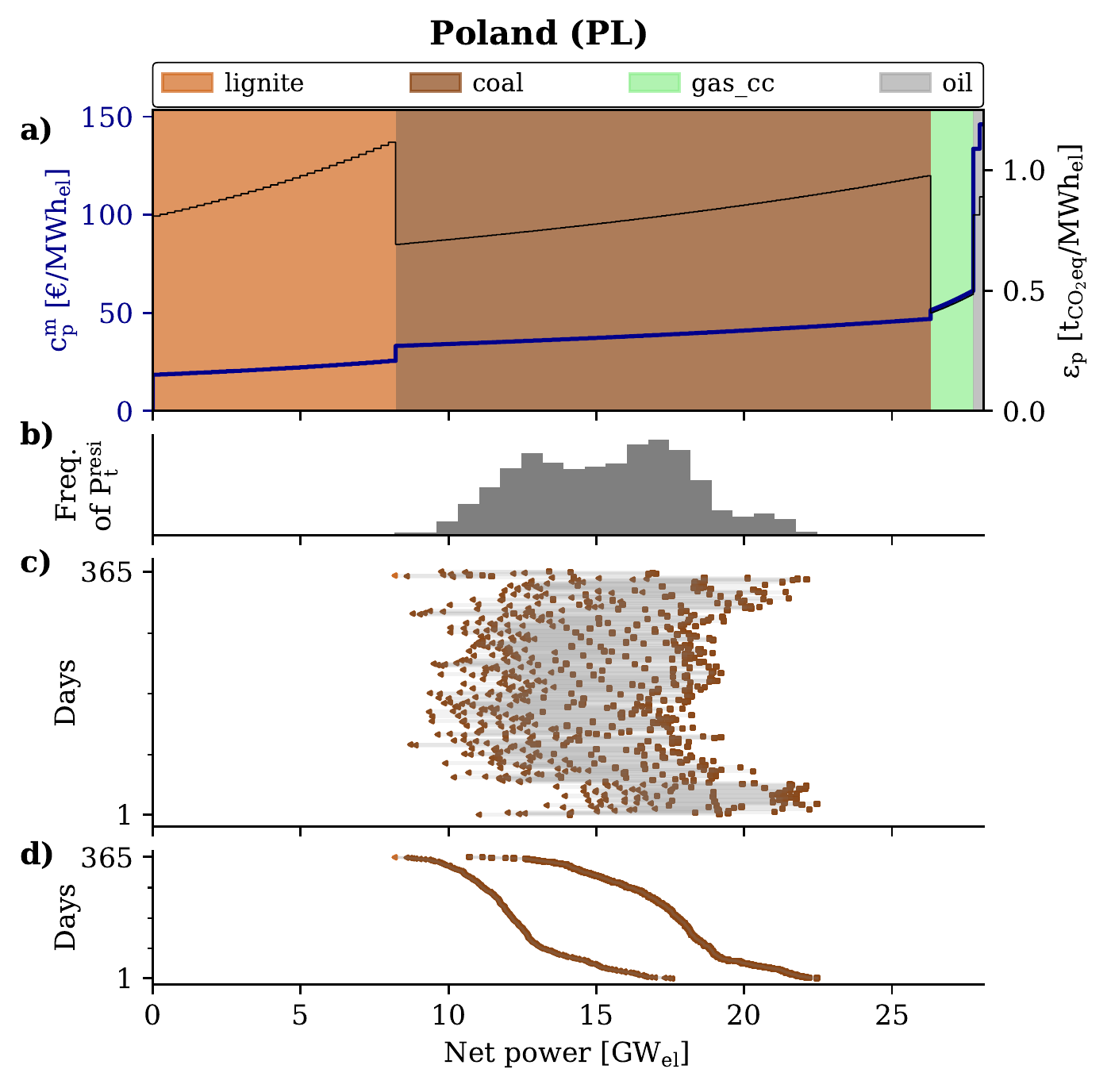}
	\includegraphics[width=0.245\linewidth]{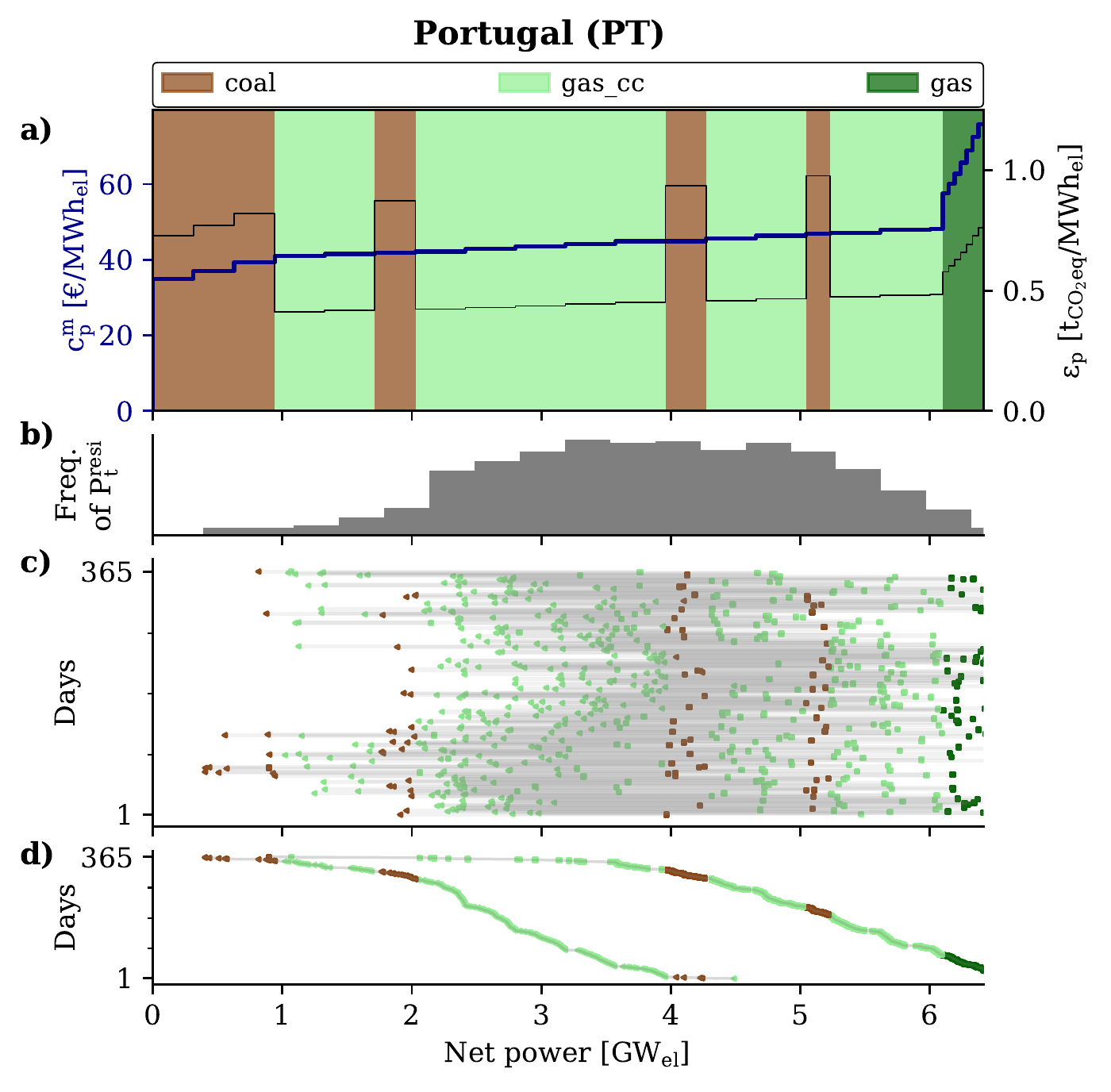}
	\includegraphics[width=0.245\linewidth]{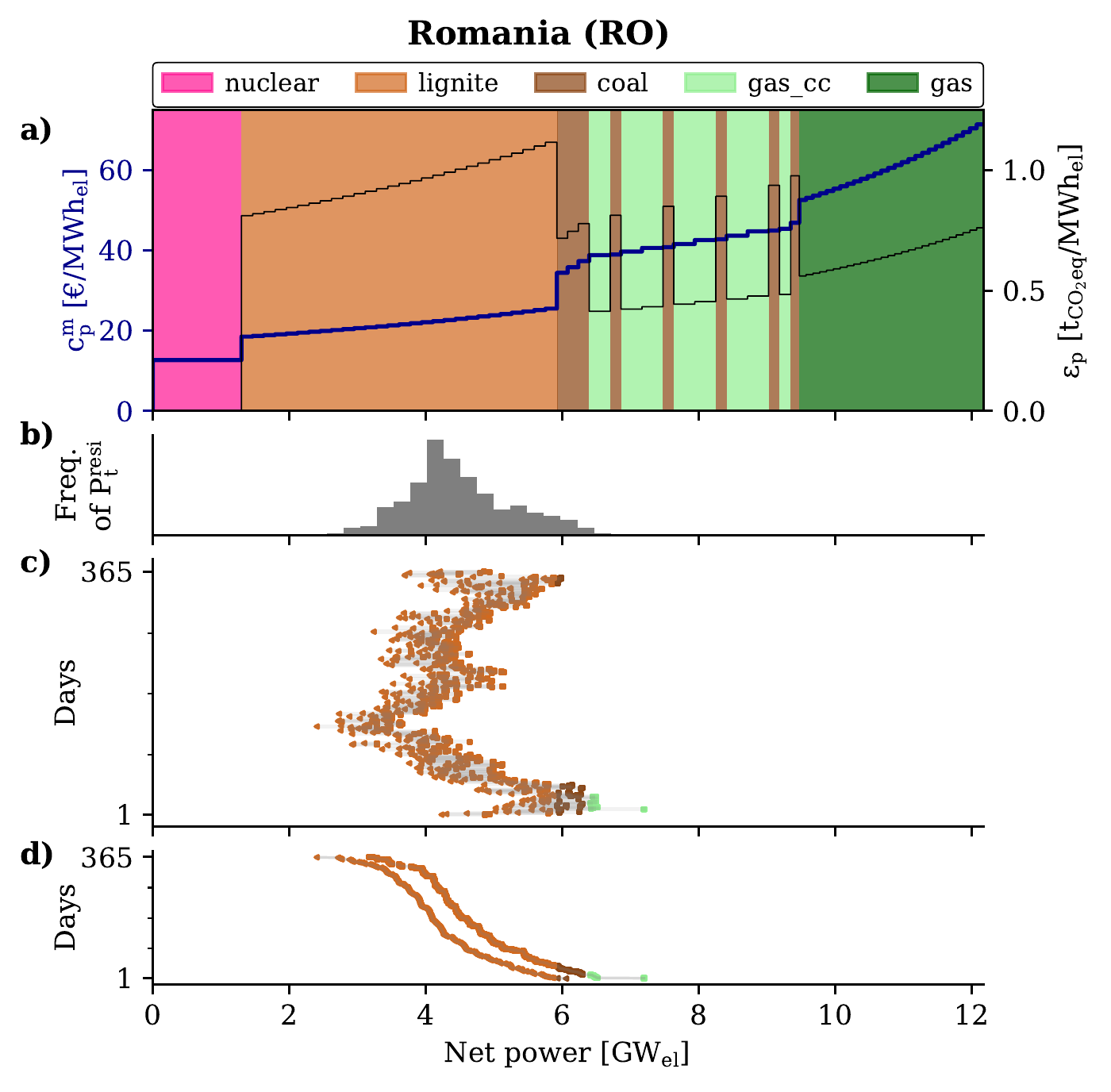}
	\includegraphics[width=0.245\linewidth]{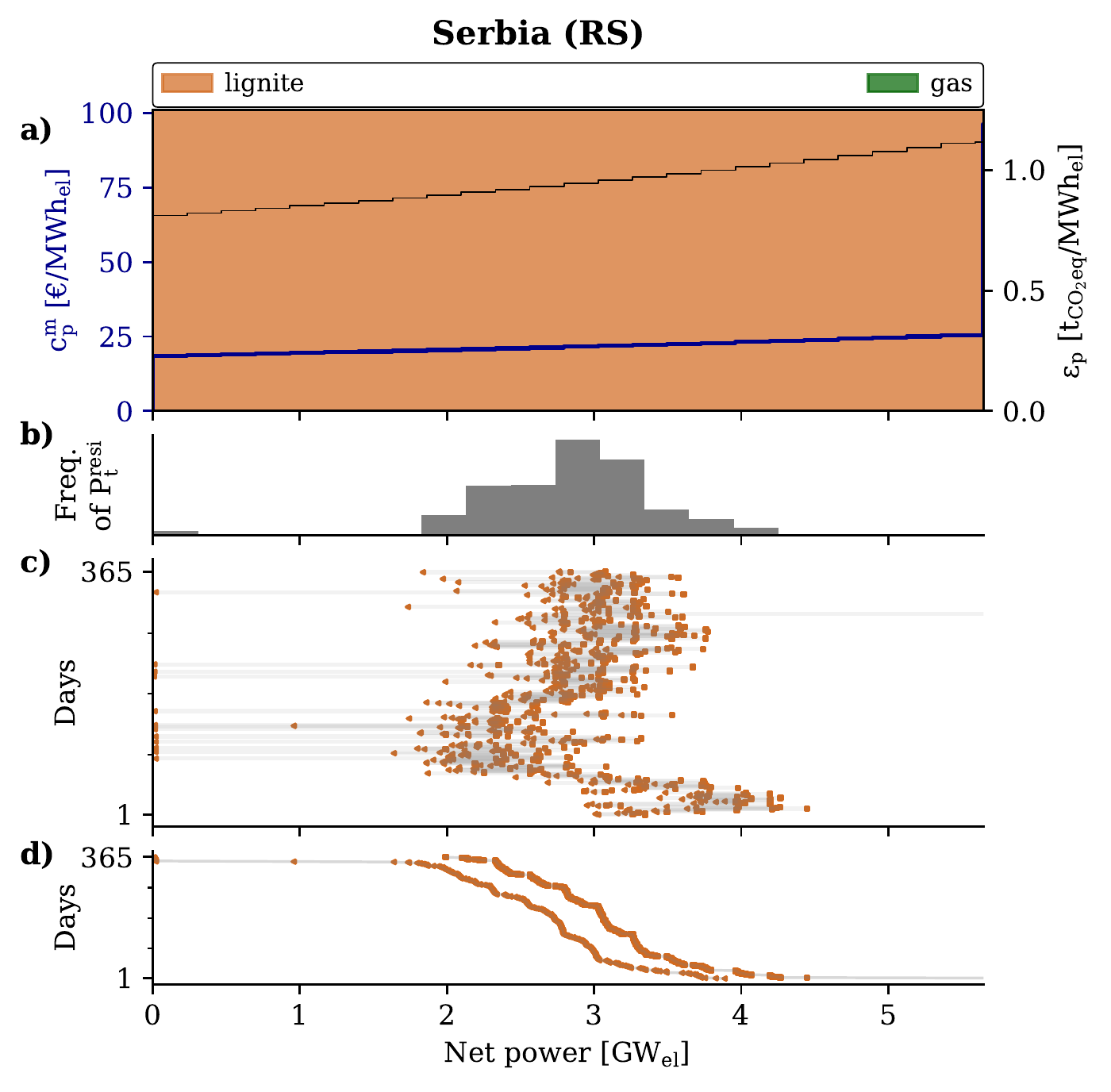}
	\includegraphics[width=0.245\linewidth]{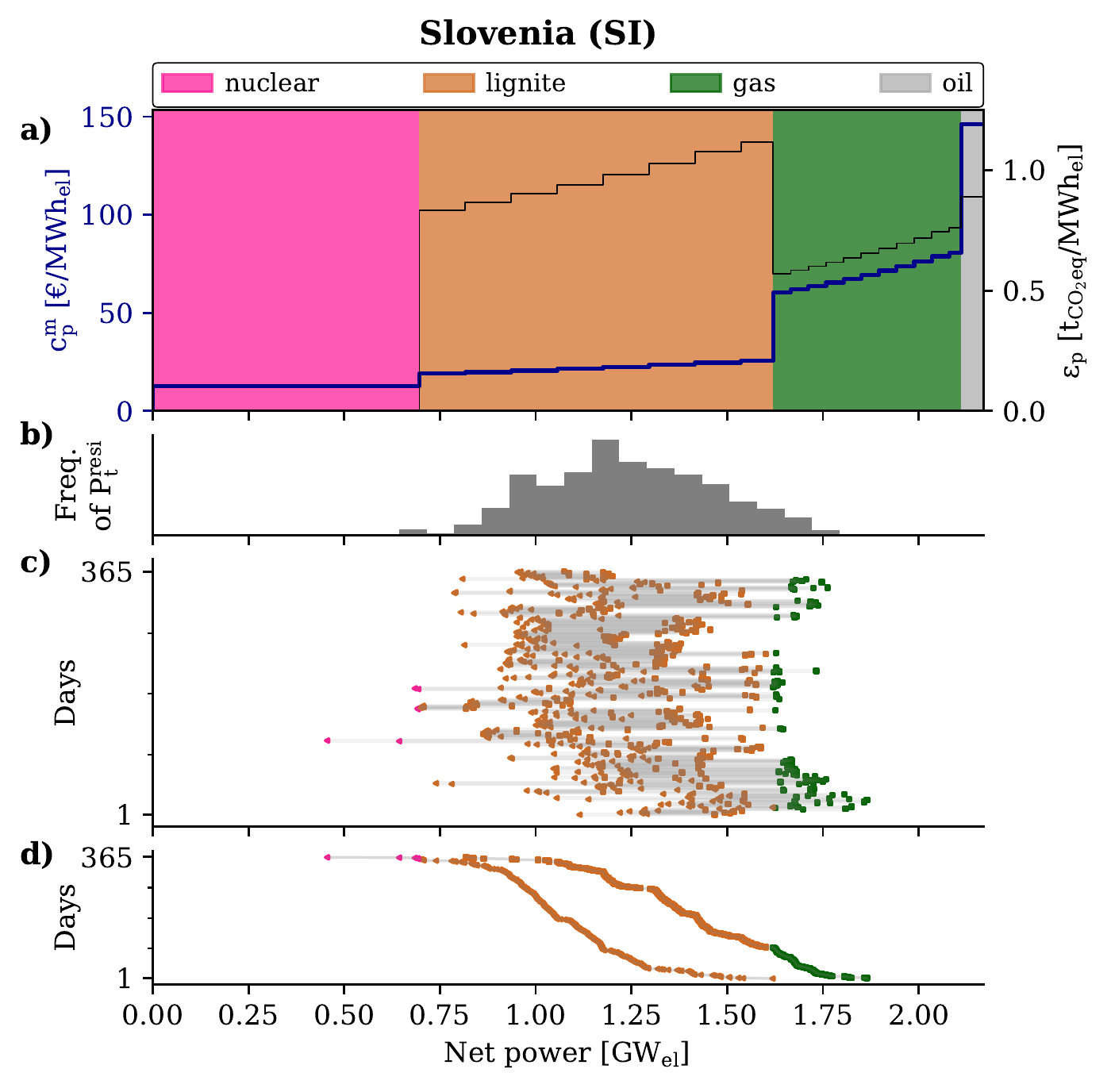}
	\caption{Merit orders and load shift analysis of analyzed European countries resulting from PWL-method for the year 2017: \textbf{a)} Merit order with marginal prices $ c_p^m $ (left) and emission intensities $ \varepsilon_p $ (right), \textbf{b)} histogram of residual load $ P_t^\mathrm{resi} $, \textbf{c)} load shifts of all days of the year (squares indicating sources and triangles indicating sinks), and \textbf{d)} inverted load duration curves of sources (right) and sinks (left)). The x-axis (net generation power in GW\textsubscript{el}) is shared across the four subplots.}
	\label{fig:mols2017}
\end{figure}

\begin{figure}[!h]
	\centering
	\textbf{2018}\par\medskip
	\includegraphics[width=0.245\linewidth]{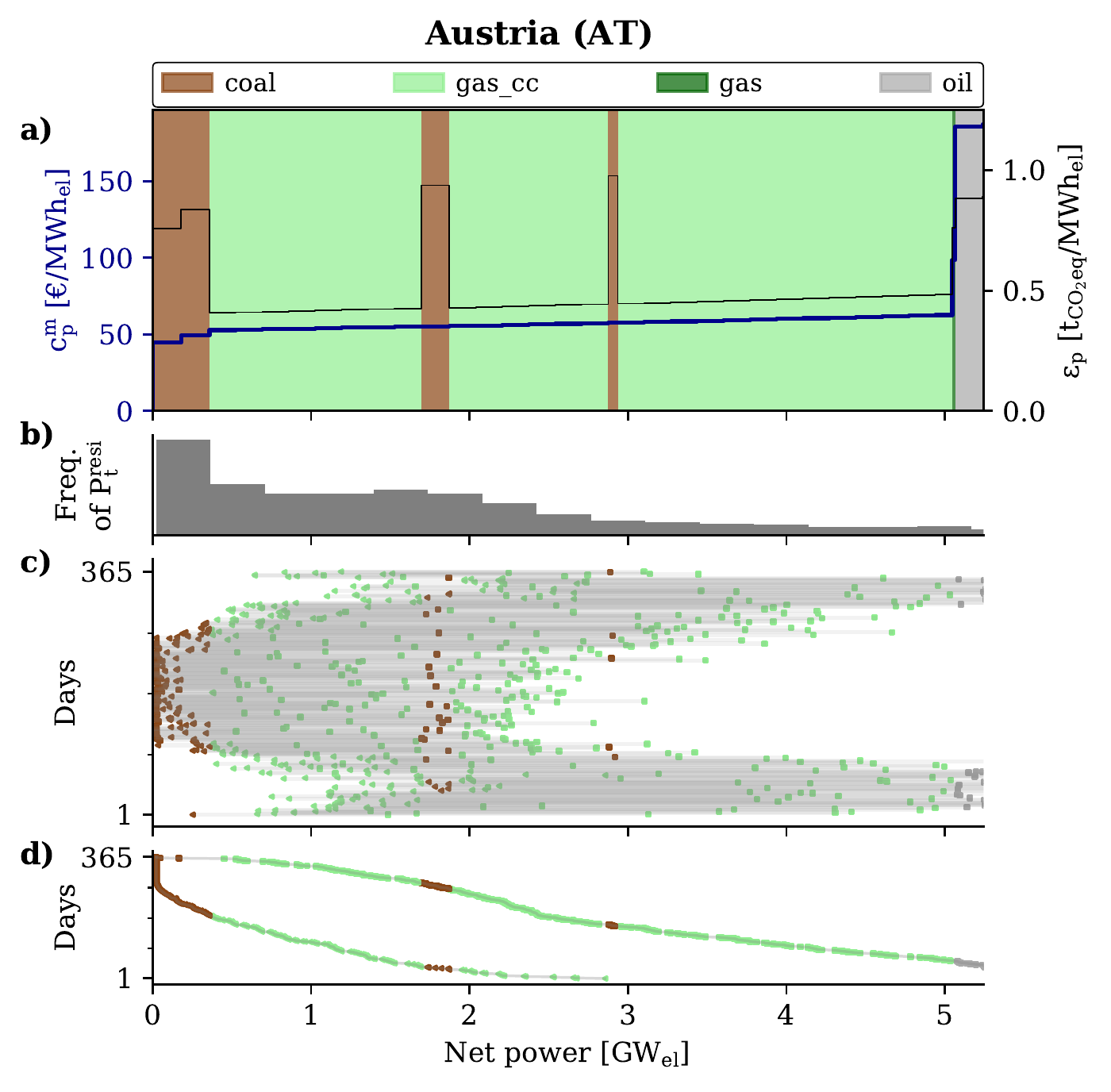}
	\includegraphics[width=0.245\linewidth]{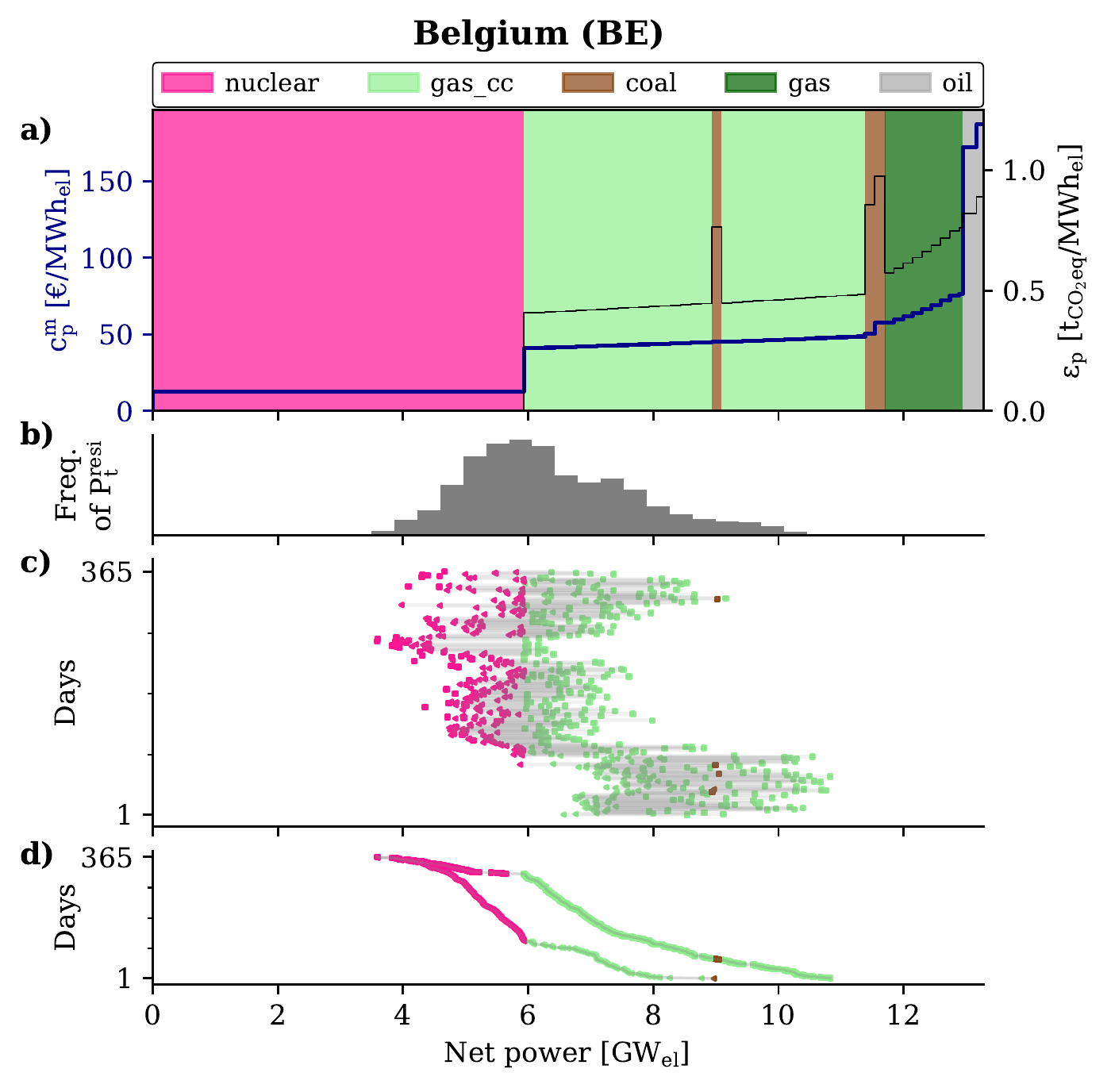}
	\includegraphics[width=0.245\linewidth]{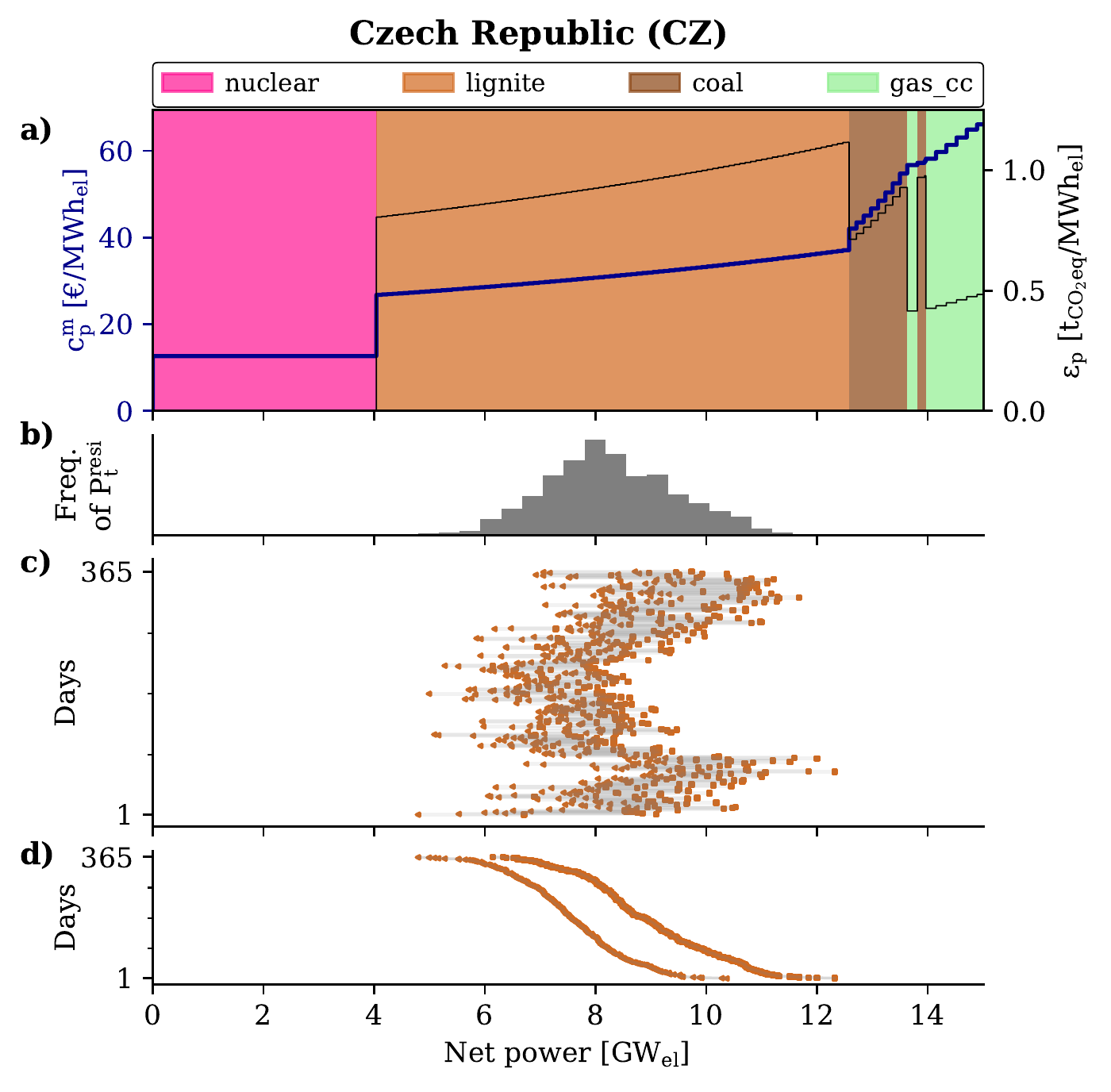}
	\includegraphics[width=0.245\linewidth]{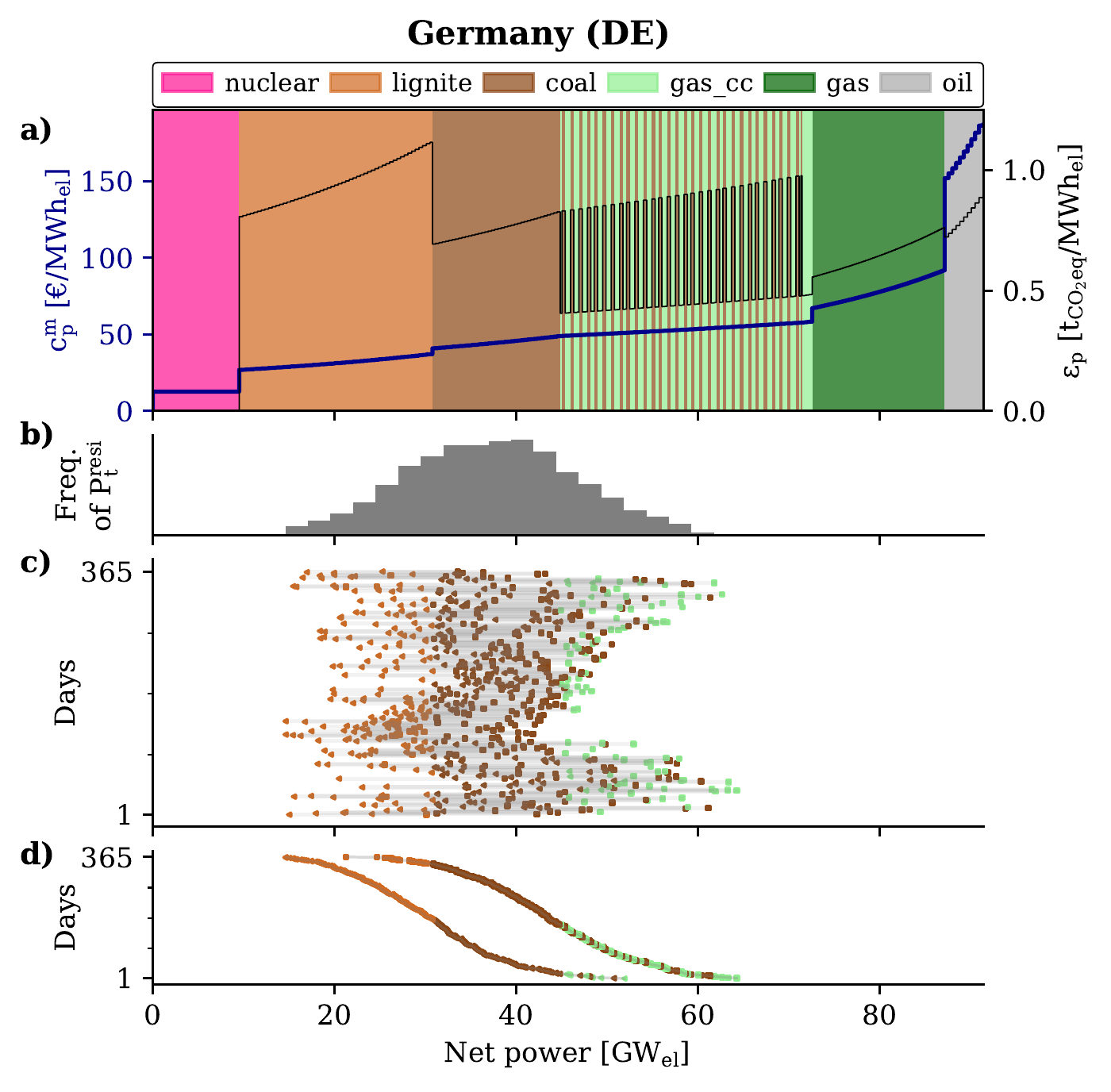}
	\includegraphics[width=0.245\linewidth]{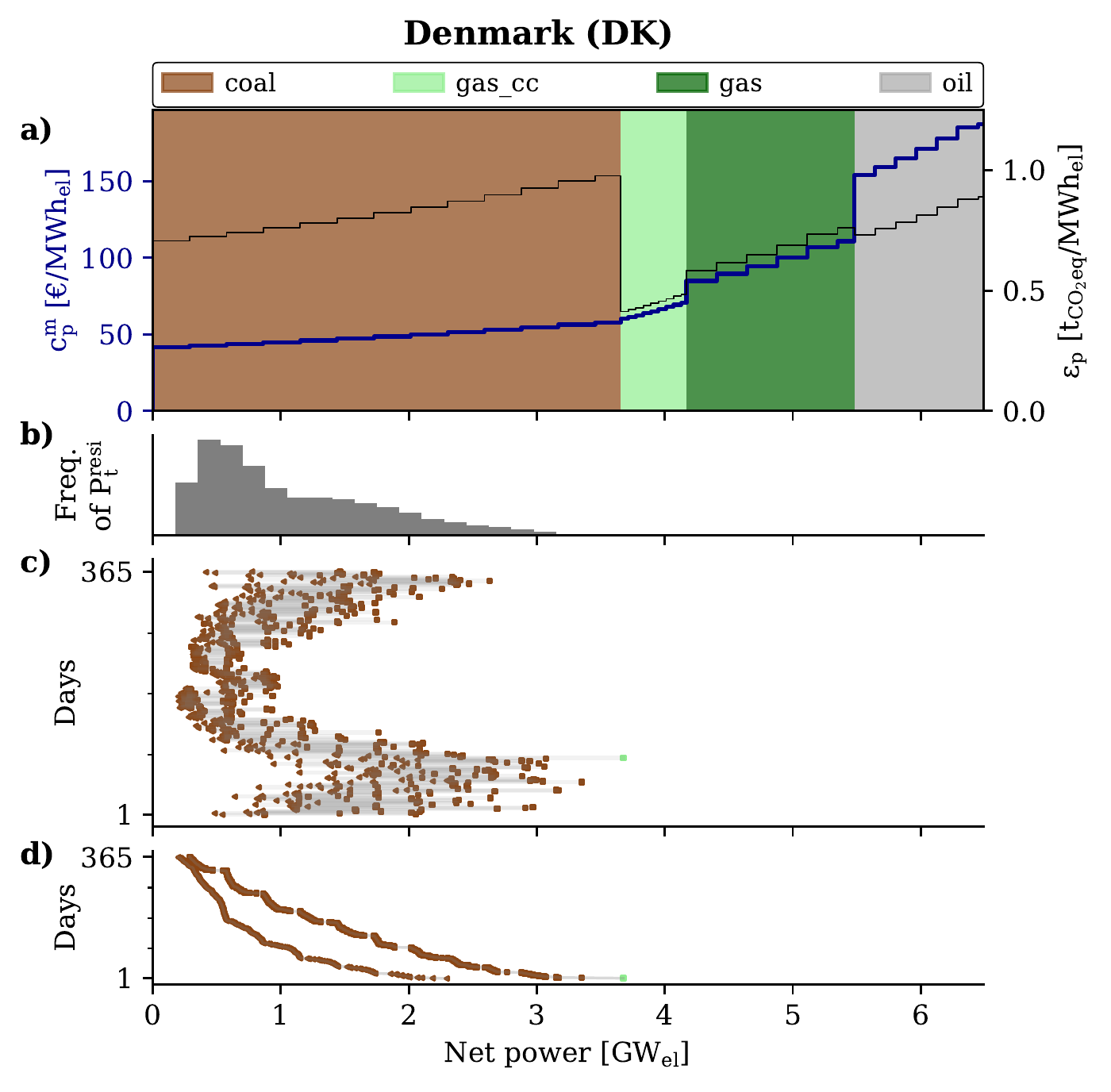}
	\includegraphics[width=0.245\linewidth]{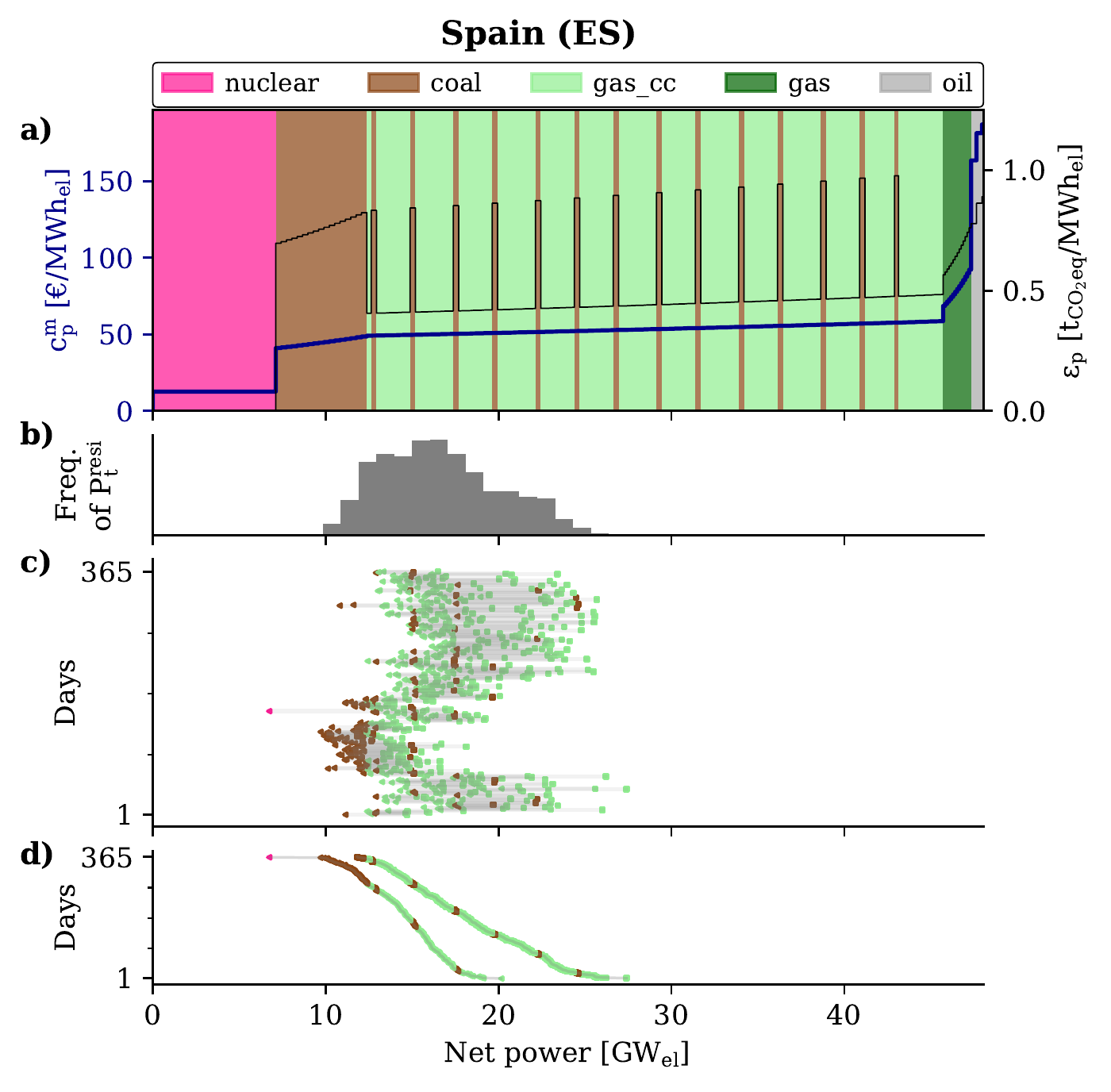}
	\includegraphics[width=0.245\linewidth]{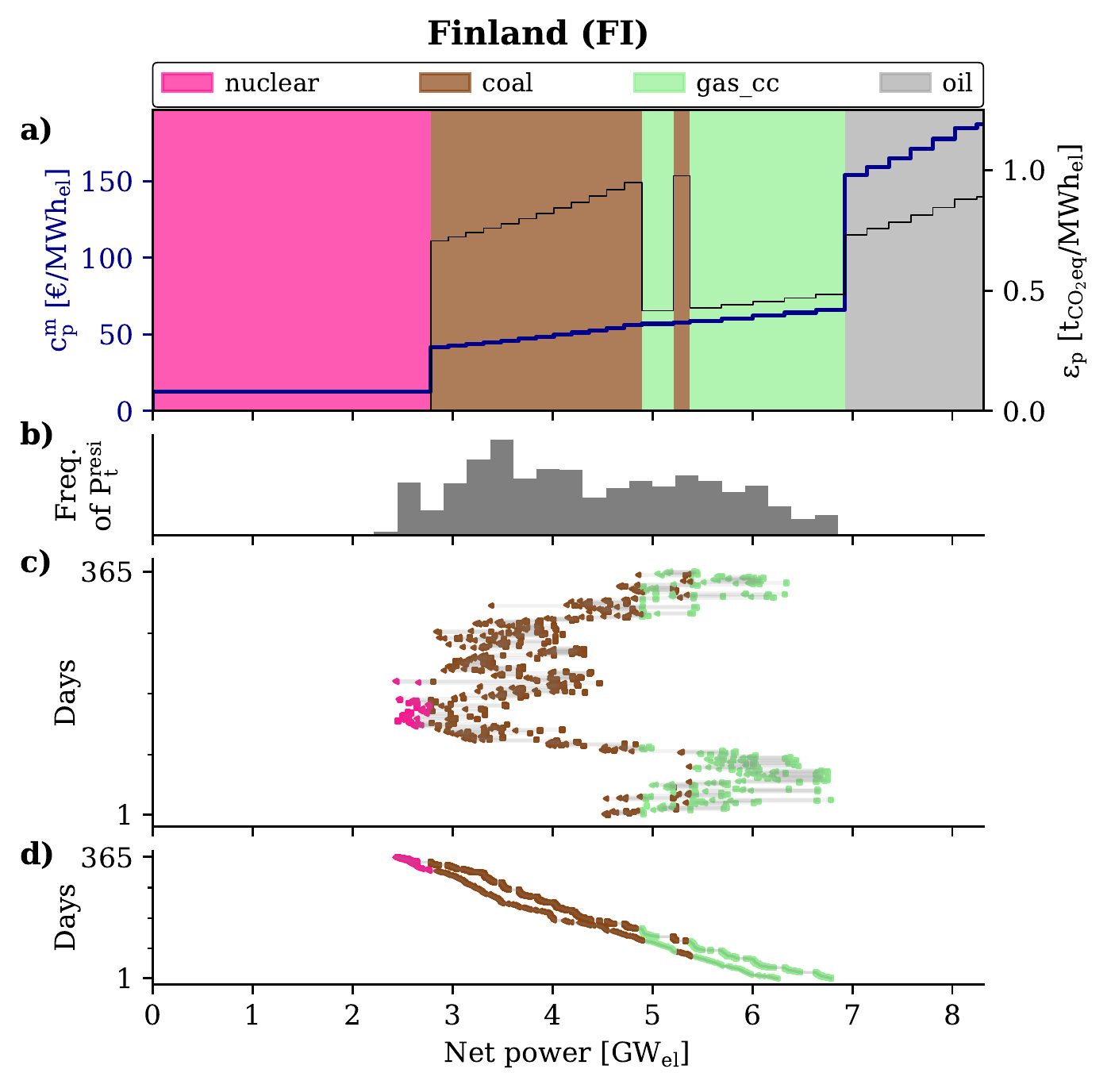}
	\includegraphics[width=0.245\linewidth]{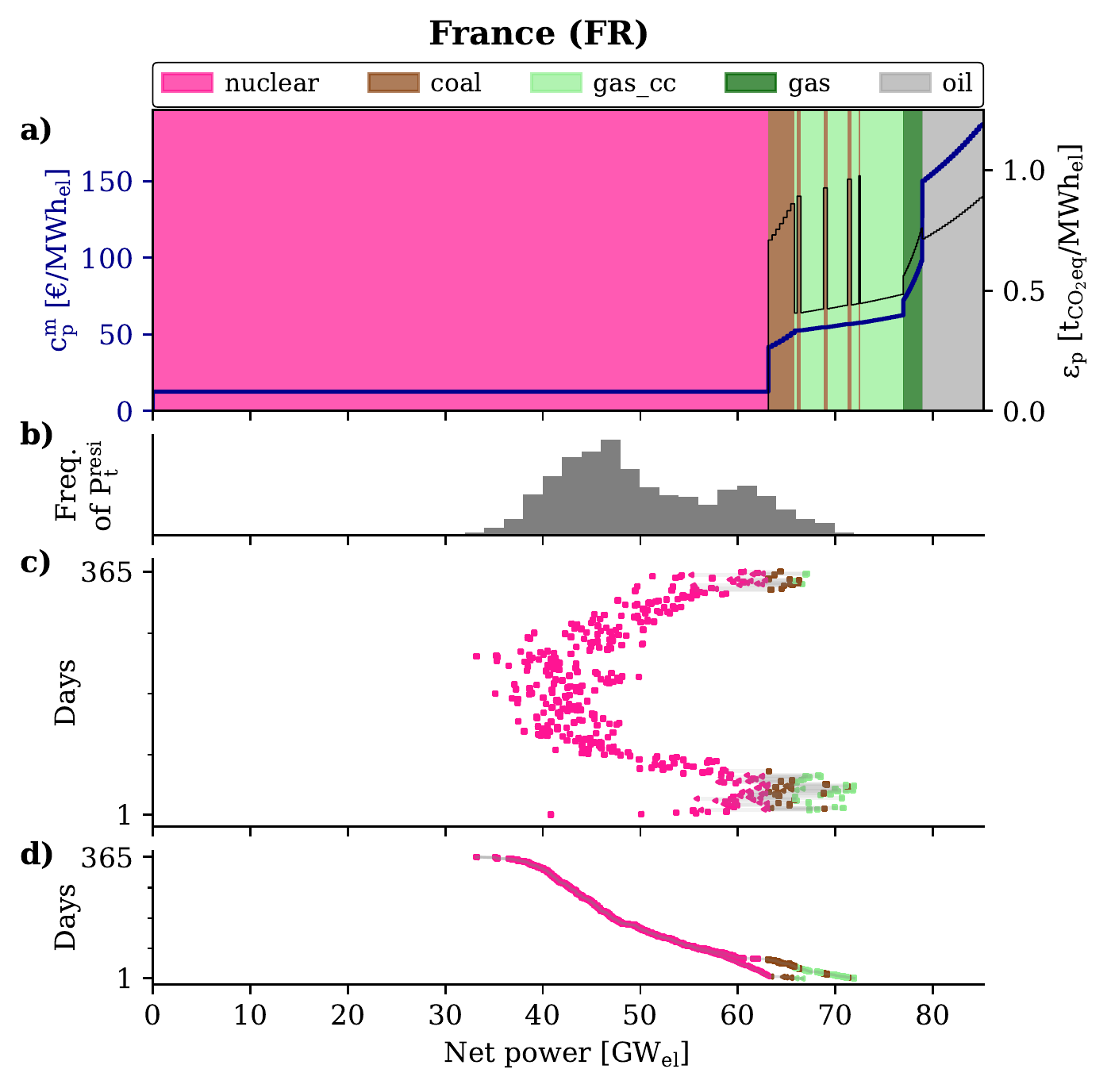}
	\includegraphics[width=0.245\linewidth]{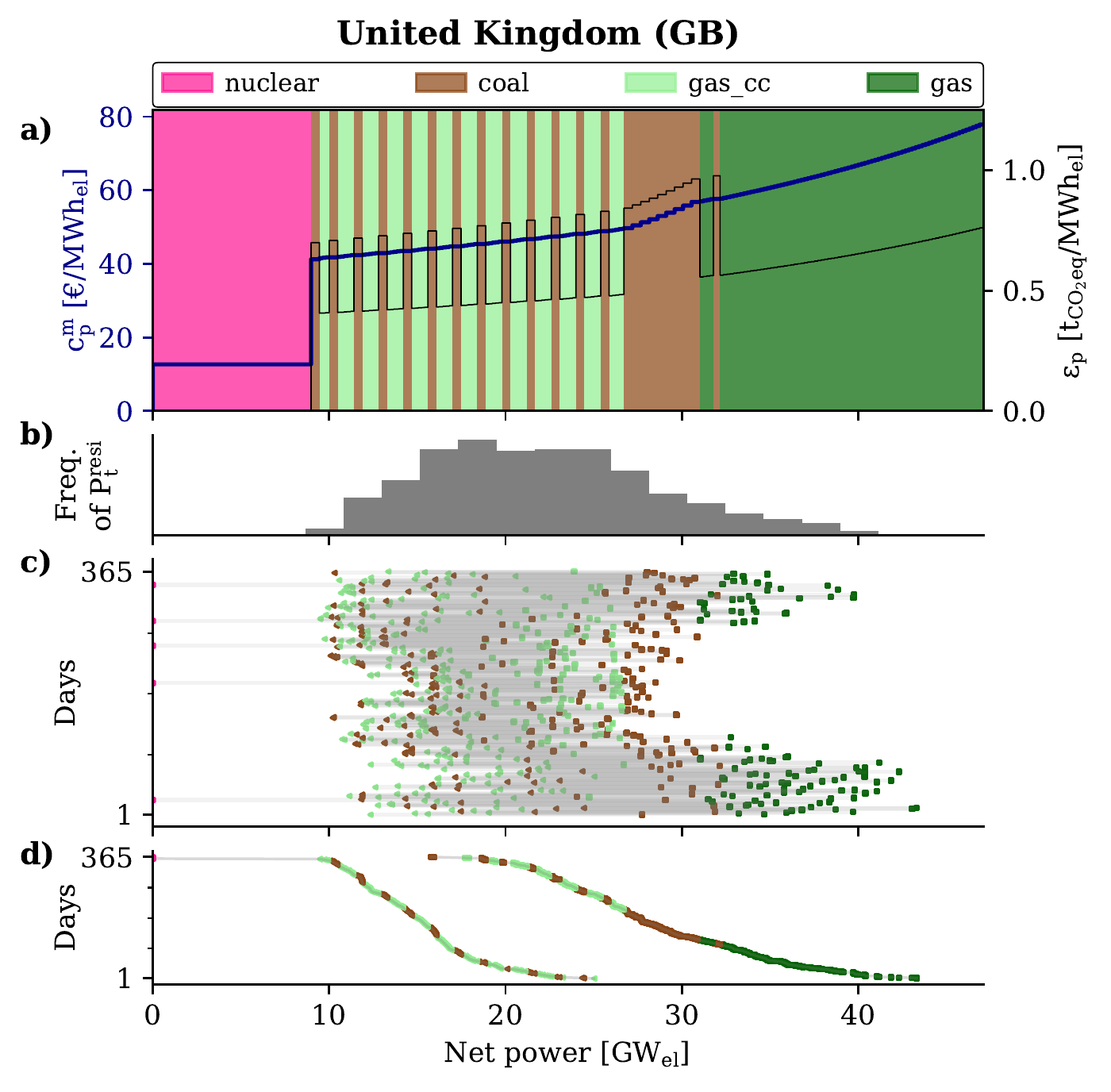}
	\includegraphics[width=0.245\linewidth]{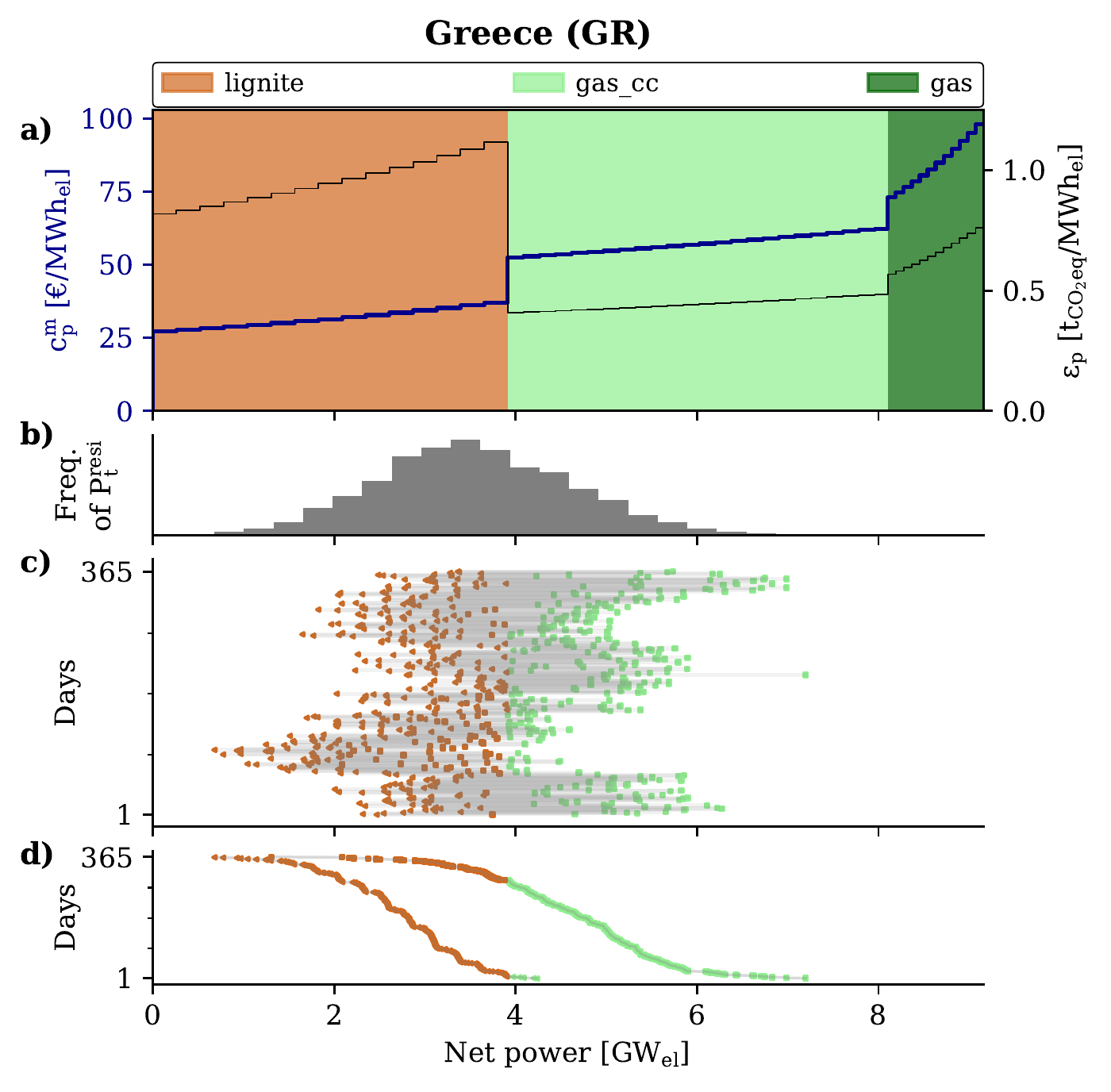}
	\includegraphics[width=0.245\linewidth]{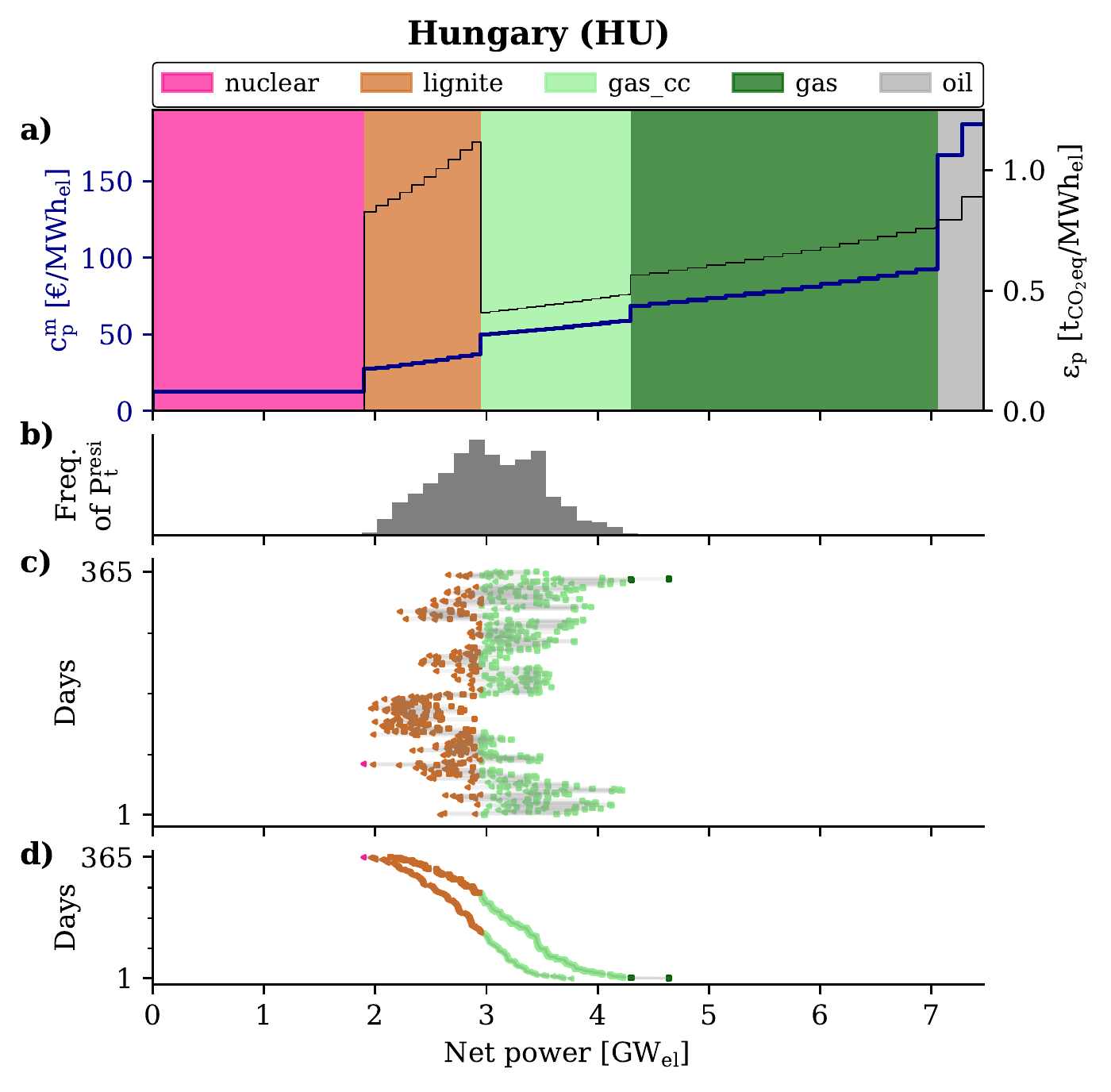}
	\includegraphics[width=0.245\linewidth]{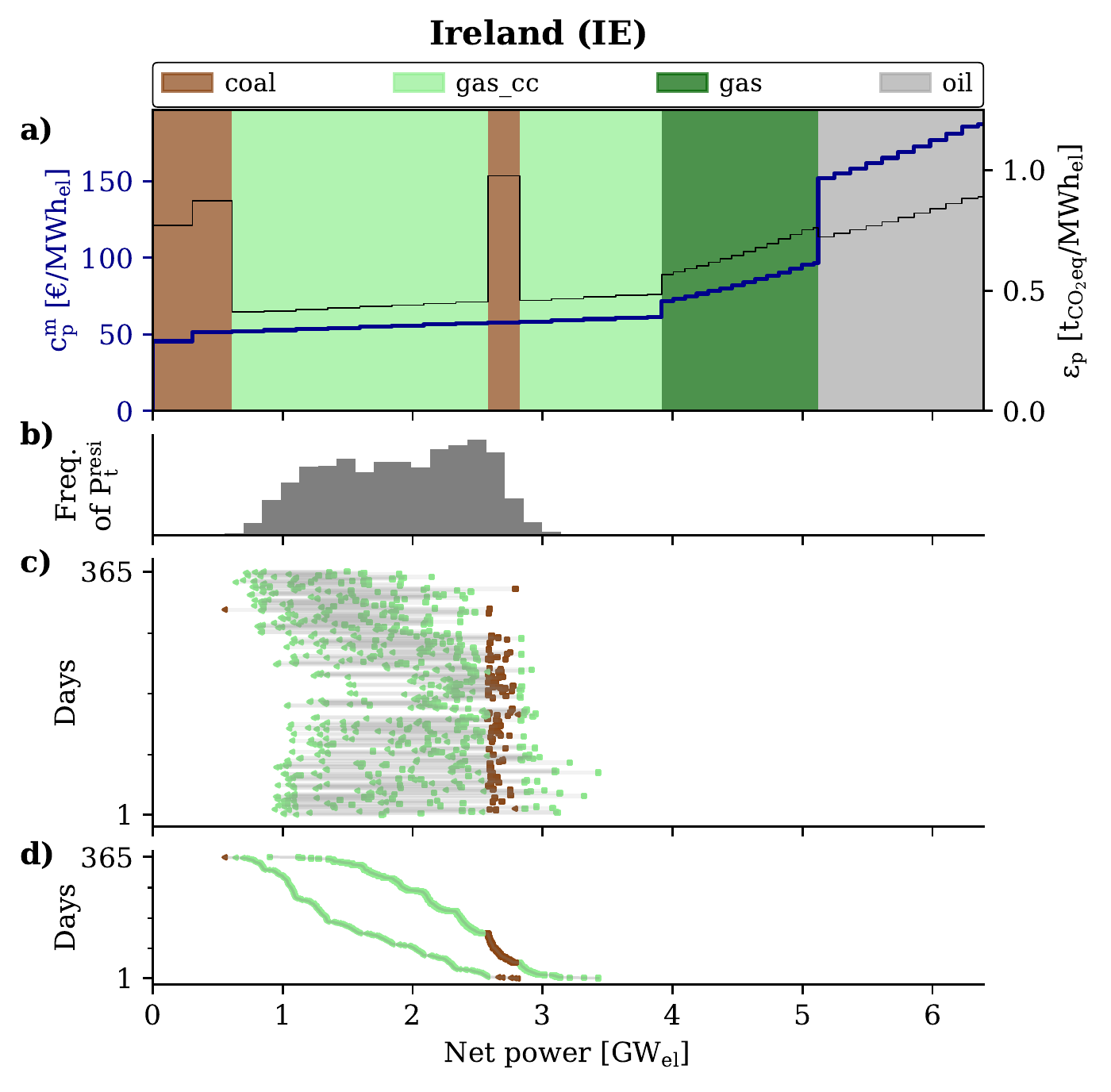}
	\includegraphics[width=0.245\linewidth]{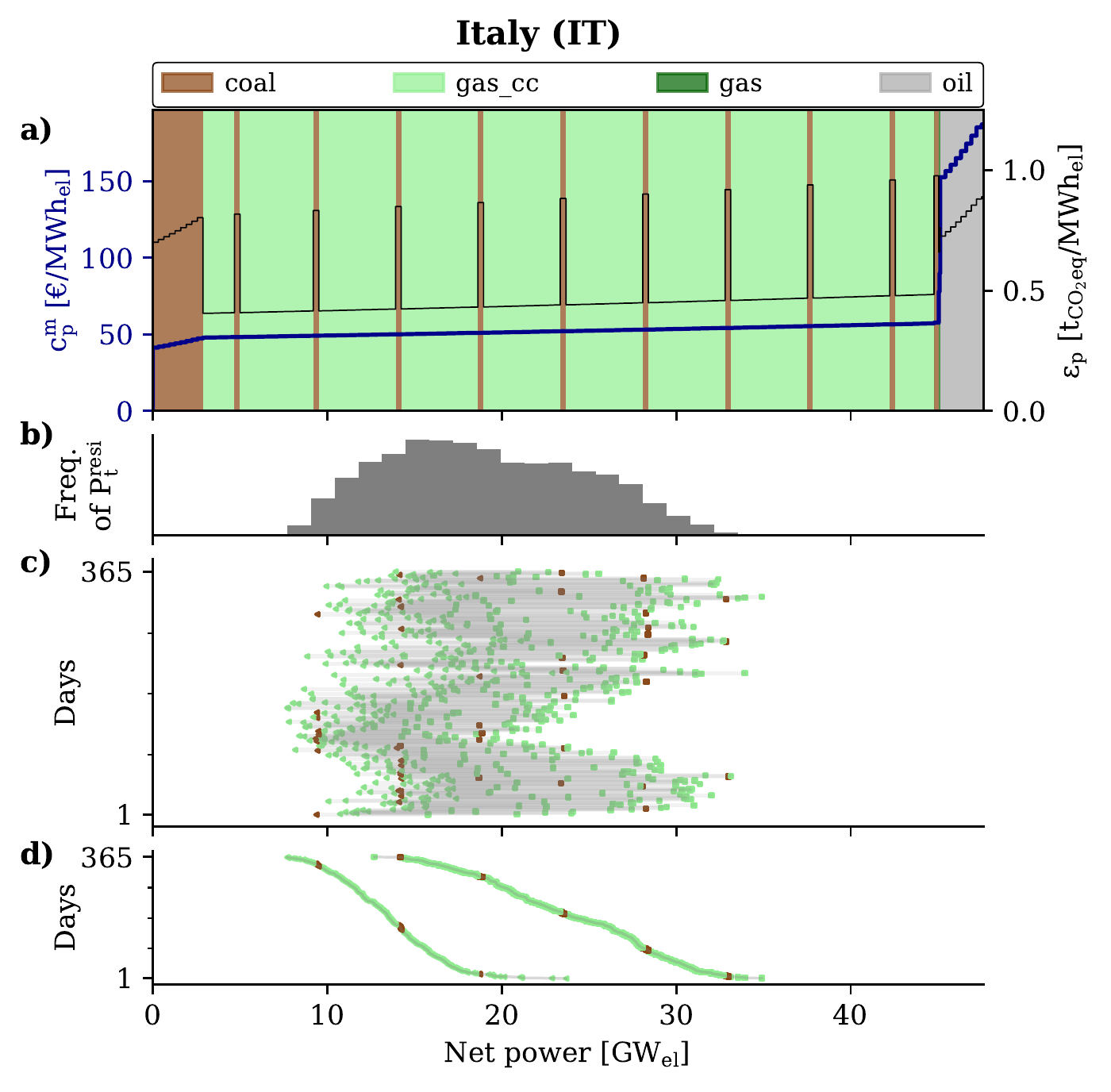}
	\includegraphics[width=0.245\linewidth]{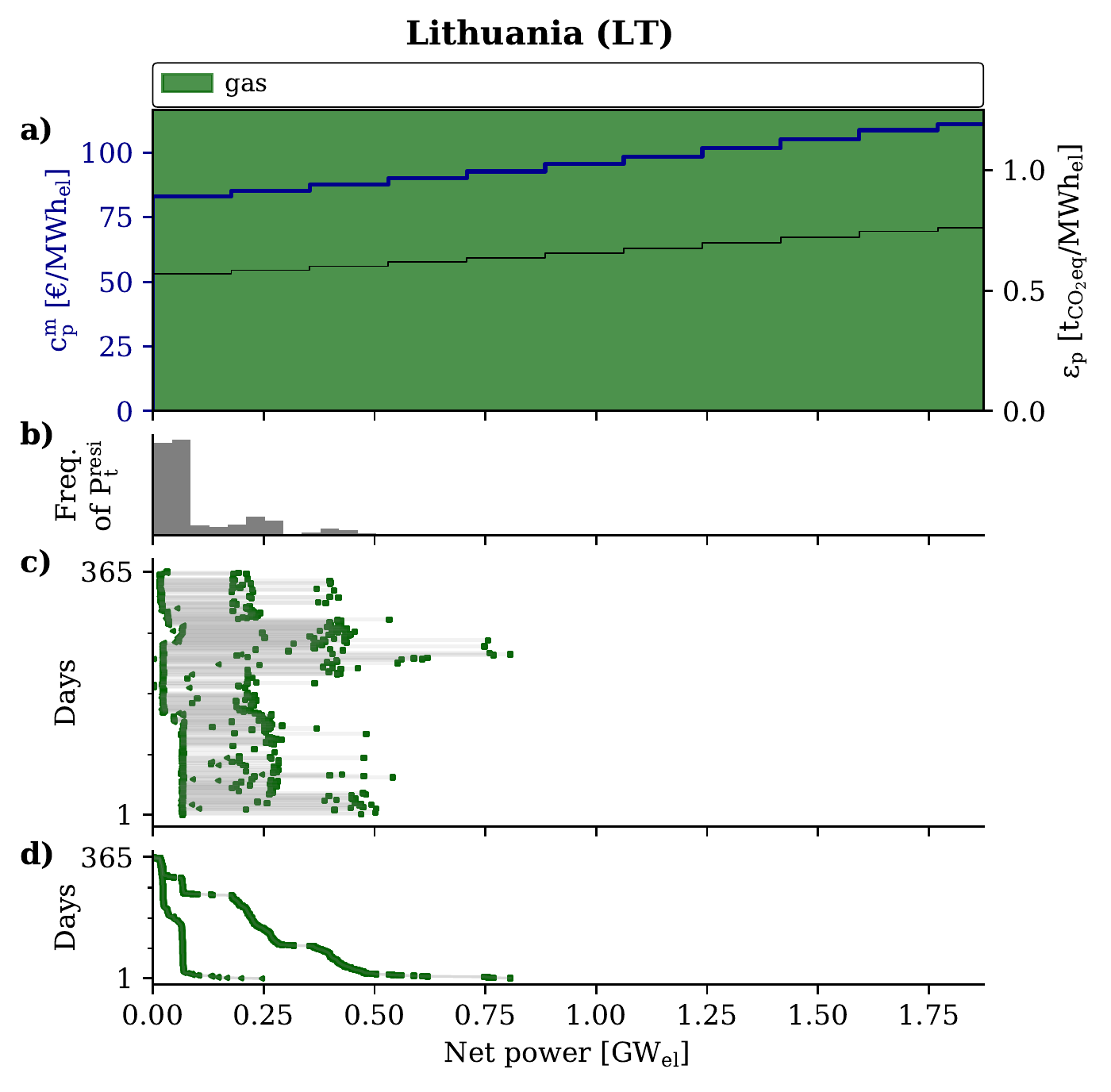}
	\includegraphics[width=0.245\linewidth]{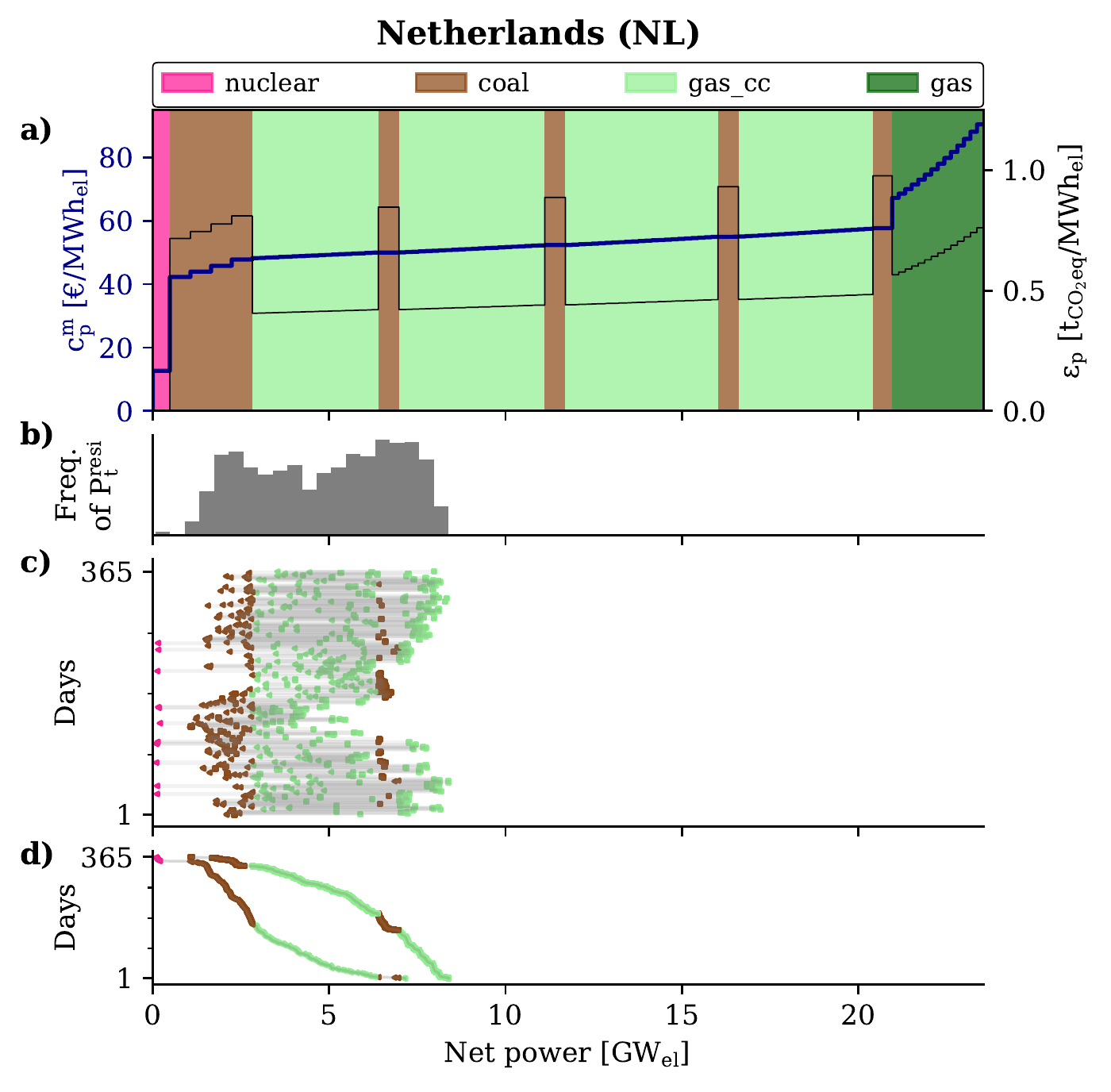}
	\includegraphics[width=0.245\linewidth]{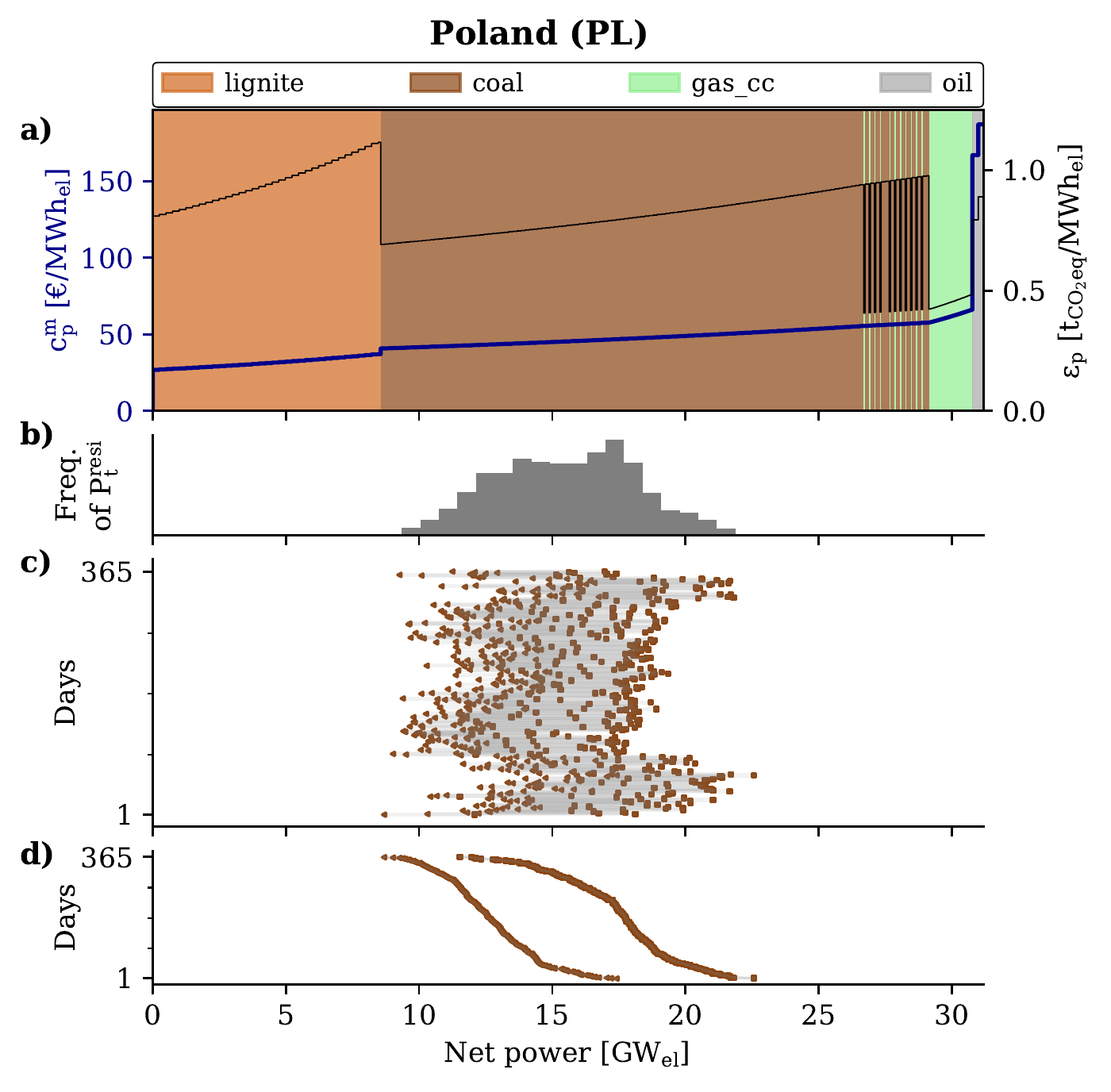}
	\includegraphics[width=0.245\linewidth]{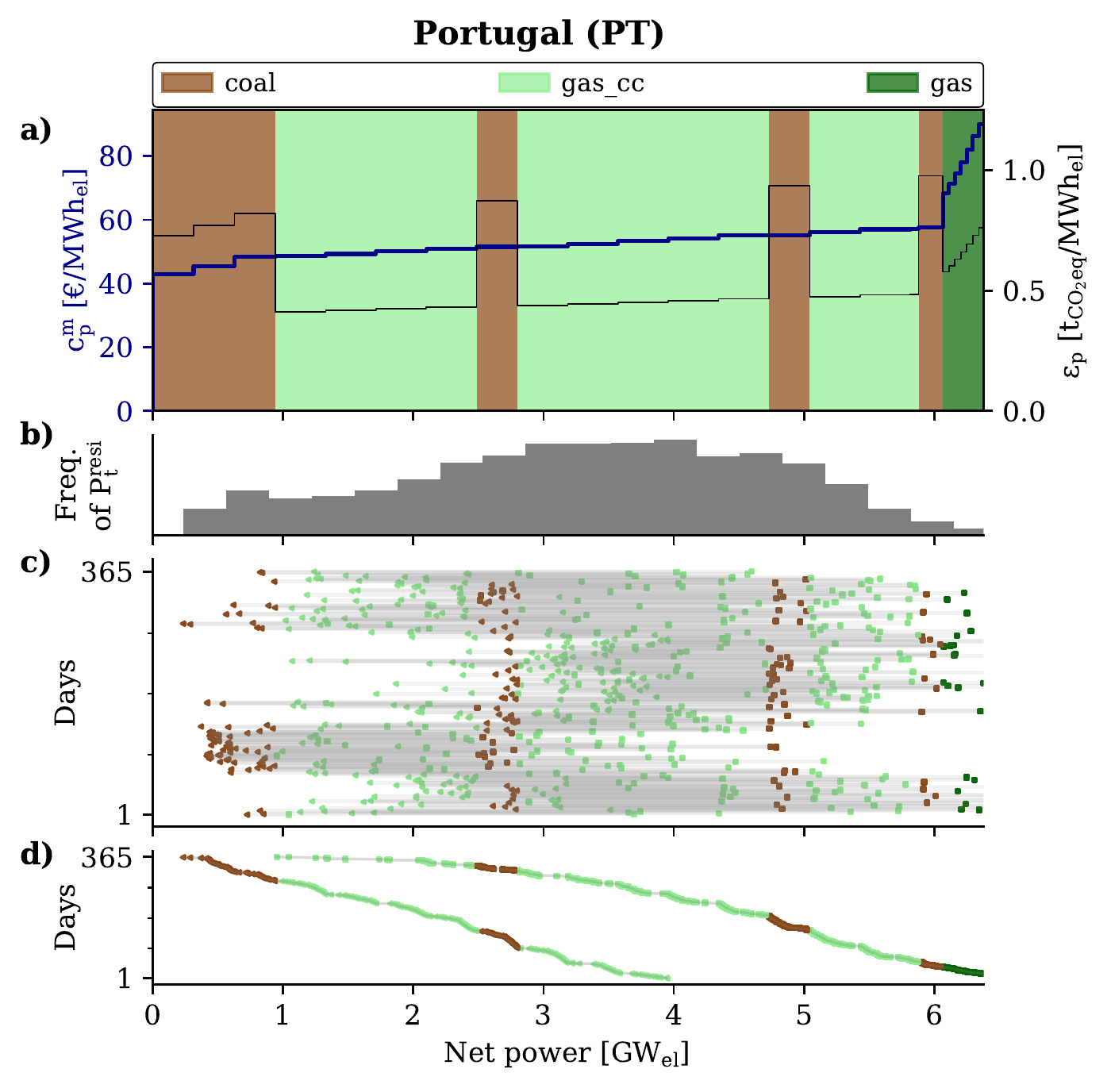}
	\includegraphics[width=0.245\linewidth]{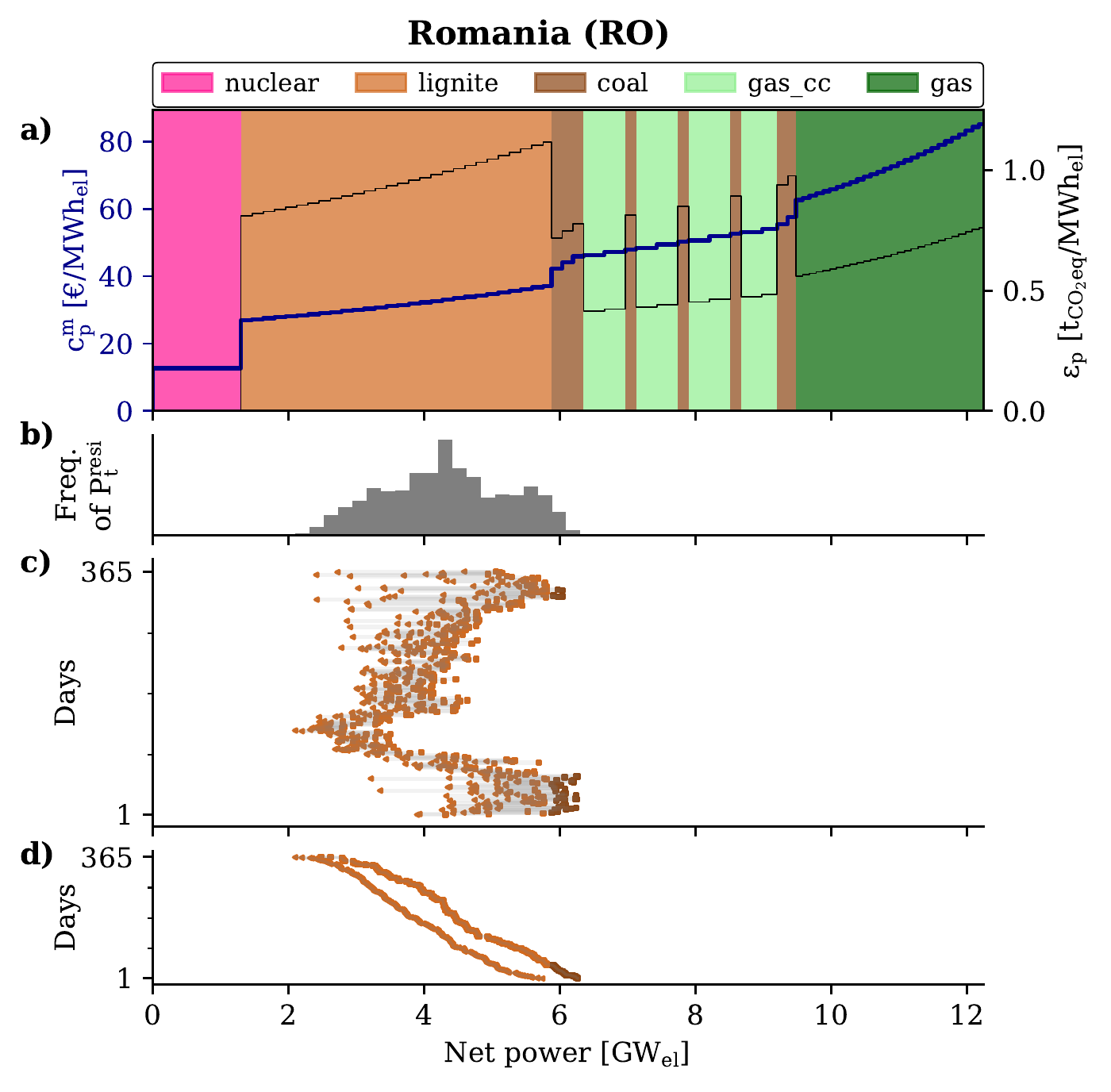}
	\includegraphics[width=0.245\linewidth]{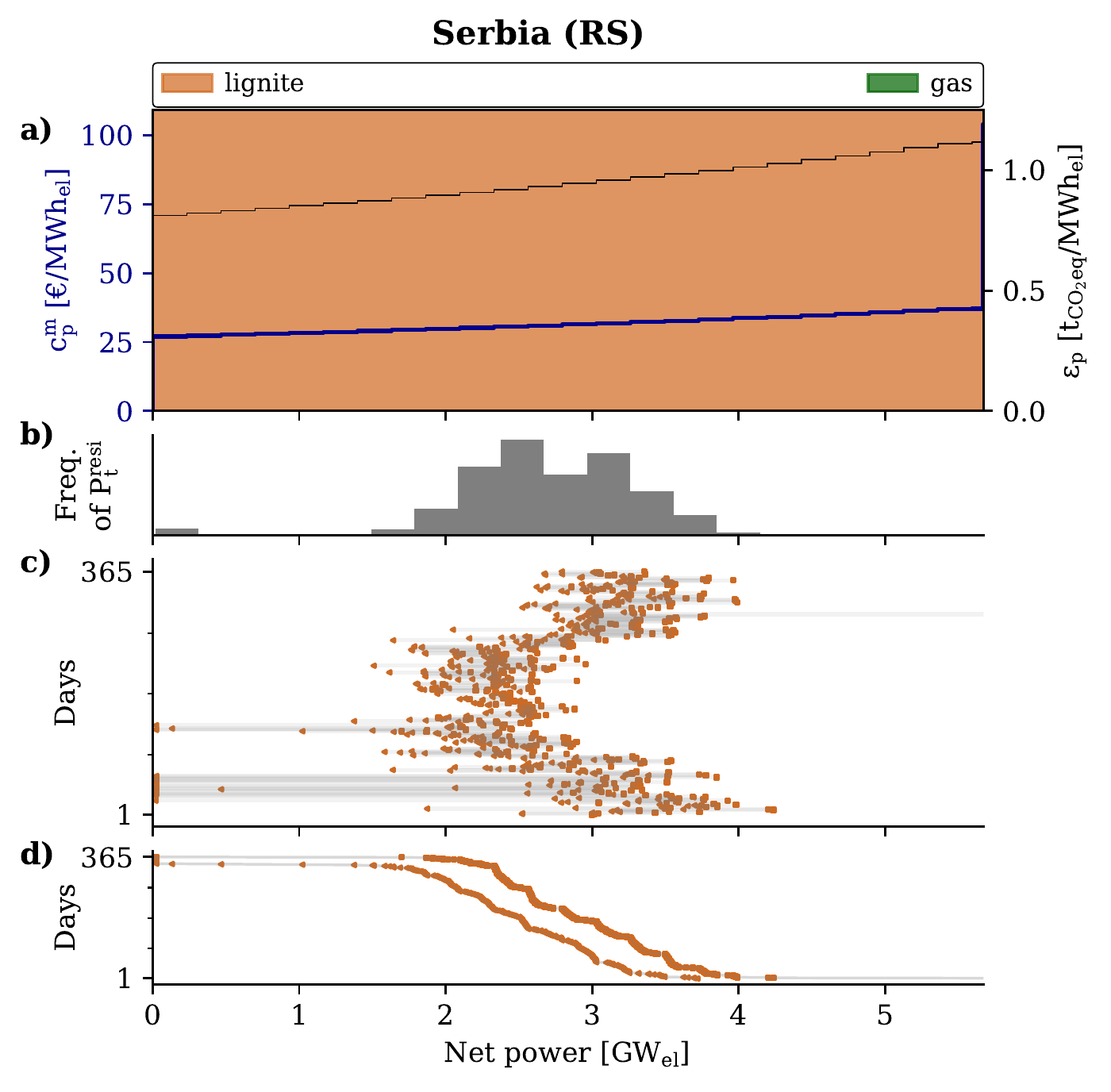}
	\includegraphics[width=0.245\linewidth]{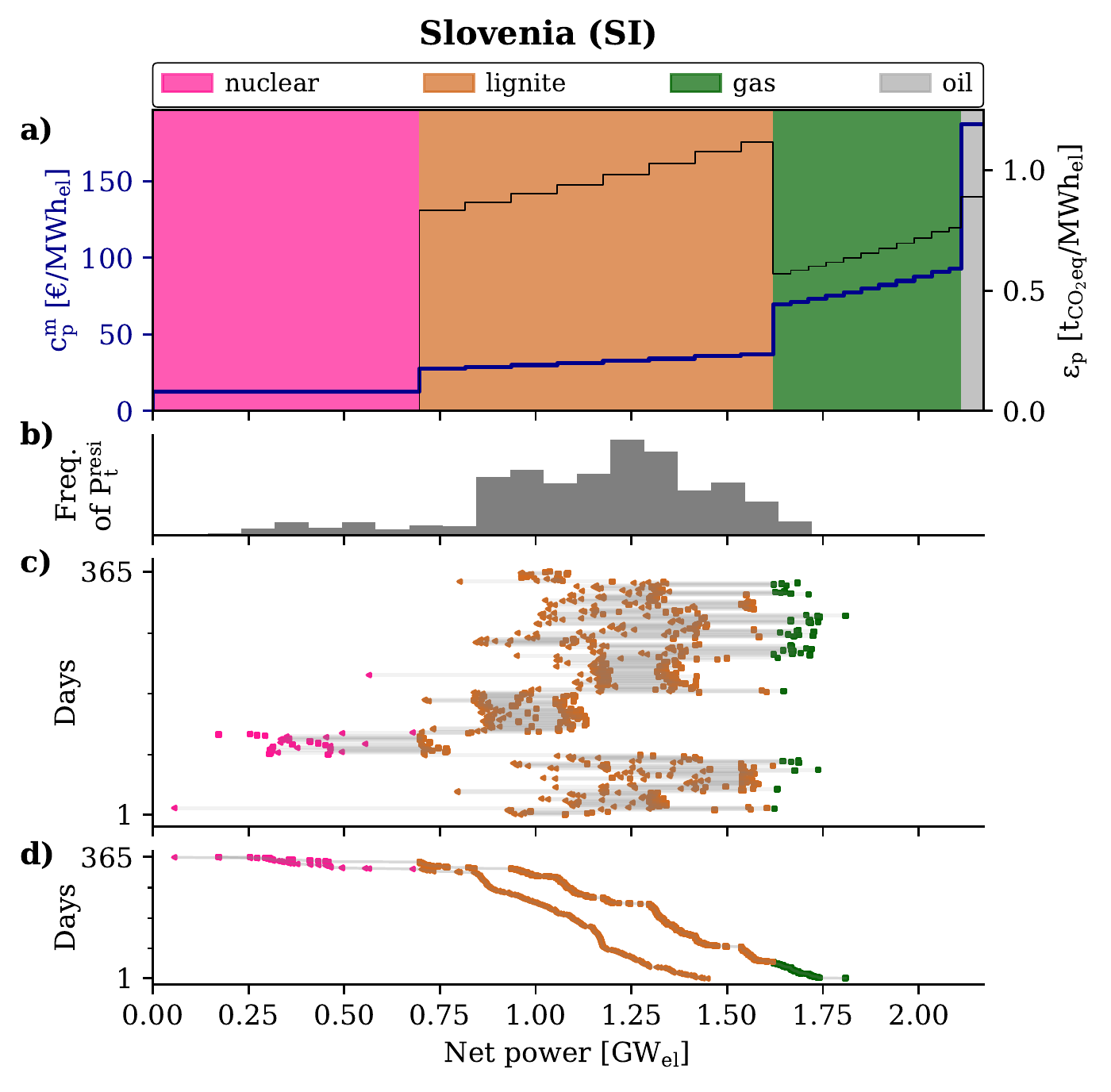}
	\caption{Merit orders and load shift analysis of analyzed European countries resulting from PWL-method for the year 2018: \textbf{a)} Merit order with marginal prices $ c_p^m $ (left) and emission intensities $ \varepsilon_p $ (right), \textbf{b)} histogram of residual load $ P_t^\mathrm{resi} $, \textbf{c)} load shifts of all days of the year (squares indicating sources and triangles indicating sinks), and \textbf{d)} inverted load duration curves of sources (right) and sinks (left)). The x-axis (net generation power in GW\textsubscript{el}) is shared across the four subplots.}
	\label{fig:mols2018}
\end{figure}

\begin{figure}[!h]
	\centering
	\textbf{2019}\par\medskip
	\includegraphics[width=0.245\linewidth]{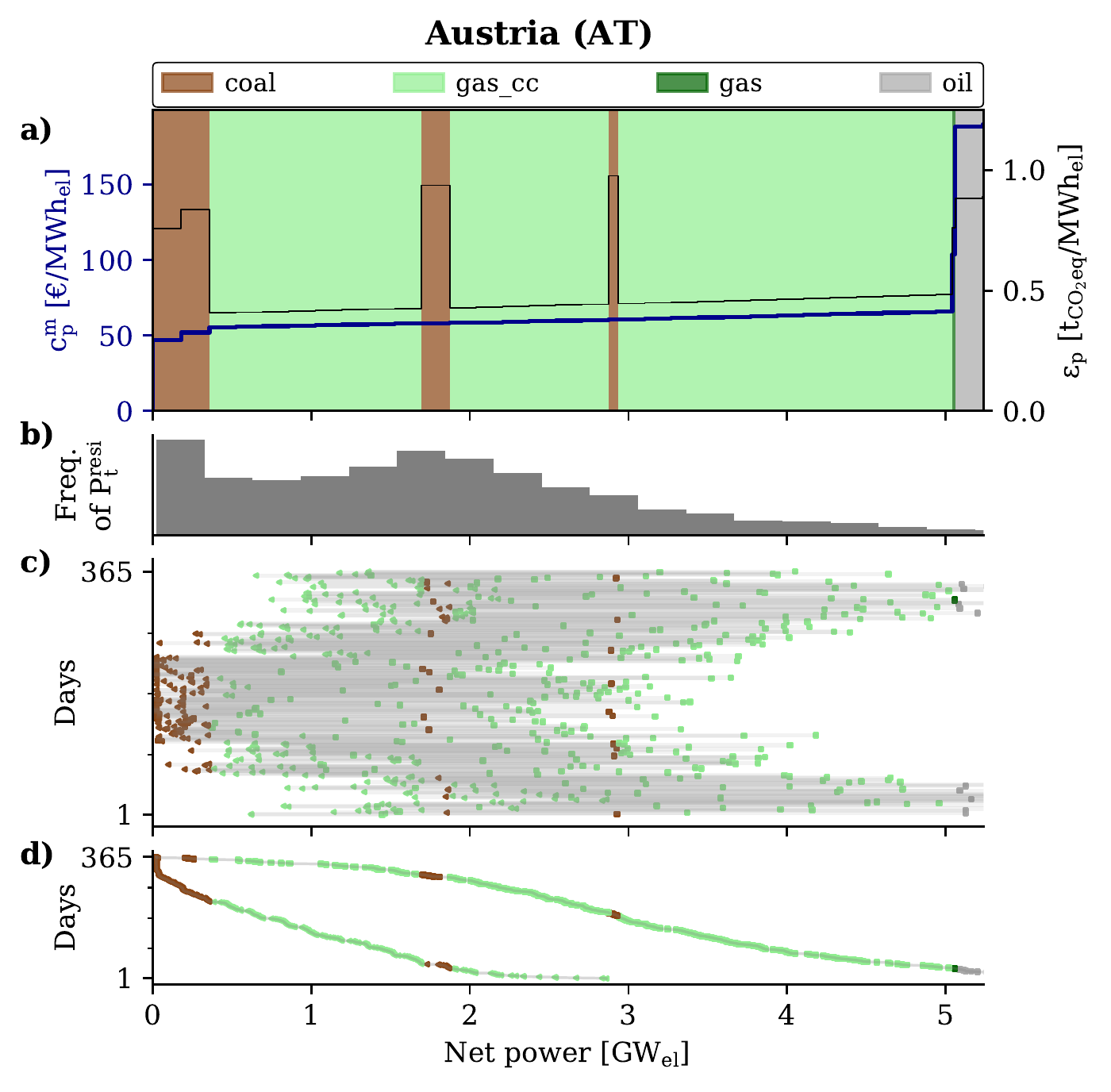}
	\includegraphics[width=0.245\linewidth]{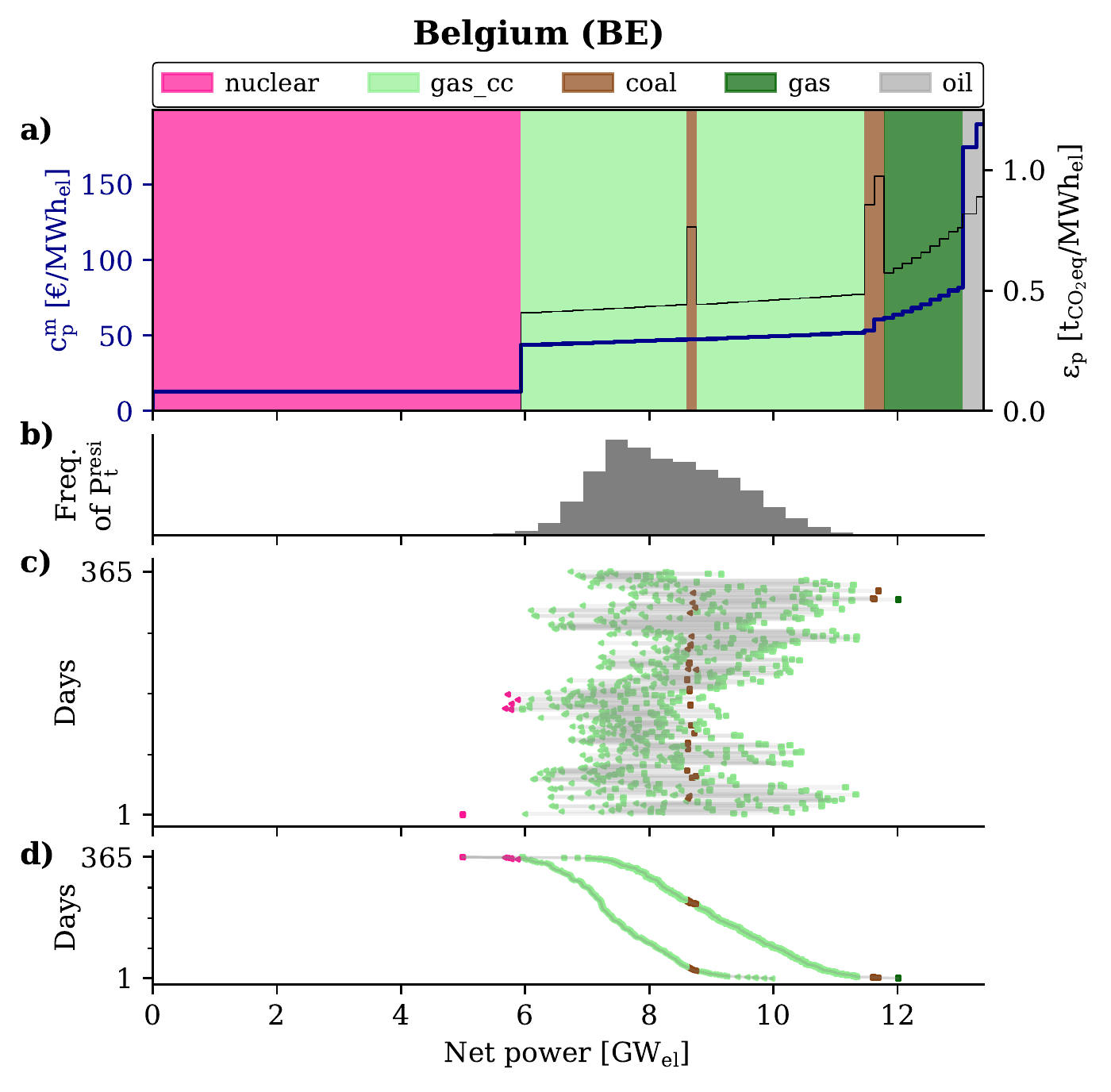}
	\includegraphics[width=0.245\linewidth]{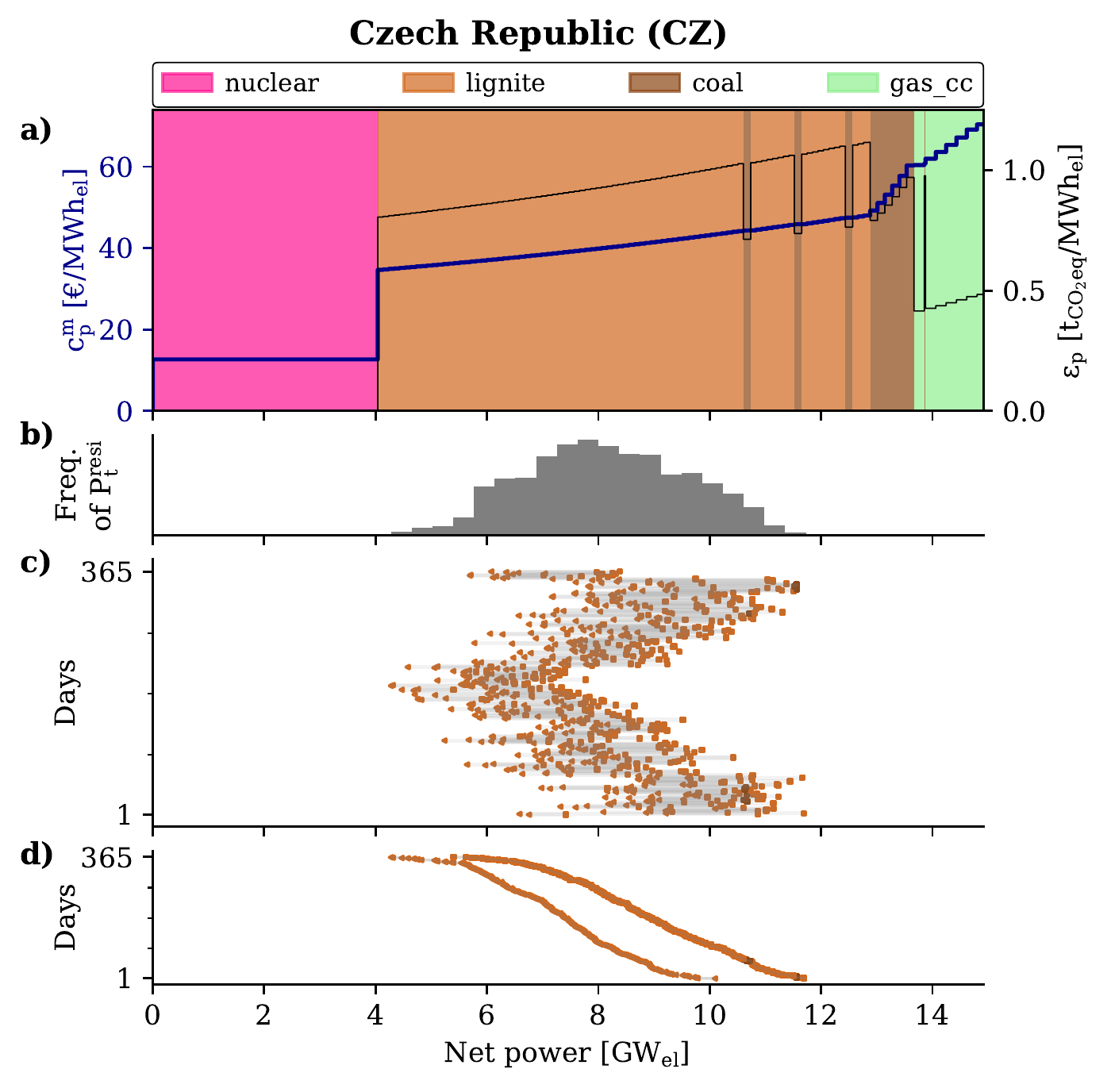}
	\includegraphics[width=0.245\linewidth]{pics/mols/mols_DE_2019}
	\includegraphics[width=0.245\linewidth]{pics/mols/mols_DK_2019}
	\includegraphics[width=0.245\linewidth]{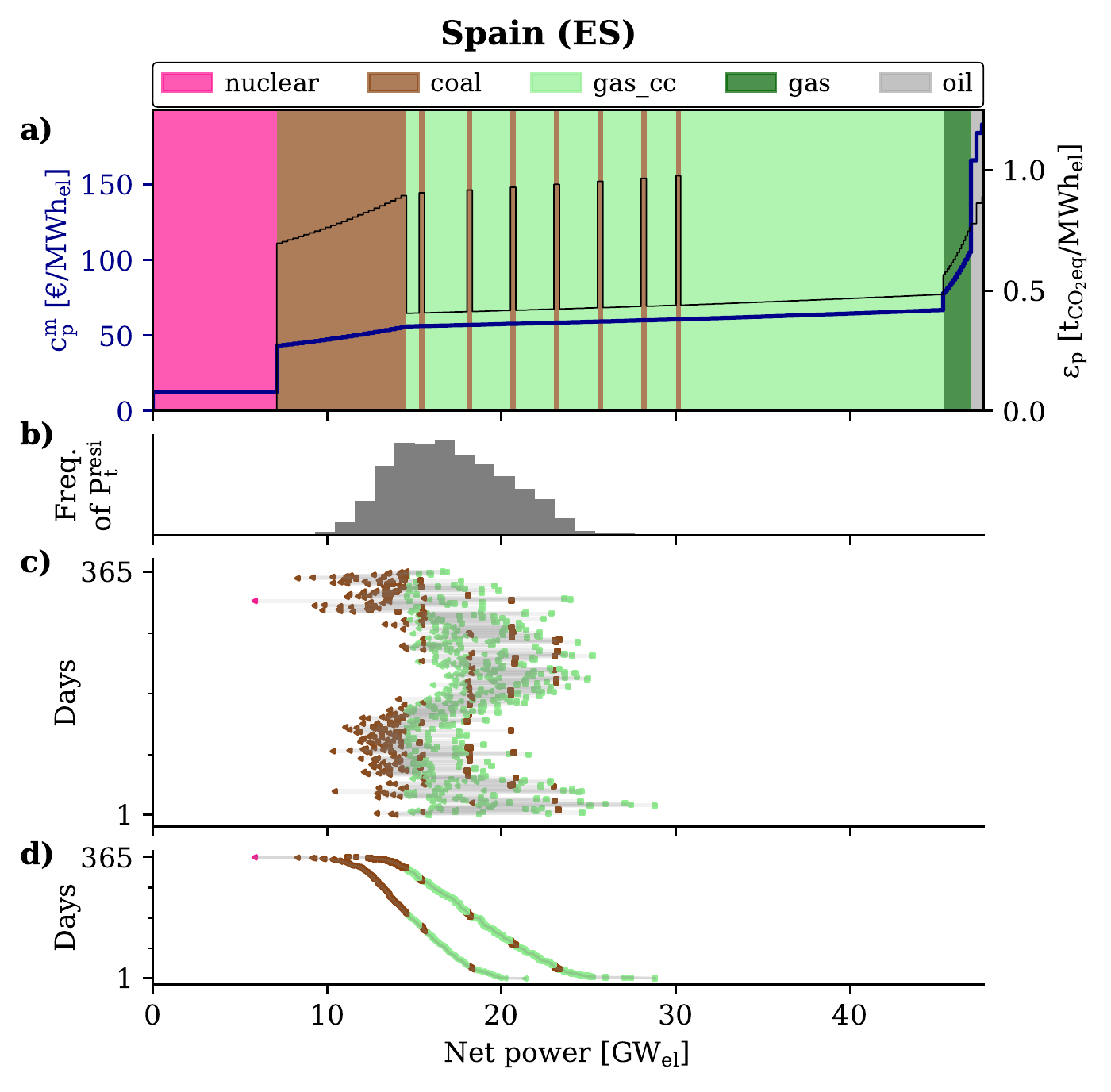}
	\includegraphics[width=0.245\linewidth]{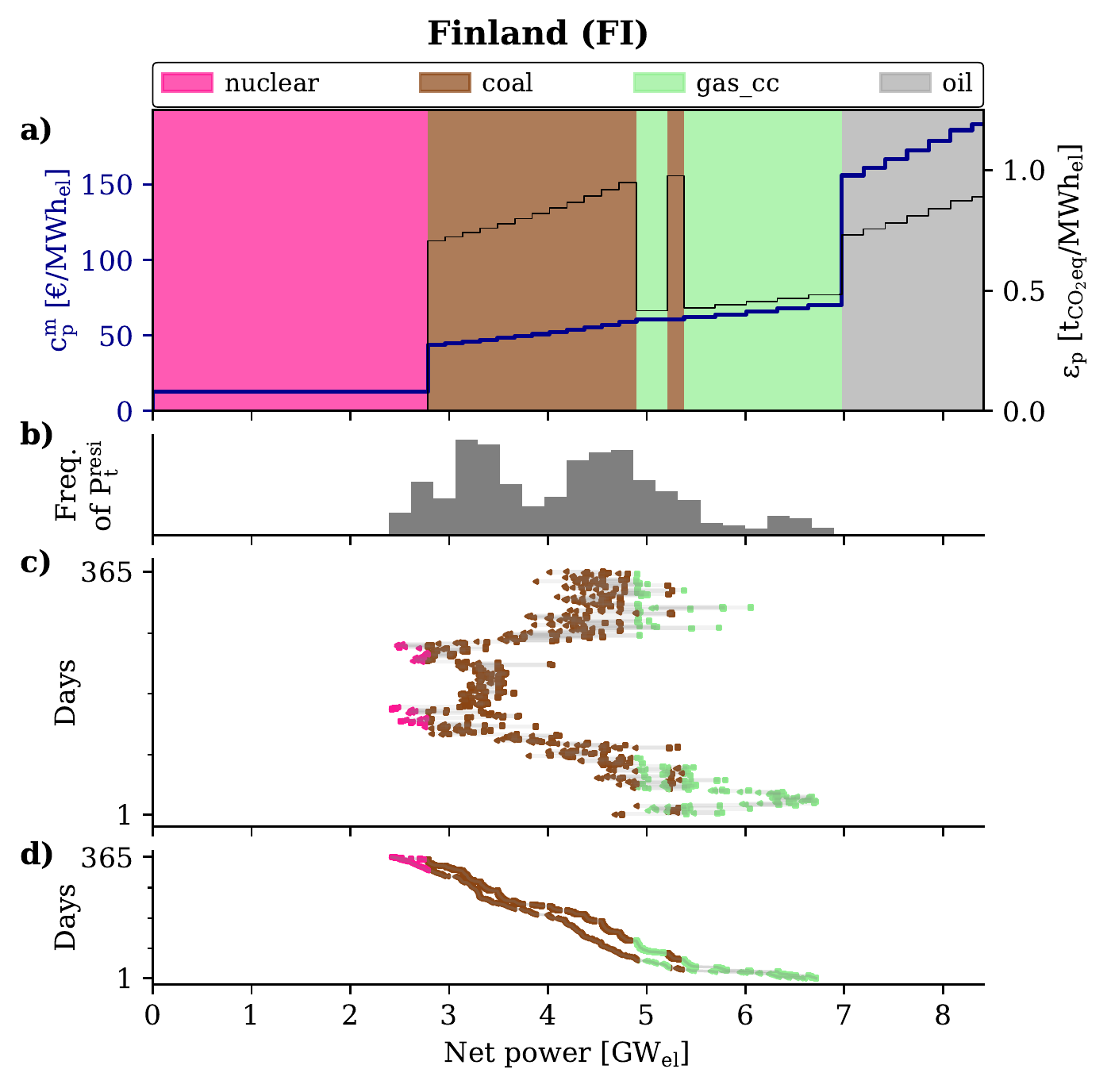}
	\includegraphics[width=0.245\linewidth]{pics/mols/mols_FR_2019}
	\includegraphics[width=0.245\linewidth]{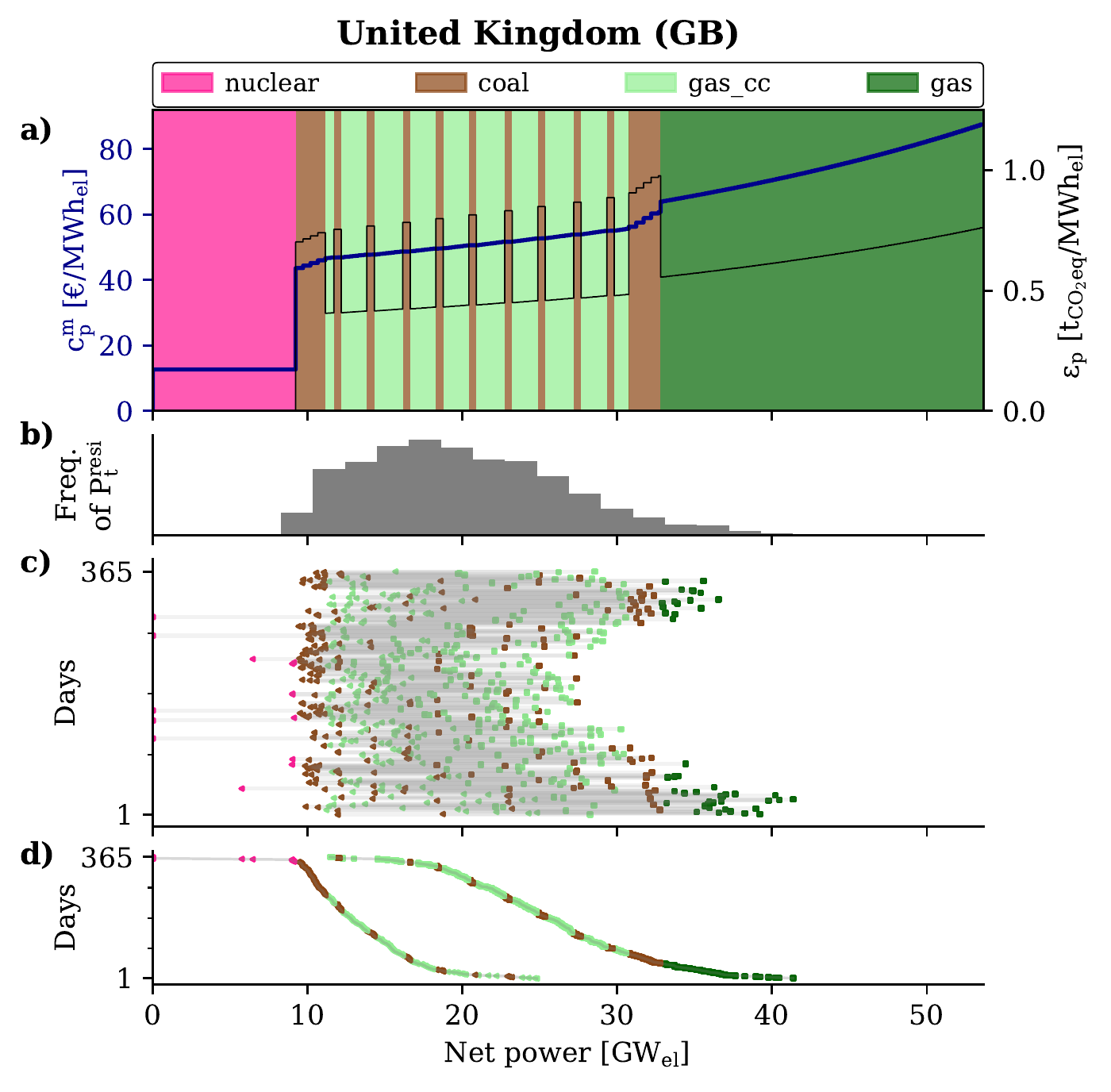}
	\includegraphics[width=0.245\linewidth]{pics/mols/mols_GR_2019}
	\includegraphics[width=0.245\linewidth]{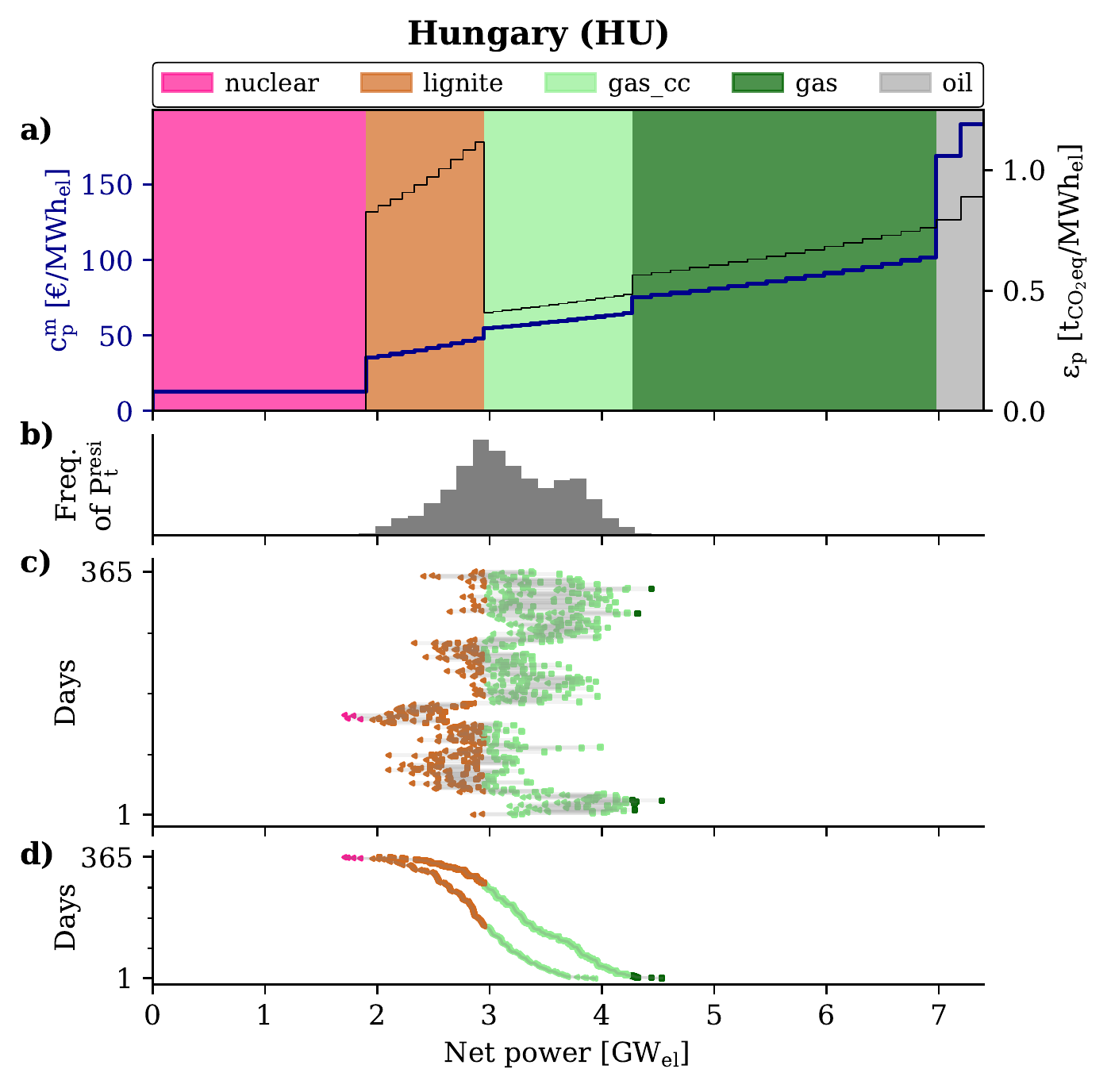}
	\includegraphics[width=0.245\linewidth]{pics/mols/mols_IE_2019}
	\includegraphics[width=0.245\linewidth]{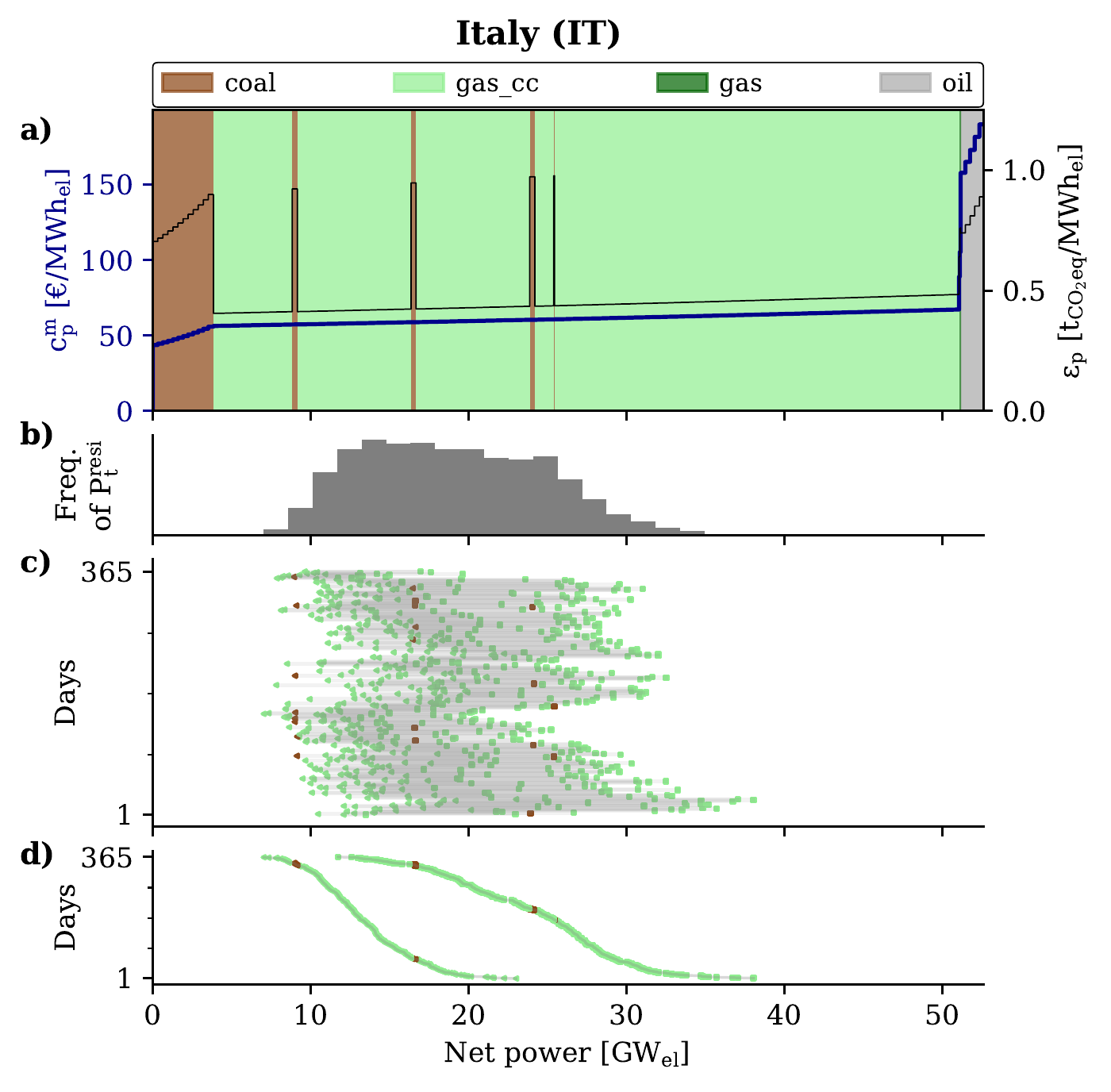}
	\includegraphics[width=0.245\linewidth]{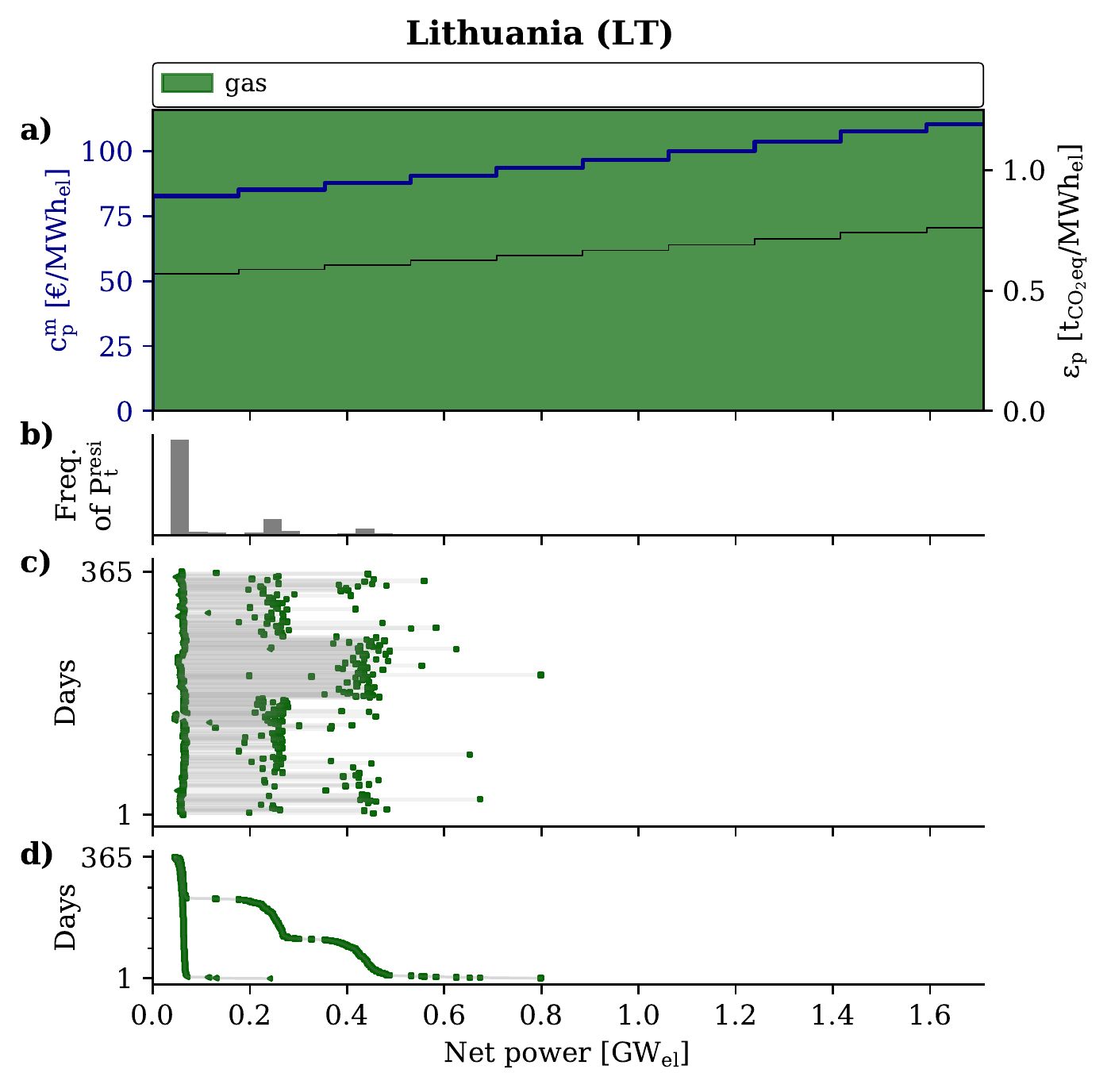}
	\includegraphics[width=0.245\linewidth]{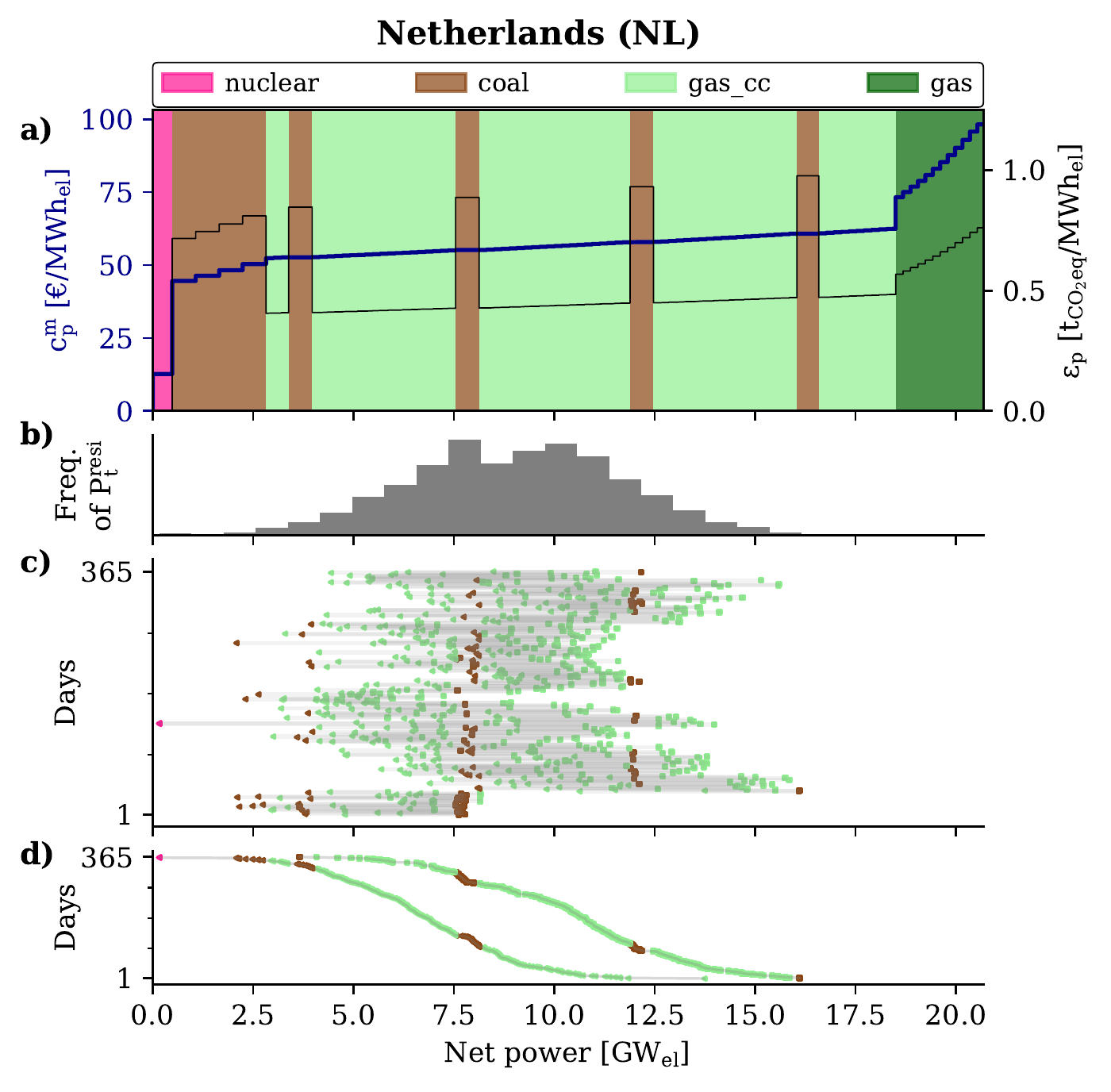}
	\includegraphics[width=0.245\linewidth]{pics/mols/mols_PL_2019}
	\includegraphics[width=0.245\linewidth]{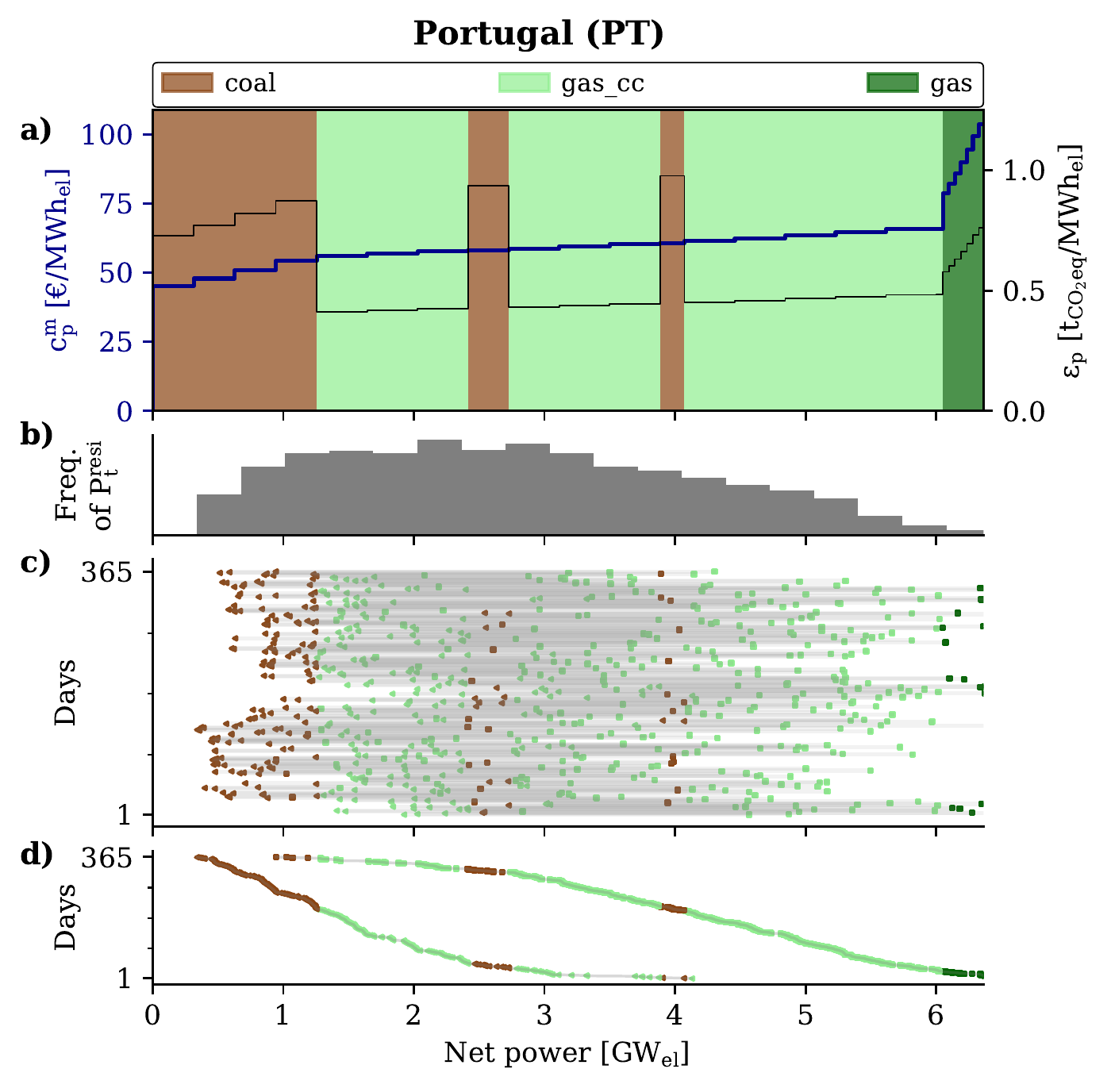}
	\includegraphics[width=0.245\linewidth]{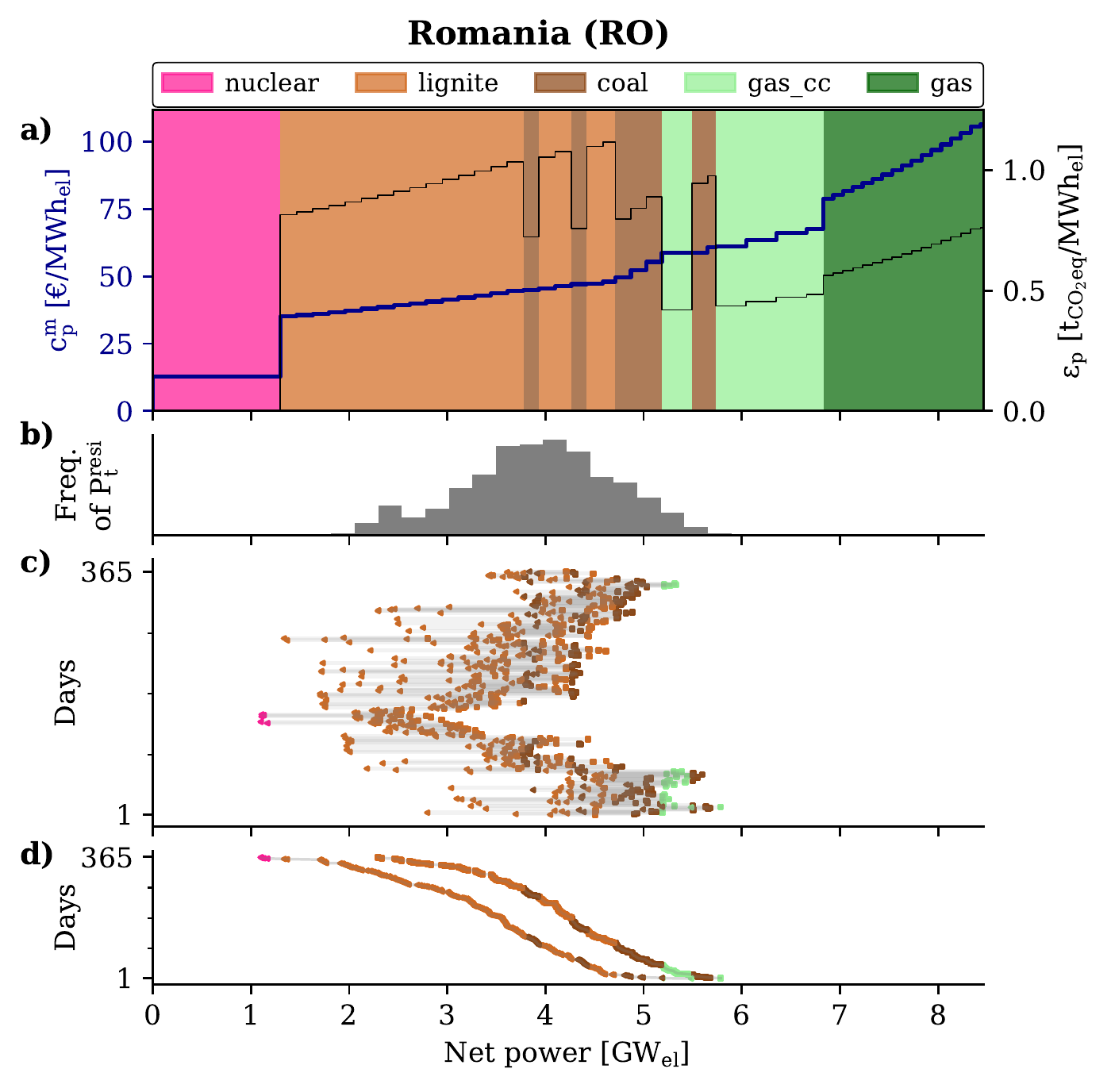}
	\includegraphics[width=0.245\linewidth]{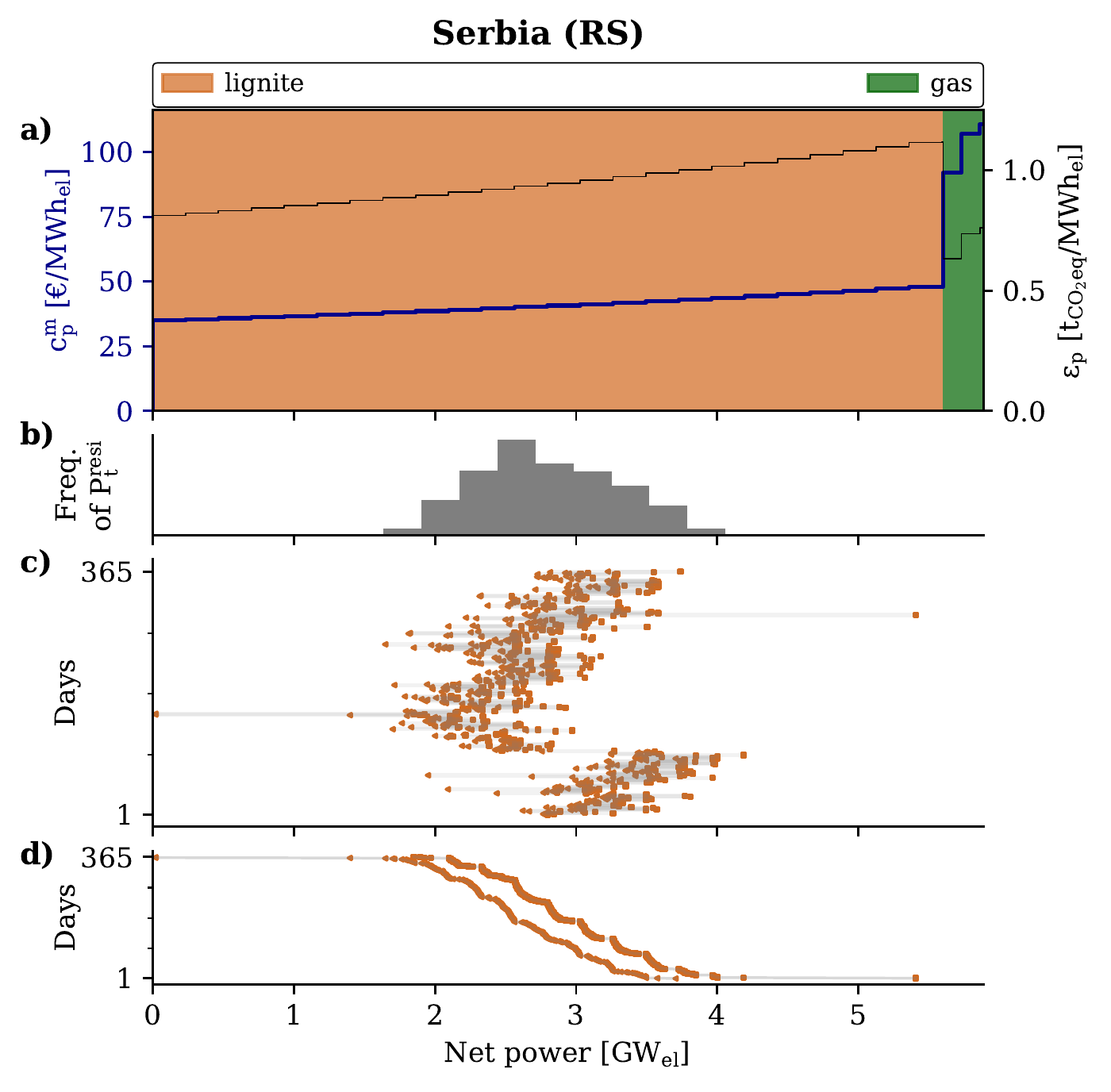}
	\includegraphics[width=0.245\linewidth]{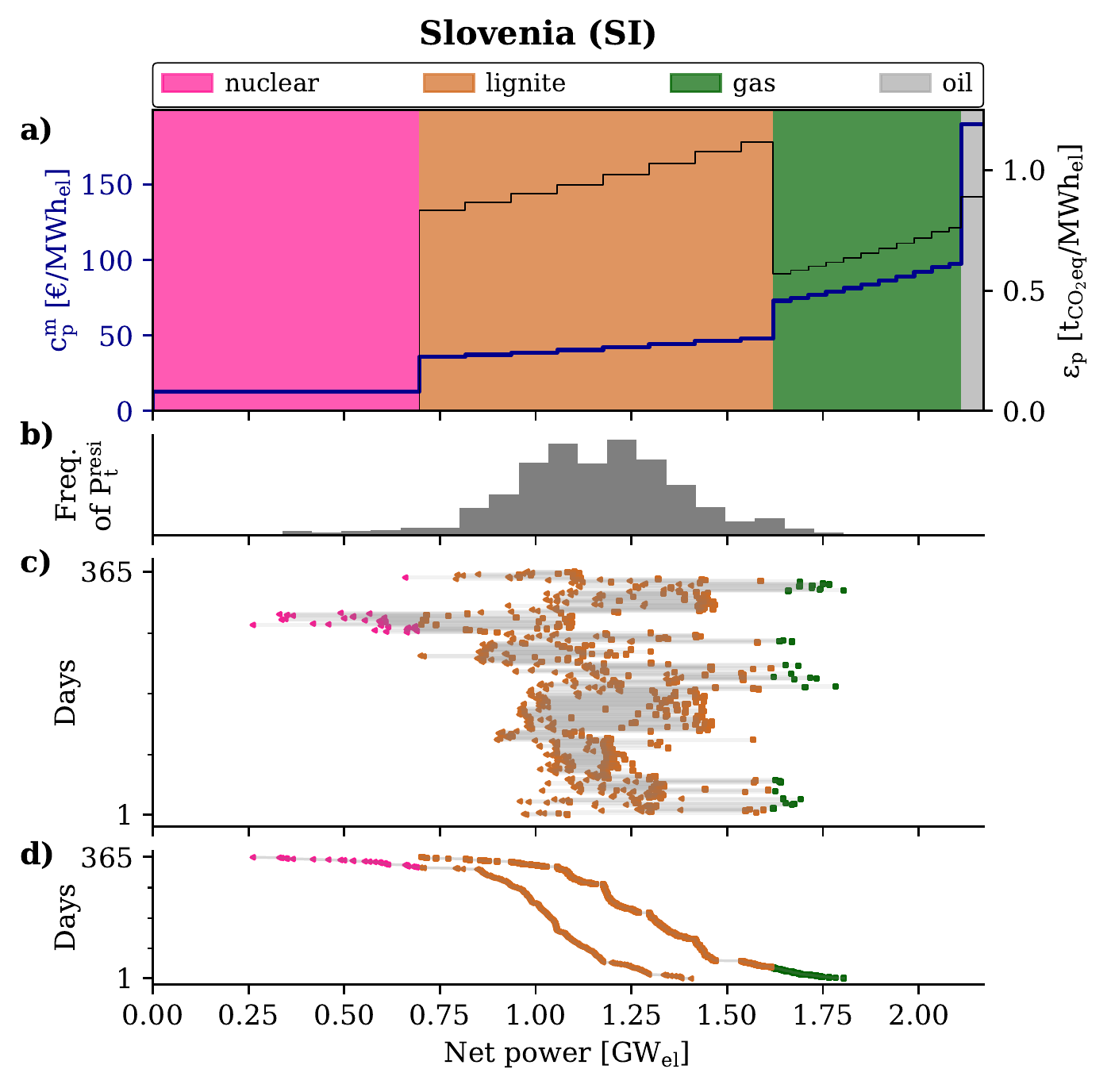}
	\caption{Merit orders and load shift analysis of analyzed European countries resulting from PWL-method for the year 2019: \textbf{a)} Merit order with marginal prices $ c_p^m $ (left) and emission intensities $ \varepsilon_p $ (right), \textbf{b)} histogram of residual load $ P_t^\mathrm{resi} $, \textbf{c)} load shifts of all days of the year (squares indicating sources and triangles indicating sinks), and \textbf{d)} inverted load duration curves of sources (right) and sinks (left)). The x-axis (net generation power in GW\textsubscript{el}) is shared across the four subplots.}
	\label{fig:mols2019}
\end{figure}

\section{Description of the remaining 14 countries}
In this section, all values refer to the year 2019 and all superlatives to the 20 countries analyzed in this paper.
Gas includes also gas\_cc.

\paragraph{Austria (AT)}
Austria has the highest RES share (74\%) after Denmark (77\%), and with 45\% the highest hydro share.
The main fossil energy sources are gas\_cc and coal.
The fluctuation of the residual load compared to the aggregated installed capacity is high and features a pronounced seasonal pattern.
Price-based load shifts led to reductions of 24\% cost and 53\% grid mix emissions but to an increase of 11\% in (marginal) emissions.
Most fuel type combinations were gas\_cc-to-gas\_cc (179) and gas\_cc-to-coal (130).

\paragraph{Belgium (BE)}
In Belgium, the strongest contributors to electricity production are nuclear (47\%), gas (27\%), wind (9\%), and solar (4\%).
Price-based load shifting led to reductions of 6\% cost, 45\% grid mix emissions, and to 7\% (marginal) emissions.
330 (90\%) load shifts were gas\_cc-to-gas\_cc.

\paragraph{Czech Republic (CZ)}
In the Czech Republic, predominant fuel types are lignite (39\%) and nuclear (35\%).
The load shifting simulation showed 354 lignite-to-lignite and 11 coal-to-lignite load shifts.
The resulting reductions were 6\% for cost, 21\% for grid mix emissions, and 5\% for (marginal) emissions.
354 (97\%) load shifts were lignite-to-lignite, 11 (3\%) were coal-to-lignite.

\paragraph{Spain (ES)}
Spain is mainly powered by gas (31\%), nuclear (22\%), and wind\_onshore (21\%).
From load shifting, the cost reduced only by 4\%, the grid mix emission reduced by 10\%, but (marginal) emission increased by 25\%.
The emission increase is mainly caused by 144 (39\%) gas\_cc-to-coal shifts.

\paragraph{Finland (FI)}
In Finland, nuclear (38\%), hydro (19\%), biomass (10\%), and wind\_onshore (9\%) are the most important fuel types.
The results of the load shifting simulation were a 10\% reduction of cost and grid mix emissions and a 4\% reduction of (marginal) emissions.
The emissions did not increase since only 51 (14\%) load shifts (gas\_cc-to-coal) were not beneficial from an emissions perspective.

\paragraph{United Kingdom (GB)}
In the United Kingdom, the main fuel types are gas (45\%), nuclear (21\%), and wind (18\%).
Price-based load shifting led to reductions of 16\% cost, 57\% grid mix emissions, and 3\% (marginal) emissions.
The most frequent fuel type combinations were gas\_cc-to-coal (114) and gas\_cc-to-gas\_cc (108).

\paragraph{Hungary (HU)}
Hungary features a low RES share of 8\%.
The most important fuel types are nuclear (51\%), gas (25\%), and lignite (14\%).
From load shifting, both cost and grid mix emission decreased by 13\%. However, (marginal) emissions increased by 31\%.
The emission increase is mainly caused by 129 gas\_cc-to-lignite shifts.

\paragraph{Italy (IT)}
Italy shows a diversified electricity supply system.
The strongest contributors are gas (44\%), hydro (13\%), wind (8\%), solar (7\%), and coal (7\%).
The load shifting simulation shows low reductions in cost (4\%), grid mix emissions (3\%), and (marginal) emissions (4\%).
The predominant fuel type combination gas\_cc-to-gas\_cc occurred 335 times (92\%).

\paragraph{Lithuania (LT)}
Lithuania is the country with the highest pumped hydro share (17\%).
Other important fuel types are wind\_onshore (39\%), biomass (12\%), and gas (11\%).
Reductions from price-based load shifting were low in the case of cost (3\%) and (marginal) emissions (3\%) but high for the case of grid mix emissions (53\%).
All load shifts were gas-to-gas.

\paragraph{Netherlands (NL)}
The Netherlands has the lowest RES share (7\%) and the highest gas share (49\%).
Besides other\_conv (21\%), coal has also a considerable share (19\%).
The results of the load shifting simulation were reductions of 5\% cost, 1\% grid mix emissions, and 5\% (marginal) emissions. 
71\% of load shifts remained within the fuel type gas\_cc.

\paragraph{Portugal (PT)}
The main fuel types in Portugal are gas (33\%), wind (27\%), hydro (11\%), and coal (11\%).
Load shifting led to a 14\% reduction of cost, a 20\% reduction of grid mix emissions, but a 22\% increase of (marginal) emissions.
The emission increase is mainly caused by 150 gas\_cc-to-coal load shifts.

\paragraph{Romania (RO)}
In Romania, the strongest contributors to electricity production are lignite (22\%), nuclear (19\%), hydro (18\%), gas (16\%)
Through load shifting, cost and grid mix emissions decreased by 11\% and 22\%, respectively while the (marginal) emissions increased by 4\%.
This slight increase of emissions is due to 112 coal-to-lignite load shifts.

\paragraph{Serbia (RS)}
Serbia features the highest lignite share (68\%).
Hydro (26\%) plays also an important role.
The effects of the price-based load shifting simulations were -3\% cost, +3\% grid mix emissions, and -3\% (marginal) emissions.
A possible explanation for the unintuitive increase of mix emissions of Serbia can be found in the paper.

\paragraph{Slovenia (SI)}
In Slovenia, predominant fuel types are nuclear (37\%), hydro (29\%), and lignite (27\%).
The load shifting simulation resulted in reductions of 19\% cost, 27\% grid mix emissions, and 13\% (marginal) emissions.
81\% of load shifts remained within the fuel type lignite.

\bibliographystyle{elsarticle-num}
\bibliography{references,ref2}